\documentclass[fleqn,usenatbib]{mnras}

\usepackage{newtxtext,newtxmath}

\usepackage[T1]{fontenc}

\DeclareRobustCommand{\VAN}[3]{#2}
\let\VANthebibliography\thebibliography
\def\thebibliography{\DeclareRobustCommand{\VAN}[3]{##3}\VANthebibliography}

\usepackage{graphicx}	
\usepackage{amsmath}	

\usepackage{subfig}
\usepackage{orcidlink}
\usepackage{breakurl}

\title[Stellar wind variability at low metallicities]{Extensive observational evidence for massive star stellar wind variability at low metallicities: implications for mass-loss rate determination}

\author[T. N. Parsons et al.]{
Timothy N. Parsons,$^{1}$\thanks{E-mail: timothy.parsons.15@ucl.ac.uk (TNP)}\orcidlink{0000-0002-7714-7823}
Raman K. Prinja,$^{1}$\orcidlink{0000-0002-5251-3743}
Derck L. Massa$^{2}$\orcidlink{0000-0002-9139-2964}
and Alex W. Fullerton$^{3}$
\\
$^{1}$Department of Physics and Astronomy, University College London, Gower Street, London WC1E 6BT, UK\\
$^{2}$Space Science Institute (SSI), 4750 Walnut Street, Suite 205, Boulder, CO 80301, USA\\
$^{3}$Space Telescope Science Institute, 3700 San Martin Drive, Baltimore, MD 21218, USA
}

\date{Accepted 2025 December 15. Received 2025 December 9; in original form 2025 August 27}

\pubyear{\the\year{}}

\begin{document}
\label{firstpage}
\pagerange{\pageref{firstpage}--\pageref{lastpage}}
\maketitle

\begin{abstract}
Mass-loss from massive stars is fundamental to stellar and galactic evolution and enrichment of the interstellar medium. Reliable determination of mass-loss rate is dependent upon unravelling details of massive star outflows, including optical depth structure of the stellar wind. That parameter introduces significant uncertainty due to the nearly ubiquitous presence of large-scale optically thick wind structure. We utilize suitable available ultraviolet spectra of 20 Large and Small Magellanic Cloud (LMC, SMC) OB stars to extend existing Galactic results quantifying uncertainty inherent in individual observations to lower metallicity environments. This is achieved by measuring standard deviations of mean optical depths of multiple observations of suitable wind-formed absorption profiles as a proportion of their mean optical depths. We confirm earlier findings that wind structure is prevalent at low metallicities and demonstrate that quantifying the consequent uncertainty is to some extent possible, despite the near-complete absence of time series UV spectroscopic observations in those environments. We find that the uncertainty inherent in any single observation of stellar wind optical depth at low metallicity is of similar magnitude to that already identified at Galactic metallicity (up to 45\% for cooler OB stars). We further demonstrate how the effect of varying narrow absorption components in wind-formed UV spectral profiles is unlikely to be properly accounted for in existing mass-loss models. We present further evidence of a binary companion to the SMC O-type giant star AzV 75. The importance of obtaining high cadence multi-epoch, or genuine time series, UV spectroscopic observations at low metallicities is highlighted.
\end{abstract}




\section{Introduction}

\subsection{Observations of massive star stellar winds}

Reliably constraining rates of mass-loss from massive stars ($M \gtrsim 8 M_\odot$) is of fundamental importance in understanding the highly significant role played by these stars in the evolution of the interstellar medium (ISM) and of their host galaxies \citep[][among others]{Vink2023, Telford2024}. Observation and reliable quantification of these processes in lower metallicity environments is essential to a better understanding of galactic evolution and of conditions prevailing at earlier cosmic epochs \citep[for example,][]{Trundle2004, Telford2021, Vink2023, Sana2024, Telford2024, Roman2025}.

These high mass stars exhibit strong stellar winds, both from continuum-originated radiation pressure and, most significantly, from the `line-driving' process in which substantial momentum (and velocity) is imparted to stellar wind material, predominantly by resonance line absorption of ultraviolet (UV) photons \citep[see, in particular,][]{Mihalas1978, Puls2000, Hubeny2015}.

The availability of large sets of UV observations of massive stars in the Milky Way either in the form of multi-epoch observations or true time series has shown that the winds of massive stars are characterized by the presence of large-scale structure and, therefore, temporal variability \citep[see, for example][]{Massa1995, Kaper1996, Prinja2002, Massa2015}. Such structure creates variability over time and is distinct from small-scale (micro-) structure which may derive from wind turbulence and is stochastic in nature. It is also a phenomenon distinct from clumping in the stellar wind \citep[see, for example][]{Oskinova2007, Sundqvist2010, Surlan2012, Brands2025}.

These observations have demonstrated that the winds of massive stars commonly show significant variability on time scales commensurate with the rotation of the star \citep[see, most recently, the discussion in][and Subsection \ref{sec:quantvar} below]{Massa2024}. The inevitably `resource hungry' nature of time series observations means however that, to date, there are very few multi-epoch or time series high resolution UV spectroscopic observations of massive stars at lower than Galactic metallicities.

We have made use of those observations which are available, at sufficiently high signal-to-noise ratios (SNR), to quantify lower limits on variability in optically-thick structure within massive star stellar winds at the lower metallicities (broadly, 0.5 $Z_\odot$ and 0.14-0.2 $Z_\odot$, respectively) found in the Large and Small Magellanic Clouds (LMC, SMC). The implications of these findings for massive star mass-loss rates derived from single observations in these environments are considered.

The presence of optically thick and large scale (macro-) structure in the winds of massive stars has been well demonstrated both through direct observation of changes in absorption patterns in P Cygni-type profiles of UV resonance lines and by comparing absorption profiles of individual elements in resonance line doublets. By the latter method, the presence of structure may be demonstrated from single-epoch observations by considering apparent departures from canonical oscillator strength ratios indicated by the radial optical depth ratios of each element of well-developed, unsaturated resonance line doublets. Where terminal wind velocities are sufficiently low, the doublet elements can be treated as decoupled, with each element's profile being fitted independently of the other (usually using the well-separated \ion{Si}{IV} 1393.76, 1402.77 \AA~feature). This technique proceeds on the basis that the apparent departure of the optical depth ratio from the oscillator strength ratio can be explained by the presence of optically thick regions partially covering the stellar disk. It is derived from \citet{Ganguly1999} and was applied and interpreted in the context of numerous individual stars in the Milky Way by \citet{Massa2008} and \citet{Prinja2010}. More recently, further evidence for the ubiquity of such large-scale structure obtained by using this technique in LMC, SMC and very low metallicity environments was presented by \citet{Parsons2024}.

Earlier observations of O-type stars in the LMC made with the \textit{Far Ultraviolet Spectroscopic Explorer} (FUSE) \citep[see,][]{Moos2000, Sahnow2000} have shown the presence of large scale structure in stellar winds of these stars, as evidenced by variability in the absorption profiles of highly-ionized species such as \ion{O}{VI} 1031.93, 1037.62 \AA~and \ion{S}{IV} 1062.66, 1072.97/1073.51 \AA~ \citep[][]{Massa2000}.

Direct evidence of variability, on a time scale of a few days (being less than one rotation period of the star), in the wind-formed UV spectral features of one SMC star, AzV 96 - spectral type B1 Iab, was shown in \citet{Parsons2024}. That result was shown by considering the individual UV spectra that together make up the high level science product (HLSP) spectrum of that star within the \textit{Hubble Space Telescope} (HST) Ultraviolet Legacy Library of Young Stars as Essential Standards (ULLYSES) spectroscopic database \citep[][]{Roman2025}\footnote{The full suite of ULLYSES spectra may be accessed at \texttt{doi:10.17909/t9-jzeh-xy14}}.

\subsection{Importance of Quantifying Short-Term Variability in Stellar Wind Optical Depths} \label{sec:quantvar}

Understanding and quantifying variability in optical depth of the stellar wind is fundamentally important for reliable determination of mass-loss rates from massive stars. The radial optical depth of the stellar wind and the derived mass-loss rate are directly related to each other, as described below.

To obtain a `best fit' wind line profile model, containing radial optical depth data, for the observed UV spectra, we utilize the `Sobolev with exact integration' (SEI) method, introduced by \citet{Hamann1981} and developed by \citet{Lamers1987}. This method employs a single application of the Sobolev approximation to derive the source function with an exact solution of the radiative transfer equation formal integral and was used, for example, by \citet{Massa2003} to model and derive radial optical depths of wind formed profiles for a large number of LMC O-type stars. The approach and method are described in more detail in \citet{Parsons2024}. A monotonically outwardly accelerating (`single $\beta$') stellar wind form is adopted whereby the wind velocity ($\varv$) at any particular radial distance (\textit{r}) is given as:

\begin{equation}
    \varv(r) = \varv_\infty \left(1-\frac{R_*}{r} \right)^\beta ,
\end{equation}
where $R_*$ is the photospheric radius of the star and $\varv_\infty$ is the terminal wind velocity.

With this constraint, we can express the radial optical depth of the stellar wind ($\tau_{\rm{rad}}$), at any distance from the photosphere in normalized velocity space ($w$), in the following manner \citep[see also][]{Massa2024}:

\begin{equation}\label{eqn:massloss}
    \tau_{\rm{rad}}(w) = C\frac{\dot{M}}{R_{*}\varv^{2}_\infty}q_{i}(w)\left(x^{2}w\frac{dw}{dx}\right)^{-1} ,
\end{equation}
where $x = r/R_{*}$, $w = \varv/\varv_{\infty}$ and \textit{C} is a constant containing atomic parameters of the species in question.

With the `single $\beta$' constraint mentioned above and additionally assuming a constant ionization fraction ($q_{i}$) throughout the stellar wind, the derived mass-loss rate ($\dot{M}$) is proportional to the radial optical depth of the stellar wind at any given location. It follows that variations in average optical depth translate into proportional variations in the mass-loss rate derived in this way \citep[see][]{Massa2024}. However, since the ionization fraction of a non-dominant species is likely, in practice, to vary considerably throughout the stellar wind and across different stars, we derive results below in terms the product of mass-loss rates and ionization fraction, $\dot{M}q_{i}$.

We can rearrange Equation \ref{eqn:massloss} in terms of $\dot{M}q_{i}$, and in which, as shown by \citet{Olson1982}, the constant term is:
\begin{equation}\label{eqn:const}
    C' = \frac{\pi e^2}{m_{e}c}  \lambda_{0}f\frac{A_E}{4\pi \mu m_H}~,
\end{equation}
where the first fraction is the classical cross section, $\lambda_{0}$ is the rest wavelength of the relevant species, $f$ the oscillator strength, $A_E$ the elemental abundance and $\mu m_{H}$ the mean molecular weight as a function of the mass of the hydrogen atom. As a result, we obtain:

\begin{equation}\label{eqn:mdotq}
    \dot{M}q_{i}(w) = \frac{\pi e^2}{m_{e}c}  \lambda_{0}f\frac{A_E}{4\pi \mu m_H}{R_{*}\varv^{2}_\infty}\tau_{\rm{rad}}(w)\left(x^{2}w\frac{dw}{dx}\right)~,
\end{equation}
which we make use of in Section \ref{sec:discuss} below to relate our observations to the question of uncertainties in mass-loss rate derivations.

Observations of Galactic OB stars using the \textit{International Ultraviolet Explorer} (IUE) \citep{Boggess1978} established that the stellar winds of massive stars exhibit near-ubiquitous variability in well-developed, unsaturated wind-formed resonance lines present in their UV spectra \citep[see, for example][]{Prinja1986, Massa1995, Kaper1996, Prinja2002}. Extensive time series UV spectroscopic observations, using IUE, of a number of these stars, have shown that this variability is modulated on timescales commensurate with the rotation periods of the stars \citep['rotational modulation';][]{Prinja1988}, with optically thick structures being observed to `migrate' from lower velocities outward to the terminal velocity of the stellar wind on time scales measurable in hours \citep[][]{Massa1995, Kaper1996, deJong2001, Prinja2002}. \citet{Massa2024} provide a more detailed description of these observational developments and of the physical characteristics of these large scale optically thick structures determinable from those observations.

FUSE observations have also provided evidence of similar wind line variability in a small number of LMC and SMC stars \citep[][]{Massa2000, Lehner2003}. More recently, the ULLYSES program \citep[][]{Roman2025} has provided higher resolution UV spectra for a large number of OB stars in the LMC and the SMC. Although not providing true time series data, many of the ULLYSES stars have been observed on more than one occasion for various reasons. Individual spectra contributing to the ULLYSES HLSP spectra are available in the Mikulski Archive for Space Telescopes (MAST) and in some cases provide sufficient suitable data to consider wind line variability in LMC and SMC OB stars, such as in the case of the SMC star AzV 96 mentioned above and briefly discussed in \citet{Parsons2024}.

A key development in \textit{quantifying} the extent of variability in stellar wind optical depths was provided by \citet{Massa2024}. Those authors considered large numbers of IUE UV time series spectra available for a selection of Milky Way OB stars. The importance of this, as discussed in that work, is that a statistical analysis of the large observational data sets available for these stars provides a robust quantification of the intrinsic uncertainty which is inherent in the measurement of a star's mass-loss rate derived from any single observation. It was found there that the proportionate $1\sigma$ variation in mass-loss rates derived from single optical depth measurements is as high as 45\% for Galactic early B-type stars, falling to 8\% for the hottest O-type stars.

There remains, however, an open question whether the inherent uncertainty in mass-loss rates deriving from large scale wind structure and variability is of similar magnitude (and, therefore, importance) in lower metallicity environments. Current efforts to extend detailed UV observations to such environments, in the ULLYSES project itself and in observations of extremely metal-poor environments \citep[e.g.][]{Telford2021, Telford2024} can be better applied to the derivation of reliable mass-loss results if we can place some constraints upon the inherent uncertainty of available single observations. It is also instructive to consider whether the relationship between effective temperature and inherent uncertainty identified by \citet{Massa2024} extends to these environments. The intention here is to take a further significant step using currently available, albeit limited, data toward that ideal. Such a demonstration is essential for further development and refinement of mass-loss estimates and recipes in these critical low-metallicity contexts.

Systematically planned high-resolution and high-cadence time series data of similar features do not yet exist for suitable stars at lower metallicities. As discussed below, a short times series for the LMC star Sk -67 166 (Spectral Type O4 If) does exist as well as lengthy multi-epoch sets of observations for the SMC stars, AzV 75 (Spectral Type O3.5 III(f)) and Sk 191 (Spectral Type B1.5 Ia). We aim, however, to demonstrate that by using even two sufficiently high resolution `snapshot' spectra with sufficiently high SNR, we can derive a credible lower limit on the intrinsic uncertainty described above. In this way, we may determine with some quantitative certainty whether the results from \citet{Massa2024} also apply at lower metallicities. 

\begin{table*}
\centering
\caption{\label{tab:Alltargets}Tabulation of LMC (first block) and SMC (second block) O and early B stars in the ULLYSES data set which are suitable for analysis as described in the main text, with spectral types and values for principal stellar physical parameters. Values of stellar radial velocity and terminal wind velocity derived for this work or, as the case may be, taken from existing literature are indicated in the table. Spectral types used in ULLYSES data releases have been updated to those recently derived by \citet{Bestenlehner2025}, based on analysis of ULLYSES and XShooter data as part of the ongoing XShootU project \citep[][]{Vink2023}.}
\setlength{\tabcolsep}{5pt}
\begin{tabular}{llrrrrrrrrrrr} \hline \hline \\
Star & Spectral type & $T_{\rm eff}$ & log $g$ & log $L$ & Radius & \textit{Ref} & $\varv\textsubscript{rad}$ & \textit{Ref} & $\varv\textsubscript{$\infty$}$ & \textit{Ref} & $\varv$ sin $i$ & \textit{Ref} \\
 & (\textit{Ref: 23}) & (K) & (cgs) & $(L/L_\odot)$ & $(R/R_\odot)$ & & (km s$^{-1}$) & & (km s$^{-1}$) & & (km s$^{-1}$) & \\[0.5ex]
\hline
VFTS 180 & O3 If* & 44,100 & 3.62 & 5.92 & 16.5 & \textit{20} & 264 & \textit{11} & 2,200 & \boldmath$x$ & 90 & \textit{20} \\
Sk -67 166 & O4 If & 40,000 & 3.6\phantom{0} & 5.98 & 18.8 & \textit{1} & 303 & \boldmath$x$ & 1,900 & \textit{13} & 97 & \textit{9} \\
VFTS 586 & O4.5 Vz & 46,300 & 3.58 & 5.42 & 7.9 & \textit{20} & 261 & \textit{11} & 2,900 & \boldmath$x$ & 78 & \textit{20} \\
Sk -70 60 & Early O(f)pe & 46,400 & [3.75] & 5.56 & 9.3 & \textit{13} & 257 & \boldmath$x$ & 2,300 & \textit{13} & 256 & \textit{9} \\
VFTS 244 & O5 III(n)fpc & 39,700 & 3.58 & 5.58 & 13.0 & \textit{20} & 244 & \textit{11} & 2,500 & \boldmath$x$ & 223 & \textit{20} \\
N11 ELS 48 & O6 Vz((f)) & 40,700 & 4.19 & 5.38 & 10.8 & \textit{6} & 305 & \textit{23} & 2,520 & \boldmath$x$ & 78 & \textit{9} \\
Sk -68 16 & O7 III(f) & 36,700 & 3.50 & 5.34 & 16.0 & \textit{13,23} & 280 & \textit{24} & 2,220 & \textit{18} & 83 & \textit{21} \\
PGMW 1363 & O8-8.5 Ifp & 31,700 & 3.12 & 5.67 & 22.6 & \textit{18} & 291 & \textit{24} & 1,510 & \textit{18} & 153 & \textit{23} \\
Sk -70 79 & B0.2 Ib & 23,700 & 2.69 & 5.63 & 30.0 & \textit{18,23} & 244 & \textit{23} & 1,450 & \textit{18} & 86 & \textit{9} \\
\hline
NGC 346 MPG 355 & ON2 III(f*) & 51,700 & 4.00 & 6.04 & 11.0 & \textit{10} & 172 & \boldmath$x$ & 2,900 & \boldmath$x$ & 130 & \textit{17} \\
AzV 388 & O3.5 V((fc))z & 43,100 & 4.01 & 5.54 & 9.8 & \textit{10} & 173 & \boldmath$x$ & 2,100 & \textit{10} & 150 & \textit{10} \\
AzV 75 & O3.5 III(f) & 38,700 & 3.51 & 5.81 & 20.5 & \textit{13,16} & 136 & \textit{23} & 2,000 & \textit{2} & 120 & \textit{10} \\
NGC 346 MPG 368 & O6 V((fc))z & 39,300 & 3.75 & 5.38 & 10.8 & \textit{10} & 195 & \boldmath$x$ & 2,100 & \boldmath$x$ & 58 & \textit{17} \\
AzV 238 & O9.5 III & 30,700 & 3.75 & 5.30 & 15.6 & \textit{13} & 197 & \boldmath{x} & 1,250 & \boldmath$x$ & 86 & \textit{17} \\
AzV 70 & O9.5 Iab & 31,300 & 3.1\phantom{0} & 5.44 & 27.9 & \textit{2} & 158 & \textit{7} & 1,800 & \boldmath$x$ & 62 & \textit{9} \\
AzV 506 & B0.5 II & 26,700 & 3.12 & 5.13 & 19.3 & \textit{14,23} & 193 & \boldmath$x$ & 425 & \boldmath$x$ & 78 & \textit{15} \\
AzV 96 & B1 Iab & 20,000 & 2.55 & 5.39 & 30.0 & \textit{4} & 166 & \textit{12} & 800 & \textit{19} & 70 & \textit{9} \\
Sk 191 & B1.5 Ia & 22,500 & 2.55 & 5.77 & 51.7 & \textit{3} & 120 & \textit{19} & 490 & \textit{22} & 118 & \textit{9} \\
NGC 330 ELS 4 & B2.5 Ib & 17,000 & 2.3\phantom{0} & 4.77 & 31.6 & \textit{7} & 158 & \textit{8} & 460 & \textit{19} & 36 & \textit{7} \\
NGC 330 ELS 2 & B3 Ib & 14,590 & 2.15 & 4.73 & 35.8 & \textit{7} & 154 & \textit{5} & 460 & \textit{19} & 14 & \textit{7} \\
\hline
\end{tabular} 
\\[1.5ex]
References: \textbf{1} \citet{Crowther2002}, \textbf{2} \citet{Evans2004c}, \textbf{3} \citet{Trundle2004}, \textbf{4} \citet{Trundle2005}, \textbf{5} \citet{Evans2006}, \textbf{6} \citet{Mokiem2007}, \textbf{7} \citet{Trundle2007}, \textbf{8} \citet{Evans2008}, \textbf{9} \citet{Penny2009}, \textbf{10} \citet{Bouret2013}, \textbf{11} \citet{Sana2013}, \textbf{12} \citet{Lamb2016}, \textbf{13} \citet{Massa2017}, \textbf{14} \citet{Castro2018}, \textbf{15} \citet{Dorigo2020}, \textbf{16} \citet{Bouret2021} \textbf{17} \citet{Dufton2019}, \textbf{18} \citet{Hawcroft2024b}, \textbf{19} \citet{Parsons2024}, \textbf{20} \citet{Hawcroft2024a}, \textbf{21} \citet{Martins2024}, \textbf{22} \citet{Bernini2024}, \textbf{23} \citet{Bestenlehner2025}, \textbf{24} \citet{Brands2025}.

(\boldmath$x$) = Derived for this work (either new results or differing from some earlier published results), ([-]) = No data available and result estimated from other parameters or comparable stars.

Note: source references listed after stellar radius data apply also to temperature, surface gravity and luminosity data, except: AzV 96 $T_{\rm eff}$ figure is derived from UV spectral fitting.
\end{table*}

\section{Target Selection, Observations and Data Reduction}

\subsection{Criteria for target selection}

Table \ref{tab:Alltargets} shows the stars considered in this work. We initially considered all LMC and SMC stars within the ULLYSES program \citep[][]{Roman2025}. In many instances, multiple observations of individual stars have contributed to the resulting HLSP spectra, or the stars have been the subject of discontinuous observations in both ULLYSES and other programmes. There are also examples of multiple observations being obtained as part of instrument calibration exercises and of genuine short time series observations obtained for other purposes.

To obtain from that data set the final list of targets suitable for analysis in this work, the target were selected based on spectra having the following characteristics:
\begin{enumerate}
    \item at least two usable observations which are at least 12 hours apart. This minimum cadence has been chosen to maximize the likelihood of finding a measurable and significant variation in profile, while retaining as many stars as possible in the sample.
    \item well-developed, unsaturated wind line profiles for at least one species. We have identified such profiles for any of the \ion{N}{V} 1238.80, 1242.80 \AA{}, \ion{Si}{IV} 1393.76, 1402.77 \AA{} or \ion{C}{IV} 1548.20, 1550.78 \AA{} resonance line doublets. Some stars exhibit suitable profiles for more than one of these species, in which case each has been considered.
    \item with limited exceptions, no obvious prior evidence of binarity nor of significant emission lines. This results in the exclusion of known high-mass X-ray binaries and other binaries \citep[for example, LMC-X-4, AzV 490 and 2dFS 163:][]{Bernini2024, Ramachandran2024}. Sk -70 60 is an emission line star, but has been included as the \ion{N}{V} and \ion{C}{IV} features are suitable for analysis. Some targets for which there is recent evidence of potential binarity are retained, for example, Sk -68 16 and AzV 70. Additionally, there are some examples included where the stellar wind characteristics traced in UV spectra of the primary star do not appear to be significantly affected by a secondary star, such as the SB1 system AzV 238 and the SMC cluster star NGC 346 MPG 368 (included although \citet{Rickard2024} conclude there is a binary companion, based on differing radial velocity measurements required to fit \ion{He}{I} and \ion{He}{II} lines in optical wavelengths). Individual cases are discussed in Subsection \ref{sec:ind_disc}. Finally, analysis of data for this work has confirmed the recent identification of AzV 75 (SMC) as a binary system, as described in Section \ref{sec:av75} below. The presence of a secondary star does significantly affect the UV resonance line profiles and the results obtained here, however this interesting system has been retained in the analysis to illustrate some unusual effects that binarity may produce in stellar wind profiles and their consequences for the analysis conducted here.
\end{enumerate}

Applying the above criteria to the ULLYSES LMC and SMC data set produced the final list of 20 suitable targets. Although target star selection was necessarily based on availability of suitable data and the presence of these characteristics, precluding a systematic selection based on spectral type or other specific physical characteristics \textit{per se}, the final set of targets covers a range of spectral types, located in a variety of environments including some well-studied massive star clusters (N11 in the LMC; NGC 330 and NGC 346 in the SMC), inner and outer regions of each galaxy and the `wing' of the SMC.

   \begin{figure*}
\includegraphics[width=\linewidth]{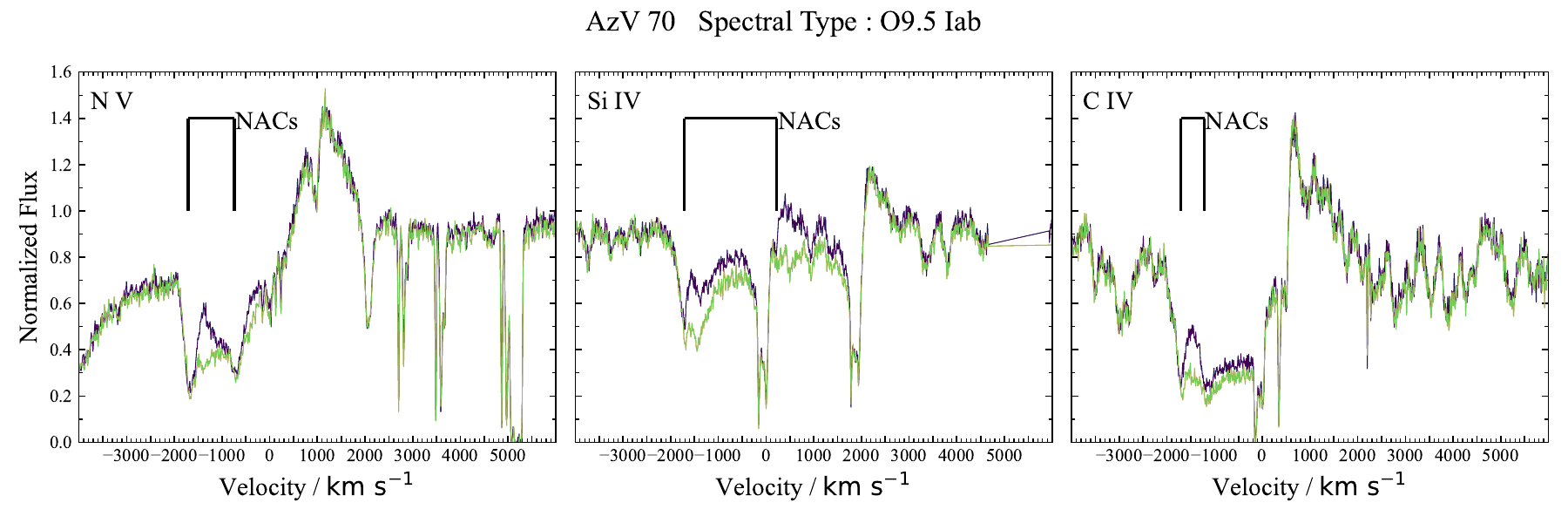} 
\caption{Comparison of two observations of the SMC star AzV 70, made at an interval of 47.954 days, the first (dark violet) spectrum obtained on 2020 June 28 and the second (light green) spectrum obtained on 2020 August 15. The significant changes in the absorption profiles of each of the \ion{N}{V}, \ion{Si}{IV} and \ion{C}{IV} resonance line profiles can be seen, in particular, the enhanced absorption toward higher stellar wind velocities in the second observation may be noted. Prominent narrow absorption component (NAC) features are also seen for each species and in each observation and these are highlighted, although the NAC for the red element of the \ion{Si}{IV} profile is obscured by the emission profile of the blue element of that doublet. Note that the separations of the NACs in each case equate to the separations of the doublet elements, confirming their nature.}
\label{fig:av70_comp_example}
   \end{figure*}

\subsection{Stellar parameters}

The relevant stellar parameters for each of the sample stars are set out in Table \ref{tab:Alltargets}. Most values are taken from existing literature however, where noted, some newly derived results are included, based upon SEI or other analysis of the UV spectra examined here. Some values of $\varv\textsubscript{$\infty$}$ derived in this way differ from existing literature. In these instances, we have cross-checked our results against any available saturated wind profiles and have also used the locations of narrow absorption components (NACs) in unsaturated profiles to derive $\varv\textsubscript{$\infty$}$ as described by \citet{Prinja1990} and \citet{Prinja1998}. For sample stars which also appear in both \citet{Bernini2024} and \citet{Parsons2024}, consistent $\varv\textsubscript{$\infty$}$ values obtained in those works (by different techniques) have been used here. As in \citet{Parsons2024}, individual stellar radial velocities derived from existing literature, or measured from the UV spectra considered here, are used rather than `bulk' LMC and SMC velocities.

Since, with very limited exceptions, the spectra have not been obtained as part of a systematic programme of time series or multi-epoch observations, the cadences of observations for each star vary considerably. Some elapsed period are up to several years.

As an illustrative example, Figure \ref{fig:av70_comp_example} shows the \ion{N}{V}, \ion{Si}{IV} and \ion{C}{IV} UV resonance line doublet features for the O9.5 Iab SMC supergiant star, AzV 70. Each of these features present well-developed but unsaturated profiles suitable for analysis. Distinct and consistent changes in the absorption profiles of each are also seen between the two observations. In each plot, the zero velocity point represents the rest wavelength (in the star's frame of reference) of the blue element of the relevant resonance line doublet. Similar spectral extracts are shown for all sample stars in Appendix \ref{allplots}. The prominent NAC features in the profiles of all of these species can readily be seen and are highlighted on each plot, the separation of each NAC for all of the plotted ions equates to the separation of the relevant doublet elements, corroborating their nature.

\subsection{Data reduction}

The data reduction process to produce model fits using the SEI method is similar to that described in detail in \citet{Parsons2024}. As there, the normalized stellar spectrum is combined with a suitable photospheric spectrum derived from the \texttt{TLUSTY} O and B star model grids \citep[described in][and subsequent guides]{Hubeny1988, Lanz2007}. Consistent identification of the UV continuum is complicated by the presence of extensive line-blanketing due to the presence of large numbers of photospheric spectral lines such as \ion{Fe}{IV}, \ion{Fe}{V} and \ion{Fe}{VI} (the so-called `iron forest'). Consistency of normalization here was ensured by identifying several short continuum regions across a large portion of the full wavelength range in each spectrum and dividing by a spline fit interpolated through these short regions. These short regions were also checked for flux consistency across individual spectra. This method produces consistency for comparisons between spectra even if the `true' continuum remains elusive.

The SEI method itself was applied with a weighted non-linear `damped' (Marquardt-Levenburg) least-squares minimization. This commences the fit at the highest velocity portion of the stellar wind and uses each velocity bin's fit to inform the fit to the next lower velocity bin. Radial optical depths of the stellar winds, $\tau_{\rm{rad}}(w)$, are those that the SEI model fitting process has derived to produce the best fit in each instance. In a number of cases, stars with suitable unsaturated wind line profiles have had to be excluded as a consequence of only one usable observation for a particular ion being available.

\section{Results}

\subsection{Optical depth variations}

The average radial optical depths of absorption features are derived in most cases for the velocity range $w$ = 0.3-0.9 (recalling that $w = \varv/\varv\textsubscript{$\infty$}$). This range includes 15 of the 21 radial velocity `bins' used in the SEI fitting process. We exclude the low velocity regions as the SEI technique begins to break down at very low wind velocities and this region is typically affected by significant ISM lines in the case of \ion{Si}{IV} and \ion{C}{IV} profiles. A small portion of the highest velocity region is also excluded to avoid effects from any uncertainty in the terminal wind velocity and of the extent of wind turbulence. For a small number of stars, a shorter range of $w$ = 0.4-0.9 was used, for reasons described in the discussions for those stars. This shorter range excludes only one further velocity 'bin' from the analysis. In addition, there are some targets with sufficiently low terminal wind velocities to allow the \ion{Si}{IV} doublet elements to be treated as decoupled from each other and individually modelled. In those examples, the same velocity range is used by reference to the rest wavelengths of each of the blue and red doublet elements, respectively.

Numerical data are set out in Table \ref{tab:Alldata}, particularly in the final column, which provides quantitative results for the observed proportional variation in average optical depth of wind-formed absorption features, $\sigma(\tau)/\tau$. As explained in Section \ref{sec:quantvar} and in \citet{Massa2024}, this provides a proxy measure of the minimum inherent uncertainty of mass-loss rates derived from any single observation (assuming, as noted, a constant ionization fraction throughout the stellar wind).
 
Results from the available data are insufficient to permit fully quantified conclusions regarding any correlation between the extent of variability observed in winds of stars across differing spectral types or effective temperatures. The results do, however, provide robust determinations of the \textit{minimum} extent of optical depth variations observed on time scales broadly commensurate, in many cases, with stellar rotation periods. The separations of observations for the stars in our final data set are still on timescales ranging from one or two months to two decades. Observed variations are therefore also likely to derive from similar processes of stellar wind variability, rather than from long-term evolutionary changes in the star itself.

Rotational periods shown in Table \ref{tab:Alldata} are approximate \textit{maximum} periods, in days, derived from the stated stellar radii and projected rotational velocities ($\varv$ sin $i$) and assuming that sin $i \approx 1$ This also assumes that the line broadening measurements producing those rotational velocities arise principally from rotational effects, not macroturbulence, the presence of which could be expected to mean true rotation periods are longer than indicated. True rotation periods may be shorter where sin $i$ $\ll$ 1. It is therefore not necessary or appropriate to introduce uncertainty measurements for this parameter, which is provided only for the purpose of approximate comparison with cadences of observations.

\begin{table*} 
\centering
\caption{\label{tab:Alldata}Tabulation of LMC (first block) and SMC (second block) data and results indicating variability of optical depths. NOTE (1): (*) indicates maximum and minimum derived average optical depths, respectively, are quoted where there are more than two observations utilized. For AzV 75, note the discussion int eh main text regarding the effects of binarity on these results. (2): Derived rotation periods are \textit{maximum} periods only, based solely upon the quoted stellar radius and $\varv$ sin $i$ values set out in Table \ref{tab:Alltargets}. The ninth column contain the results for the derived lower bounds on proportionate uncertainty in average radial optical depths of any single observation and averaged across the same wind velocity range used to obtain average radial optical depth results. The final three columns show the product of derived mass-loss rates and ionization fraction for the relevant ion for two observations of each target, together with the associated uncertainty inherent in that result. (3): ($^\varnothing$) indicates that AzV 388 displays very prominent and varying NAC features. The effect of this on these results is discussed in the main text.}
\begin{tabular}{llrrrlccrrrc} \hline \hline \\
Star & Spec. type & Rot. Per. & Obs. Cadence & Obs. & Ion & \multicolumn{2}{c}{Ave. Opt. Depth}  & $\sigma(\tau)/\tau$ & \multicolumn{2}{c}{$\dot{M}q_{i}/10^{-9}$} & $\sigma(\dot{M}q_{i})/10^{-10}$ \\
 & & (d., max.) & (d., §=mean) & ($\#$) & & ($\tau_{\rm{obs1}}$) & ($\tau_{\rm{obs2}}$) & & \multicolumn{2}{c}{(M$_{\odot}$yr$^{-1}$:obs$_{\rm{1,2}})$} & (M$_{\odot}$yr$^{-1}$) \\[0.5ex]
\hline
VFTS 180 & O3 If* & 9.3 & 1264.637\phantom{0} & 2 & \ion{Si}{IV} & 0.80 & 0.81 & < 0.01 & 0.802 & 0.812 & 0.071 \\
Sk -67 166* & O4 If & 9.8 & 0.880§ & 10 & \ion{Si}{IV} & 1.80 & 1.46 & 0.06 & 1.533 & 1.244 & 2.043 \\
VFTS 586 & O4.5 Vz & 5.1 & 952.753\phantom{0} & 2 & \ion{C}{IV} & 1.61 & 1.86 & 0.10 & 0.391 & 0.452 & 0.431 \\
Sk -70 60* & Early O(f)pe & 1.8 & 76.028§ & 3 & \ion{N}{V} & 2.19 & 2.62 & 0.13 & 2.399 & 2.871 & 3.337 \\
 & & & & 3 & \ion{C}{IV} & 1.39 & 1.54 & 0.08 & 0.249 & 0.277 & 0.198 \\
VFTS 244 & O5 III(n)fpc & 3.0 & 1288.255\phantom{0} & 2 & \ion{Si}{IV} & 0.29 & 0.32 & 0.05 & 0.296 & 0.326 & 0.212 \\
N11 ELS 48 & O6 Vz((f)) & 7.0 & 1.051\phantom{0} & 2 & \ion{N}{V} & 1.80 & 1.70 & 0.04 & 2.749 & 2.597 & 1.075 \\
 & & & & 2 & \ion{C}{IV} & 1.04 & 1.00 & 0.02 & 0.261 & 0.251 & 0.071 \\
Sk -68 16 & O7 III(f) & 8.1 & 2.311\phantom{0} & 2 & \ion{N}{V} & 1.29 & 1.11 & 0.10 & 2.265 & 1.949 & 2.234 \\
 & & & & 2 & \ion{Si}{IV} & 0.60 & 0.49 & 0.14 & 0.594 & 0.485 & 0.771 \\
 & & & & 2 & \ion{C}{IV} & 2.99 & 2.84 & 0.04 & 0.863 & 0.819 & 0.311 \\
PGMW 1363 & O8-8.5 Ifp & 7.5 & 171.310\phantom{0} & 2 & \ion{N}{V} & 1.01 & 1.09 & 0.11 & 1.159 & 1.251 & 0.651 \\
Sk -70 79 & B0.2 Ib & 17.6 & 0.857\phantom{0} & 2 & \ion{Si}{IV} & 4.89 & 5.26 & 0.03 & 3.871 & 4.164 & 2.072 \\
\hline
NGC 346 MPG 355 & ON2 III(f*) & 4.3 & 8060.059\phantom{0} & 2 & \ion{C}{IV} & 0.47 & 0.48 & 0.02 & 0.064 & 0.065 & 0.007 \\
AzV 388$^\varnothing$ & O3.5 V((fc))z & 3.3 & 2135.151\phantom{0} & 2 & \ion{C}{IV} & 0.63 & 0.68 & 0.05 & 0.040 & 0.043 & 0.021 \\
AzV 75* & O3.5 III(f) & 8.6 & 201.456§ & 25 & \ion{C}{IV} & 4.33 & 1.73 & 0.33 & 0.519 & 0.207 & 2.206 \\
NGC 346 MPG 368 & O6 V((fc))z & 9.1 & 8000.632\phantom{0} & 2 & \ion{C}{IV} & 0.61 & 0.51 & 0.12 & 0.042 & 0.036 & 0.042 \\
AzV 238 & O9.5 III & 8.4 & 23.434\phantom{0} & 2 & \ion{C}{IV} & 0.59 & 0.59 & < 0.01 & 0.021 & 0.021 & < 0.001\phantom{0} \\
AzV 70 & O9.5 Iab & 22.8 & 47.954\phantom{0} & 2 & \ion{N}{V} & 1.37 & 1.64 & 0.13 & 1.103 & 1.321 & 1.541 \\
 & & & & 2 & \ion{Si}{IV} & 0.40 & 0.72 & 0.40 & 0.182 & 0.328 & 1.032 \\
 & & & & 2 & \ion{C}{IV} & 1.44 & 2.05 & 0.24 & 0.190 & 0.271 & 0.573 \\
AzV 506 & B0.5 II & 12.5 & 6.135\phantom{0} & 2 & \ion{C}{IV} & 0.75 & 0.64 & 0.11 & 0.004 & 0.003 & 0.778 \\
AzV 96 & B1 Iab & 21.68 & 4.827\phantom{0} & 2 & \ion{Si}{IV} & 2.20 & 3.46 & 0.32 & 0.212 & 0.334 & 0.863 \\
 & & & & 2 & \ion{C}{IV} & 3.32 & 3.92 & 0.12 & 0.093 & 0.110 & 0.120 \\
Sk 191* & B1.5 Ia & 22.2 & 87.057§ & 11 & \ion{C}{IV} & 4.61 & 3.70 & 0.15 & 0.084 & 0.067 & 0.120 \\
NGC 330 ELS 4* & B2.5 Ib & 44.4 & 945.674§ & 4 & \ion{Si}{IVb} & 1.30 & 0.86 & 0.12 & 0.044 & 0.029 & 0.106 \\
 & & & & 4 & \ion{Si}{IVr} & 0.74 & 0.61 & 0.05 & 0.055 & 0.045 & 0.068\\
 & & & & 4 & \ion{C}{IV} & 0.60 & 0.49 & 0.05 & 0.006 & 0.005 & 0.008 \\
NGC 330 ELS 2 & B3 Ib & 129.3 & 10.967\phantom{0} & 2 & \ion{Si}{IVb} & 0.29 & 0.64 & 0.53 & 0.011 & 0.024 & 0.092 \\
 & & & & 2 & \ion{Si}{IVr} & 0.29 & 0.36 & 0.37 & 0.024 & 0.030 & 0.041 \\
 & & & & 2 & \ion{C}{IV} & 0.59 & 1.16 & 0.47 & 0.007 & 0.013 & 0.046 \\
\hline
\end{tabular} 
\\[1.5ex]
\end{table*}

The limited data available do not permit definite quantitative confirmation at lower metallicities of the extent of inherent uncertainty of stellar wind optical depth measurements. However, consistent with the result in \citet{Massa2024}, significant variation of this nature can clearly be observed in both LMC and SMC stars. We discuss in the following Section, the extent to which we may also conclude that the extent of short time scale optical depth variability in winds of hotter stars is smaller than that for relatively cooler stars. We also consider further the hypothesis that the \textit{extent} of variability itself changes with varying metallicity.

The figures presented in Appendix \ref{allplots} overplot wind line profiles for each star in the target sample for the two or more usable observations available and for each of the \ion{N}{V}, \ion{Si}{IV} and \ion{C}{IV} resonance line doublets (except for the latest spectral types where the \ion{N}{V} feature is not present). Saturated profiles are included as these provide a further check on measured terminal wind velocities using the $\varv_{\infty} = \varv\textsubscript{black}$ method described by \citet{Prinja1990}. References in the text to `outward' migration of features within the spectra refer to movements from slower to faster regions of the stellar wind. The zero velocity point in each instance refers to the location of the rest wavelength, in the star's frame of reference, of the blue element of the relevant resonance line doublet feature plotted. Examination of these plots readily shows complex and, often significant, variations in the depths and shapes of wind-formed absorption profiles for many stars in our sample.

The SEI model fits for each star and ion species fitted to obtain average optical depths of wind-formed absorption features are shown in Appendix \ref{allSEI}. We present there the model fits for each observation of a particular ion species side-by-side to enable ready comparison of the observed line profiles and best-fit radial optical depth profiles. Where more than two suitable observations exist, those from which the highest and lowest average radial optical depths have been derived are shown. Each of these plots indicates the location of the rest wavelength for the blue and red elements of the relevant doublet feature. The key physical wind parameters produced by the fitting process are also shown and NAC features are indicated, where identifiable. These plots are accompanied by a smoothed plot of radial optical depths derived from the model fit and of the photospheric contribution to the observed profile. In a small number of cases, otherwise suitable \ion{N}{V} profiles have not been fitted as the lack of a distinct `blue edge' due to extensive Lyman-$\alpha$ absorption compromises the ability to produce a credible model fit.

Appendix \ref{alltauradcomp} sets out plots of the SEI-derived radial optical depths for each observation of a star against each other to illustrate how the variations in optical depths are manifested across different velocities in the stellar wind. In the case of stars with more than two observations available, the plots are limited to, at most, four observations for clarity.

The final three columns of Table \ref{tab:Alldata} set out results derived using the average radial optical depth results obtained in this work and the relationship shown in Equation \ref{eqn:mdotq} for the product of mass-loss rate and ionization fraction, $\dot{M}q_{i}$. Uncertainties for those results are also provided.

\subsection{Prevalence of NACs at low metallicities}

\begin{table}
    \centering
    \caption{Wind velocity location of NAC features. Where no value is given, existence of NACs is unclear or doubtful.}
    \setlength{\tabcolsep}{5pt}
    \begin{tabular}{lllr} \hline \hline \\
     & & & NAC \\
    Star & Spectral type & Ion(s) & Central Vel. \\
     & & & (km s$^{-1}$) \\[0.5ex]
    \hline
    VFTS 180 & O3 If* & - & - \\
    Sk -67 166 & O4 If & \ion{Si}{IV} & 1,765 \\
    VFTS 586 & O4.5 Vz & \ion{C}{IV} & 2,840 \\
    Sk -70 60 & Early O(f)pe & \ion{N}{V}, \ion{C}{IV} & 2,185 \\
    VFTS 244 & O5 III(n)fpc & - & - \\
    N11 ELS 48 & O6 Vz((f)) & \ion{N}{V}, \ion{C}{IV} & 2,420 \\
    Sk -68 16 & O7 III(f) &  \ion{N}{V}, \ion{C}{IV} & 2,175 \\
    PGMW 1363 & O8-8.5 Ifp & \ion{N}{V} & 1,450 \\
    Sk -70 79 & B0.2 Ib & \ion{Si}{IV} & 1,380 \\
    \hline
    NGC 346 MPG 355 & ON2 III(f*) & \ion{C}{IV} & 2,785 \\
    AzV 388 & O3.5 V((fc))z & \ion{C}{IV} & 2,015 \\
    AzV 75 & O3.5 III(f) & \ion{C}{IV} & 1,900 \\
    NGC 346 MPG 368 & O6 V((fc))z & \ion{C}{IV} & 1,995 \\
    AzV 238 & O9.5 III & \ion{C}{IV} & 1,200 \\
    AzV 70 & O9.5 Iab & \ion{N}{V}, \ion{Si}{IV}, \ion{C}{IV} & 1,710 \\
    AzV 506 & B0.5 II & \ion{C}{IV} & 395 \\
    AzV 96 & B1 Iab & \ion{Si}{IV}, \ion{C}{IV} & 760 \\
    Sk 191 & B1.5 Ia & - & - \\
    NGC 330 ELS 4 & B2.5 Ib & \ion{Si}{IV}, \ion{C}{IV} & 435 \\
    NGC 330 ELS 2 & B3 Ib & \ion{Si}{IV}, \ion{C}{IV} & 440 \\
    \hline
    \end{tabular}  
    \label{tab:NAC}
\end{table}

Visual reference to the overplotted spectra in Appendix \ref{allplots} and to the SEI-derived model fits in Appendix \ref{allSEI} readily illustrates the nearly universal occurrence of NACs in our sample, including across multiple species where suitably unsaturated profiles exist. This confirms earlier observations of this phenomenon at low metallicity \citep[e.g., for LMC and SMC B supergiants,][]{Parsons2024} and provides further evidence of the continued prevalence of these features in such environments and across OB spectral types and luminosity classes. In some instances the terminal wind velocity and separation of doublet elements means that the NAC for the red element of a doublet is partly or wholly obscured by interstellar absorption features from the blue element. In those cases, the theoretical red element NAC location is still noted in cases where the blue element NAC is clearly present. Where necessary, these have been used to obtain measurements of $\varv_{\infty}$ in Table \ref{tab:Alltargets}.

Table \ref{tab:NAC} sets out the central velocity within the stellar wind of the NAC features for all relevant stars in our sample.

The existence, and persistence, of NACs is understood as a key consequence of the presence and evolution of broad discrete absorption components (DACs) across the stellar wind. The corollary of the observed ubiquity of NACs at low metallicities is, therefore, that DACs can also be considered to be common in these environments. This in turn implies that distinct, and evolving, structure - not merely clumping - is likely to be a common feature throughout all regions of stellar winds at low metallicities \citep[see also][]{Massa2015}.

A feature of some observed NACs is that these features may change in depth and width even in the absence of any significant variation in the `continuum' segment of a P Cgyni absorption profile. This is discussed in relation to AzV 388 below (Subsection \ref{sec:388disc}), where this is most clearly seen. Similar characteristics are observed to varying extents in some other targets (e.g. AzV 70, Sk -70 60). This behaviour may arise from the unseen migration of a DAC occurring largely between observations and only observed at the stage of merging into a persistent NAC. We consider below the significance of this phenomenon for the accuracy and precision of determinations of mass-loss rates.

   \begin{figure*}
\includegraphics[width=\linewidth]{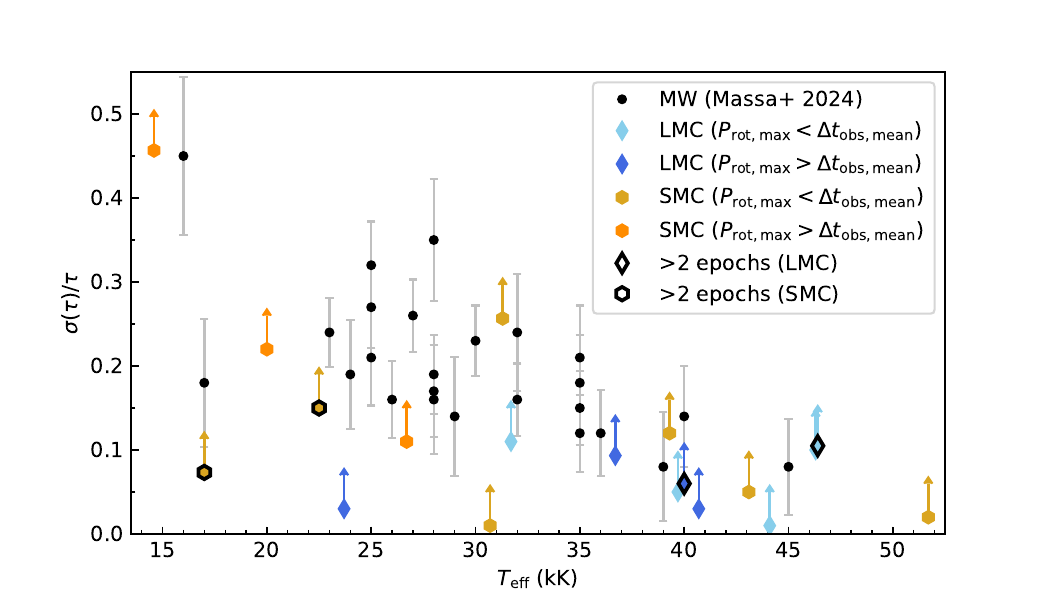} 
\caption{Comparison of the optical depth inherent ncertainty results from this work against those for Galactic stars presented by \citet{Massa2024}. The plot shows derived \textit{lower bounds} on $\sigma(\tau)/\tau$ against stellar effective temperature for the LMC and SMC targets considered in this work. Uncertainties on the results for Galactic stars are also from \citet{Massa2024}. Targets with observations at a cadence longer than the stellar rotation period are distinguished from those where the cadence is shorter than the stellar rotation period. Targets with more than two observations used are highlighted. The result for AzV 75, although included in Table \ref{tab:Alldata}, is not plotted here due to the significant effect that the binary nature of AzV 75, discussed below, has on the inherent uncertainty derived in this work.}
\label{fig:massaplot}
   \end{figure*}

\section{Discussion}\label{sec:discuss}

\subsection{Mass-loss uncertainties, comparison with Galactic results}

The lower bounds on variability derived here (ninth column of Table \ref{tab:Alldata}) support the conclusion that, for the latest O stars and for early B stars in the lower metallicity environments of the LMC and SMC, the extent of wind line variability is commensurate with, and not significantly smaller than, that observed in the Galactic sample examined by \citet{Massa2024}. As in that work, short time scale variability is quantified by measuring the standard deviation of mean radial optical depths as a proportion of the overall `mean of mean' optical depths: $\sigma(\tau)/\tau$.

We find, for example, a lower bound on the proportionate degree of observed variability at the $1\sigma$ level of 13-40\% for the O9.5 Iab SMC star, AzV 70 and 37-53\% for the B3 Ib SMC star, NGC 330 ELS 2. A similar lower bound (12-32\%) is observed for the B1 Iab SMC star, AzV 96. Significant proportionate variability in average optical depth, of the order of 15\%, is observed for the B1.5 Ia SMC star Sk 191. That star exhibits other unusual spectral features which add further uncertainty to mass-loss rate determinations, as discussed below. The lesser extent of variability observed in the earlier O stars in the present sample can be seen. Despite the absence of extensive time-series or multi-epoch observations for these stars, we may still conclude that these figures also represent a lower bound on variability and are not inconsistent with the results found by \citet{Massa2024}. The O3.5 III(f) target, AzV 75 shows a similarly large measured variability in wind optical depth as that shown by some stars of later spectral types. As discussed in Section \ref{sec:av75} below, that result is complicated by the previously unreported binary nature of AzV 75.

Although the data available for individual LMC and SMC stars are much less numerous than for the Galactic stars studied by \citet{Massa2024}, plotting our `lower bound' results against the results from that work (see Figure \ref{fig:massaplot}) highlights in a similar fashion the inverse relationship between the inherent uncertainty of a single average wind optical depth measurement and stellar effective temperature. The results here generally show equivalent or lower inherent uncertainties of single observations to those obtained by \citet{Massa2024} for Galactic stars. Even with our limited data providing only lower bounds on these results, Figure \ref{fig:massaplot} shows that a similar trend of increased inherent uncertainty at lower effective temperatures also applies to lower metallicity environments.

The limited extent of our available data does not permit a definite conclusion as to whether the magnitude of inherent uncertainty of an optical depth measurement derived from a single observation is dependent upon metallicity in addition to stellar effective temperature. Nevertheless, it can be seen from Figure \ref{fig:massaplot} that a comparable degree of inherent observational uncertainty does persist at lower metallicities.

The final 3 columns of Table \ref{tab:Alldata} quantitatively relate optical depth uncertainties to mass-loss rates by providing measures of the product of mass-loss rate and ionization fraction ($\dot{M}q_i$), derived using Equation \ref{eqn:mdotq}. To obtain those results, we use elemental abundances from \citet{Asplund2009} (scaled by 0.5 for LMC stars and by 0.2 for SMC stars), oscillator strengths from \citet{Morton1991} and the method described in greater detail by \citet{Olson1982} and \citet{Massa2003} to obtain a mass-loss rate multiplied by ion fraction ($\dot{M}q_{i}$), averaged across the same normalized velocity region of the stellar wind for which average radial optical depth has also been measured. This calculation is carried out for each observation and a resulting inherent lower-bound uncertainty for $\dot{M}q_i$ derived from any single observation is obtained.

It is noted that direct quantification of mass-loss rates from these methods requires accurate quantification of ionization fractions. This is only possible when considering a dominant ion, where $q_i \simeq1$, which is not the case for any of \ion{N}{V}, \ion{Si}{IV} or \ion{C}{IV}. For these species, the ionization fraction can be expected to vary significantly with stellar effective temperature, location within the stellar wind and the extent of inhomogeneity in the wind itself. The overall ionization fraction of a species within the wind of any single star should not however significantly vary between two observations that are reasonably proximate. Our results for $\sigma(\dot{M}q_{i})$ therefore do allow us to relate the observed variability in average radial optical depths directly to uncertainties in mass-loss rates derived from single observations.

\subsection{Comments on individual stars}\label{sec:ind_disc}

\subsubsection{LMC: VFTS 180}

As may be expected for an O3 supergiant, the \ion{N}{V} and \ion{C}{IV} profiles are saturated. There is a wind-formed \ion{Si}{IV} feature, albeit weak, consistent with the early spectral type. Almost no variation between the two observations is discernible. The very prominent ISM lines have caused difficulty in obtaining accurate SEI model fits in lower velocity regions. The computation of average optical depth is therefore limited to the normalized velocity range $w = 0.4-0.9$, rather than extending to $w = 0.3$. This represents a loss of only one velocity 'bin' and avoids the poor fit in the lower velocity region of the first observation. 

\subsubsection{LMC: Sk -67 166}

The multiple observations of this star are discussed in Section \ref{sec:multi} below.

\subsubsection{LMC: VFTS 586}

The two spectra for this star show distinct differences in optical depth in the higher velocity regions of the stellar wind. What appear to be multiple NAC-type features in the first observation have merged (or otherwise been superseded) in the second observation by a broad, deep absorption trough extending across more than 1,000 km s$^{-1}$. The NAC feature close to the terminal wind velocity remains.

\subsubsection{LMC: Sk -70 60}

The \ion{N}{V} and \ion{C}{IV} features for this emission line star both display very prominent NACs. Some minor variation is observed in the \ion{C}{IV} feature between observations, principally at higher velocities. The \ion{N}{V} feature shows some reduction in the intensity of the absorption element at mid-range velocities from the second to the third observation, despite these being separated by only approximately 3 hours. We have included that additional observation in this instance as an example of very short-term variability being evident along with variability observed over the longer separation of the first 2 observations.

\subsubsection{LMC: VFTS 244}

Very little variation between the two observed \ion{Si}{IV} profiles is evident, although there is some increased optical depth in the second observation at high wind velocities.

\subsubsection{LMC: N11 ELS 48}

Very little variation is observed between the spectra in either of the \ion{N}{V} or \ion{C}{IV} features. This is reflected in there being no significant variation in average optical depths. Visually, there is an outward migration of enhanced optical depth discernible between the two NACs in the \ion{C}{IV} profile, resulting in narrowing of these features near the terminal wind velocity. This is consistent with the expected behaviour of DACs and can be identified in the radial optical depth plots in Figure \ref{fig:11048_tauradcomp}. This change is less evident in the \ion{N}{V} profile.

\subsubsection{LMC: Sk -68 16}

The migration of a region of reduced optical depth close to the terminal velocity of the stellar wind, is observable in the \ion{N}{V} profiles, corresponding with a narrowing of the prominent NAC feature observable near the terminal wind velocity. The cadence of the two observations (being certainly shorter than a single stellar rotation) suggests this is indeed an outward migration of the feature during a single stellar rotation. This is not repeated in the \ion{C}{IV} spectra, but this may be a result of the \ion{C}{IV} absorption approaching saturation toward higher wind velocities. Consistent variation in the extent of average wind optical depth obtained from SEI model fits to the spectra is seen from both the \ion{N}{V} and \ion{Si}{IV} features. The near-saturated \ion{C}{IV} profile understandably shows a much smaller degree of variation. The NAC features, seen so clearly in the \ion{N}{V} profile, are barely discernible in the \ion{C}{IV} profile which, conversely, displays a very flat absorption trough. This may be due to approaching saturation, although the flux within the \ion{C}{IV} absorption does not reach zero. That may indicate the presence of a cooler unseen companion; however, there is no clear evidence of this from radial velocity measurements of the \ion{N}{III} 1183.03, 1184.55 \AA{} and \ion{O}{IV} 1338.61 \AA{} photospheric lines. \citet{Brands2025} similarly found no clear evidence of binarity from their radial velocity measurements. This star would be a worthwhile subject for obtaining further multi-epoch spectra for measuring any radial velocity changes. 

\subsubsection{PGMW 1363}

The two spectra of this late O type supergiant show a distinct variation in the upper portion of the blue edge of the absorption profiles for each of the \ion{Si}{IV} and \ion{C}{IV} features. These profiles, although saturated, are shown in Figure \ref{fig:1363_compare} to illustrate this. The change is not observable in the \ion{N}{V} feature probably due to contamination by the red wing of the Lyman-$\alpha$ absorption feature. The observed variation in blue edge profile is similar to that seen in some of the spectra of the LMC star Sk -67 166 discussed in Section \ref{sec:multi} below. The changing blue edge slope suggests varying macro-turbulence in the wind around the terminal wind velocity. This is in contrast to the variability in the location of the entire blue edge (not always accompanied by a change in slope) evident in the spectra of Sk 191, also discussed further in Section \ref{sec:multi}.

Both the \ion{N}{V} profile and the red element of the \ion{Si}{IV} profile show the emergence of a distinct NAC feature in the second observation in contrast to an earlier broad DAC feature (this is not evident in the blue element of the \ion{Si}{IV} doublet due to saturation). The optical depth plots in Figure \ref{fig:1363_tauradcomp} illustrate this change.

\subsubsection{LMC: Sk -70 79}

The two available spectra are separated by 20.5 hours. The absorption profiles are unusually flat across the entire velocity range, but with prominent NAC features close to the terminal wind velocity in the \ion{Si}{IV} absorption profiles. The \ion{C}{IV} absorption is saturated for much of its width. There is evidence of a wind-formed \ion{N}{V} absorption profile, but this is minimally visible and heavily affected by the red wing of the Lyman-$\alpha$ absorption. The very flat \ion{Si}{IV} absorption profile is difficult to fit using SEI techniques adopted here. It may be that a single wind acceleration parameter ($\beta$) does not adequately reflect the characteristics of the stellar wind of this star, although it is unclear why this should be so in this instance. Credible single $\beta$ values do not fully reproduce the emission peaks observed, while larger values for $\beta$ create significant problems with fitting the low velocity portion of the flat absorption profile. For this reason, the optical depth comparison set out in Table \ref{tab:Alldata} for this star uses only the normalized range $w = 0.4-0.9$, rather than extending to $w = 0.3$.

There is a distinct change in the \ion{Si}{IV} absorption profile in the short interval between the two observations. A region of enhanced optical depth seen in the first observation has migrated outward in that time, to be incorporated into the NAC which persists near the terminal wind velocity. The SEI fits for each observation suggest a relatively optically thick stellar wind (despite the \ion{Si}{IV} absorption not being saturated) throughout the full radial velocity range.

This star exhibits strong wind-formed \ion{Si}{IV} and \ion{C}{IV} profiles and the most suitable \texttt{TLUSTY} photospheric profile also indicates a significantly lower surface gravity than previous literature has suggested. The recent reclassification of this star by \citet{Bestenlehner2025} as a supergiant (rather than a giant) of spectral type B0.2 Ib is strongly supported by these factors and is adopted here.

\subsubsection{SMC: NGC 346 MPG 355}

The two observations of this star (also known as NGC 346 SSN 9) are separated by approximately 22 years. Minor changes in the \ion{C}{IV} profiles between each observation are observed. The small scale of the changes may be due to a lack of significant variation in the wind profile over short time scales, or it may be that the observations are, by chance, of similar features within the stellar wind. Based on the findings of \citet{Massa2024}, we may expect little variation on rotational time scales in the wind of a very hot, early O-type star. Our observations are consistent with this. Measurements based on the prominent NAC feature of the \ion{C}{IV} profile and the blueward extent of saturation of the \ion{N}{V} profile (see Fig. \ref{fig:355_compare}) both suggest a terminal wind velocity figure of 2,900 km s$^{-1}$, lower than the figure derived by \citet{Hawcroft2024b}, but higher than that derived by \citet{Rickard2022}.

The variation in the high velocity portion of the \ion{C}{IV} absorption profiles is not replicated in \ion{N}{V} due to saturation of that portion of those absorption profiles. \citet{Rickard2022} considered these two observations, together with a third observation made approximately one hour after the second (but here combined into a single HASP spectrum). Those two successive observations suggest the outward migration of the red edge of the NAC feature over a timescale of hours (see Figure 7 of that work).

\subsubsection{SMC: AzV 388}\label{sec:388disc}

The \ion{C}{IV} feature exhibits very prominent NACs which are significantly deeper and wider in the second observation. This behaviour is also observed in the \ion{N}{V} absorption profile of both observations. The cadence of the two observations is far longer than the rotational period of the star, so little else can be observed regarding the development of these NACs. Observations to establish whether this behaviour is observed on rotational time scales would be highly instructive for greater understanding of how migrating DAC features asymptotically merge into NACs at low metallicity in a manner similar to that already observed in Galactic O stars \citep[see, for example:][among others]{Henrichs1988, Prinja1992, Kaper1994, Kaper1996}.

The changing width of the NACs, as well-illustrated in these spectra, presents a challenge to existing methods of determining mass-loss rates based upon fitting to the overall shape of the resonance line absorption features. Such fitting, in the absence of using the SEI-based methods applied here will not probe the NAC features in any detail. As shown by \citet{Prinja2013}, accurately reflecting NAC optical depths is of great importance in estimating true mass-loss rates. The prominence of these NACs highlight this need. The fact that these features also change significantly in width will have a substantial effect on actual mass-loss rates since the integrated optical depth of the whole absorption feature is increased as the NAC becomes wider, despite the absorption `continuum' remaining unchanged. This is highlighted by the comparative optical depth plots in Figure \ref{fig:388_tauradcomp}.

The dramatic effect of the prominent and varying NACs in this instance can be seen by comparing the results for average optical depth and $\sigma(\tau)/\tau$ set out in Table \ref{tab:Alldata} with the same results if we include in our calculation the full extent of the NAC feature out to $w = 1.0$. As expected, this produces increased average optical depth measurements of 0.85 and 1.24 for Observations 1 and 2, respectively. All other things being equal, this alone would produce a significantly increased estimate of mass-loss rate, in relation to which, the observation by \citet{Prinja2013} that it is the NAC features from which the best estimates of true mass-loss rates are derived, may be recalled. Importantly for the conclusions here, the inherent uncertainty lower bound increases from 0.06 to 0.26. The NAC features here are unusually prominent and variable, as noted, however this observation suggests that in some cases, even the hottest O-type stars may exhibit stellar wind structure which significantly varies apparent mass-loss rates based on single observations.

\subsubsection{SMC: AzV 75}

The multiple observations of this star are discussed in Section \ref{sec:multi} below.

\subsubsection{SMC: NGC 346 MPG 368}

Observations of this star (also known as NGC 346 SSN 15), provide two UV spectra separated by approximately 22 years. This mid O-type main sequence star exhibits a discernible variation in the radial optical depths of its stellar wind, as traced by the observed \ion{C}{IV} profiles. The principal variation is the narrowing of the NAC features in the second observation although the remainder of the profile is largely unchanged. This narrowing results in a significant increase in the overall average optical depth of the \ion{C}{IV} absorption feature in the second observation. A similar change is seen between the two \ion{N}{V} profiles, despite the incompleteness of the second observation.

The NAC features in both of the \ion{N}{V} and \ion{C}{IV} profiles indicate a terminal wind velocity of 2,100 km s$^{-1}$, consistent with the model-derived result of \citet{Rickard2022}.

It should be noted that observations of this star's optical spectrum considered by \citet{Rickard2024} indicate that this is actually a binary system. That does not appear to affect the relevant portions of the UV spectrum in a manner that significantly alters the conclusion here that the observed variation is the result of variability in the stellar wind of the primary star.

\subsubsection{SMC: AzV 238}

Little change is observed in the \ion{C}{IV} profile between two observations separated by approximately 3 rotations of the star. It is of course possible that the observations happen to coincide with an integer number of rotations such that the same features are presented to the observer in each. The principal discernible change is the steepness of the blue edge of the unsaturated absorption feature, likely indicating varying levels of macro-turbulence in the stellar wind \citep[e.g.][]{Prinja1992}. Only a very minor change in the average optical depth of the absorption feature is observed. Based on NAC locations, we find $\varv_{\infty}$ = 1,250 km s$^{-1}$, slightly higher than, but within uncertainty bounds of, the result in \citet{Puls1996}, itself stated there as a likely underestimate.

Examining UV photospheric lines provides evidence of radial velocity variations, supporting the conclusion that AzV 238 is indeed a binary system \citep[][]{Dufton2019, Sana2024, Shenar2024}, although, consistent with its SB1 designation, binarity does not appear to affect the limited variation in the optical depth of the \ion{C}{IV} absorption in these two observations.  \citet{Sana2024} also measure slightly larger radial velocities from more recent optical data, compared to the UV-derived result provided here in Table \ref{tab:Alldata}.

\subsubsection{SMC: AzV 70}

Each of the \ion{N}{V}, \ion{Si}{IV} and \ion{C}{IV} wind-formed profiles are well-developed and unsaturated. The absorption profile for \ion{Si}{IV} is relatively shallow compared to the others, as expected for a star of this spectral type (O9.5 Iab). All of these features show distinct variations in profile between the two observations, with significantly increased optical depth in the stellar wind, particularly at higher velocities, apparent in the second observation. The separation of the two observations is of an order commensurate with the rotation of the star, indicating significant variation in the line of sight wind structure over the star's rotational time scale. We observe a lower limit of 13-40\% in the proportional variation in average radial optical depths of the absorption features for the species mentioned above for this late O-type supergiant. The greatest variation is observed, for each species, in the higher velocity portions of the stellar wind, approaching the NAC features, which are very prominent in each case.

The \ion{C}{IV} absorption profile is very flat-bottomed. \citet{Backs2024} suggested that this, together with a broadened \ion{He}{I} profile may be evidence of the presence of a cooler companion. Considering the individual UV spectra suggests that this alone is unlikely to be the explanation for the \ion{C}{IV} absorption profile shape: the profile remains flat-bottomed in the second observation, but with a higher `floor' and both observations clearly show the presence of prominent NACs, indicating that the flat-bottomed \ion{C}{IV} absorption is not the result of a saturated profile being `infilled' by a companion's flux (see Figure \ref{fig:7060_c4_SEI}). AzV 70 is in the Binarity at Low Metallicity (BLOeM) data set \citep[][]{Bodensteiner2025}, but observations to date in that programme have not identified it as a binary system \citep[][]{Bestenlehner2025a}.

\subsubsection{SMC: AzV 506}

Little variation is observed between the two \ion{C}{IV} profiles illustrated. The cadence of the observations is such that we may, coincidentally, be observing the star at similar points in its rotation on each occasion. The other ions do not show well-developed wind profiles. The fitted profiles obtained in this instance do, however, show variation. The very small optical depth of the \ion{C}{IV} feature and significant contamination by interstellar absorption lines of the mid-velocity portion of the stellar wind combine to decrease the precision of results in this instance.

\subsubsection{SMC: AzV 96}

Variability in the \ion{Si}{IV} profile was shown and discussed in \citet{Parsons2024}. Fig. \ref{fig:96_compare} also illustrates the similar variation shown in the \ion{C}{IV} profile. The extent of variation is smaller than in the \ion{Si}{IV} profile. This appears to be a consequence of the deep \ion{C}{IV} absorption approaching saturation. In both cases however, the second observation is characterized by a region of significantly increased optical depth in the stellar wind at mid-range velocities. The cadence of these observations is a significant proportion of, but less than, a single rotation of the star, even if the projection angle is small, indicating a significant degree of wind line variability on rotationally-modulated timescales.

\subsubsection{SMC: Sk 191}

The multiple observations of this star are discussed in Section \ref{sec:multi} below.

\subsubsection{SMC: NGC 330 ELS 4}

Although the \ion{Si}{IV} absorption profile shows significant variation in optical depth of the stellar wind across a wide range of velocities, particularly in the blue element of that doublet, this is not replicated to the same degree in the \ion{C}{IV} profiles. The \ion{C}{IV} absorption profiles are much shallower than those of \ion{Si}{IV}, some variation is present, but is less evident.

The low terminal velocity for this star's stellar wind allows us to treat the two elements of the \ion{Si}{IV} doublet as effectively decoupled \citep[see][]{Prinja2010, Parsons2024}. We may therefore use the technique illustrated in those works to measure any variation over time of the apparent ratio of the radial optical depths required to produce a model fit to each doublet absorption feature. A significant variation over time from the canonical ratio of the oscillator strengths of the two doublet elements may be taken to indicate a change in the prevalence of optically thick large scale structure in the stellar wind \citep[see also][]{Ganguly1999, Massa2008}. Across the four available observations of this star, we do indeed see such a variation in the value of this ratio, namely: 1.85 and 1.76 in the observations separated by 11 months in 2001 and 2002 and 1.36 and 1.42 in the observations from 2009, separated by 5 days. The substantial variation between each of the two pairs of observations separated by approximately 7-8 years suggests an increase in the degree of optically thick structure in the line of sight stellar wind over that time scale. This may represent a steady increase over that time period or may be the cumulative result of shorter-term increases and decreases. The mean of these values, 1.60 $\pm{0.14}$ (at the 1$\sigma$ level), accords with the result derived by \citet{Parsons2024} using the single ULLYSES HLSP spectrum (1.60 $\pm{0.30}$).

The passage of optically thick structure through the stellar wind may be observed in Figure \ref{fig:3304_tauradcomp}. The movement of these features is especially clear in the comparisons over time of radial optical depths for each element of the \ion{Si}{IV} resonance line doublet.

\subsubsection{SMC: NGC 330 ELS 2}

There is a significant increase in optical depth of the wind across most of the velocity range outside the NAC feature and the region dominated by interstellar absorption lines. This change appears consistently in both the \ion{Si}{IV} and \ion{C}{IV} absorption profiles. The extent of variation in average optical depth observed here is consistent with that observed by \citet{Massa2024} for some of the Galactic early B supergiant stars in their sample. In this example, we observe proportionate variations in optical depths of 37\% and 53\% for each ion observed, at the 1$\sigma$ level.

We can again treat the two elements of the \ion{Si}{IV} doublet as effectively decoupled due to the low terminal wind velocity. For the second observation of this star, we derive a mean optical depth ratio of 1.78, in accordance with the ULLYSES HLSP-derived value of 1.80 $\pm{0.38}$ determined by \citet{Parsons2024}. The first observation produces a much lower value of 1.40, however the very low optical depth values derived from the SEI fits (0.29 and 0.21) render this single data point of limited significance. 

\subsection{Targets with several multi-epoch spectra or a time series}\label{sec:multi}

Three of the stars in our sample have several, or more, suitable spectra. In the case of Sk -67 166, this represents a short, genuine, time series observation over a period of 9 days. There are 21 observations of AzV 75 considered, comprised of one earlier STIS spectrum and 20 COS spectra, those having been obtained at intervals of a few to several months in most instances. In the case of Sk 191, there are 12 observations, mostly each separated by periods of 2-3 months. We discuss each star's results individually below.

\subsubsection{LMC: Sk -67 166}

\begin{figure}
\begin{center}
 \includegraphics[width=\linewidth]{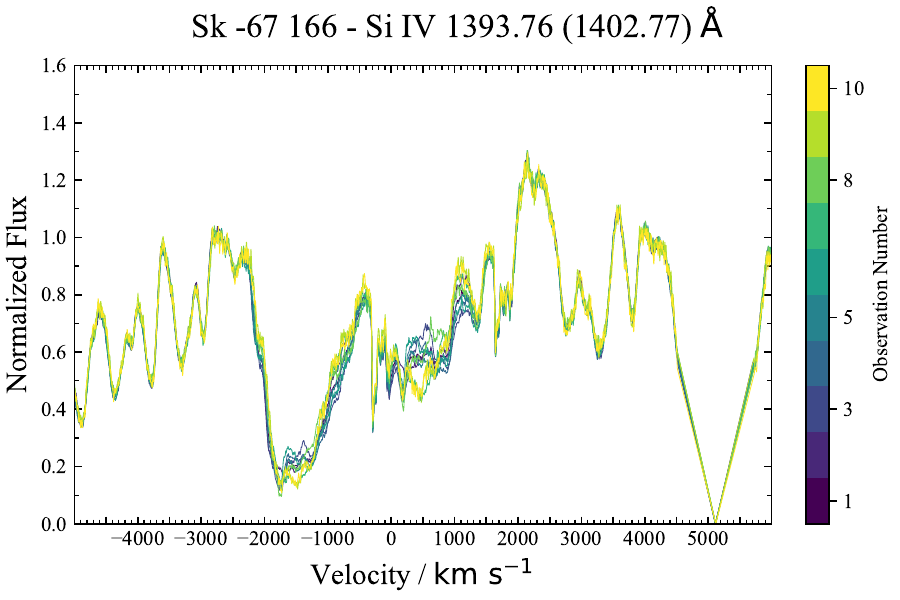} 
 \caption{Comparison of 10 HASP COS UV spectra for the LMC star Sk -67 166, spectral type O4 If, concentrating on the \ion{Si}{IV} resonance line doublet. Plotted in the rest frame of the star (rest wavelength of the blue element of the doublet is at zero velocity). Considerable variation in the overall absorption profile shapes and the simultaneous increases and decreases in optical depth in different regions of the stellar wind can be observed. The average time between each observations is approximately 21 hours.}
 \label{fig:67166_compare_si4}
\end{center}
\end{figure}

This LMC blue supergiant star (spectral type O4 If) provides the sole example of a genuinely high-cadence time series observation of variations in the stellar winds of a massive star at low metallicity. The star was observed using the E140M grating with the STIS instrument on HST for a period of 9 days (HST proposal 16304, PI: J. Chisholm) \citep{Chisholm2020}. This period was designed to cover two successive complete rotation periods of the star. It should be noted that the observing proposal states the rotation period as 4 days, rather than the 9.8 day \textit{maximum} figure in Table \ref{tab:Alldata}.

The observing program obtained 42 separate UV spectra over its 9-day duration with a view to establishing whether observed stochastic oscillations of this blue supergiant correspond to changes in its stellar wind or instead result from internal gravity waves. We have utilized the 10 Hubble Advanced Science Product (HASP) co-added spectra available within the ULLYSES data set to provide sufficient signal-to-noise ratio reliably to observe variations within the absorption profiles of the \ion{Si}{IV} wind profiles. This approach smooths out the details of potential variations on time scales much shorter than one day, but still allows observation of a complete cycle of observations of structure throughout the stellar wind across 1-2 rotations of the star. A more detailed consideration of the movement of structure through the stellar wind on shorter times scales of a few hours forms part of a follow-up work examining time series observations of this star (Parsons, et al., in prep.). Spectra for this star also show a well-developed \ion{C}{IV} wind-formed profile which is, however, saturated and so unsuitable for analysis for the purposes of this work.

The observed variations of the well-developed, but unsaturated \ion{Si}{IV} resonance line doublet feature make this an ideal subject for quantification of the effects of rotationally modulated wind variability upon estimates of massive star mass-loss at lower than solar metallicity.

We derive from NAC locations in each of the spectra shown in Fig. \ref{fig:67166_compare} and measurement of \ion{C}{IV} saturated profile widths, a terminal wind velocity of 1,900 km s$^{-1}$, which is higher than in some previous litersture.

Examination of the overplotted spectra of the \ion{Si}{IV} resonance line doublet shows significant variation in the intensity of absorption across a range of wind velocities. DAC features are seen in Figure \ref{fig:67166_tauradcomp} to move through the wind over a timescale commensurate with the apparent rotation of the star. This is consistent with observations at Galactic metallicities \citep[see, for example][]{Massa1995, Kaper1996, Prinja2002, Massa2015} and provides strong evidence that similar behaviour is observed at LMC metallicity. This behaviour will be examined in greater detail in a further work focused on the full set of FUSE, COS and STIS data available for this star (Parsons et al., in prep.).

The variation in average optical depth (see Table \ref{tab:Alldata}) derived from fitting the available observations of the \ion{Si}{IV} absorption profile appears small in comparison to the obvious physical changes in that profile. This is at least partly explained by the fact that simultaneous increases and decreases in the optical depth at different radial velocities, while showing clear evidence of wind structure, are cancelling out changes in the overall average optical depth across the whole stellar wind. This can be observed in Figure \ref{fig:67166_compare_si4}. This in itself provides further evidence that the rapid migration of regions of enhanced optical depth outward through the stellar wind seen in many Milky Way stars is also observed at lower metallicity in the LMC.

\subsubsection{SMC: AzV 75}\label{sec:av75}

\begin{figure}
\begin{center}
 \includegraphics[width=\linewidth]{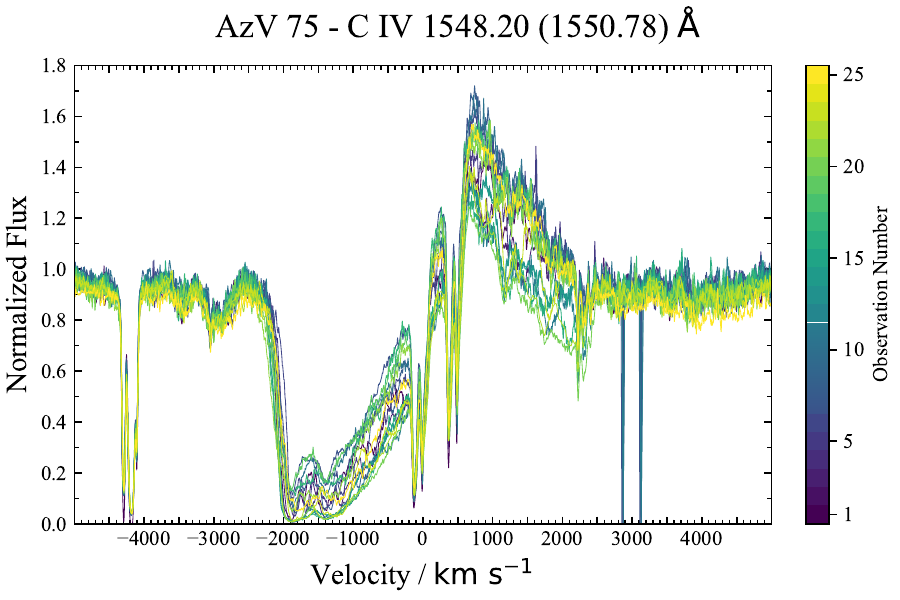} 
 \caption{Comparison of one STIS and 24 COS UV spectra for the SMC star AzV 75, spectral type O3.5 III(f) highlighting the \ion{C}{IV} resonance line doublet. Plotted in the rest frame of the star (rest wavelength of the blue element of the doublet is at zero velocity). The considerable variation in the blue edge of the absorption feature, in the shape of the absorption and, most unusually, in the emission features can be clearly observed. Note that not all observations extend to this portion of the spectrum, so the applicable colours are skipped in the sequence to maintain consistency and comparability with other plots.}
 \label{fig:75_compare_select}
\end{center}
\end{figure}

\begin{figure}
\begin{center}
 \subfloat[ ]{\includegraphics[width=\linewidth]{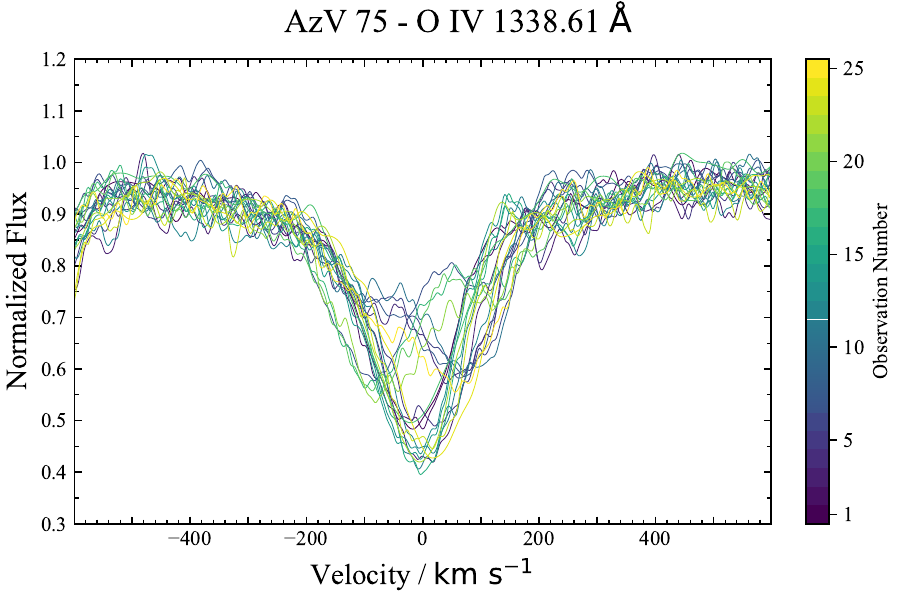}}
 
 \subfloat[ ]{\includegraphics[width=\linewidth]{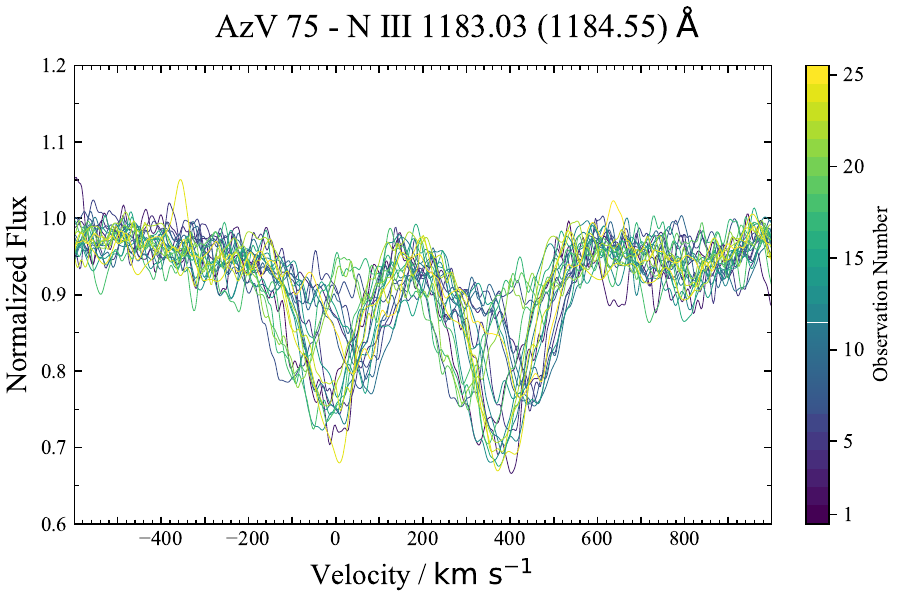}}
 \caption{Time-ordered comparisons of multiple UV spectra of (a) the excited \ion{O}{IV} 1338.61 \AA{} and (b) the excited \ion{N}{III} 1183.03, 1184.55 \AA{} photospheric lines observed in AzV 75. The colour sequence of the observations is the same in each plot and matches that in Figure \ref{fig:75_compare_select}, however some observations do not extend to the wavelength of the \ion{N}{III} feature and/or the \ion{O}{IV} feature, so the applicable colours in the sequence are skipped to maintain consistency and comparability.}
 \label{fig:75_o4_n3}
\end{center}
\end{figure}

AzV 75 has been the subject of repeated observations using COS over an extended period between 2012 and the present day, as well as an earlier STIS observation in 1999. The COS observations form an ongoing series of regular observations for instrument calibration purposes. The ULLYSES HLSP spectrum is formed from a number of these observations, however additional spectra, obtained as part of continuing calibration checks, have continued to be added to the MAST archive since the final ULLYSES data release, the most recent at the time of writing being dated 2025 July 11. We have plotted in Figure \ref{fig:75_compare} the single STIS spectrum and 24 COS spectra, all being flux-calibrated data products from the available HASP suite of spectra. An enlarged image of the \ion{C}{IV} feature for these observations is shown in Figure \ref{fig:75_compare_select} to highlight the extreme variability to be observed. Most of the spectra show very clear and deep NACs, although in some, these NACs are at, or very close to, saturation. There is also a prominent \ion{N}{V} feature, which shows a lesser degree of variation in the absorption profiles between observations, but consistently shows prominent NACs.

A highly unusual and clearly observable feature of note is the emission profile of the \ion{N}{V} feature and, to a greater extent, the \ion{C}{IV} feature (see Figure \ref{fig:75_compare}). Both sets of emission profiles show distinctly different strengths in different observations (refer also to the \ion{C}{IV} profile plots in Figure \ref{fig:75_compare_select}). This emission variation far exceeds any variation in the continuum level, by nearly an order of magnitude, as noted by \citet{Patel2024}. The spectra showing enhanced emission also exhibit shallower \ion{C}{IV} and, to a lesser extent, \ion{N}{V} absorption profiles. The reduced emission in some \ion{C}{IV} spectra also appears to uncover additional absorption features to the red side of the emission peak.

Since the emission profile is produced by scattered and retransmitted photons from the entire stellar atmosphere, we would not expect to observe such a variation from a single star occurring repeatedly over such a relatively short timescale. AzV 75 was excluded from the recent study by \citet{Pedersen2025} of stellar stochastic low-frequency oscillations on the basis that the data used for that work, from the Transiting Exoplanet Survey Satellite (TESS) \citep{Ricker2014}, showed that AzV 75 is an eclipsing binary system. AzV 75 had not been identified as a binary system in previous literature, including the recent comprehensive `pipeline' results for ULLYSES stars presented by \citet{Bestenlehner2025}. Reference to the relevant TESS data, and additional available photometric data, confirms the eclipsing binary nature of this system first identified by \citet{Pedersen2025}.

Presented with this unusual variability in multiple UV spectra, combined with the apparent lack of optical spectral evidence for binarity in previously published literature, we may nevertheless use the available UV spectra for this purpose. The extensive series of papers in \textit{The Observatory}, commencing with \citet{Stickland1987}, shows the efficacy of using UV spectroscopic data for examining binary systems, particularly where the components are of comparable luminosities or where optical lines are blended. Here, it is largely a case of making best use of available data, but we see that the UV spectra do provide clear results. We show in Figure \ref{fig:75_o4_n3} the excited \ion{O}{IV} 1338.61 \AA{} photospheric feature of AzV 75. The overplotting of 25 observations in a single, consistent rest frame (confirmed by reference to the \ion{C}{II} 1335 \AA{} ISM lines in the spectra), using the systemic radial velocity for AzV 75 given in Table \ref{tab:Alltargets} shows 3 distinct `phases' in the observations. First, a deep absorption feature at the \ion{O}{IV} rest wavelength. Secondly, we observe in several of the spectra a much broader profile, which has in each case a deep and a shallow absorption component. In some of these, we see deep absorption blue-shifted by up to approximately 100 km s$^{-1}$ and shallow absorption red-shifted by up to approximately 90 km s$^{-1}$. Conversely, in the remainder of those spectra, we see an almost mirror image with a blue-shifted shallow absorption component and red-shifted deep absorption. Each of these patterns is consistently repeated in both elements of the excited \ion{N}{III} 1183.03, 1184.55 \AA{} profiles also shown in Figure \ref{fig:75_o4_n3}. These distinctive features are also consistent with the conclusion that AzV 75 is a binary system with the secondary star contributing significantly to the total flux observed. The excited \ion{C}{III} complex around 1176 \AA{} (not illustrated here) also shows evidence of blue and red shifts consistent with the photospheric lines described.

Binarity, of course, complicates the measurement of stellar wind optical depth variations. Potential wind interactions may significantly affect stellar winds. The available observations do, however, show that the observed profiles fall broadly into 2 categories; those where the possible component stars appear to be observed at a phase at or approaching quadrature (on either side of conjunction) and those closer to stellar conjunction, where radial velocity differences are smaller. The former (where we observe the 2-component red- and blue-shifted \ion{O}{IV} absorption) correspond to \ion{N}{V} and \ion{C}{IV} profiles showing higher emission peaks and shallower absorption troughs. Those observations showing a single, deep \ion{O}{IV} absorption at the rest wavelength, correspond to \ion{N}{V} and \ion{C}{IV} profiles with lower emission peaks and deeper absorption troughs. Figure \ref{fig:75_compare} also shows that no significant wind-formed \ion{Si}{IV} feature is present. This is to be expected from an early O-type star. The lack of this feature suggests also that the secondary star is of a spectral type which does not exhibit a strong wind-formed \ion{Si}{IV} profile, but is nevertheless sufficiently hot to show prominent excited \ion{O}{IV} photospheric lines, suggesting that the secondary star may be a mid-O giant or, possibly, a mid/late-O dwarf.

The blue edge of the \ion{C}{IV} absorption and, to a lesser degree, the Lyman-$\alpha$ affected \ion{N}{V} absorption blue edge show considerable variation in slope and apparent terminal wind velocity. The higher terminal wind velocities appear to coincide broadly with those observations at or close to primary eclipse/stellar conjunction, with lower terminal velocities corresponding to observations made at, or close to, quadrature, either side primary eclipse/conjunction. Comparison of the dates of the UV observations examined here with the dates of apparent primary eclipses seen in TESS and other optical data confirm this. The apparent variations in $\varv_{\infty}$ therefore appear to be explained by the radial velocity variations of the primary star. Detailed consideration of this, along with derivation of orbital parameters forms the subject of a separate work (Parsons \& Pauli, in prep.).

Table \ref{tab:Alldata} quantifies the variation in observed average optical depths, suggesting a proportionate uncertainty in any single measure of at least 33\%. Indeed, the fact that there are other spectra containing saturated \ion{C}{IV} profiles strongly indicates a degree of variation even larger than this measure of uncertainty itself indicates. Further consideration of the effect of binarity and possible wind interaction is merited and is considered below and in the above-mentioned forthcoming work.

Almost all of the spectra of AzV 75 plotted in Figure \ref{fig:75_compare} show distinct NAC profiles in their \ion{N}{V} and \ion{C}{IV} features. These can be observed to change depth and width on the timescales of these observations. Further observation of AzV 75 is certainly warranted.

\subsubsection{SMC: Sk 191}

\begin{figure}
\begin{center}
 \subfloat[]{\includegraphics[width=\linewidth]{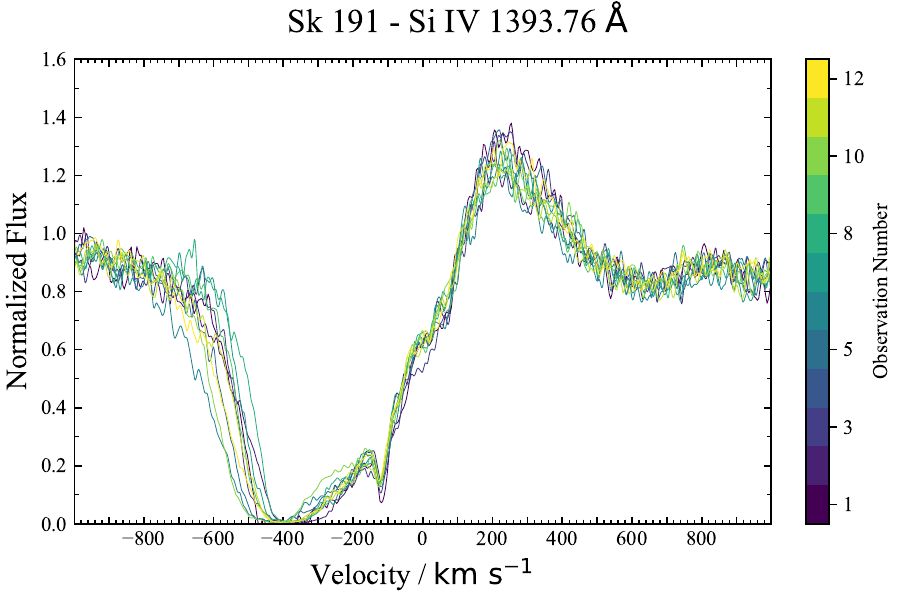}}
 
 \subfloat[]{\includegraphics[width=\linewidth]{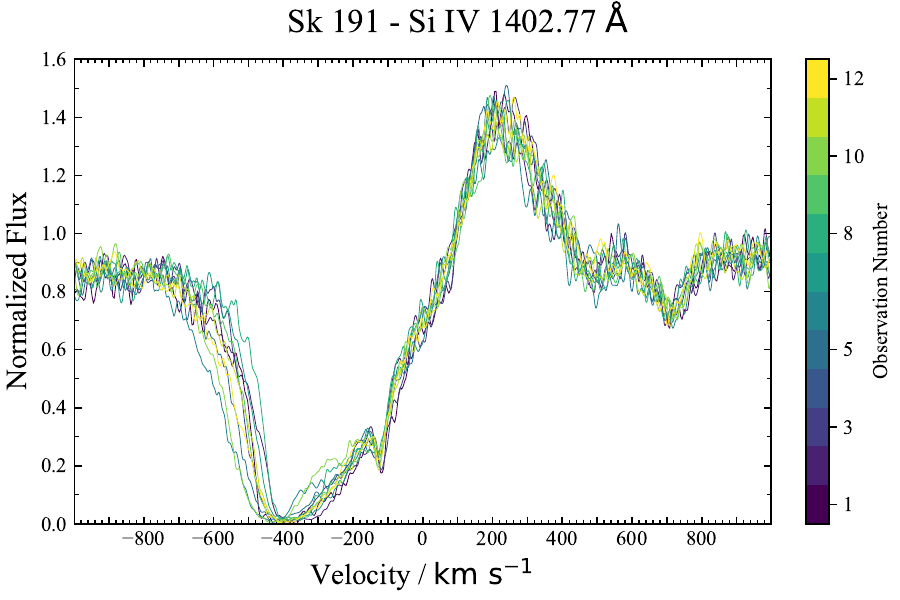}}
 \caption{Comparison of 11 observed spectra of the SMC star Sk 191, spectral type B1.5 Ia highlighting (a) the blue element and (b) the red element of the \ion{Si}{IV} resonance line doublet. Plotted in the rest frame of the star. The considerable variation in the blue edge of the absorption feature, and in the presence or otherwise of a deep absorption feature of changing width, visible in some of the unsaturated profiles for the red element of the doublet, is noticeable over time scales of 2-3 months between observations made between 2009 and 2012 (the first observation was made several years earlier, dated 2001 October 20). The time-ordered colour sequence is the same for each plot, as shown in the colour bars.}.
 \label{fig:191_si4blue_red_compare}
\end{center}
\end{figure}

\begin{figure}
\begin{center}
 \includegraphics[width=\linewidth]{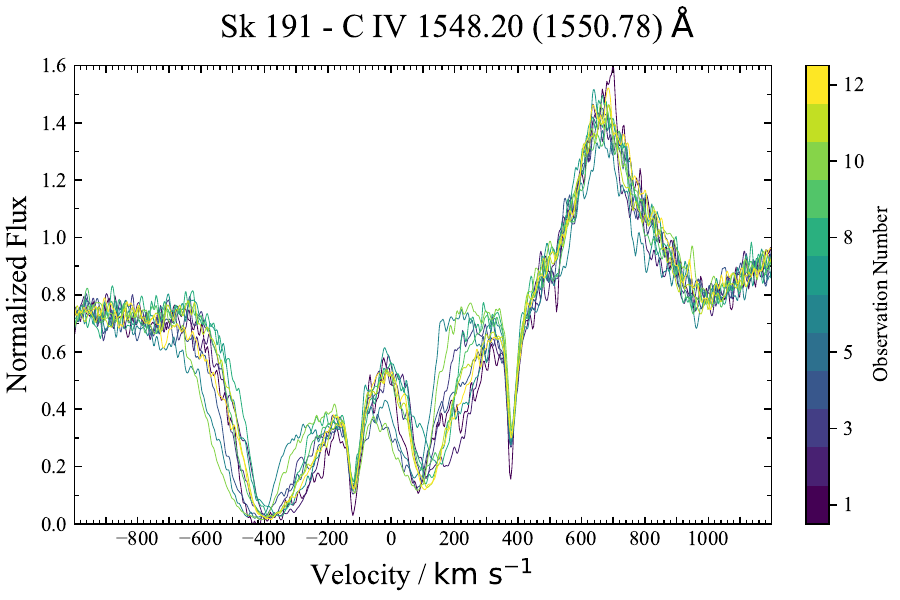} 
 \caption{Comparison of 11 spectra for the SMC star Sk 191, highlighting the detail of variations observable across the full range of velocities within the \ion{C}{IV} resonance line profile. Plotted in the rest frame of the star (rest wavelength of the blue element of the doublet is at zero velocity). The considerable variation in the blue edge of the absorption feature, and in the shape of the absorption features can be clearly observed. The colour sequence of plotted observations is the same as in Figure \ref{fig:191_si4blue_red_compare}).}
 \label{fig:191_c4_compare}
\end{center}
\end{figure}

Sk 191 was observed on multiple occasions with the G130M and G160M gratings using the COS instrument on HST as part of an internal/external wavelength monitoring exercise for that instrument. The programs (HST proposals 11997, 12425 and later, PI: C. Oliveira) \citep[][and others]{Oliveira2009, Oliveira2010} consisted of determining the centroids of several absorption lines spread throughout the COS spectrum and then comparing them to the centroids derived for the same lines in the STIS data and tracking how the centroids of COS lines changed over different visits. The result is a library of high-resolution spectra separated at approximately 3 month intervals which, while not constituting a true time series on rotational time scales, provides multi-epoch observations of well-developed \ion{Si}{IV} and \ion{C}{IV} wind lines for this SMC star of spectral type B1.5 Ia. The \ion{Si}{IV} feature is saturated and therefore of limited use for this work. That for \ion{C}{IV} is however unsaturated in some observations, thereby providing an opportunity to explore variability in the wind line absorption profiles at SMC metallicity over an extended period.

All of the wind-formed lines visible in the UV spectra of this star obtained as part of these calibration programmes show significant variability in the blue edges of their absorption features. The emission features remain consistent across all observations for each ion observed. The \ion{C}{IV} 1548.20, 1550.78 \AA~profile in particular displays prominent NACs close to the terminal wind velocity in both elements of this doublet. Figure \ref{fig:191_si4blue_red_compare} shows the considerable variation to be seen in the entire blue edge of each element of the \ion{Si}{IV} absorption feature (saturated in some observations). In those observations where the \ion{Si}{IV} absorption is not saturated, the visible NAC features in their entirety appear shifted to larger or smaller velocities. In some instances, however, the width of the NACs themselves varies.

The observations of the high velocity portions of the stellar wind of Sk 191 are similar in many respects to those observed by \citet{Prinja1992} in relation to the extensively studied Galactic O supergiant $\zeta$ Puppis insofar as the blue edge of the wind-formed absorption profile (of both of the \ion{Si}{IV} and \ion{C}{IV} resonance lines in this instance) appears to shift significantly. Here, however, many of the observed blue edge variations in the wind-formed profiles do not appear to vary significantly in \textit{slope}. Such variation would be expected if it is the result of changing conditions of turbulence at the terminal velocity of the stellar wind. It seems, therefore, that this changing feature cannot only be the result of turbulence and shocks at this location but appears to be a true variation in the terminal velocity of the stellar wind, as traced by the saturated absorption feature \citep[as also noted by][]{Bernini2024}. There is no evidence of binarity presented in current literature nor has a similar degree of apparent $\varv_{\infty}$ variation been identified in another B-type supergiant. The true nature and cause of this variation is worthy of further investigation.

This variable blue edge feature also has the consequence that performing suitable SEI fitting to the observed spectrum using a single value for the terminal velocity of the stellar wind is rendered very difficult. We have utilised the figure of 490 km s$^{-1}$, consistently with the figure derived by \citet{Bernini2024} using the ULLYSES HLSP spectrum on the basis that this corresponds well with the value of the blue edge of the saturated absorption profile in many of those spectra showing saturated profiles ( refer to the blue element of the \ion{Si}{IV} resonance line profiles shown in Figure \ref{fig:191_si4blue_red_compare}). Figures \ref{fig:191_si4blue_red_compare} and \ref{fig:191_c4_compare} both however show that measuring terminal wind velocity from the blueward extent of saturation, $\varv_{\rm black}$, gives results varying between approximately 420 and 490 km s$^{-1}$. Reference to the examples of SEI-derived fitted spectra for Sk 191 in Figure \ref{fig:191_c4_SEI} shows that, in some cases, a complete fit to all of the features of the absorption profile is not possible. The changing profiles suggest not only genuine variations in $\varv_{\infty}$, but also changing levels of turbulence near that velocity over relatively short time scales. This perhaps suggests non-isotropic conditions in the ISM surrounding the star.

\section{Conclusions}

\subsection{Variability and wind structure}
As has been demonstrated in earlier work, we can observe variability in optical depths of stellar winds at low metallicities. Importantly, we show here that just two `snapshot' spectra of sufficient resolution are enough to provide an approximate lower bound on the extent of variability. If additional spectra are obtained, a more precise approximation to the actual extent of variability can be obtained. Observations at cadences less than the rotational period of the star will help to provide a dynamic description of the motion of structure through stellar winds.

We confirm the near-ubiquity of large scale, optically thick structure in the winds of stars at low metallicities and provide lower limits on the extent of variability of optical thickness of those winds. This permits us to provide credible lower bounds on the inherent uncertainty of any single measurement of average optical depth derived from a single observation, as set out in the ninth column of Table \ref{tab:Alldata} and illustrated in Figure \ref{fig:massaplot}. Those lower bounds are consistent with the findings of \citet{Massa2024} at Galactic metallicities and show a similar inverse relationship between inherent uncertainty and stellar effective temperature. This, in turn, provides similar information regarding uncertainties in mass-loss rates.

We can conclude, even from this limited sample, that the extent of variability in the optical depths of stellar winds over time scales commensurate with the rotation of stars at lower (LMC, 0.5 $Z_\odot$ and SMC, 0.14 - 0.2 $Z_\odot$) metallicities is similar to, and may indeed be greater than, that observed at Galactic metallicity. We may further conclude that the warning provided by \citet{Massa2024} regarding the intrinsic error in deriving mass-loss rates, or testing mass-loss recipes, from any single observation of a massive star applies with at least equal force when considering stars in low metallicity environments.

The limited data available do not permit definite conclusions regarding a possible correlation between metallicity and the extent of wind variability, but it appears that the phenomenon of variability arising from the presence of large-scale stellar wind structure is not significantly reduced at lower metallicities. We do show (see Figure \ref{fig:massaplot}) that at low metallicities, the inverse relationship between effective temperature and the extent of wind optical depth variability observed by \citet{Massa2024} for massive stars in the Milky Way is also likely to apply to these stars in the lower metallicity environments of the LMC and SMC.

It is therefore evident that studies of stellar wind properties and derivations of mass-loss rates, particularly for late O and early B type stars need to keep in mind the inherent uncertainties associated with the results. Without multi-epoch or time series UV spectral data, determinations of stellar mass-loss rates at low metallicities, especially for these cooler and more numerous massive stars, currently based almost exclusively on single epoch spectra, remain subject to an uncertain degree of variability.

The results demonstrated here highlight the desirability of obtaining genuine time series UV spectral observations of some exemplar stars of differing spectral types at SMC or even lower metallicities covering approximately two or more complete rotations of those stars. Such observations will constrain to a statistically significant degree that intrinsic error in such environments and, in addition to better constraining mass-loss models and formulae, will contribute significantly to the ongoing research into the precise origins of rotationally modulated stellar wind structure. Genuine time series data, even for representative examples of 4 or 5 different spectral types of SMC O and early B stars will provide detailed information about the behaviour of large scale wind structure at low metallicity and provide greater ability to interpret mass-loss results obtained from existing single or multi-epoch data.

Consistently with earlier studies, we also observe that NAC features are nearly universal in wind-formed profiles in the UV spectra of massive stars even at low metallicites. We see in the wind of some stars very prominent NACs that also vary in width and depth on relatively short time scales. These changes are most evident in the spectra of AzV 388 and may be seen despite the absence of any obvious change to the shape or depth of absorption profiles in other regions of the stellar wind. As discussed in Subsection \ref{sec:388disc}, including the full extent of those features significantly increases not only the calculated average radial optical depth results (and the mass-loss rate conclusions to be drawn from those), but significantly increase the inherent uncertainty of that result derived from any single observation. In the case of AzV 388, we have shown that fully accounting for NAC profile changes increases the observed proportionate extent of variability in average radial optical depth by a factor of 5, therefore affecting mass-loss rates for the star to a significant extent. Such changes are unlikely to be accounted for in existing mass-loss models based upon broadly fitting the overall shape of those absorption profiles without necessarily probing the detailed shape and depth of NACs. As explained by \citet{Prinja2013} modelling and accounting for NACs is particularly important in this context and is essential for both accurate and precise determination of mass-loss rates. AzV 388 highlights that, even in the case of a very hot early O-type star, we may observe significant variability in optical depth profiles.

\subsection{Further evidence of binarity: AzV 75}
We also present further evidence for the binary nature of the SMC star, AzV 75, in particular, multi-epoch UV spectroscopic data, which provide additional information about this system. The extreme variability in the emission profiles of P Cygni features in the UV spectra presented here, as well as the variability in photospheric absorption lines such as the excited \ion{O}{IV} and \ion{N}{III} lines all point clearly to the same conclusion of binarity. The eclipsing nature of this system, as first disclosed by \citet{Pedersen2025} is indeed evidenced by TESS light curve data and is further demonstrated by other optical photometric data (including from additional recent TESS observations). This finding may also partly explain earlier differences in the spectral classification of AzV 75: O3.5 III(f) \citep{Bestenlehner2025}; O5 III(f+) \citep{Walborn2000} and O5.5 I(f) \citep{Bouret2021}. Follow-up observations to determine the properties of this interesting system are warranted. Currently available data permit a number of key parameters to be determined.  Initial investigation indicates highly elliptical orbits and a long orbital period. A further work will present more complete details and characterization of this system (Parsons \& Pauli, in prep.).

\section*{Acknowledgements}

Based on observations obtained with the NASA/ESA Hubble Space Telescope, retrieved from the Mikulski Archive for Space Telescopes (MAST) at the Space Telescope Science Institute (STScI). STScI is operated by the Association of Universities for Research in Astronomy, Inc. under NASA contract NAS 5-26555.

This research has made use of the ``Aladin sky atlas'' developed at CDS, Strasbourg Observatory, France \citep{Aladin2000}.

The \texttt{STARLINK} software \citep{Currie2014} is currently supported by the East Asian Observatory.

The authors wish to express their thanks to the anonymous referee for the very helpful comments and recommendations made following review of the draft of this work.

\section*{Data Availability}

The spectral data used in this work may be obtained from the Mikulski Archive for Space Telescopes (MAST) at the Space Telescope Science Institute (STScI), accessible at: \texttt{https://mast.stsci.edu/portal/Mashup/Clients/Mast/\hspace{0pt}Portal.html}
in particular, the ULLYSES data contained therein.

\texttt{DIPSO} spectral analysis software used in this work is available as part of the \texttt{STARLINK} package \citep[][]{Currie2014}, accessible at: \texttt{https://starlink.eao.hawaii.edu/starlink}.

\texttt{TLUSTY} model grids and associated materials are accessible at: \texttt{http://tlusty.oca.eu/}, with thanks to Ivan Hubeny and Thierry Lanz.

The SEI-fitting code and results therefrom will be made available upon reasonable request to the corresponding author of this work.



\bibliographystyle{mnras}
\bibliography{bibliography2} 



\appendix

\section{Spectral Plots for Target Sample}\label{allplots}

The plots presented here overplot available observations (in most cases, two, for each star in the target sample), to demonstrate in a visual manner the variations observed in key resonance line profiles. In each case, the time-ordered colour sequence runs from dark (violet) to light (green-yellow). Time lapse between each observation is indicated in the caption to each plot. Where more than two observations are plotted,the colour cycle and observation cadences are indicated by a colour bar on the relevant plot. The zero velocity point in each plot is the rest wavelength of the blue element of the relevant doublet feature in the star's frame of reference.

   \begin{figure*}
\includegraphics[width=\linewidth]{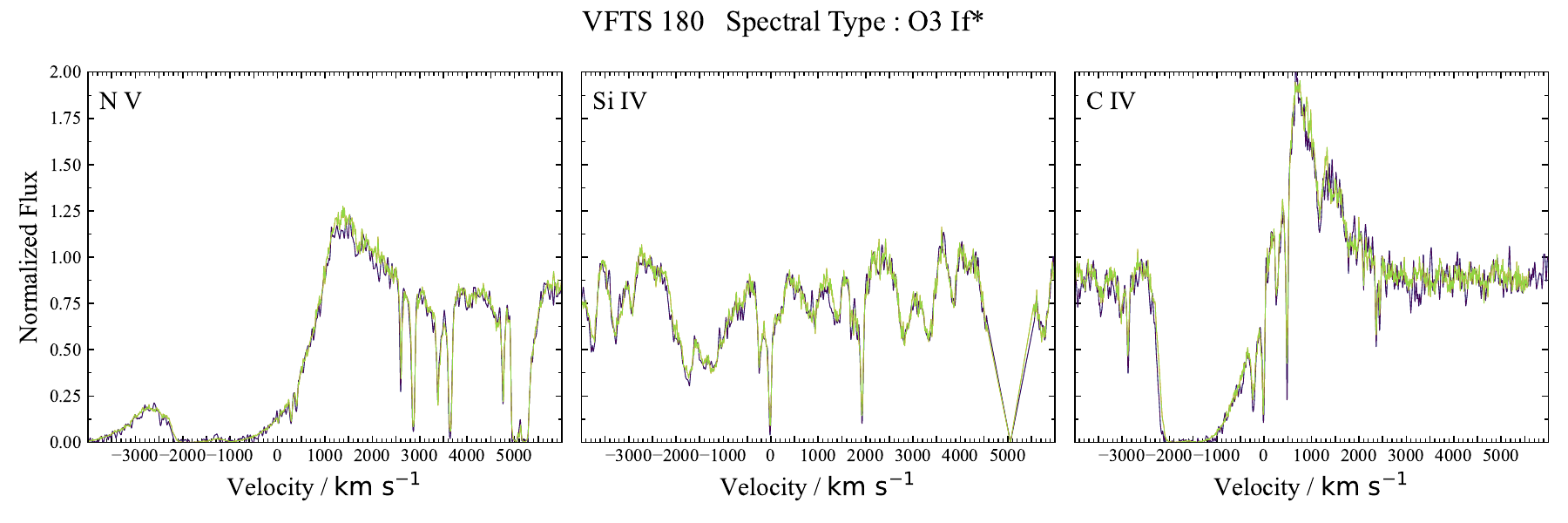} 
\caption{Comparison of two observed spectra, separated by $\Delta(t)= 1264.637$ days, for the LMC star VFTS 180. Plots respectively show each of the \ion{N}{V} 1238.80, 1242.80 \AA{}, \ion{Si}{IV} 1393.76, 1402.77 \AA{} and \ion{C}{IV} 1548.20. 1550.78 \AA{} UV resonance line doublet profiles. Observation 1: dark violet, observation 2: light green. Saturated profiles, here and elsewhere in this Appendix, are included to assist in confirming $\varv_{\infty}$ used in this work.}
\label{fig:180_compare}
    \end{figure*}

   \begin{figure*}
\includegraphics[width=\linewidth]{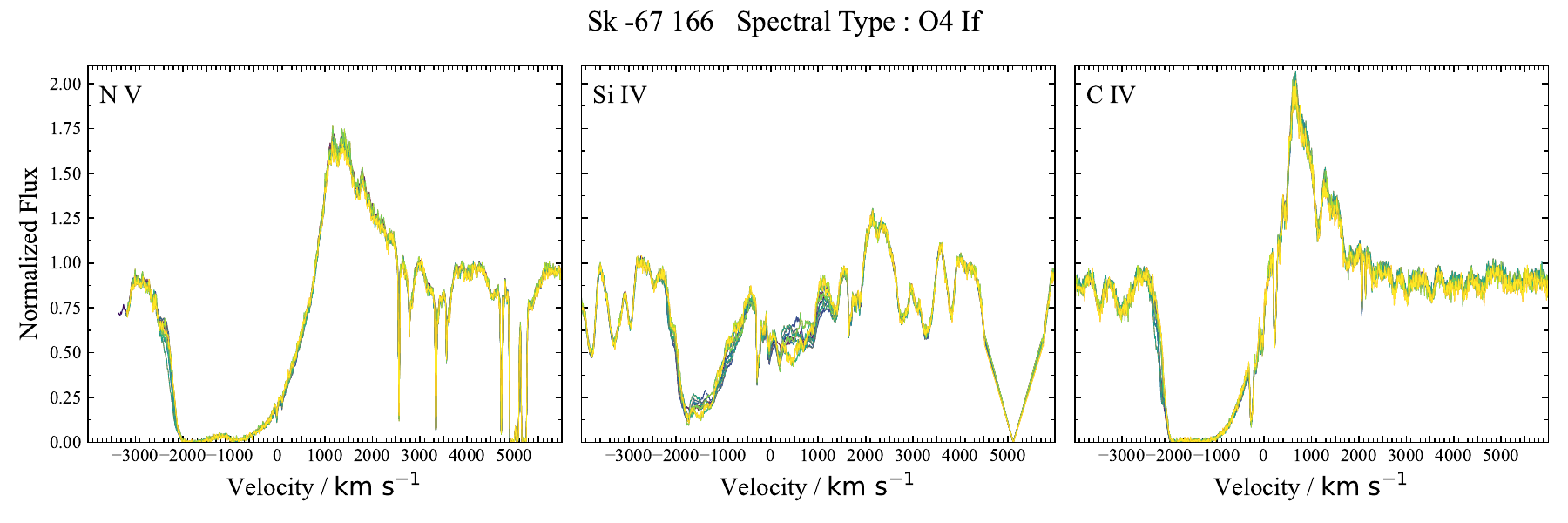} 
\caption{Comparison of 10 observed spectra, separated by an average $\Delta(t)= 0.880$ days, for the LMC star Sk -67 166. Plots respectively show each of the \ion{N}{V} 1238.80, 1242.80 \AA{}, \ion{Si}{IV} 1393.76, 1402.77 \AA{} and \ion{C}{IV} 1548.20. 1550.78 \AA{} UV resonance line doublet profiles. Colour sequence runs from dark to light in time order, as shown in the colour bar in the main text, Figure \ref{fig:67166_compare_si4}.}
 \label{fig:67166_compare}
    \end{figure*}

   \begin{figure*}
\includegraphics[width=\linewidth]{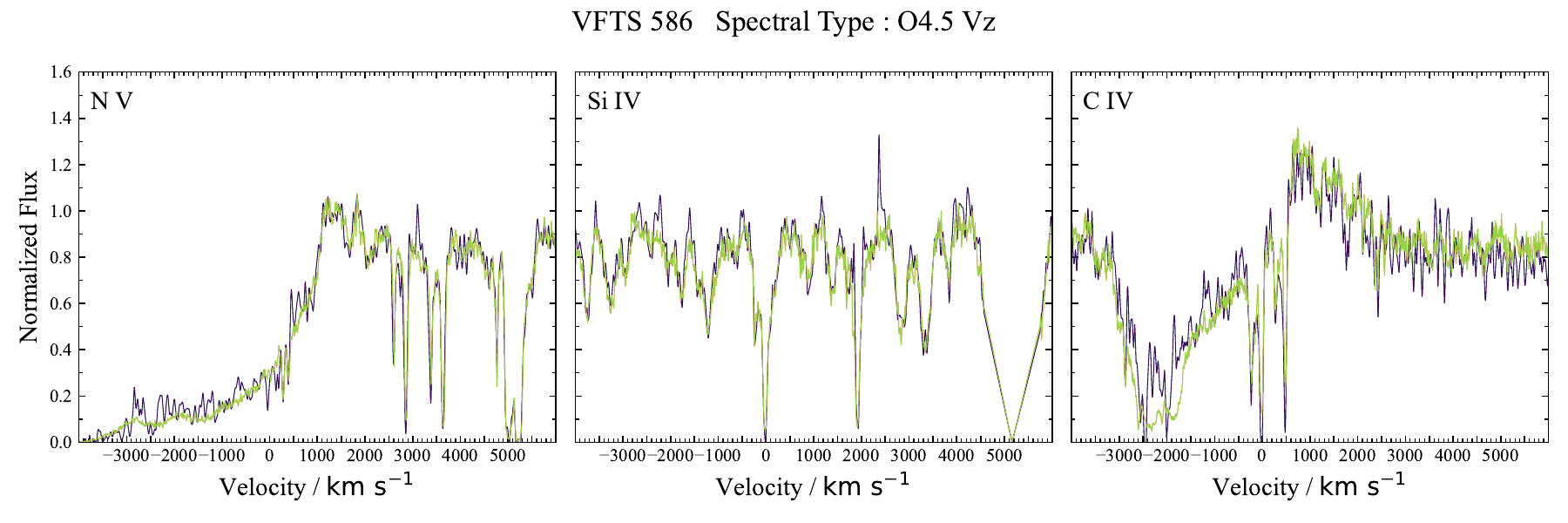} 
\caption{Same as Fig. \ref{fig:180_compare}, for the LMC star VFTS 586 $\Delta(t)= 952.753$ days.}
\label{fig:586_compare}
    \end{figure*}

   \begin{figure*}
\includegraphics[width=\linewidth]{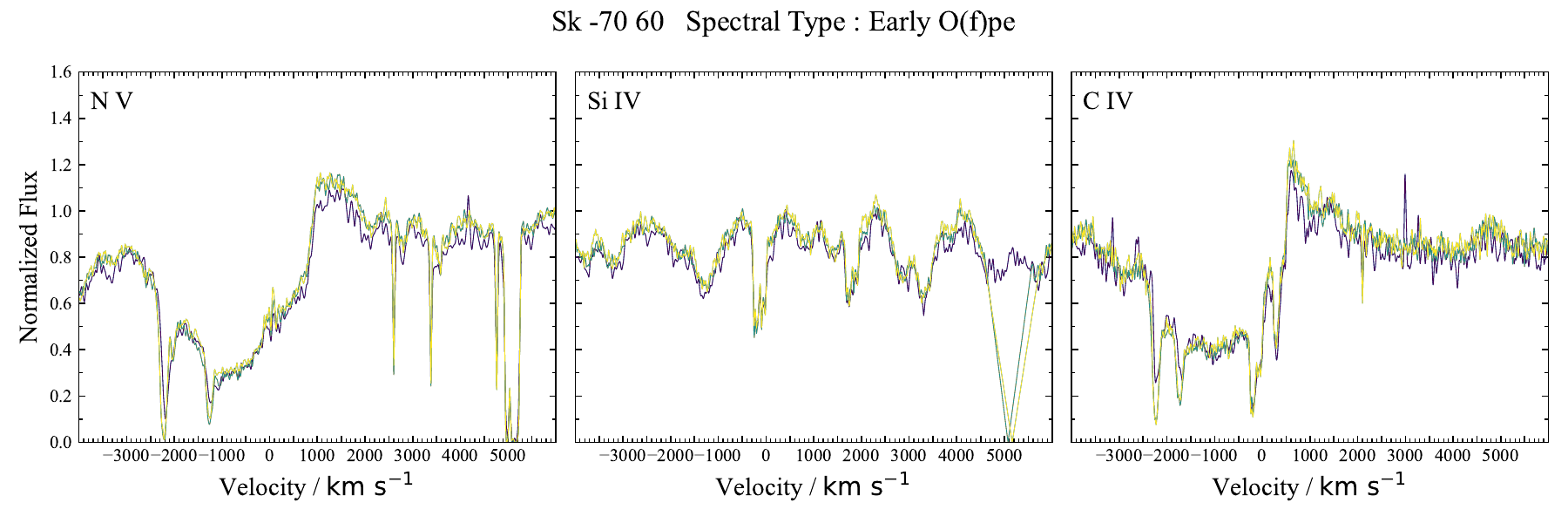} 
\caption{Same as Fig. \ref{fig:180_compare}, for the LMC star Sk -70 60, $\Delta(t) = 152.056$ days, with a third observation, plotted in yellow, separated from the second observation by approximately 3 hours.}
\label{fig:7060_compare}
    \end{figure*}

    \begin{figure*}
\includegraphics[width=\linewidth]{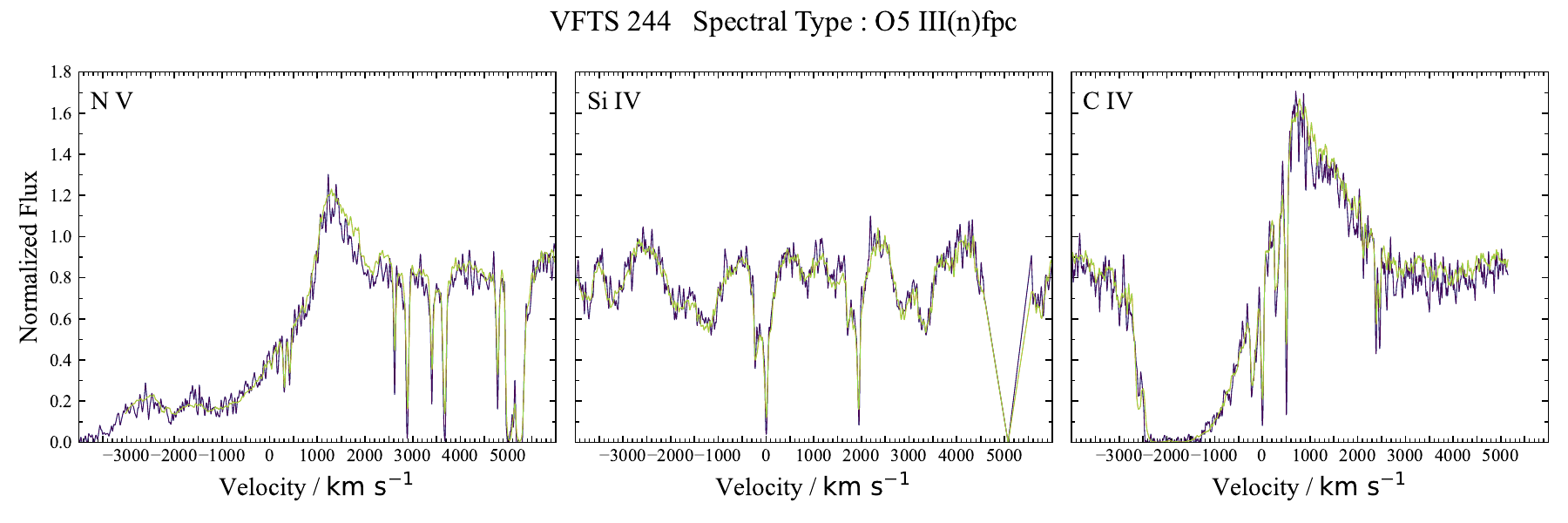} 
\caption{Same as Fig. \ref{fig:180_compare}, for the LMC star VFTS 244, $\Delta(t) = 1288.255$ days.}
\label{fig:244_compare}
    \end{figure*}

   \begin{figure*}
\includegraphics[width=\linewidth]{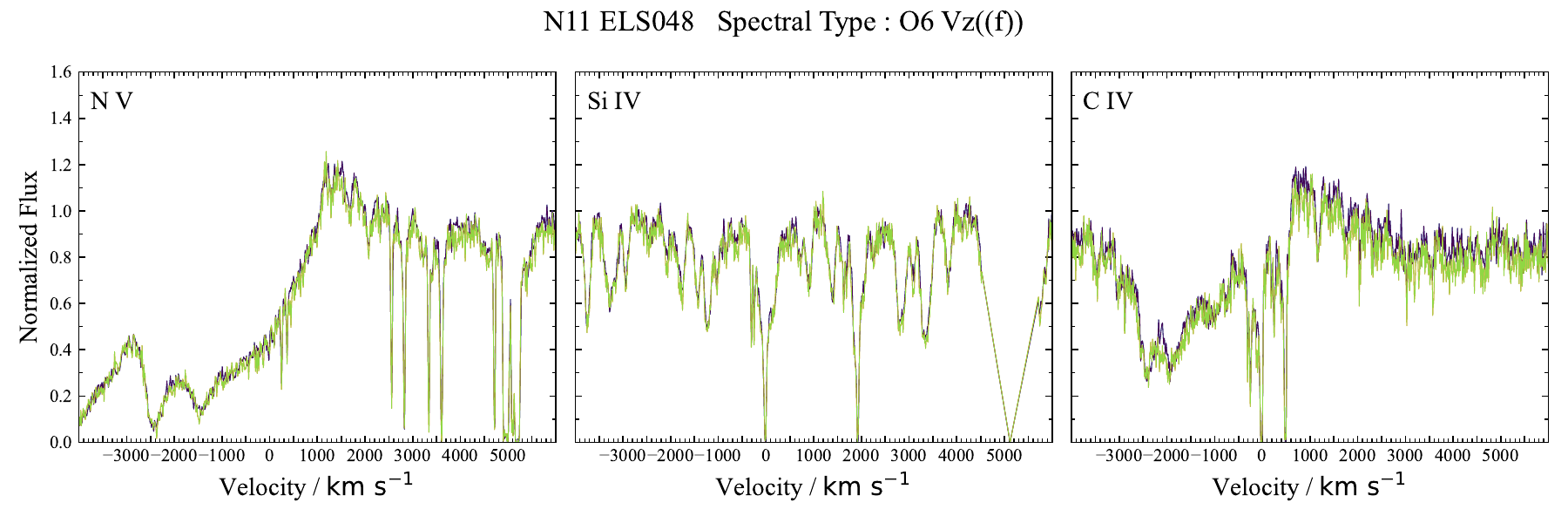} 
\caption{Same as Fig. \ref{fig:180_compare}, for the LMC star N11 ELS048, $\Delta(t) = 1.051$ days.}
\label{fig:11048_norm_compare_eps}
    \end{figure*}

   \begin{figure*}
\includegraphics[width=\linewidth]{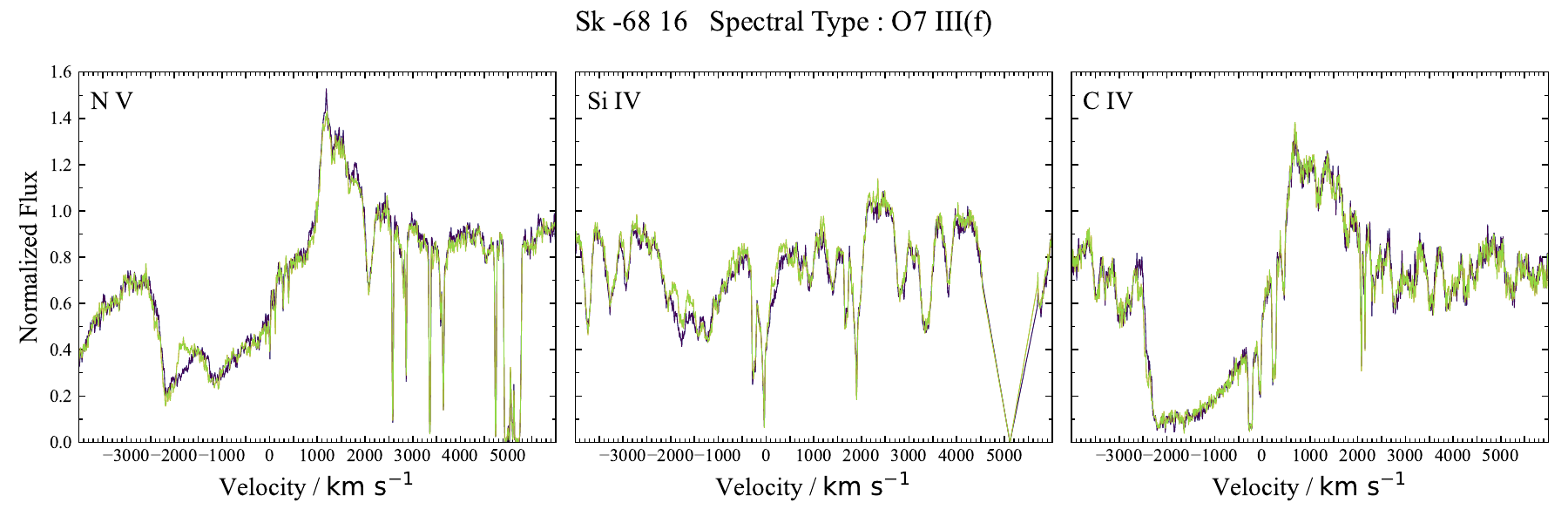} 
\caption{Same as Fig. \ref{fig:180_compare}, for the LMC star Sk -68 16, $\Delta(t) = 2.311$ days.}
\label{fig:6816_compare}
    \end{figure*}

   \begin{figure*}
\includegraphics[width=\linewidth]{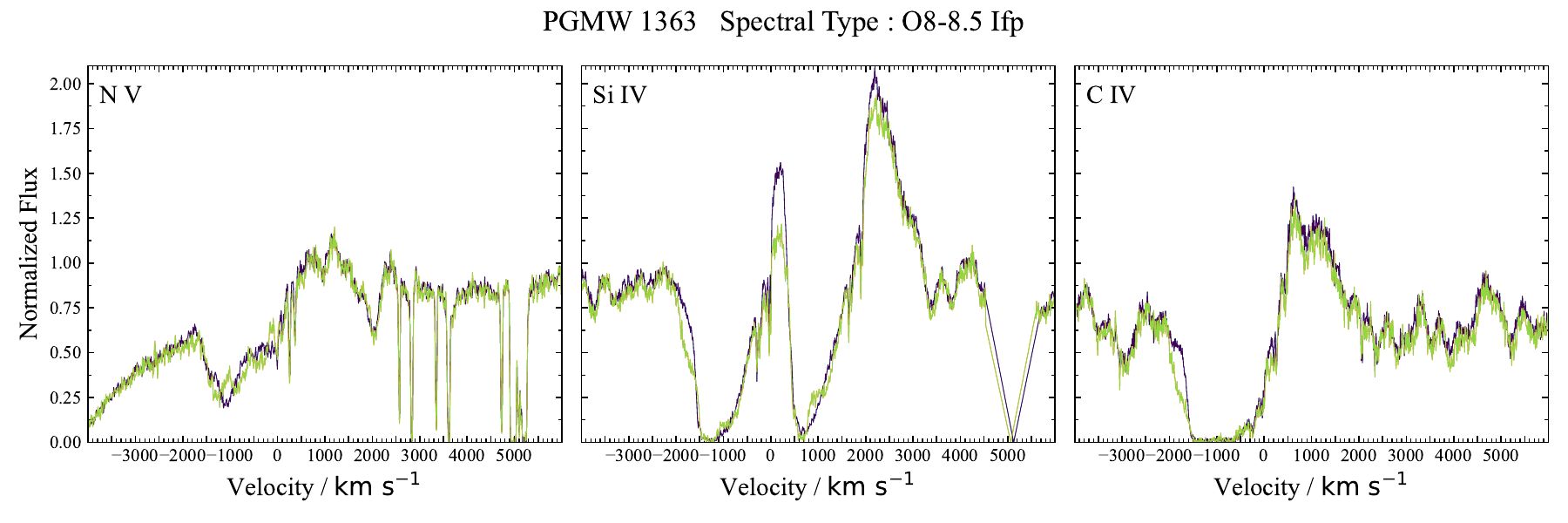} 
\caption{Same as Fig. \ref{fig:180_compare}, for the LMC star PGMW 1363, $\Delta(t) = 171.310$ days. The saturated \ion{Si}{IV} and \ion{C}{IV} spectra highlight the distinct and unusual change observed in the shape of the upper portions of the blue edges of each of the absorption profiles. The change is consistent across both of those profiles, but is not evident in the \ion{N}{V} profile as it is largely obscured by the red wing of the Lyman-$\alpha$ absorption.}
\label{fig:1363_compare}
    \end{figure*}

   \begin{figure*}
\includegraphics[width=\linewidth]{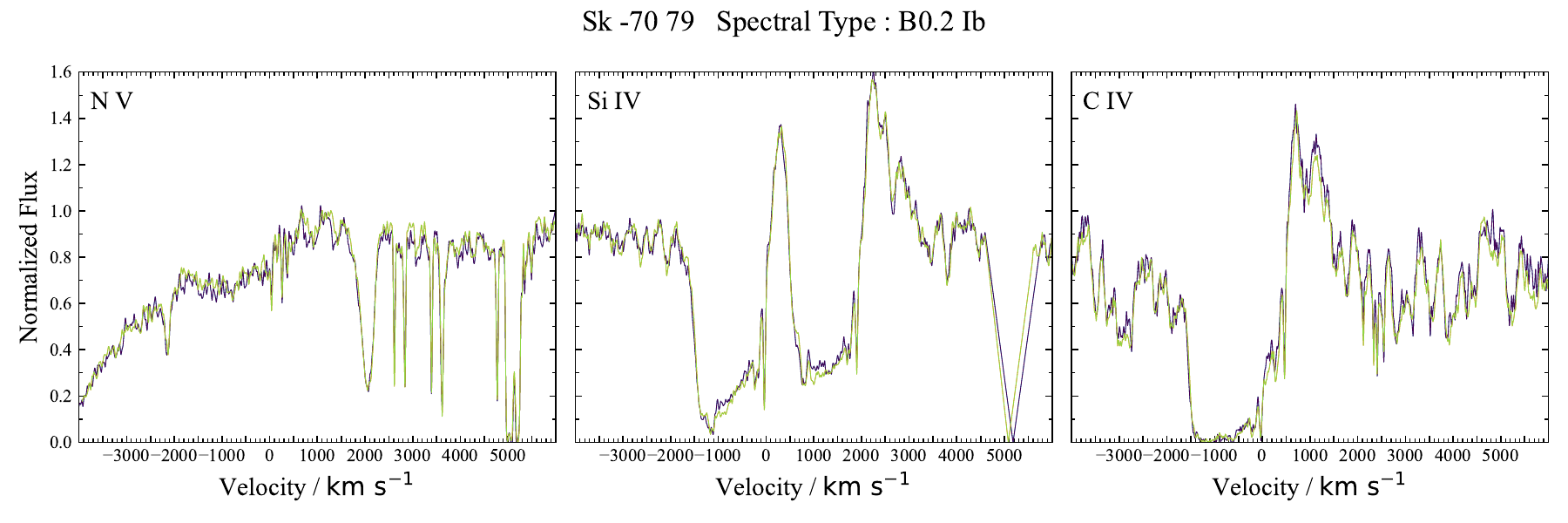} 
\caption{Same as Fig. \ref{fig:180_compare}, for the LMC star Sk -70 79, $\Delta(t) = 0.857$ days.}
\label{fig:7079_compare}
    \end{figure*}

   \begin{figure*}
\includegraphics[width=\linewidth]{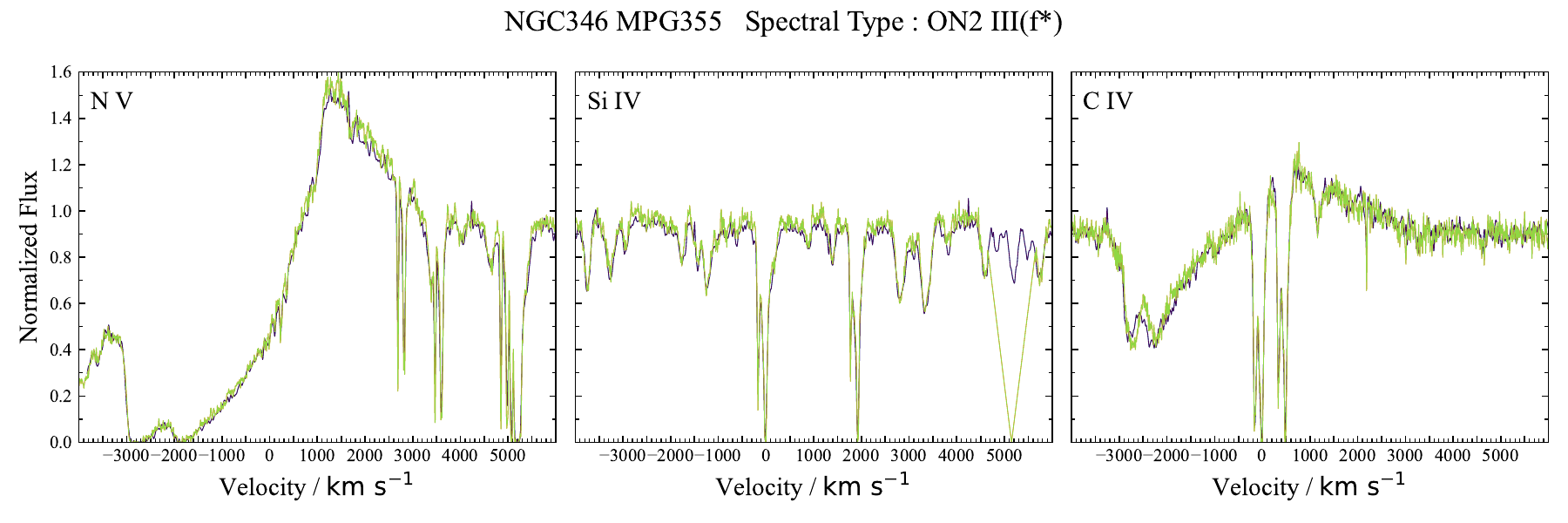} 
\caption{Same as Fig. \ref{fig:180_compare}, for the SMC star NGC 346 MPG 355, $\Delta(t) = 8060.059$ days.}
\label{fig:355_compare}
    \end{figure*}

   \begin{figure*}
\includegraphics[width=\linewidth]{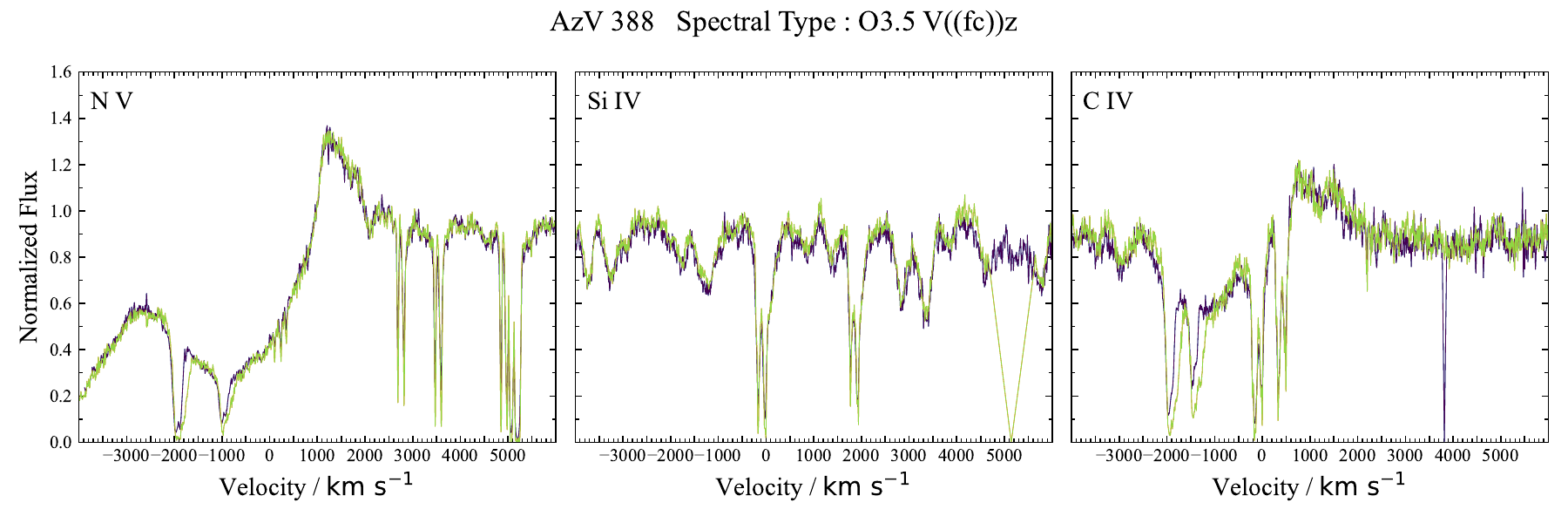} 
\caption{Same as Fig. \ref{fig:180_compare}, for the SMC star AzV 388, $\Delta(t) = 2135.151$ days.}
\label{fig:388_compare}
    \end{figure*}

   \begin{figure*}
\includegraphics[width=\linewidth]{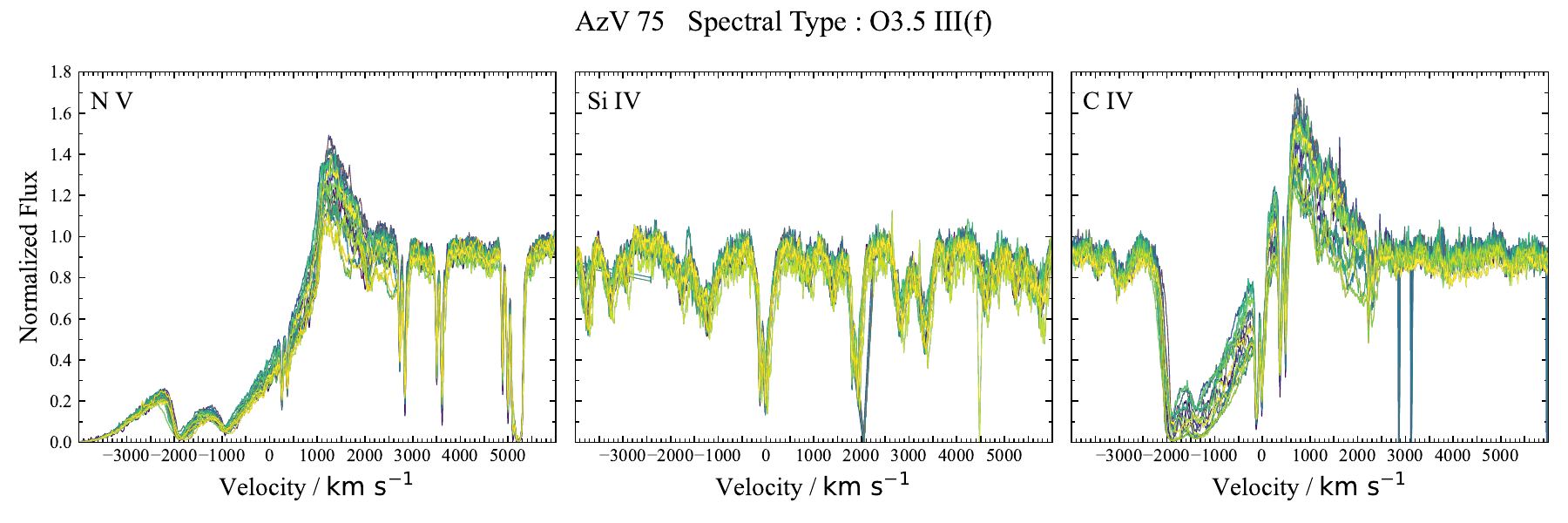} 
\caption{Comparison of 1 STIS and 24 COS spectra, separated by an average $\Delta(t)= 201.456$ days, for the SMC star AzV 75. Colour sequence runs from dark to light in time order as shown in the colour bar in the main text, Figures \ref{fig:75_compare_select} and \ref{fig:75_o4_n3}.}
\label{fig:75_compare}

    \end{figure*}

   \begin{figure*}
\includegraphics[width=\linewidth]{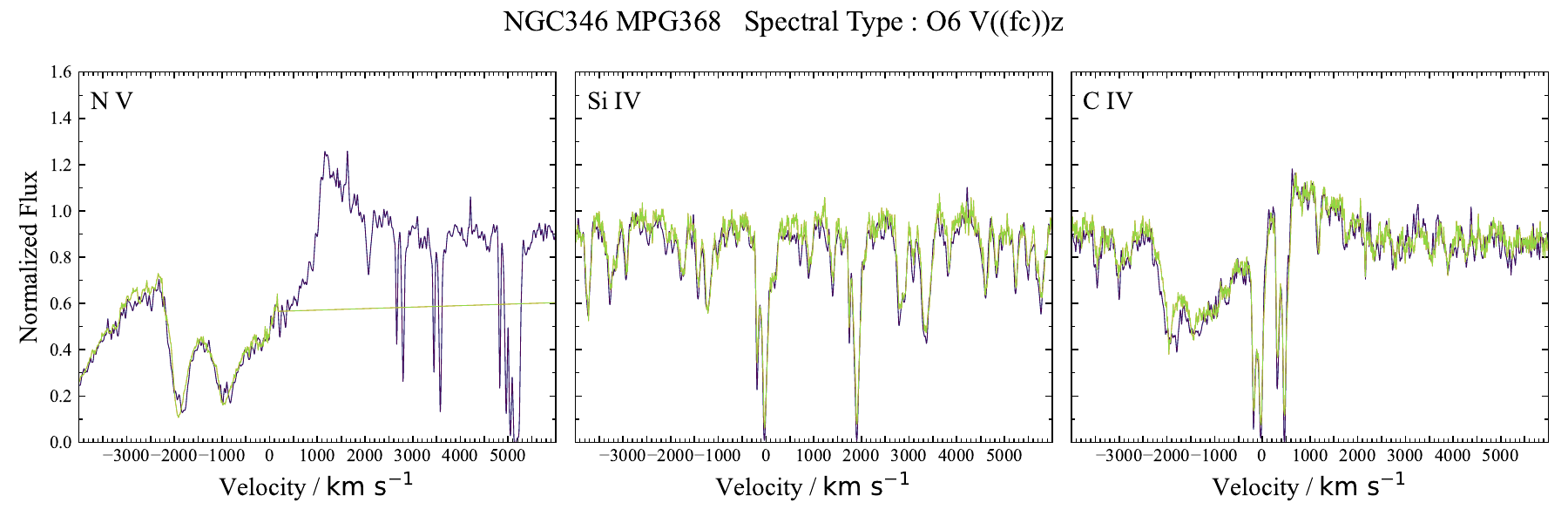} 
\caption{Same as Fig. \ref{fig:180_compare}, for the SMC star NGC 346 MPG 368, $\Delta(t) = 8000.632$ days. Note that the data for the second observation only incompletely cover the \ion{N}{V} feature but is presented to confirm the variations seen in the high velocity portion of the wind also seen in the \ion{C}{IV} feature.}
\label{fig:368_compare}
    \end{figure*}

   \begin{figure*}
\includegraphics[width=\linewidth]{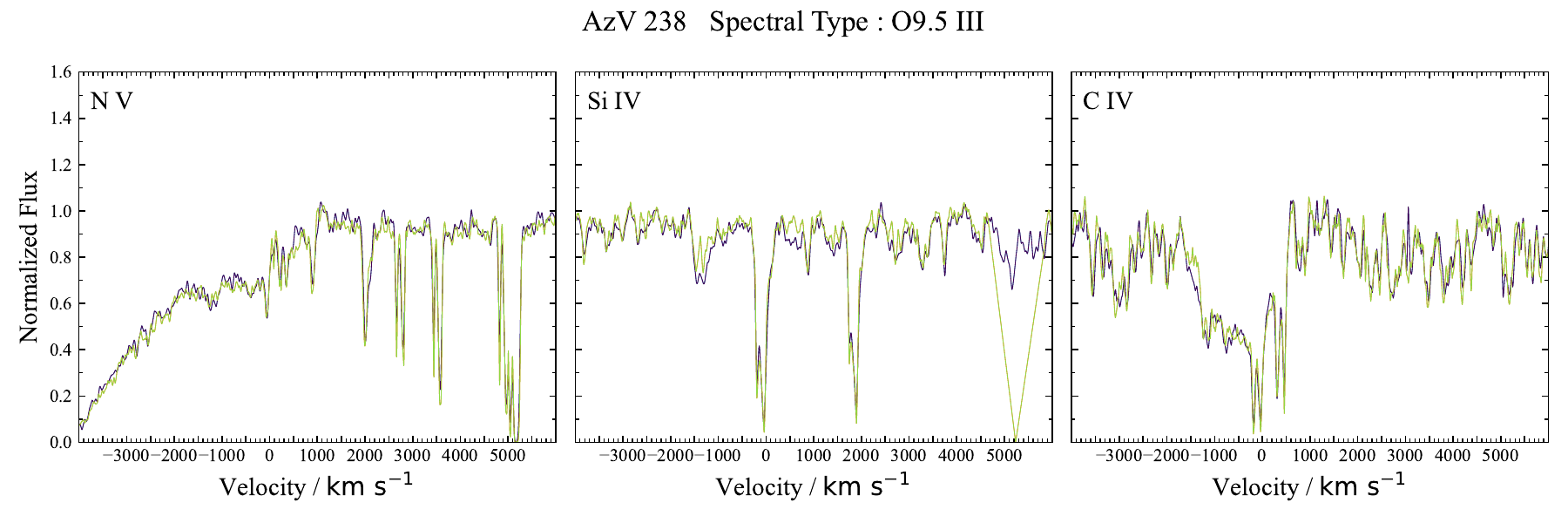} 
\caption{Same as Fig. \ref{fig:180_compare}, for the SMC star AzV 238, $\Delta(t) = 23.434$ days..}
\label{fig:238_compare}
    \end{figure*}

   \begin{figure*}
\includegraphics[width=\linewidth]{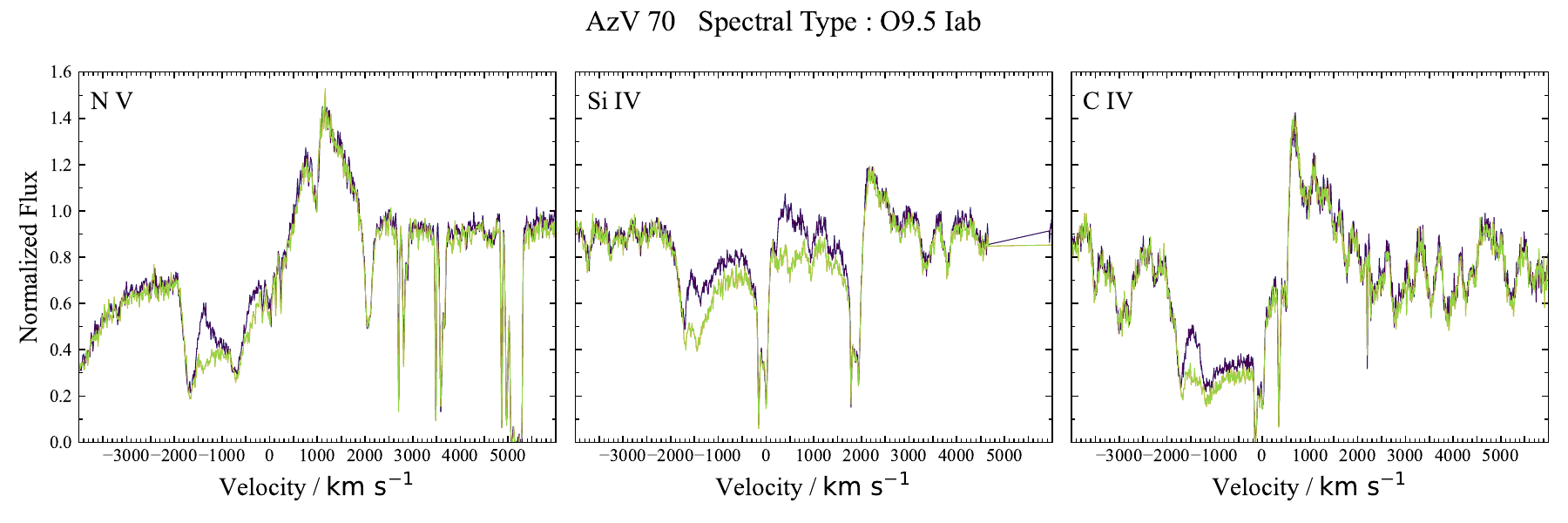} 
\caption{Same as Fig. \ref{fig:180_compare}, for the SMC star AzV 70, $\Delta(t) = 47.954$ days.}
\label{fig:70_compare}
    \end{figure*}

   \begin{figure*}
\includegraphics[width=\linewidth]{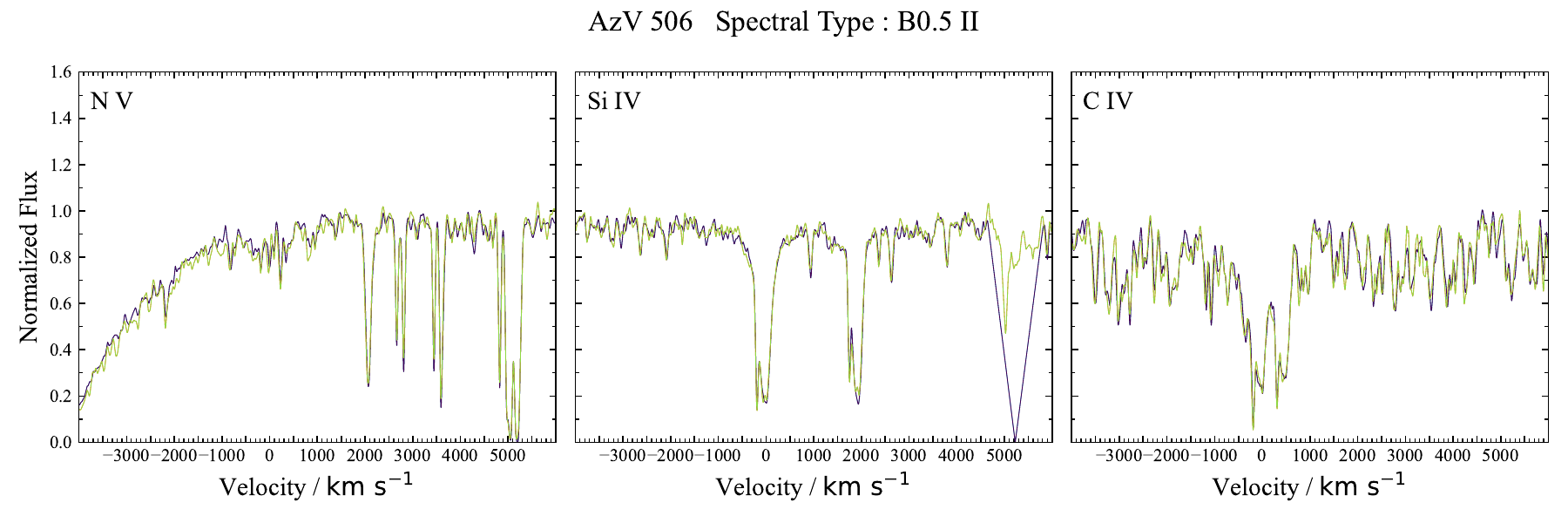} 
\caption{Same as Fig. \ref{fig:180_compare}, for the SMC star AzV 506, $\Delta(t) = 6.135$ days.}
\label{fig:506_compare}
    \end{figure*}

   \begin{figure*}
\includegraphics[width=\linewidth]{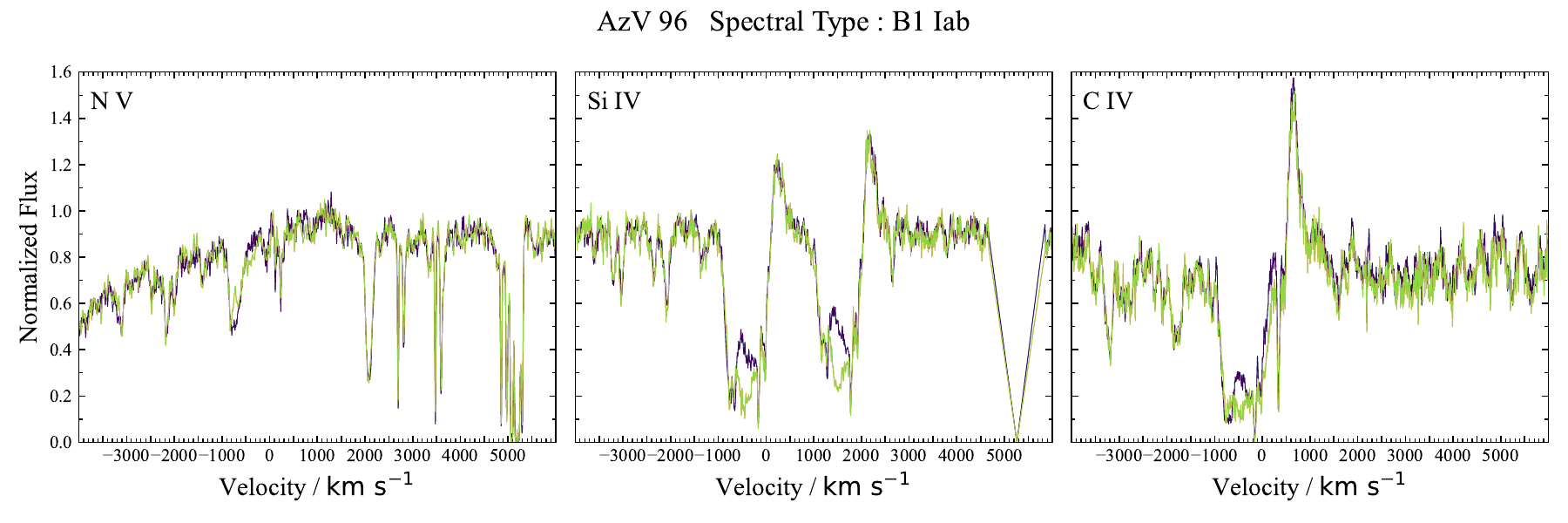} 
\caption{Same as Fig. \ref{fig:180_compare}, for the SMC star AzV 96, $\Delta(t) = 4.827$ days.}
\label{fig:96_compare}
    \end{figure*}

   \begin{figure*}
\includegraphics[width=0.69\linewidth]{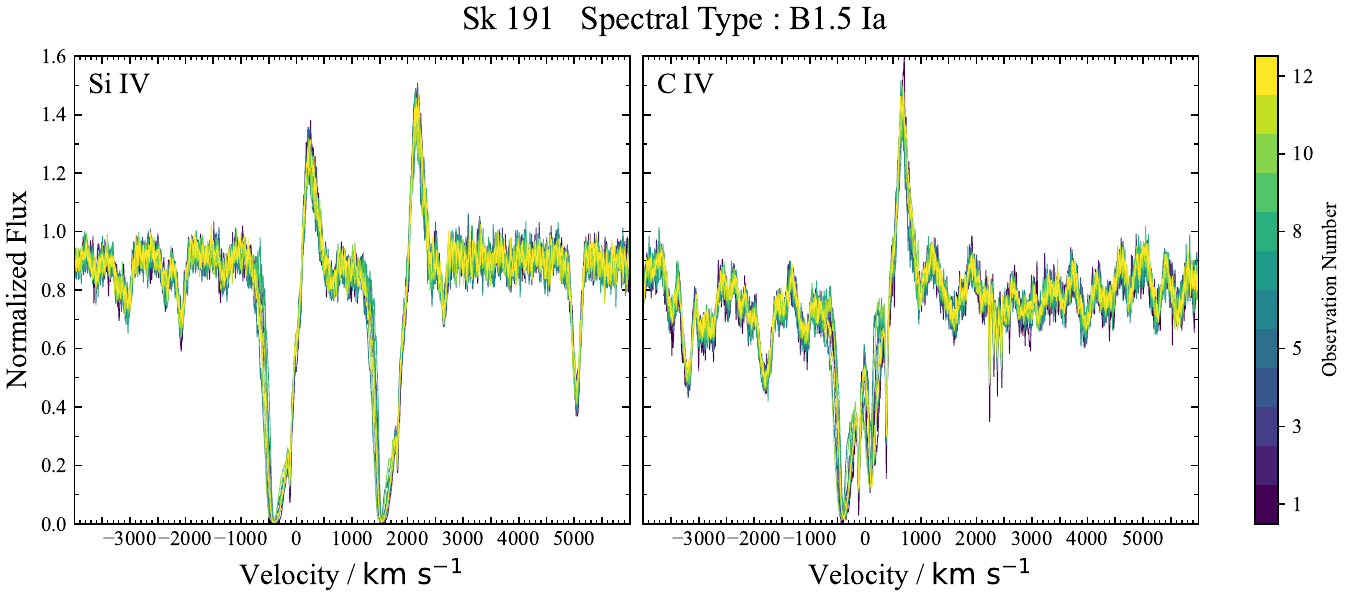} 
\caption{Comparison of 12 observed spectra, separated by an average $\Delta(t)= 87.057$ days, for the SMC star Sk 191. Plots respectively show each of the \ion{Si}{IV} 1393.76, 1402.77 \AA{} and \ion{C}{IV} 1548.20. 1550.78 \AA{} UV resonance line doublet profiles. Colour sequence runs from dark to light in time order, as shown by the colour bar.}
\label{fig:191_compare}
    \end{figure*}

   \begin{figure*}
\includegraphics[width=0.69\linewidth]{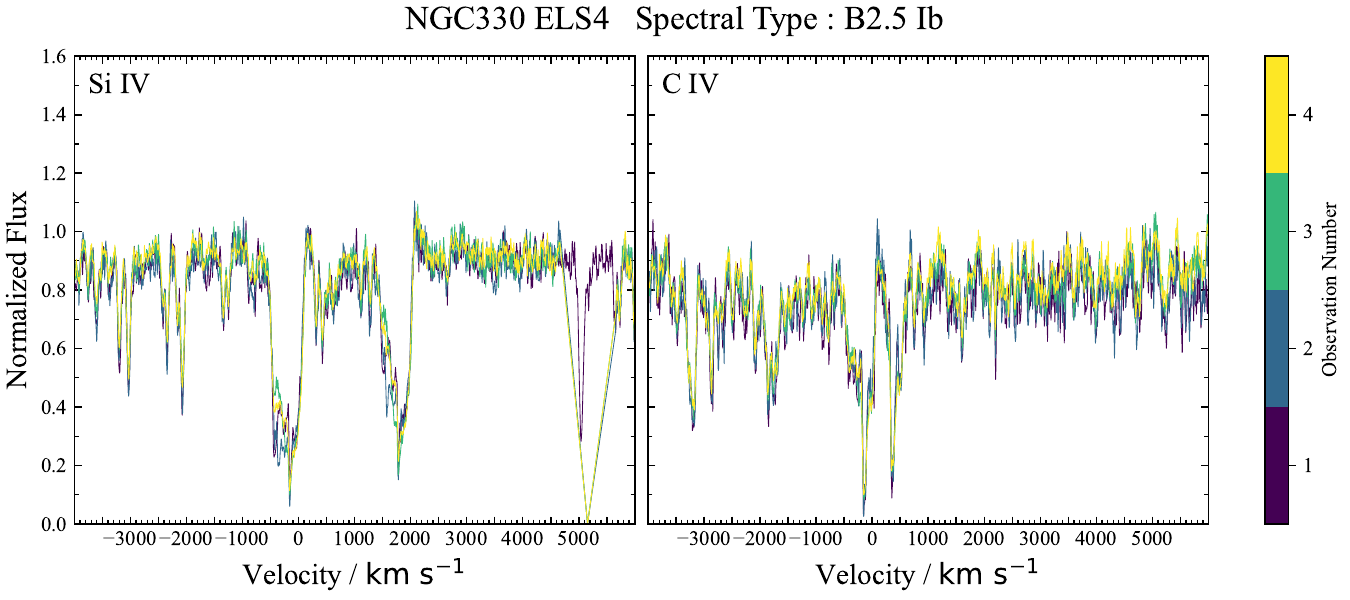} 
\caption{Comparison of four spectra, separated by an overall elapsed time between first and last observation of $\Delta(t)= 2837.012$ days (between Observations 1 and 2, $\Delta(t)= 340.840$ days and between Observations 3 and 4, $\Delta(t)= 4.864$ days), for the SMC star NGC 330 ELS4. Plots respectively show each of the \ion{Si}{IV} 1393.76, 1402.77 \AA{} and \ion{C}{IV} 1548.20. 1550.78 \AA{} UV resonance line doublet profiles. The colour bar shows the time-ordered colour sequence of the plotted spectra.}
\label{fig:3304_compare}
    \end{figure*}

   \begin{figure*}
\includegraphics[width=0.69\linewidth]{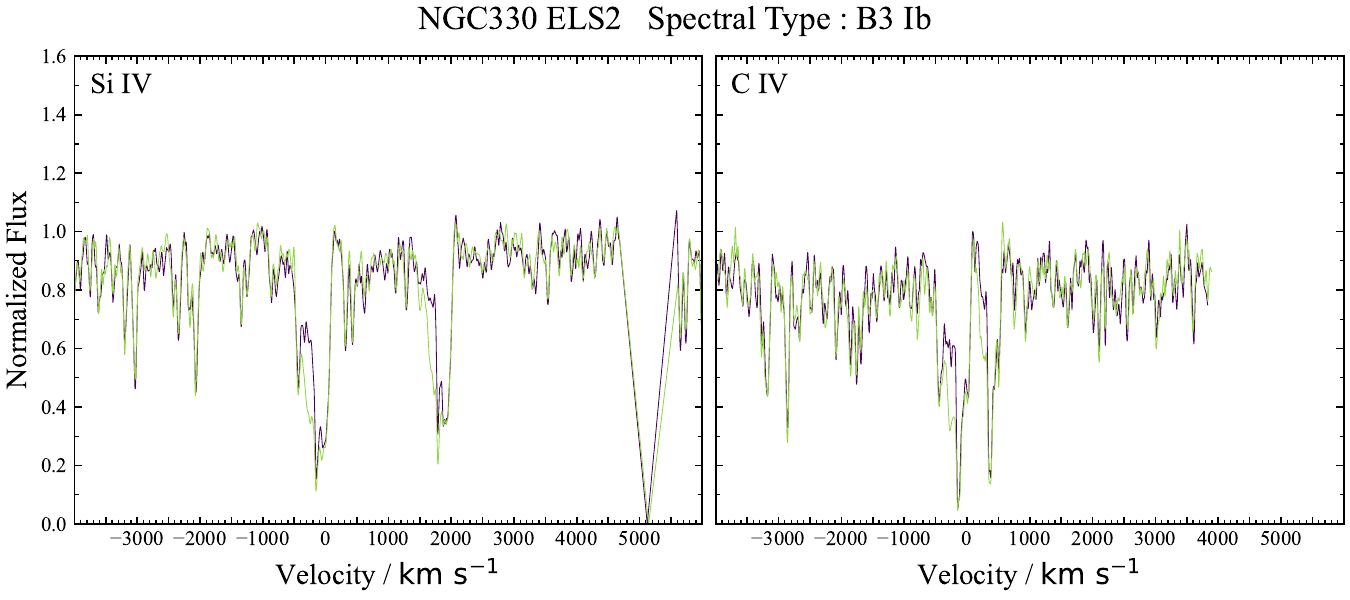} 
\caption{Same as Fig. \ref{fig:3304_compare} for the SMC star NGC 330 ELS2, $\Delta(t) = 10.967$ days. Two observations only (dark violet, light green).}
\label{fig:3302_compare}
    \end{figure*}

\clearpage

\section{SEI Fits} \label{allSEI}

The following plots show SEI fits for two observations of each target star. For stars where more than two observations are available, those two with the extremes of derived average radial optical depths are shown.

Rest wavelengths for each blue element and each red element of the plotted doublet features are indicated, respectively, by a blue bar and a red bar on the lower edge of each plot. The main upper panel of each Figure shows the SEI fit to the observed spectrum, produced using the method described in the main text. The lower left panel plots the 21 radial optical depths derived from that fitting process and the lower right panel shows the photospheric contribution to the fit, derived from the relevant selected TLUSTY photospheric grid model spectrum.

The regions used for deriving average radial optical depths are highlighted by shading in the main plot, and by the coloured section of the lower left plot, in each Figure. NAC locations are also highlighted in each main plot, where identifiable.

   \begin{figure*}
\begin{center}
 \subfloat[ ]{\includegraphics[width=3.34in]{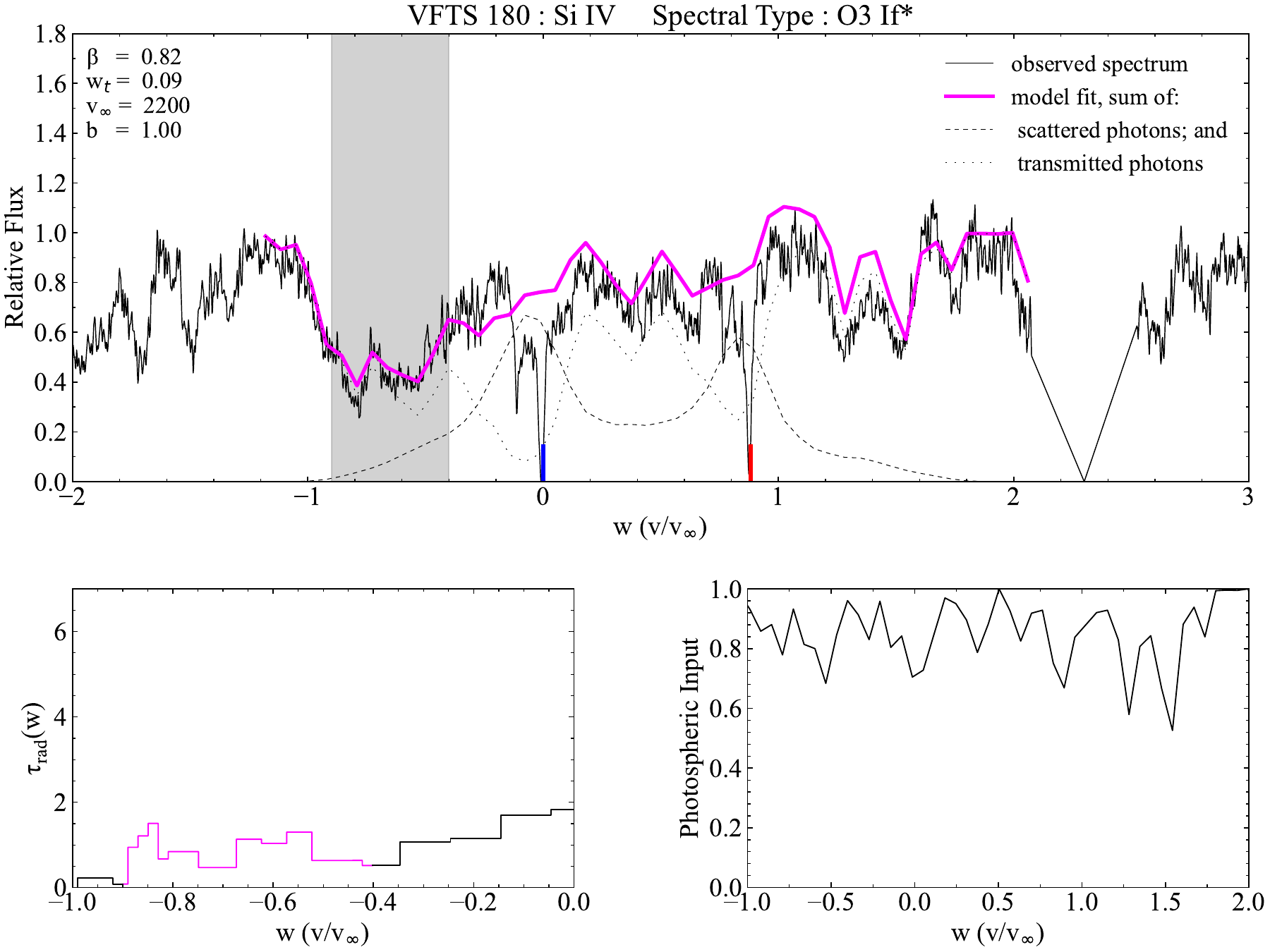} }
 \qquad
 \subfloat[ ]{\includegraphics[width=3.34in]{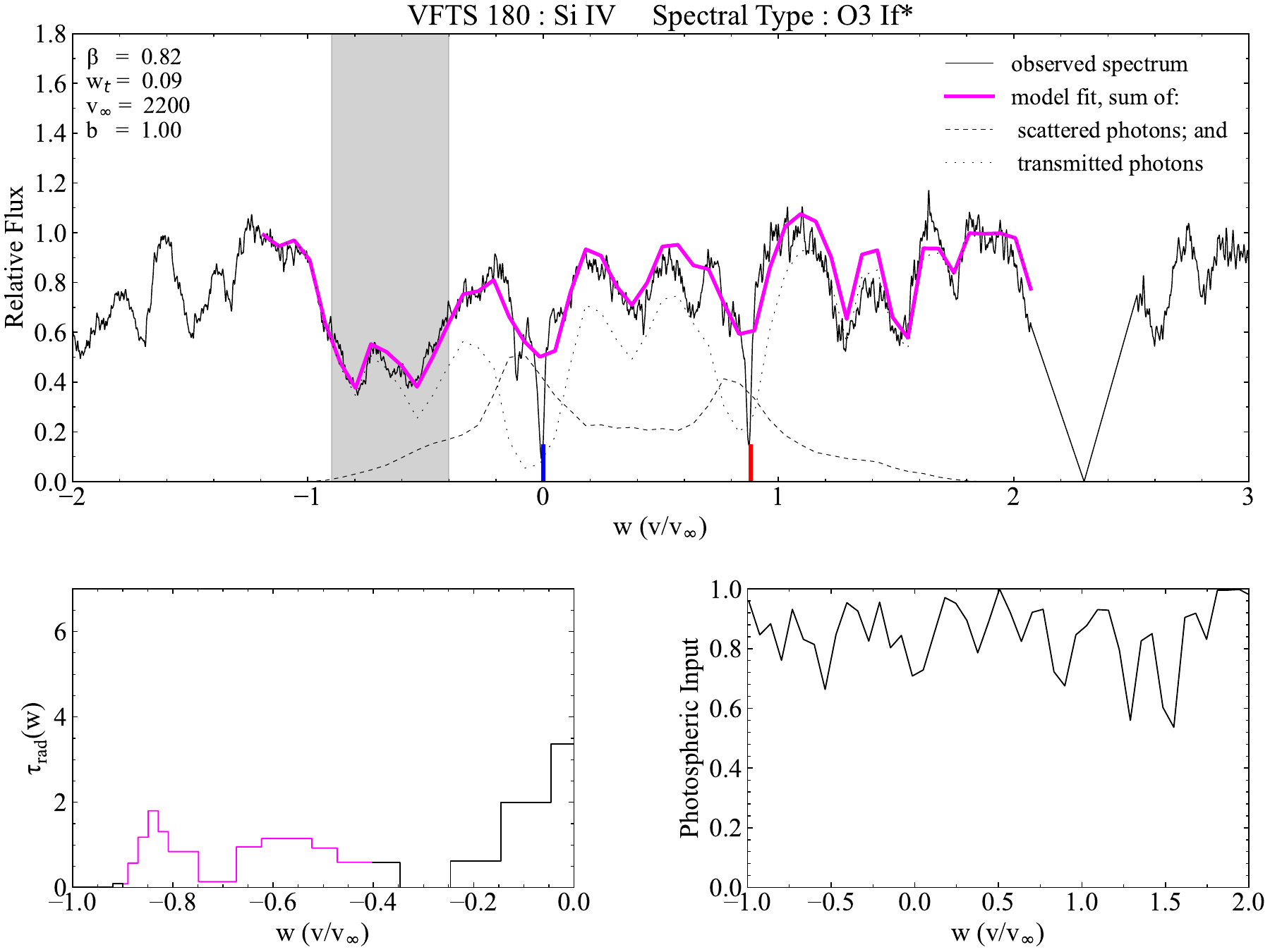} }
 \caption{SEI-derived model fits for the \ion{Si}{IV} doublet feature for two observations, dated (a) 2019 November 14 and (b) 2023 May 02, of the LMC star VFTS 180.}
 \label{fig:180_si4_SEI}
\end{center}
    \end{figure*}

   \begin{figure*}
\begin{center}
 \subfloat[ ]{\includegraphics[width=3.34in]{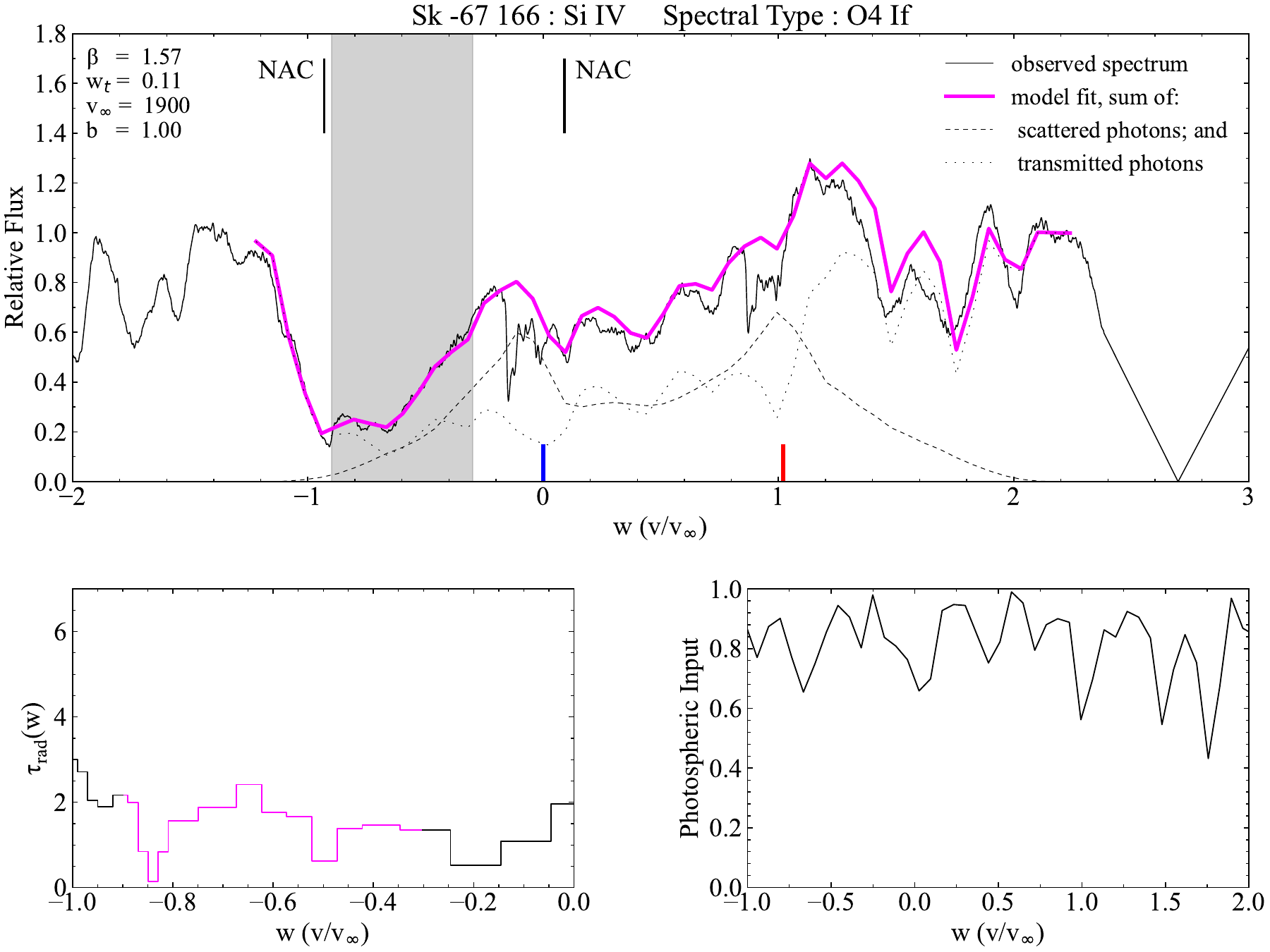} }
 \qquad
 \subfloat[ ]{\includegraphics[width=3.34in]{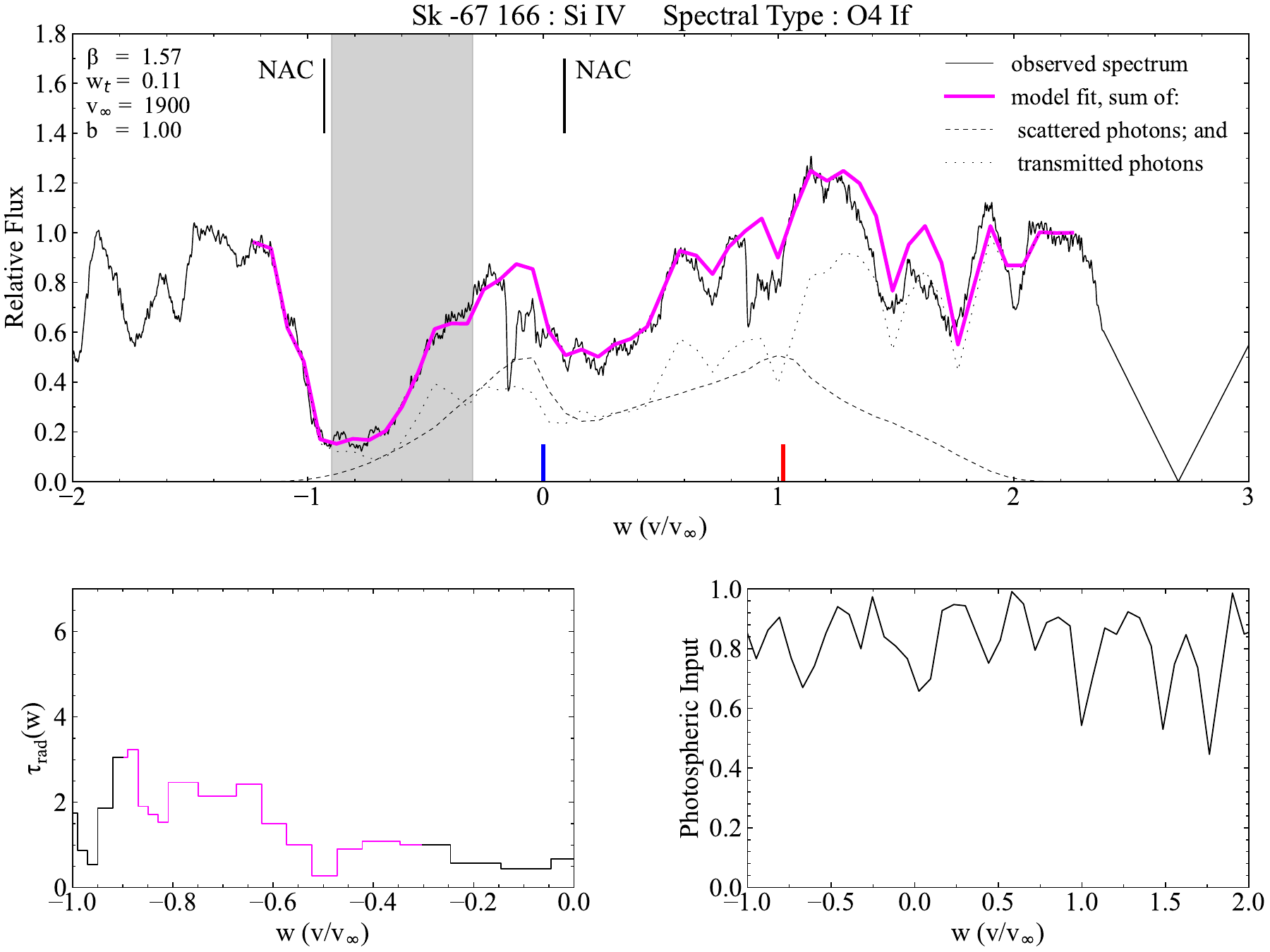} }
 \caption{SEI-derived model fits for the (a) and (b) \ion{Si}{IV} doublet feature for two observations, dated (a) 2021 January 16 and (b) 2021 January 20, of the LMC star Sk -67 166. These plots are two of the observations available showing the highest and lowest average radial optical depths in the stellar wind.}
 \label{fig:67166_si4_SEI}
\end{center}
    \end{figure*}

   \begin{figure*}
\begin{center}
 \subfloat[ ]{\includegraphics[width=3.34in]{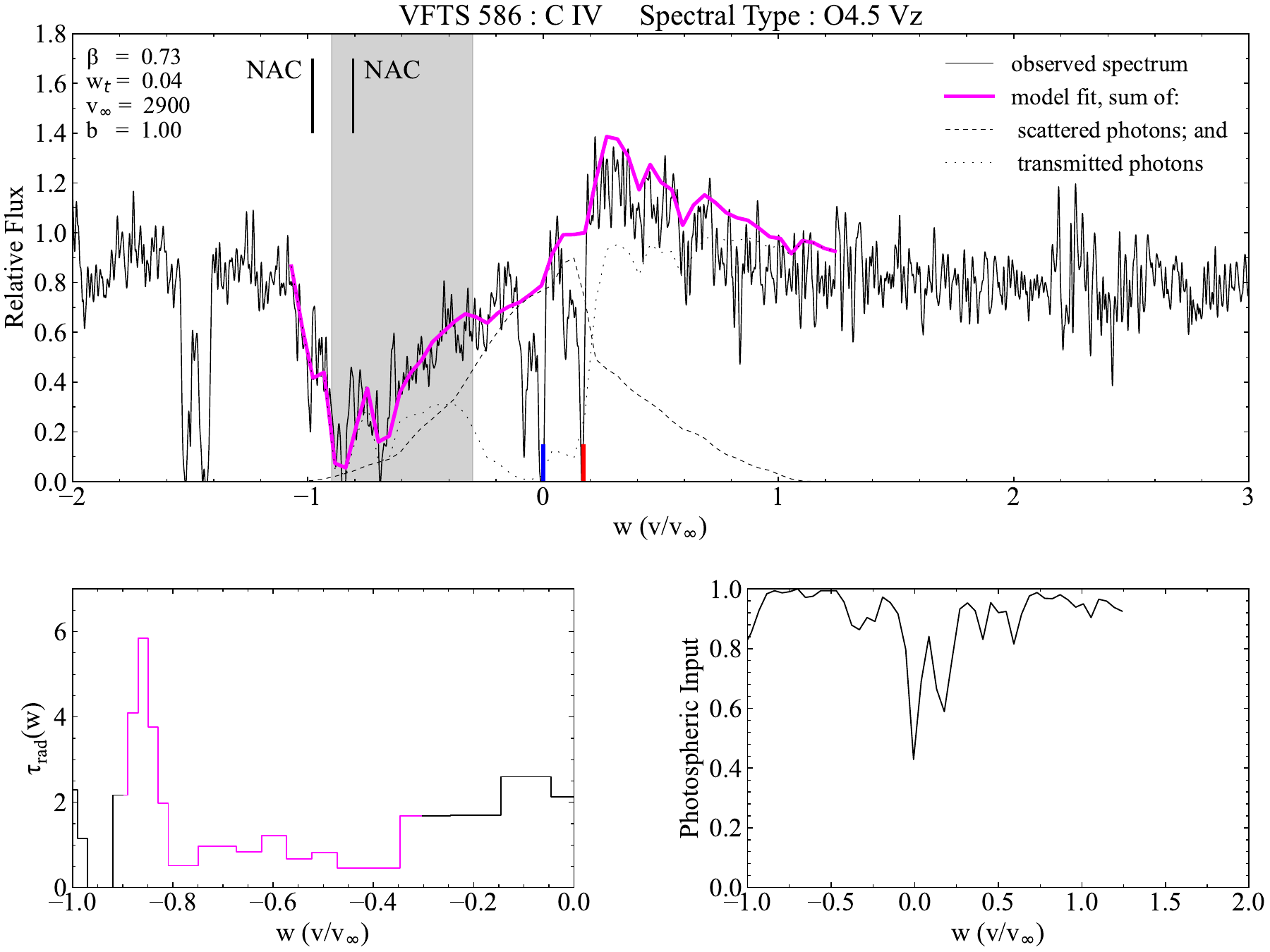} }
 \qquad
 \subfloat[ ]{\includegraphics[width=3.34in]{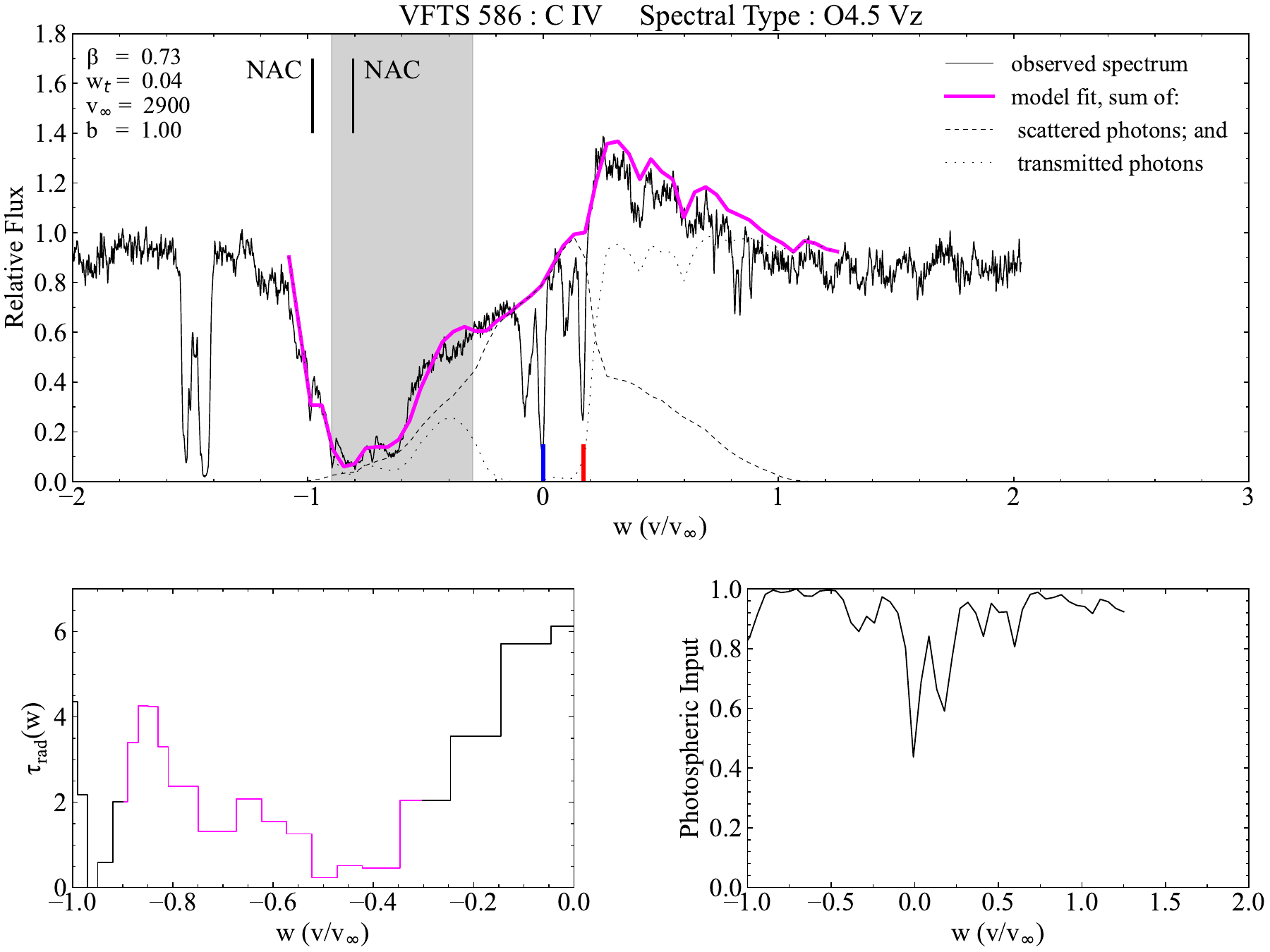} }
 \caption{SEI-derived model fits for the \ion{C}{IV} doublet feature for two observations, dated (a) 2020 February 26 and (b) 2022 October 06, of the LMC star VFTS 586.}
 \label{fig:586_si4_SEI}
\end{center}
    \end{figure*}

   \begin{figure*}
\begin{center}
 \subfloat[ ]{\includegraphics[width=3.34in]{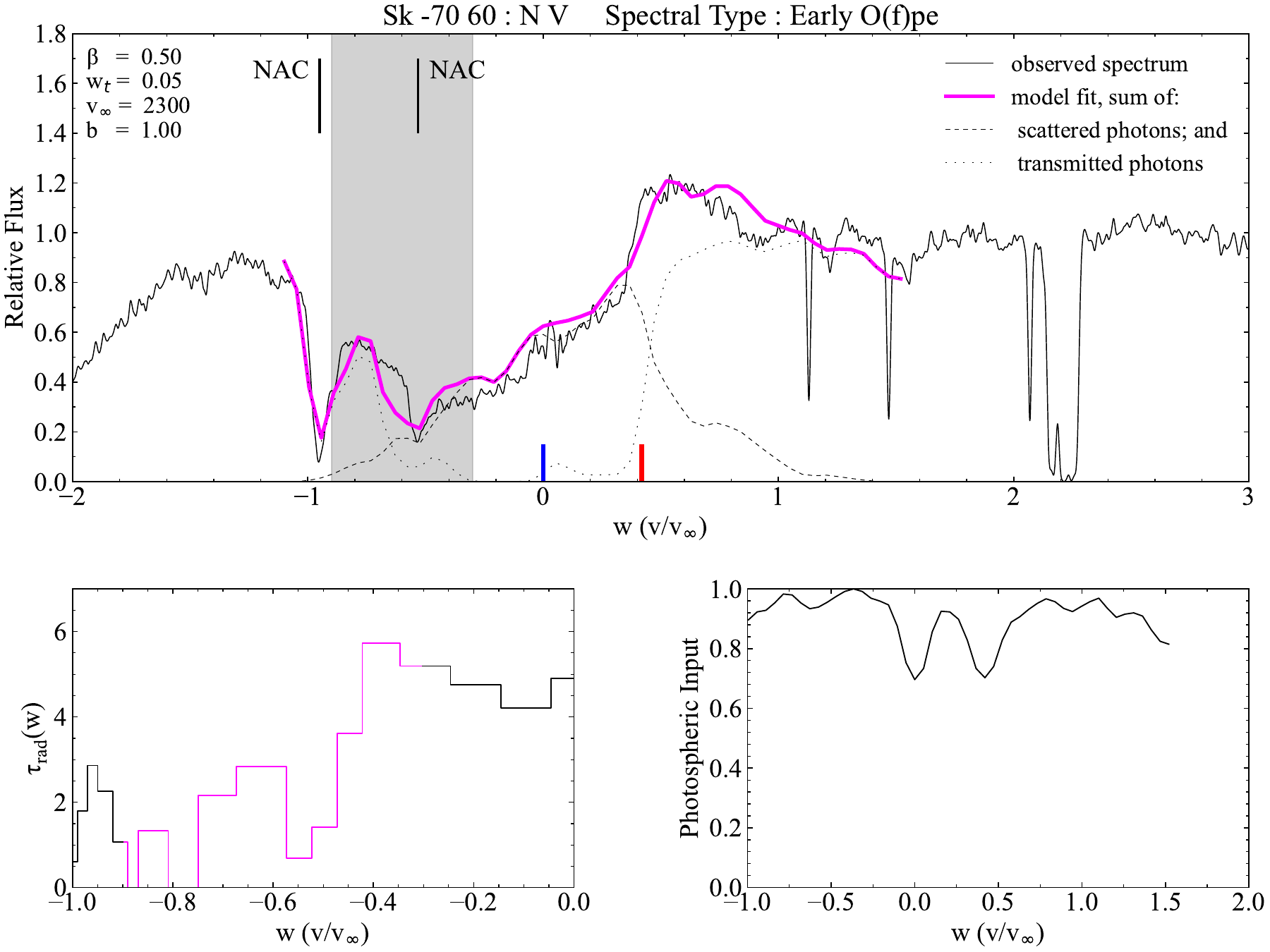} }
 \qquad
 \subfloat[ ]{\includegraphics[width=3.34in]{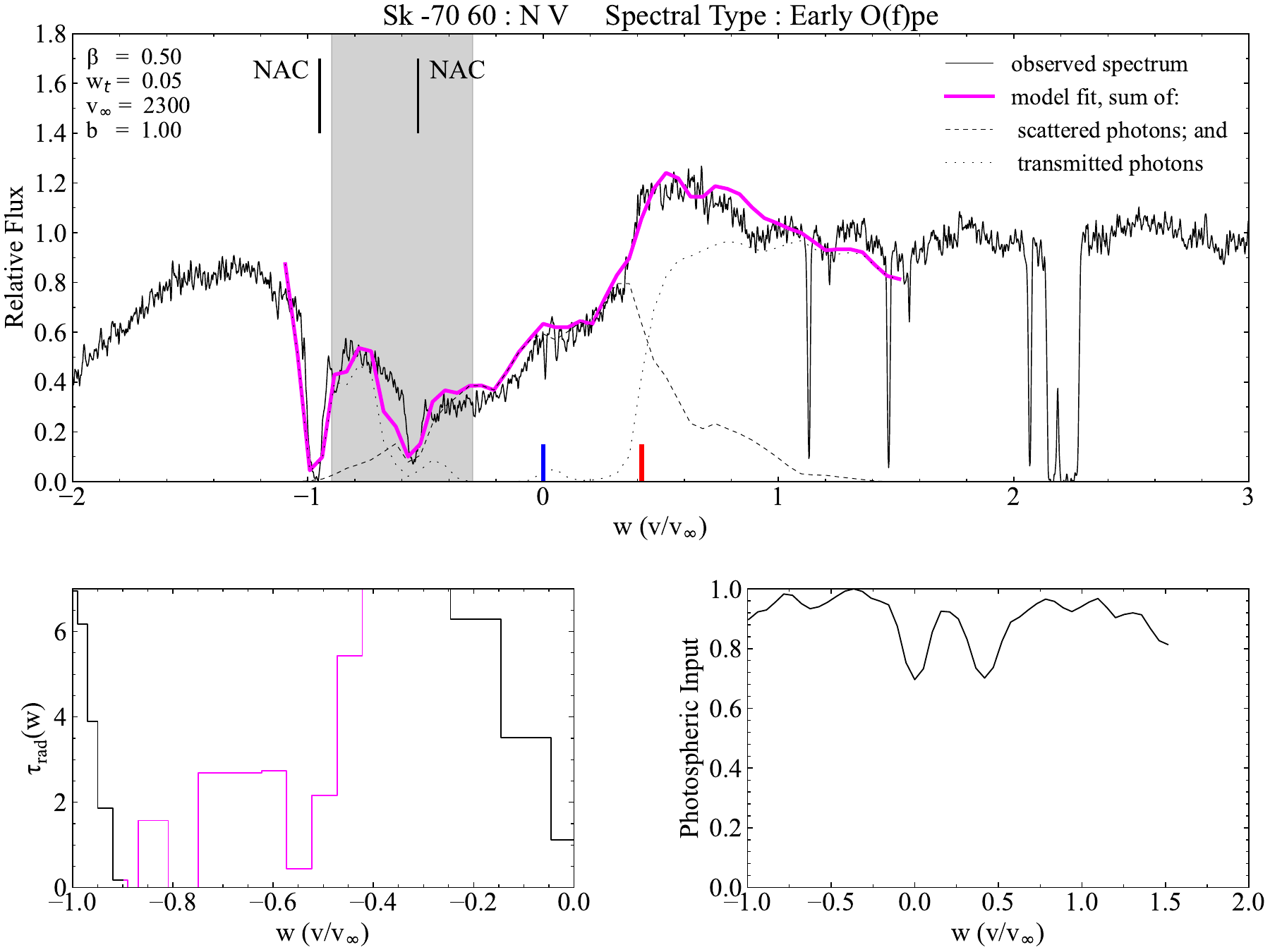} }
 \qquad
 \subfloat[ ]{\includegraphics[width=3.34in]{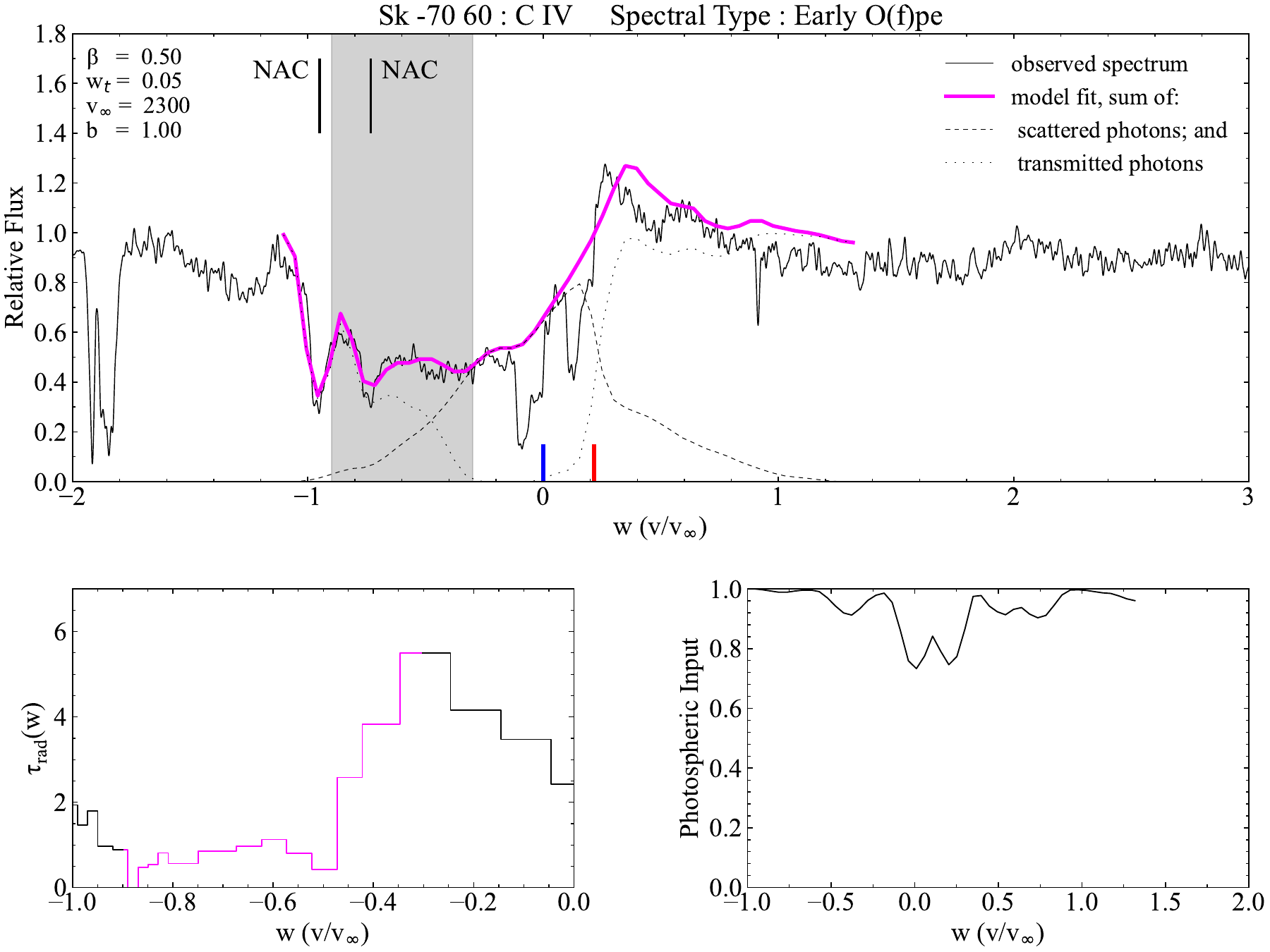} }
 \qquad
 \subfloat[ ]{\includegraphics[width=3.34in]{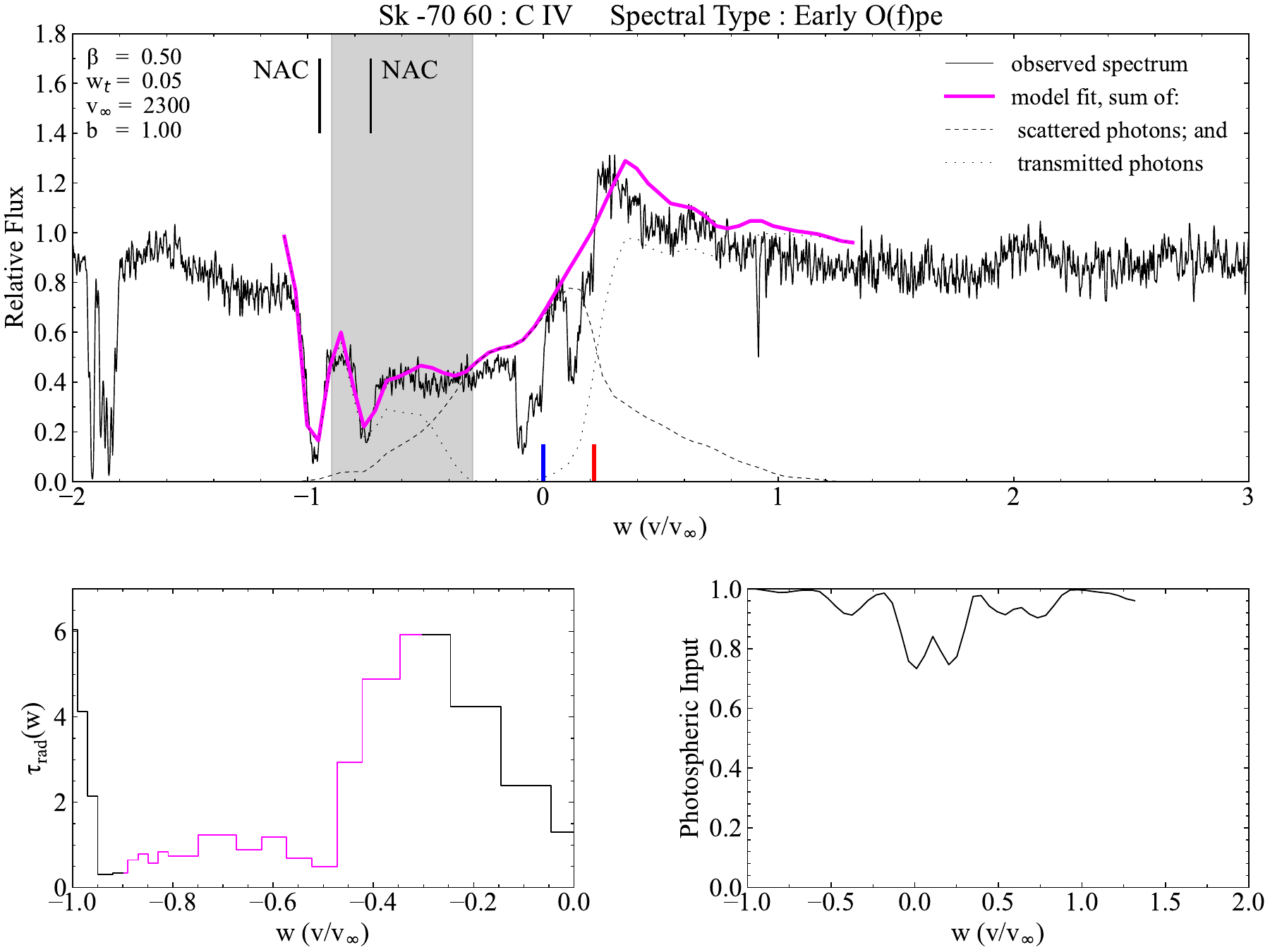} }
 \caption{SEI-derived model fits for (a) and (b) the \ion{N}{V} and (c) and (d) the \ion{C}{IV} doublet feature for two observations, dated (a) 2021 August 28 and (b) 2022 January 26, of the LMC star Sk -70 60.}
 \label{fig:7060_c4_SEI}
\end{center}
    \end{figure*}

   \begin{figure*}
\begin{center}
 \subfloat[ ]{\includegraphics[width=3.34in]{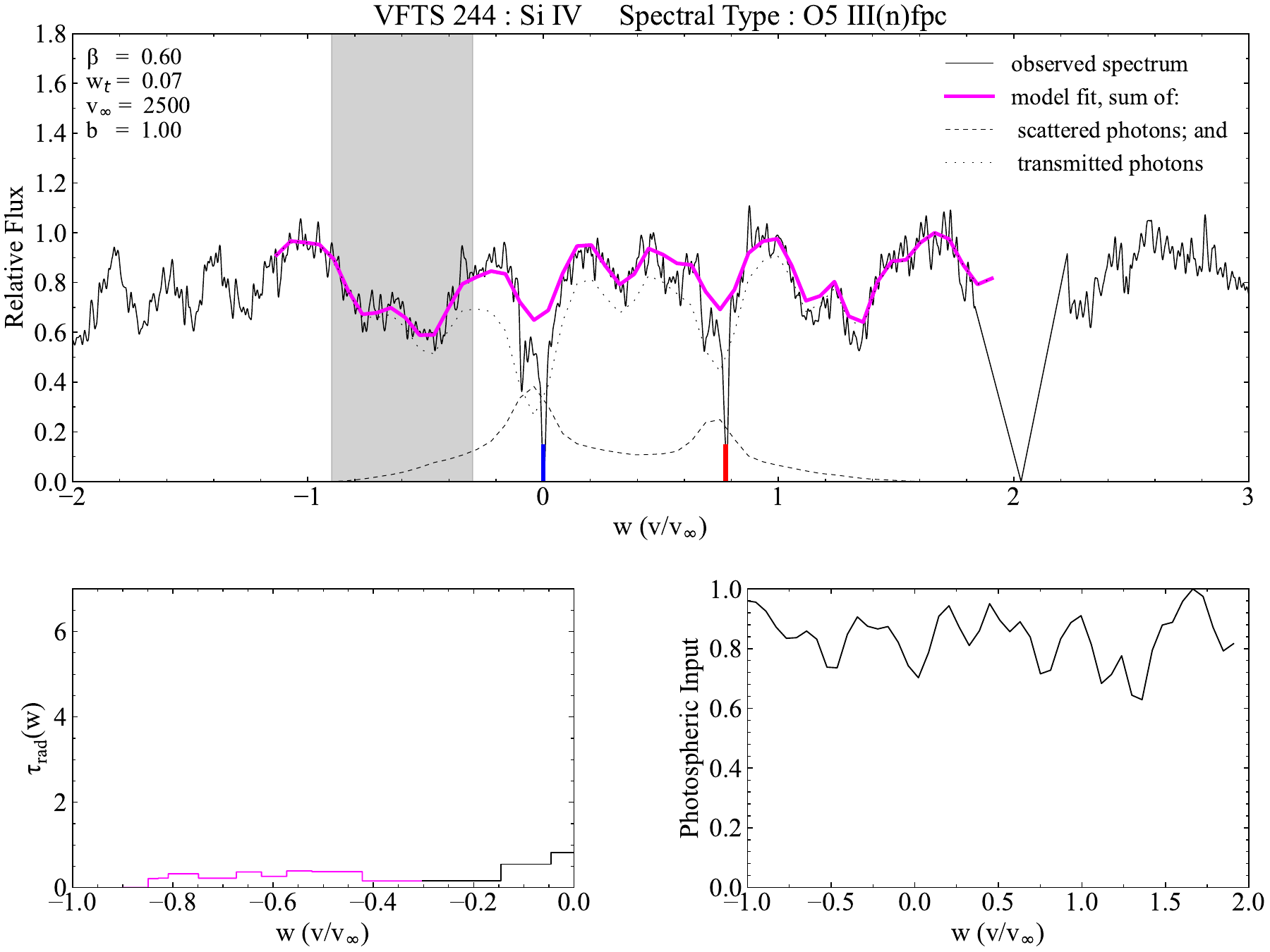} }
 \qquad
 \subfloat[ ]{\includegraphics[width=3.34in]{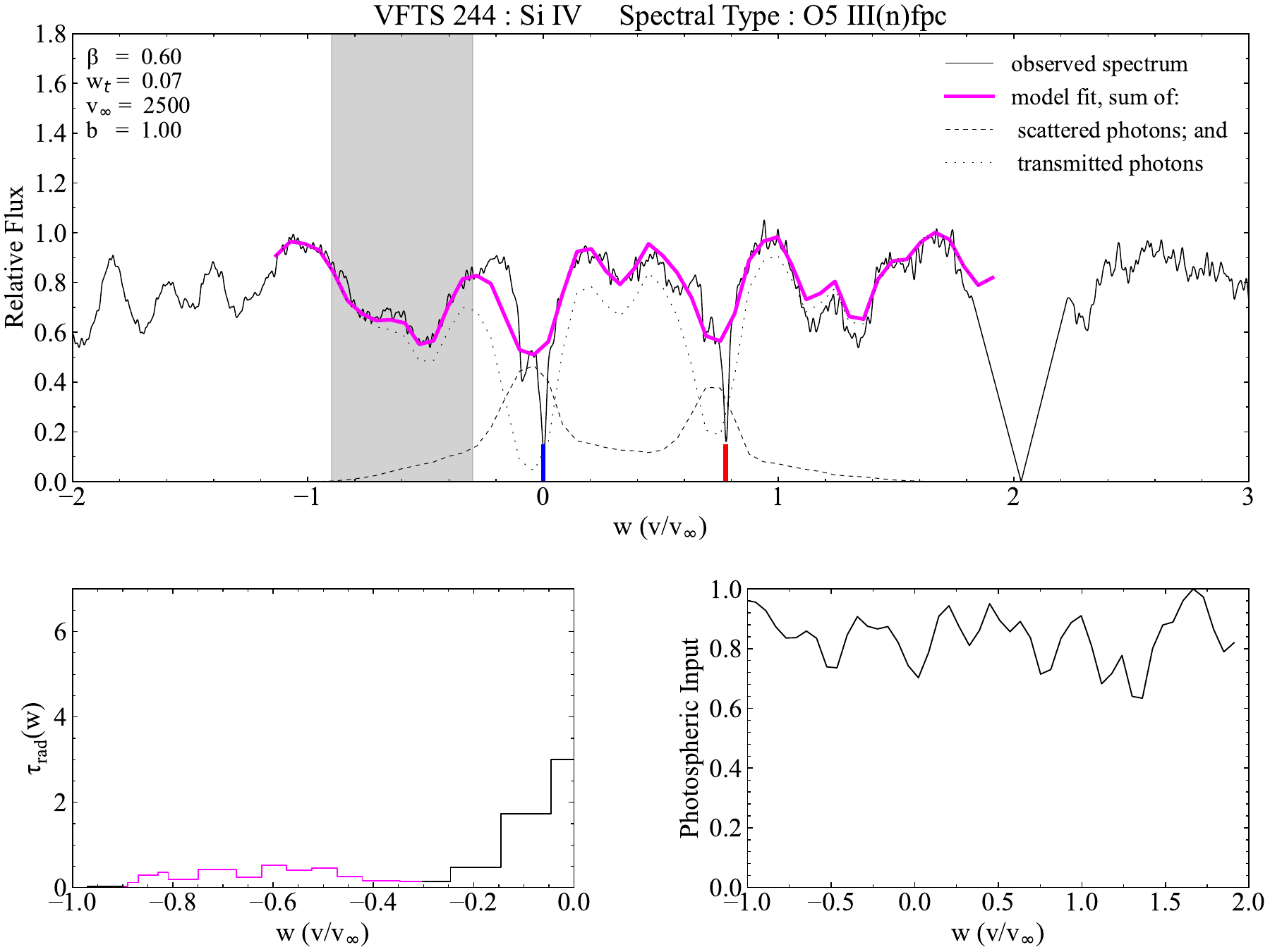} }
 \caption{SEI-derived model fits for the \ion{Si}{IV} doublet feature for two observations, dated (a) 2019 November 17 and (b) 2023 May 29, of the LMC star VFTS 244.}
 \label{fig:244_si4_SEI}
\end{center}
    \end{figure*}

   \begin{figure*}
\begin{center}
 \subfloat[ ]{\includegraphics[width=3.34in]{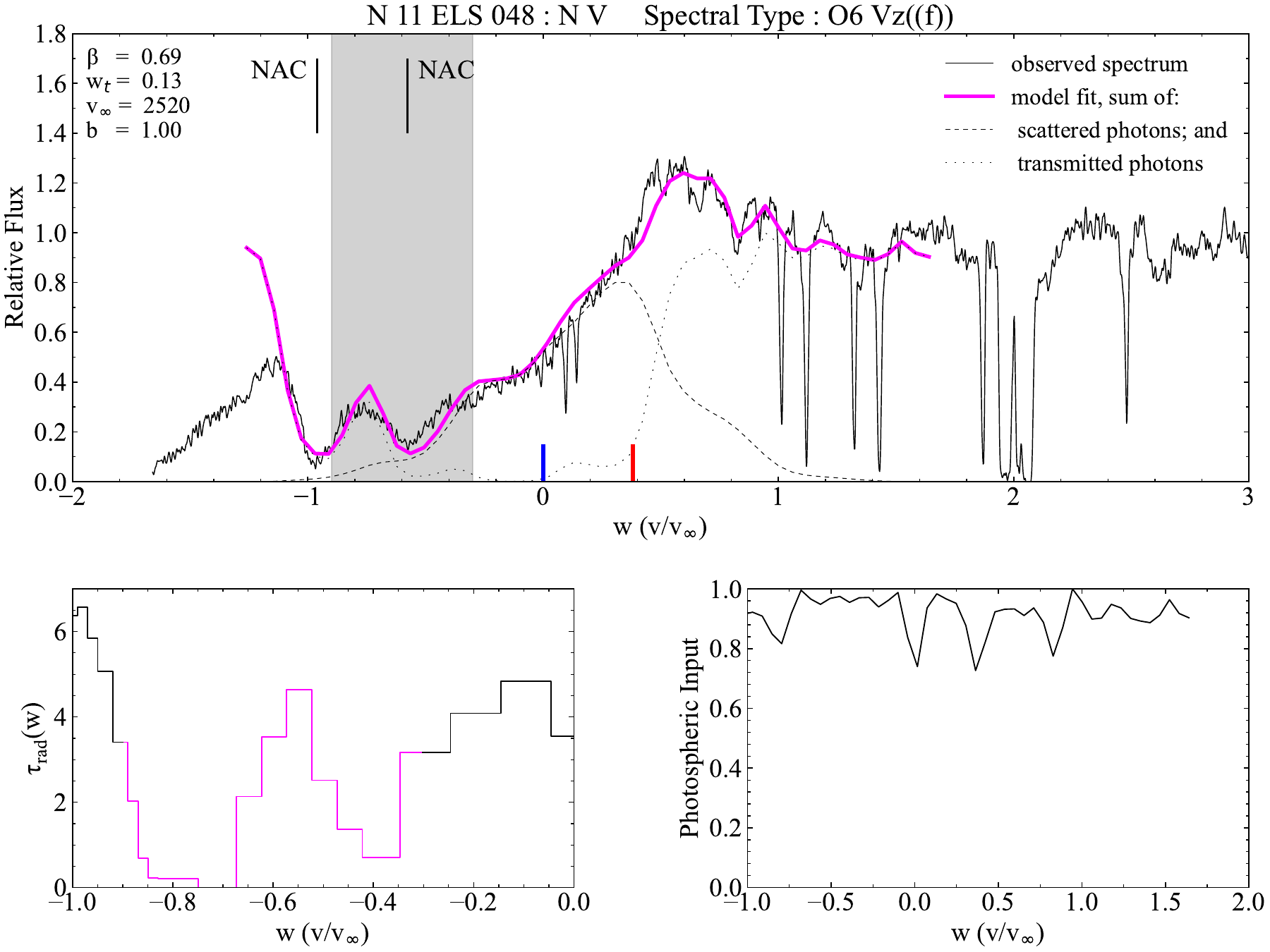} }
 \qquad
 \subfloat[ ]{\includegraphics[width=3.34in]{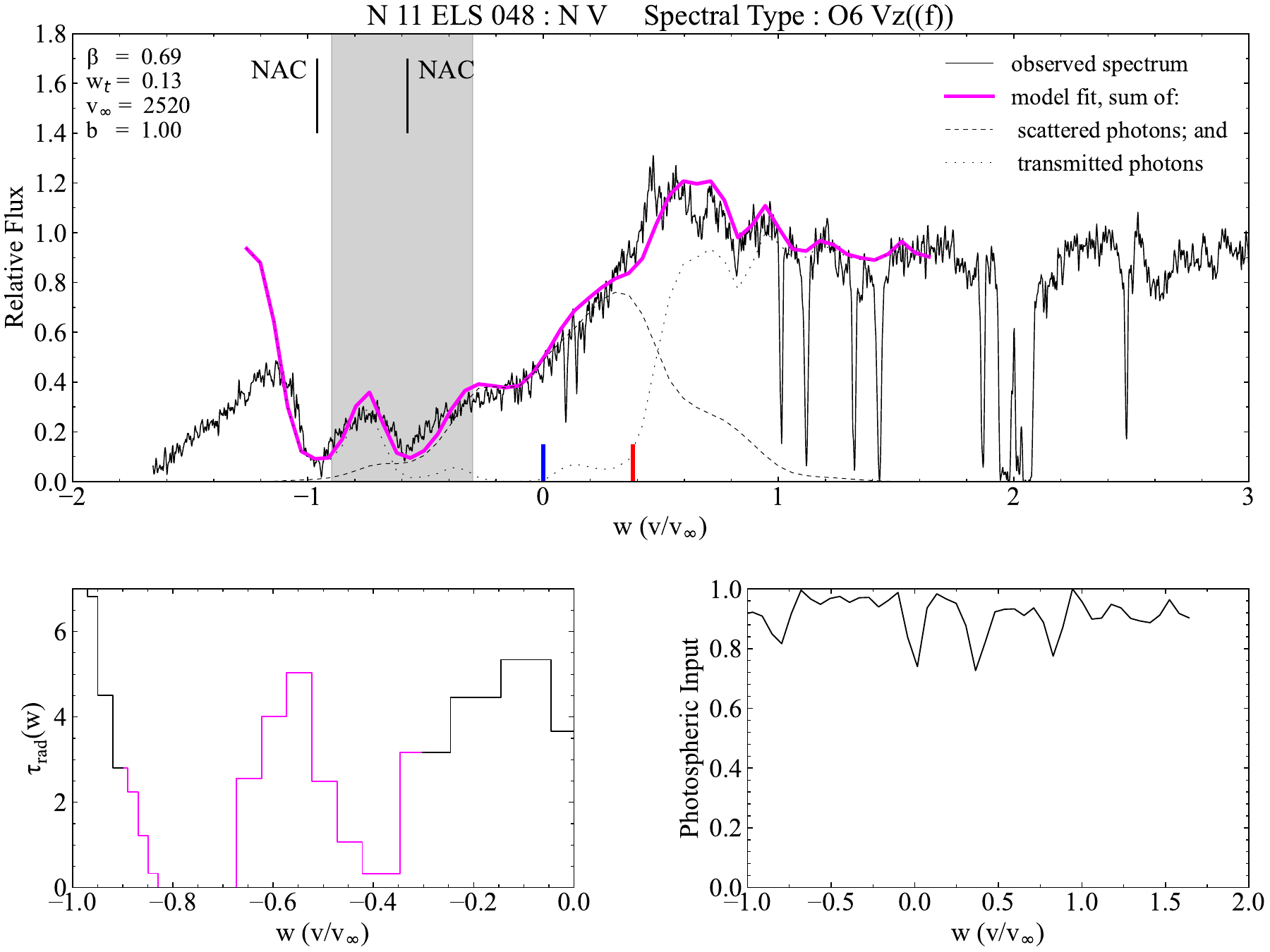} }
 \qquad
 \subfloat[ ]{\includegraphics[width=3.34in]{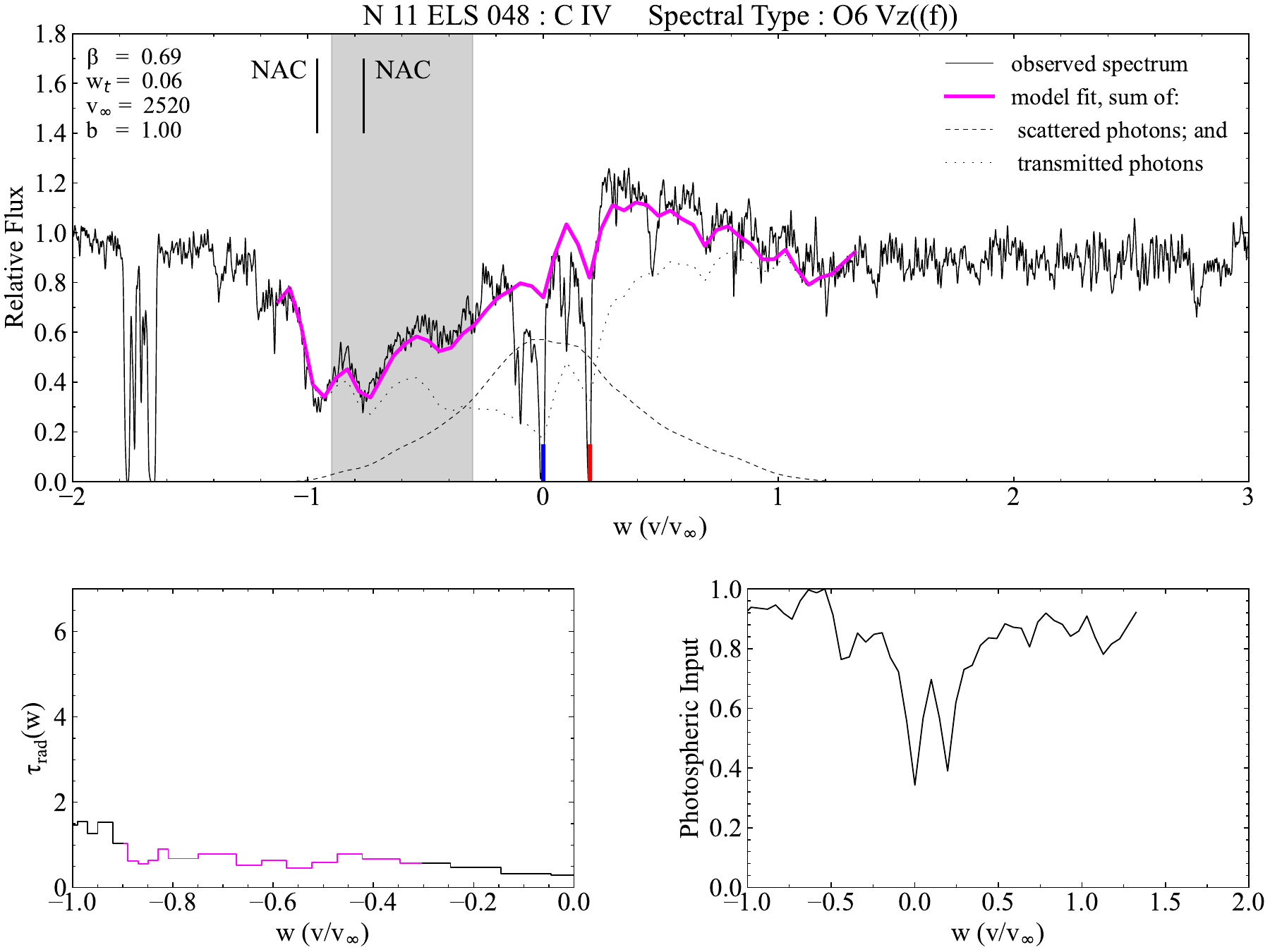} }
 \qquad
 \subfloat[ ]{\includegraphics[width=3.34in]{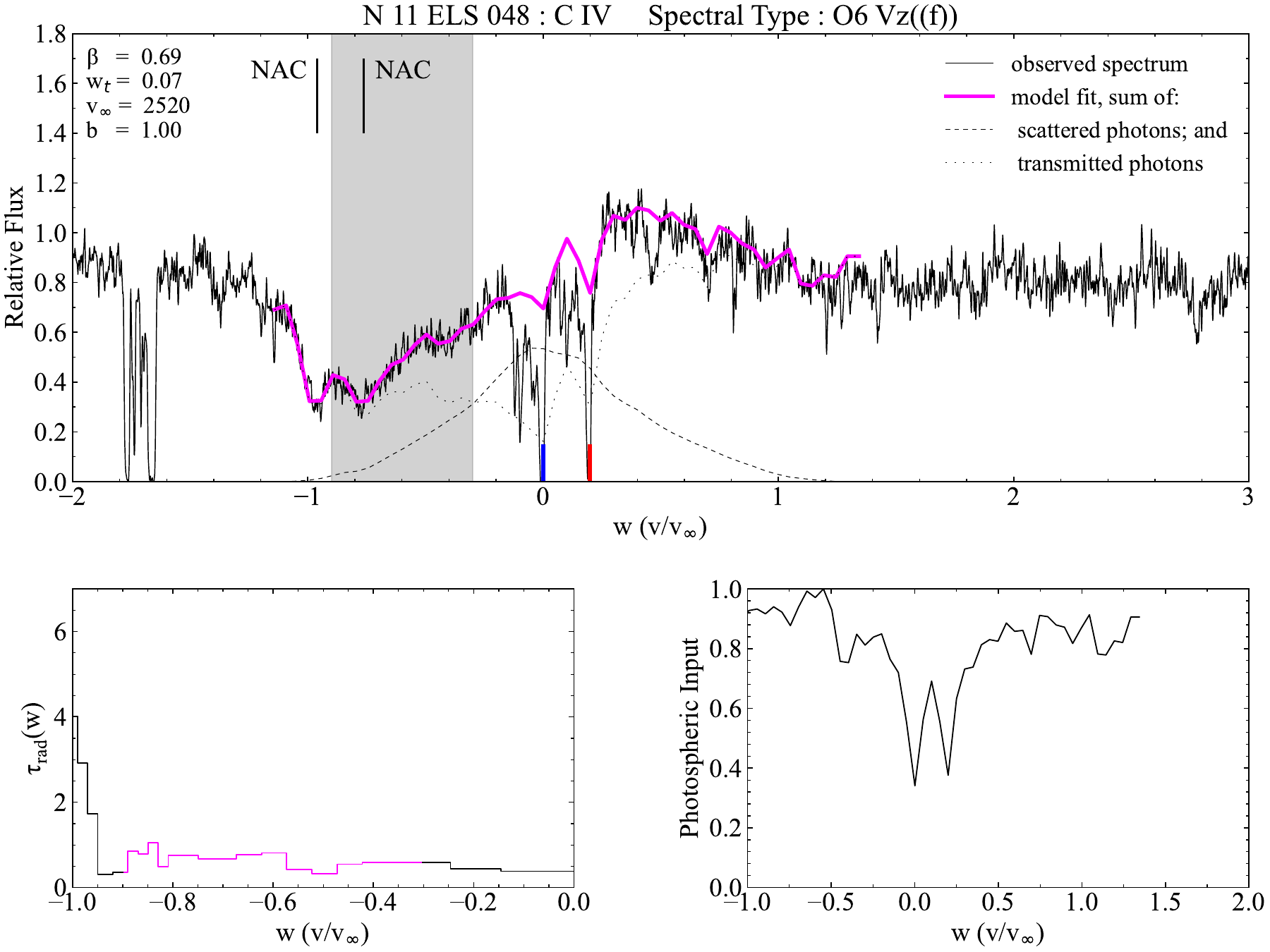} }
 \caption{SEI-derived model combined fit for (a) and (b) the \ion{N}{V} and (c) and (d) the \ion{C}{IV} doublet feature for two observations, dated (i) 2023 May 08 and (ii) 2023 May 09, of the LMC star N11 ELS 048.}
 \label{fig:11048_c4_SEI}
\end{center}
    \end{figure*}

   \begin{figure*}
\begin{center}
 \subfloat[ ]{\includegraphics[width=3.34in]{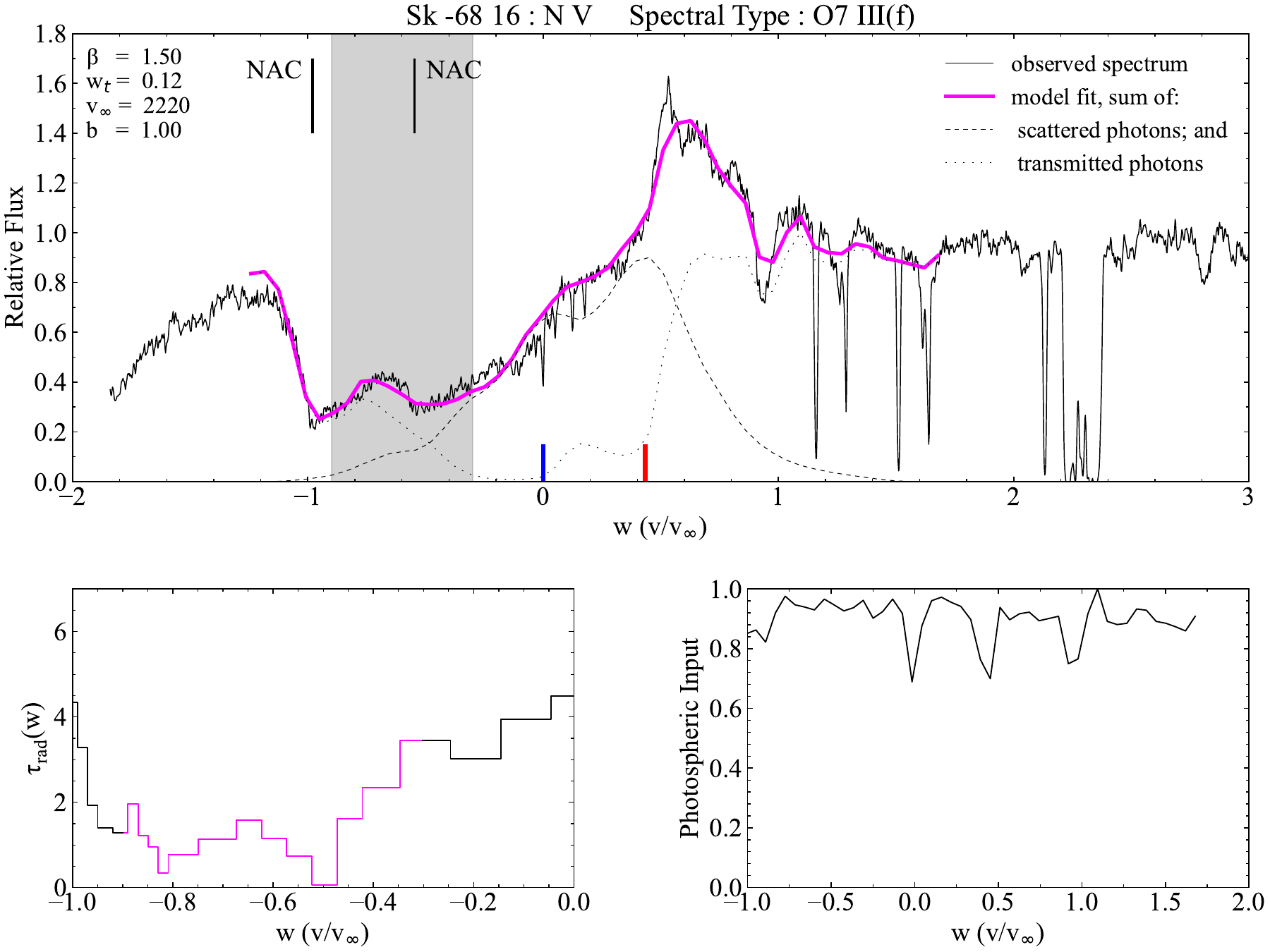} }
 \qquad
 \subfloat[ ]{\includegraphics[width=3.34in]{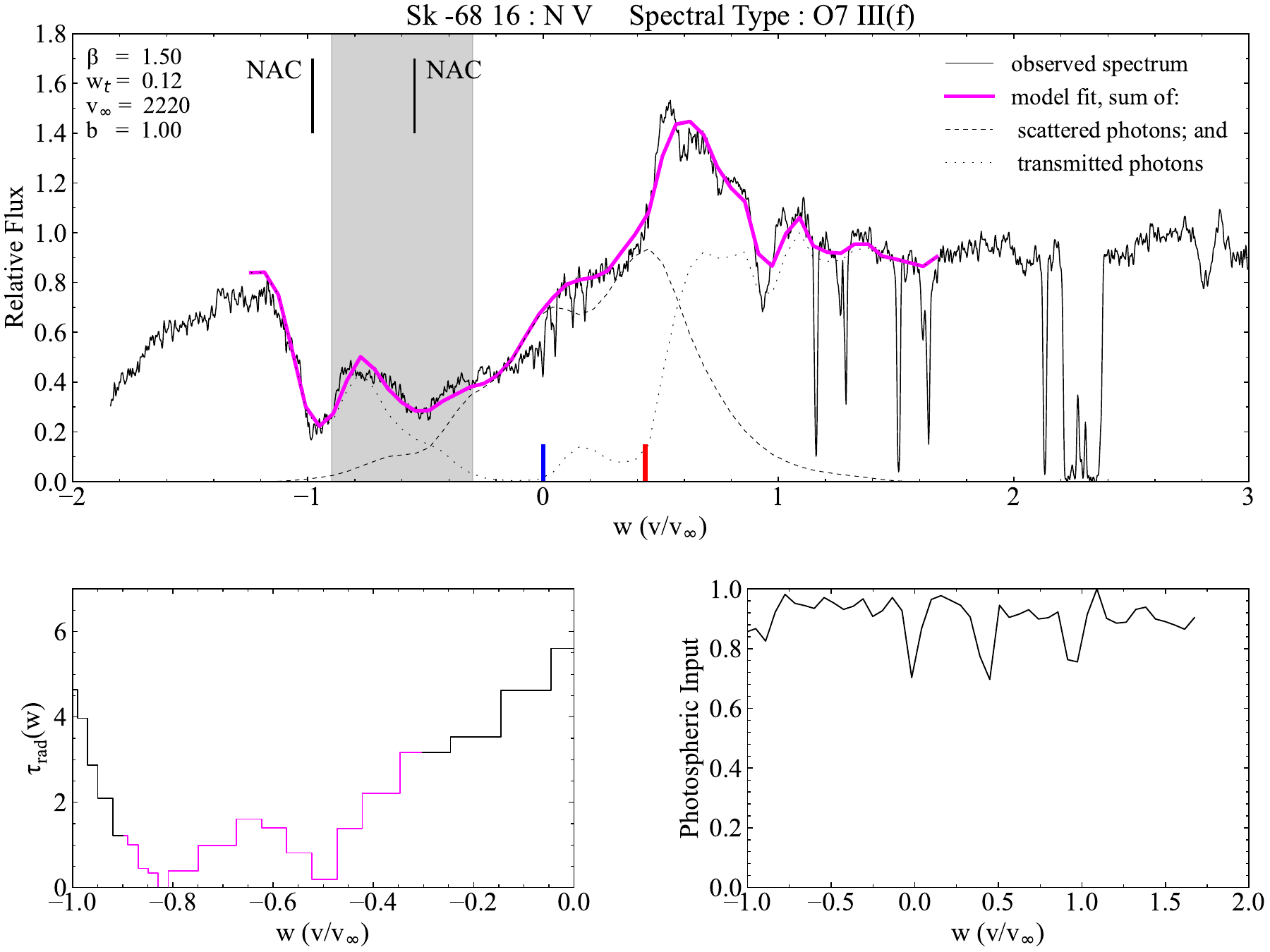} }
 \qquad
 \subfloat[ ]{\includegraphics[width=3.34in]{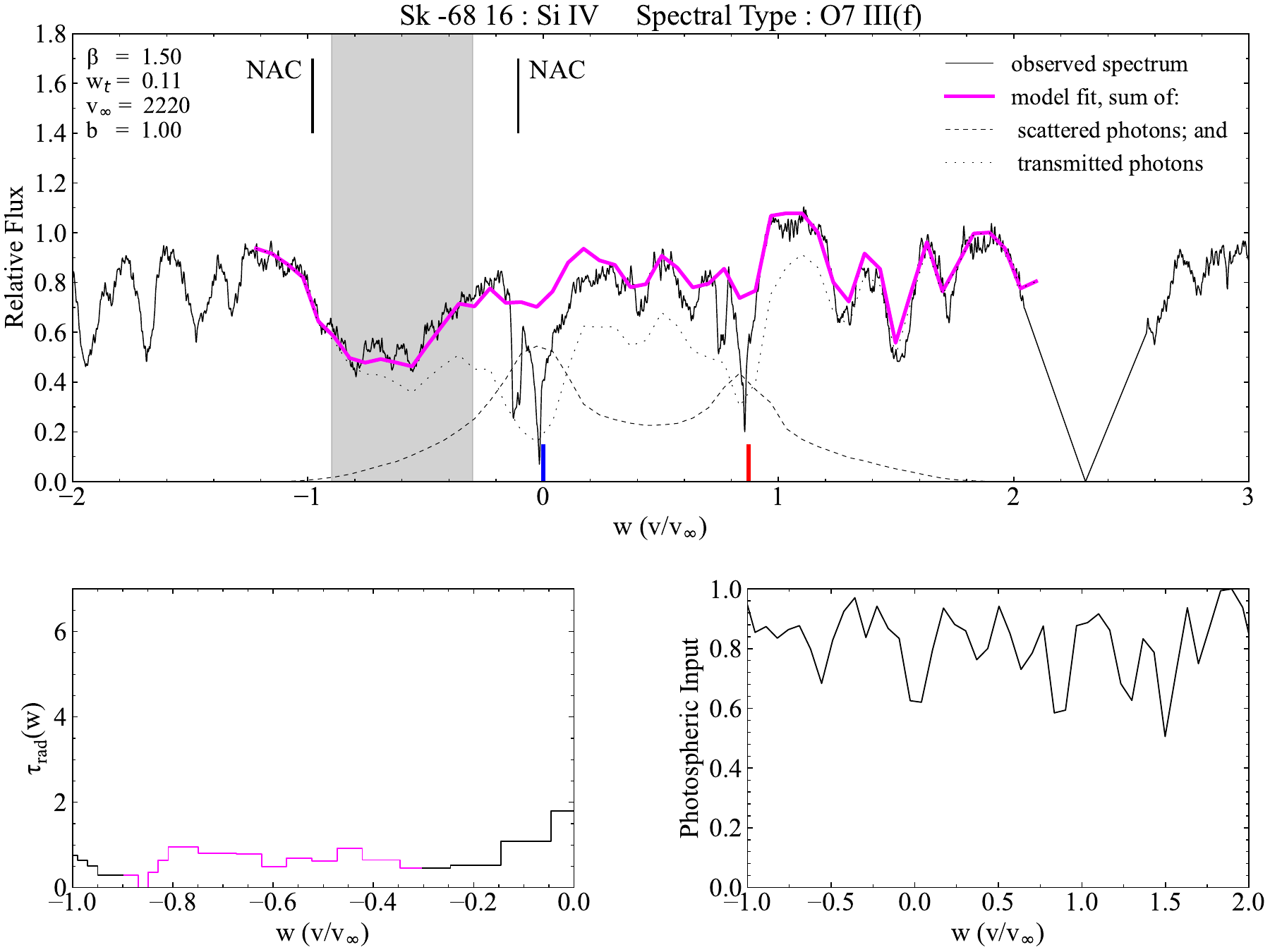} }
 \qquad
 \subfloat[ ]{\includegraphics[width=3.34in]{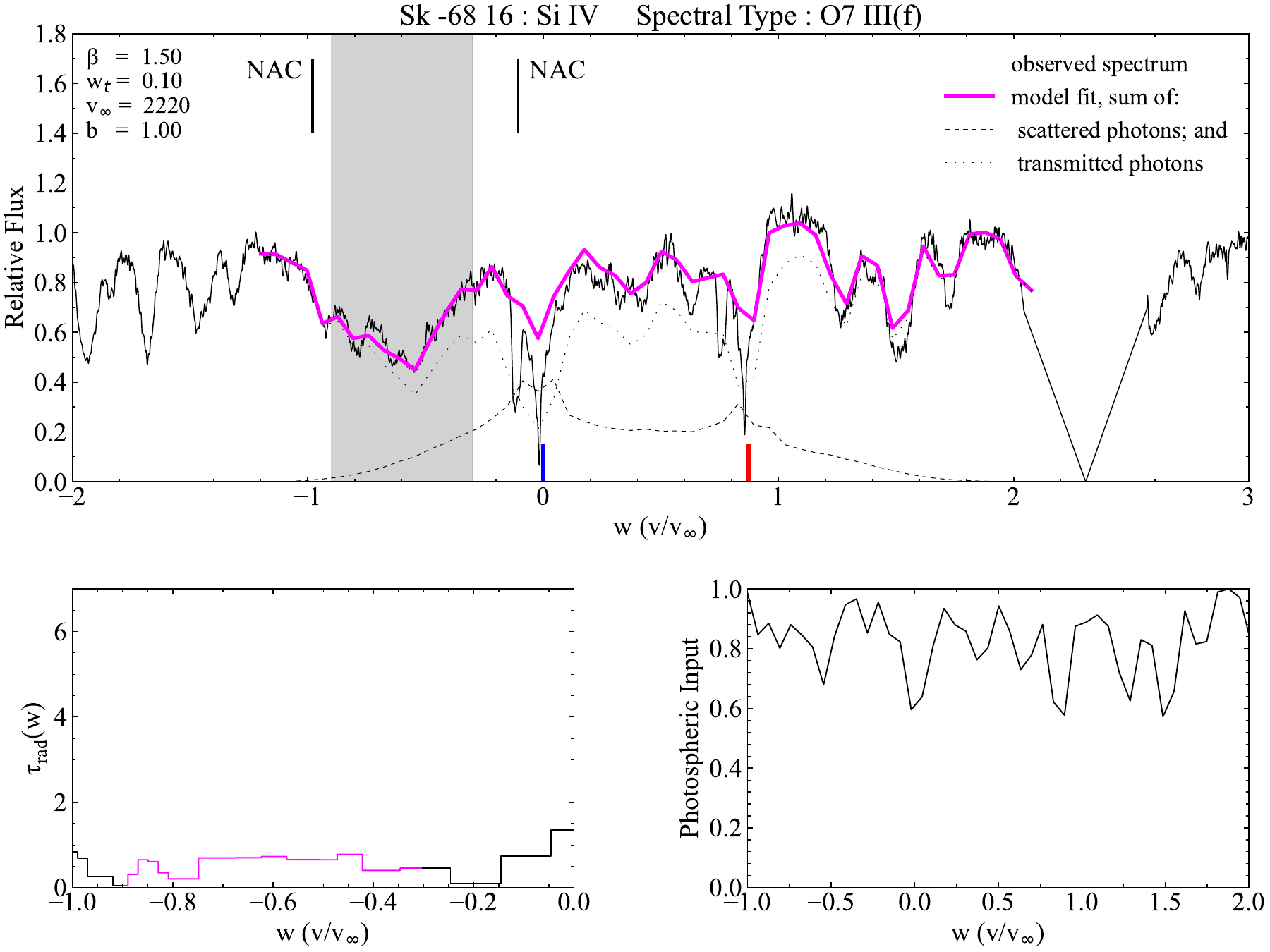} }
 \qquad
 \subfloat[ ]{\includegraphics[width=3.34in]{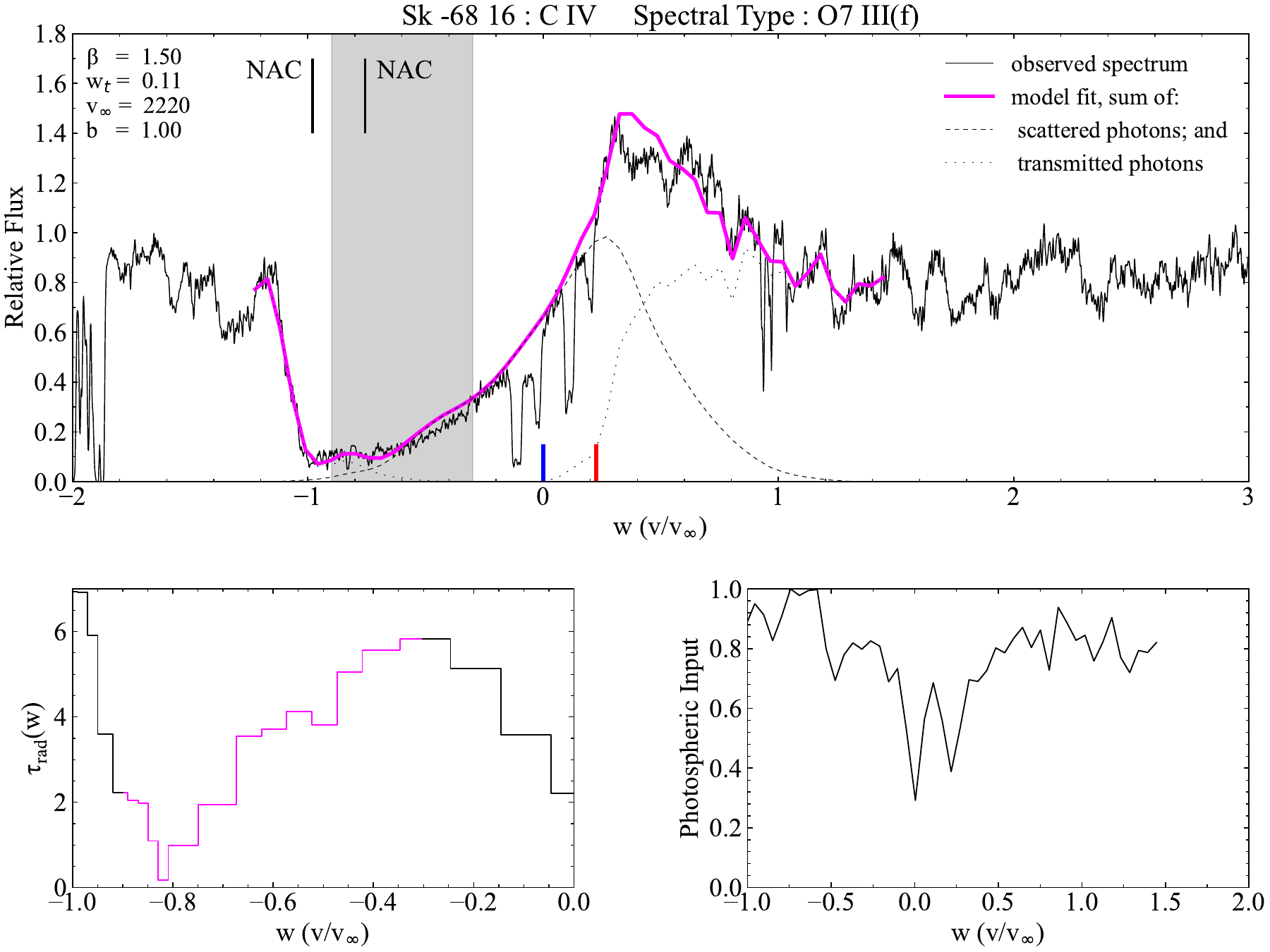} }
 \qquad
 \subfloat[ ]{\includegraphics[width=3.34in]{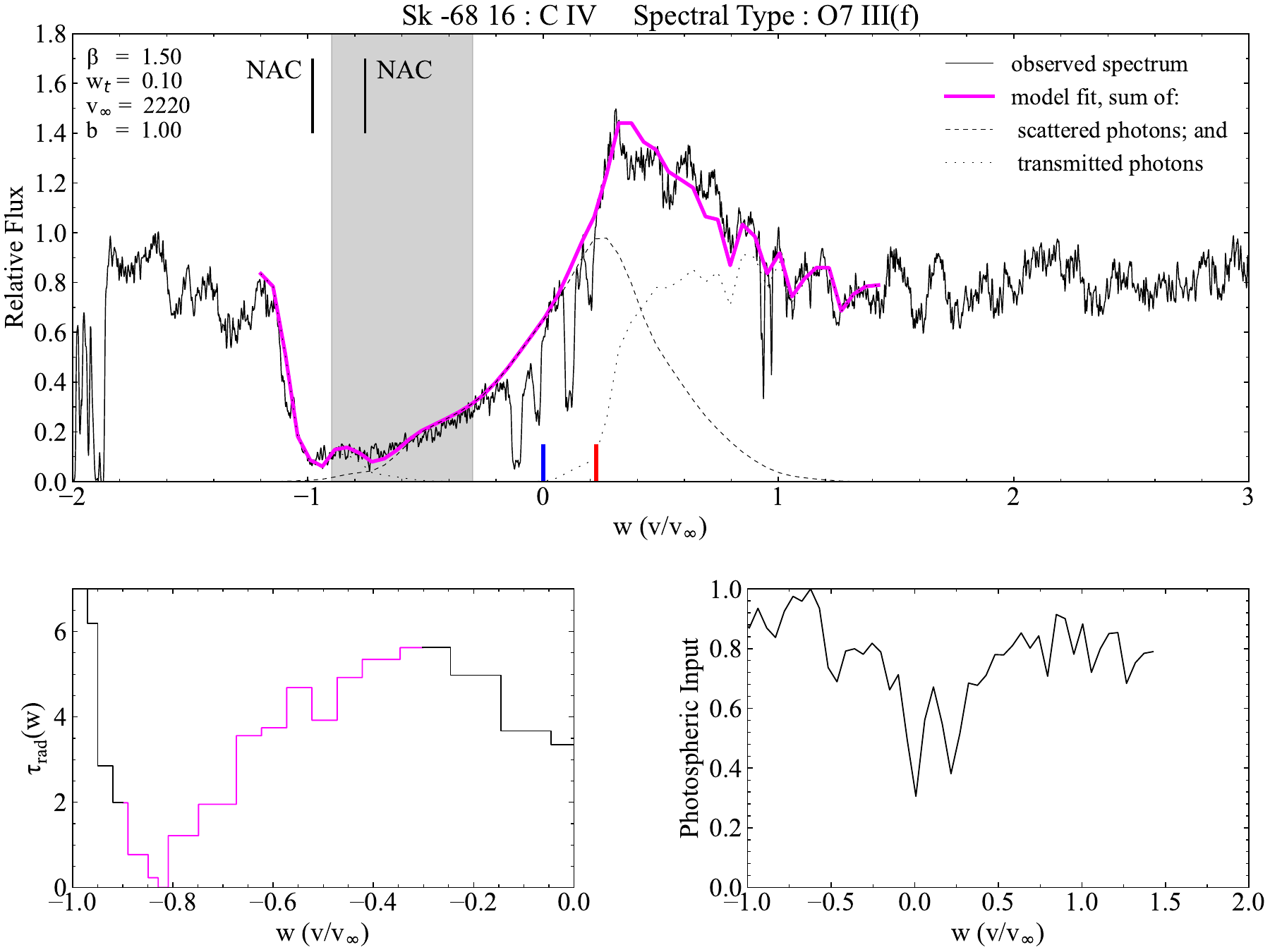} }
 \caption{SEI-derived model fits for (a) and (b) the \ion{N}{V}, (c) and (d) the \ion{Si}{IV} and (e) and (f) the \ion{C}{IV} doublet feature for two observations, dated (a) 2020 October 22 and (b) 2020 October 24, of the LMC star Sk -68 16.}
 \label{fig:6816_n5_c4_SEI}
\end{center}
    \end{figure*}

   \begin{figure*}
\begin{center}
 \subfloat[ ]{\includegraphics[width=3.34in]{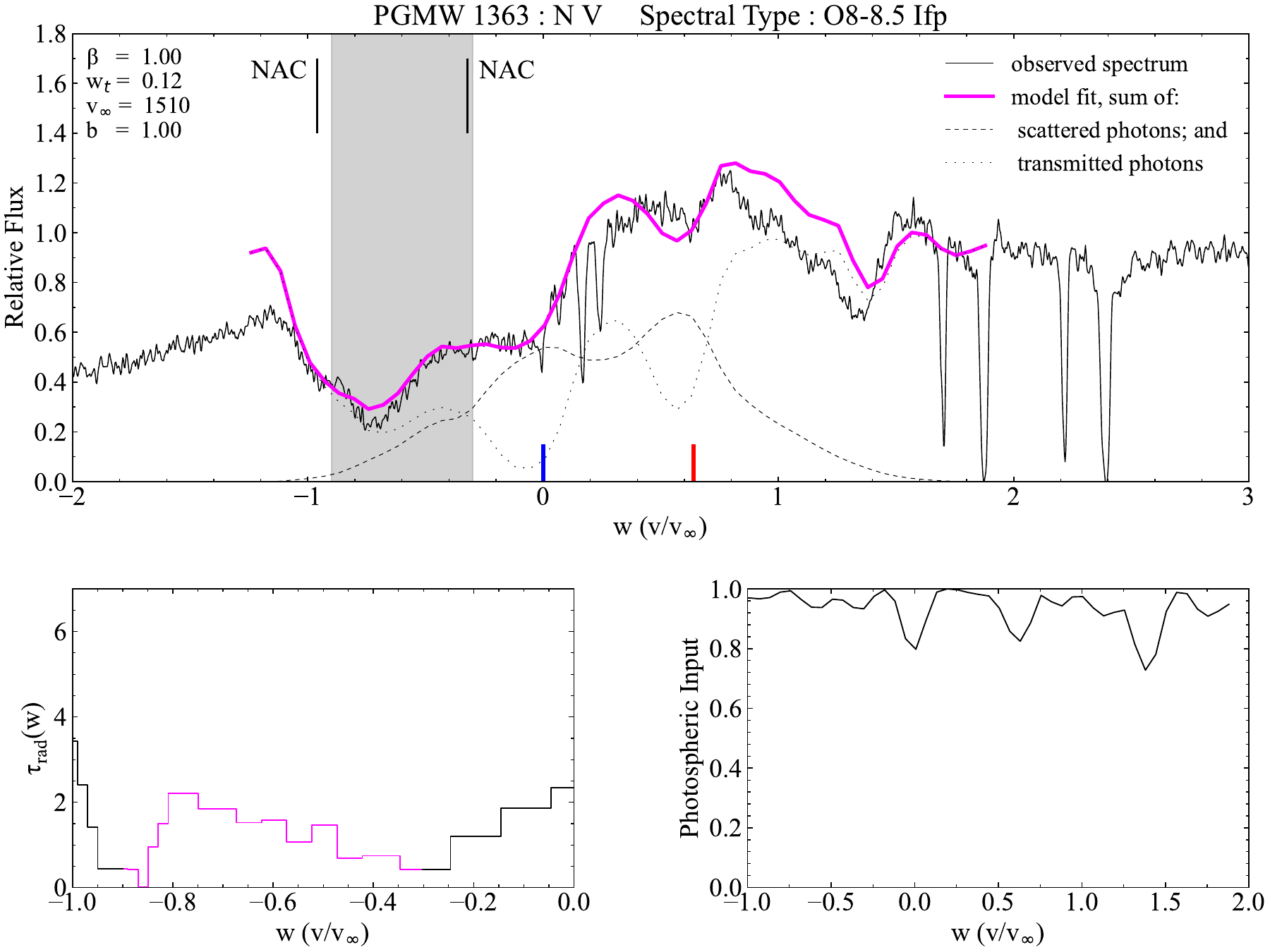} }
 \qquad
 \subfloat[ ]{\includegraphics[width=3.34in]{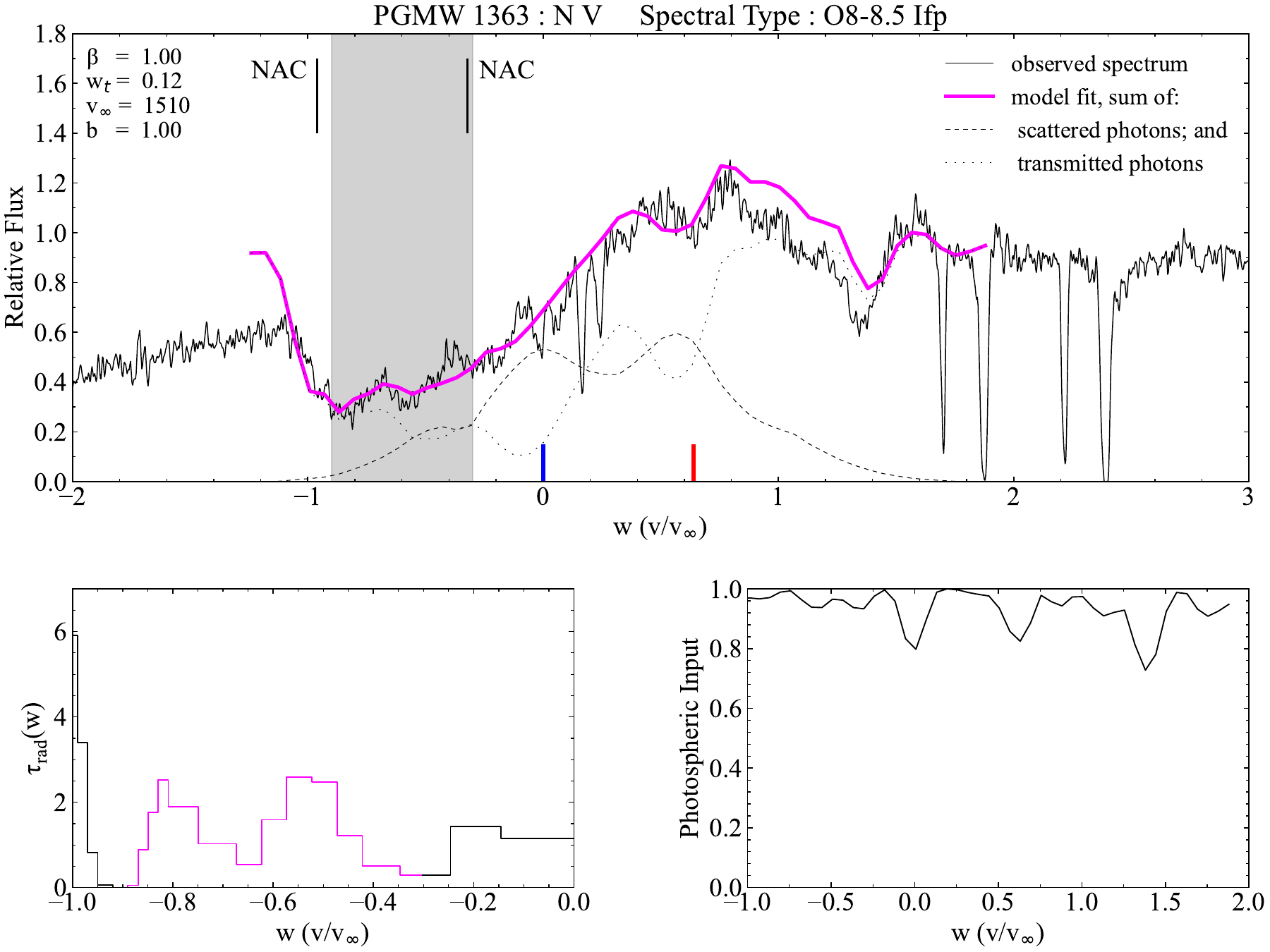} }
 \caption{SEI-derived model combined fit for both components of the \ion{N}{V} doublet feature for two observations, dated (a) 2021 February 15 and (b) 2021 April 08, of the LMC star PGMW 1363.}
 \label{fig:1363_n5_SEI}
\end{center}
    \end{figure*}

   \begin{figure*}
\begin{center}
 \subfloat[ ]{\includegraphics[width=3.34in]{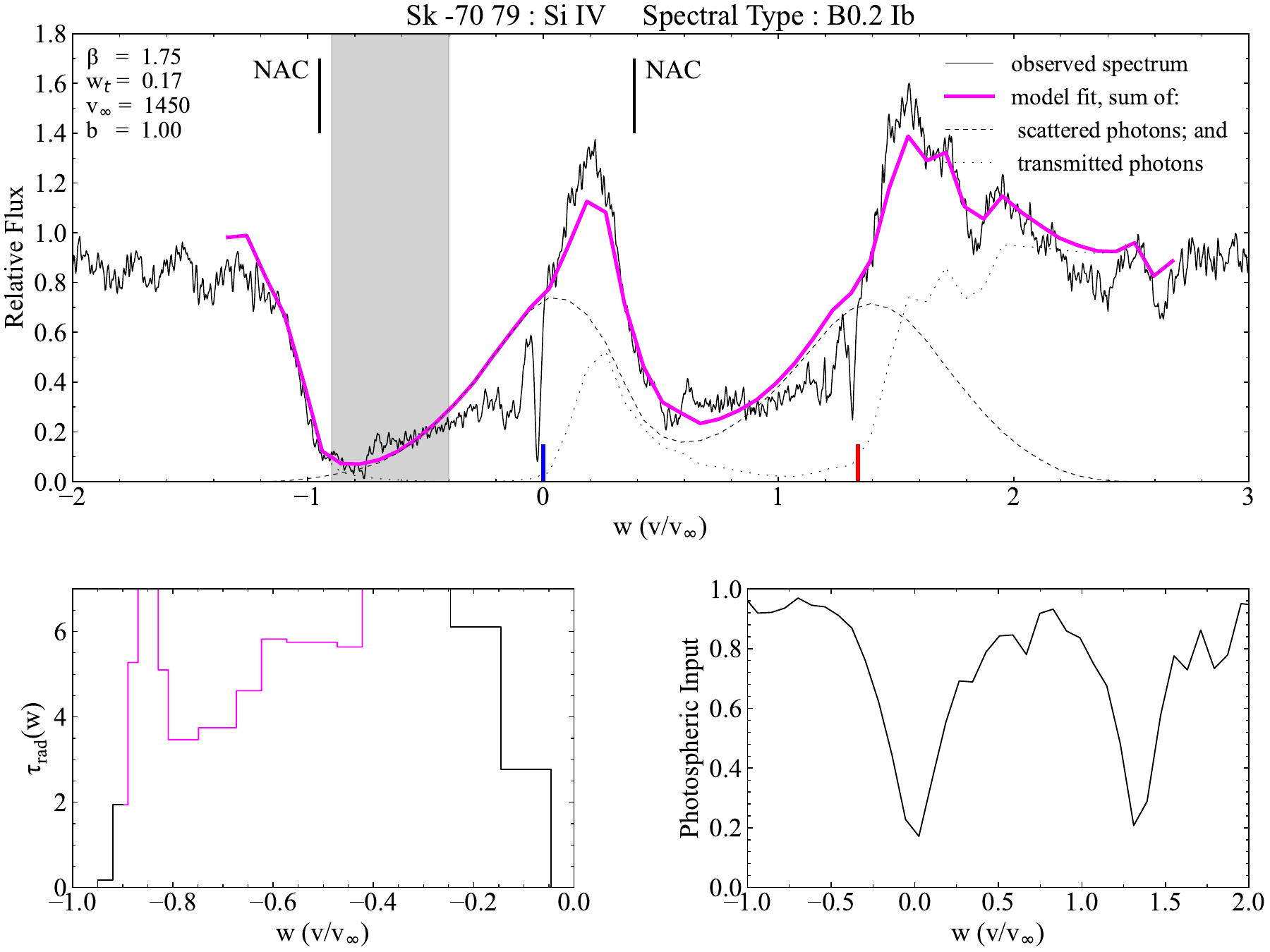} }
 \qquad
 \subfloat[ ]{\includegraphics[width=3.34in]{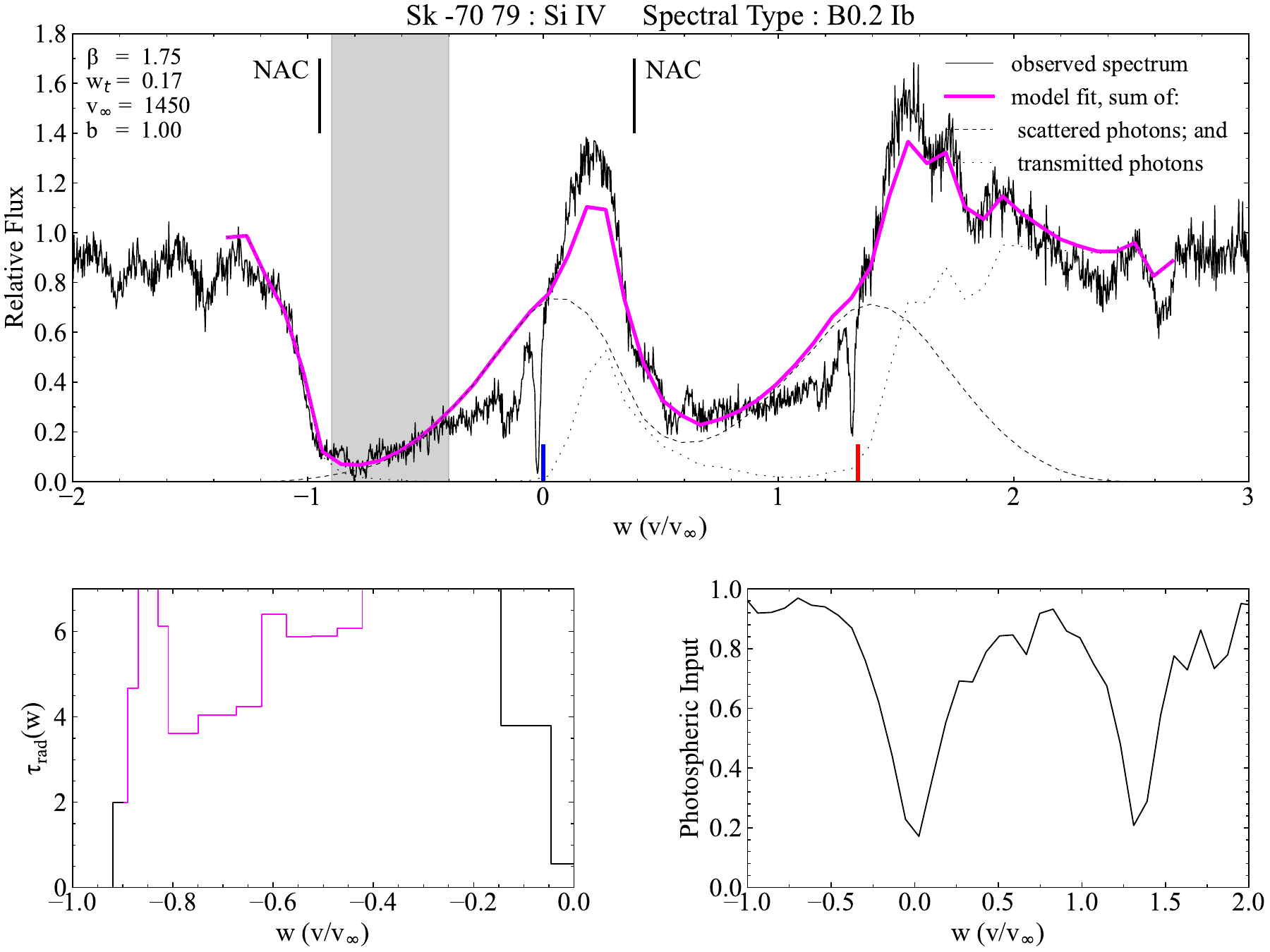} }
 \caption{SEI-derived model fits for the \ion{Si}{IV} doublet feature for two observations, dated (a) 2017 June 11 and (b) 2017 June 12, of the LMC star Sk -70 79. The difficulty in obtaining an accurate model fit for lower wind velocities in these flat, near-saturated absorption profiles may be noted and is discussed in the main text.}
 \label{fig:7079_si4_SEI}
\end{center}
    \end{figure*}

   \begin{figure*}
\begin{center}
 \subfloat[ ]{\includegraphics[width=3.34in]{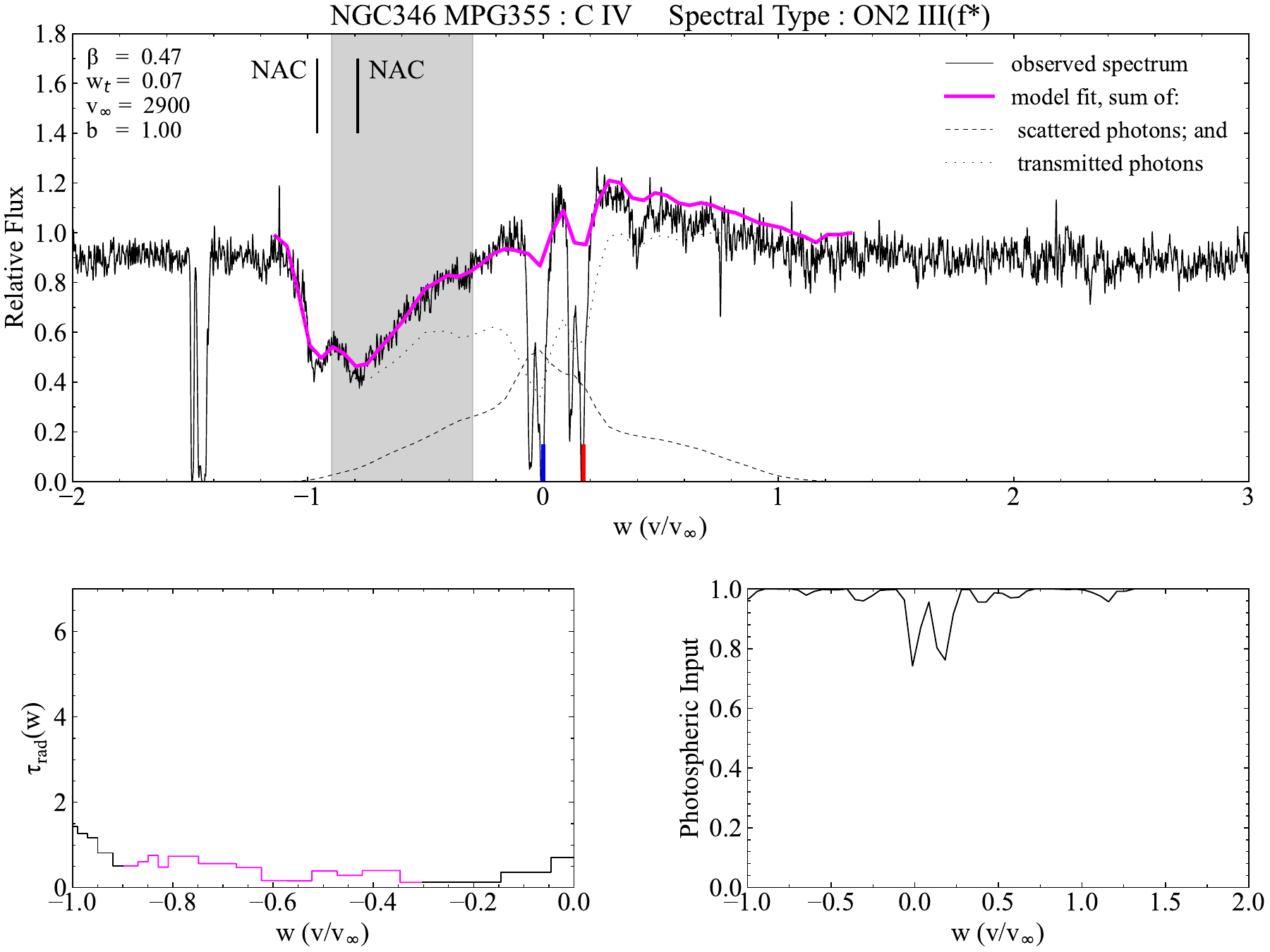} }
 \qquad
 \subfloat[ ]{\includegraphics[width=3.34in]{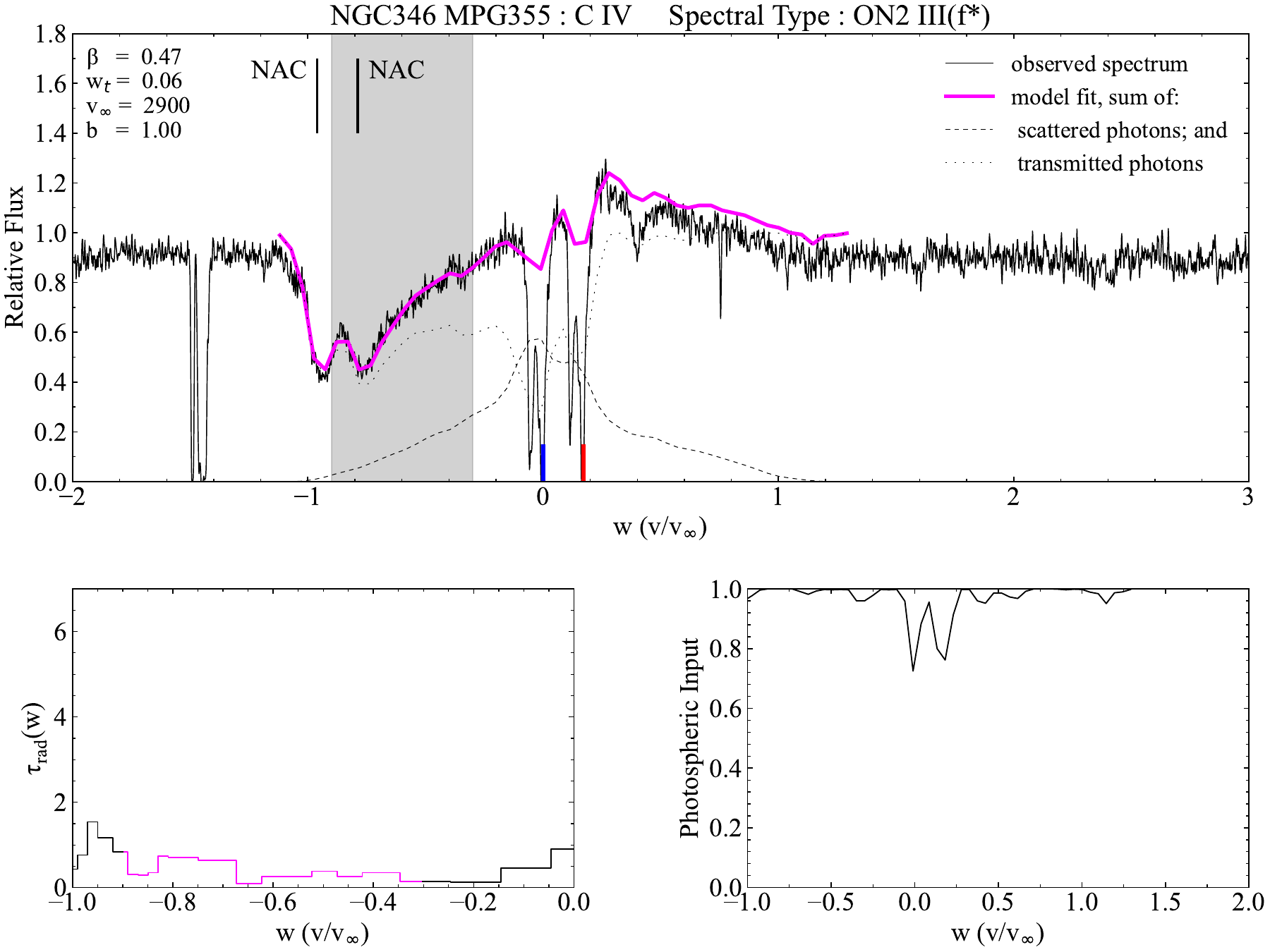} }
 \caption{SEI-derived model fits for the \ion{C}{IV} doublet feature for two observations, dated (a) 1998 August 08 and (b) 2020 September 01, of the SMC star NGC 346 MPG 355.}
 \label{fig:355_c4_SEI}
\end{center}
    \end{figure*}

   \begin{figure*}
\begin{center}
 \subfloat[ ]{\includegraphics[width=3.34in]{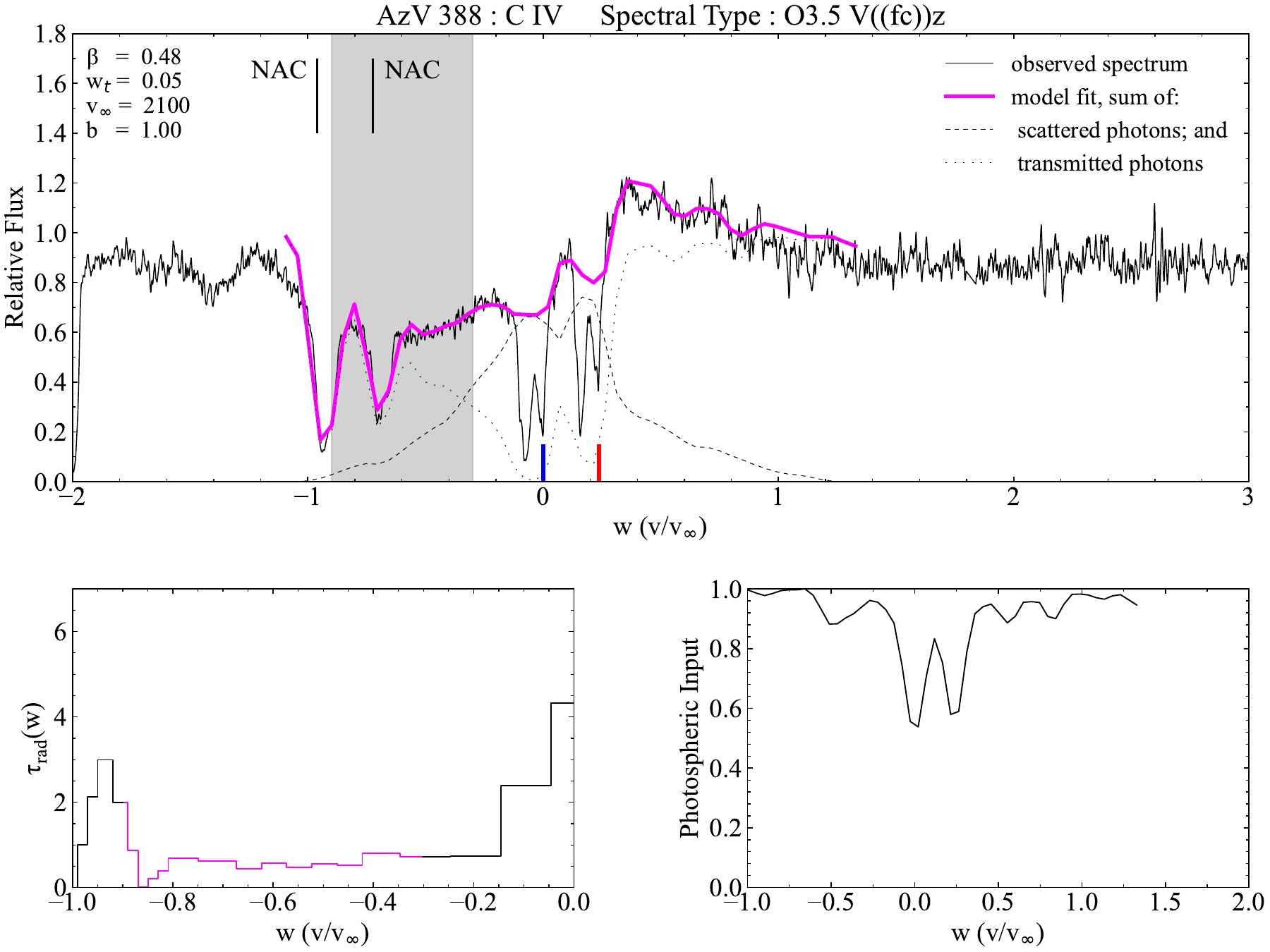} }
 \qquad
 \subfloat[ ]{\includegraphics[width=3.34in]{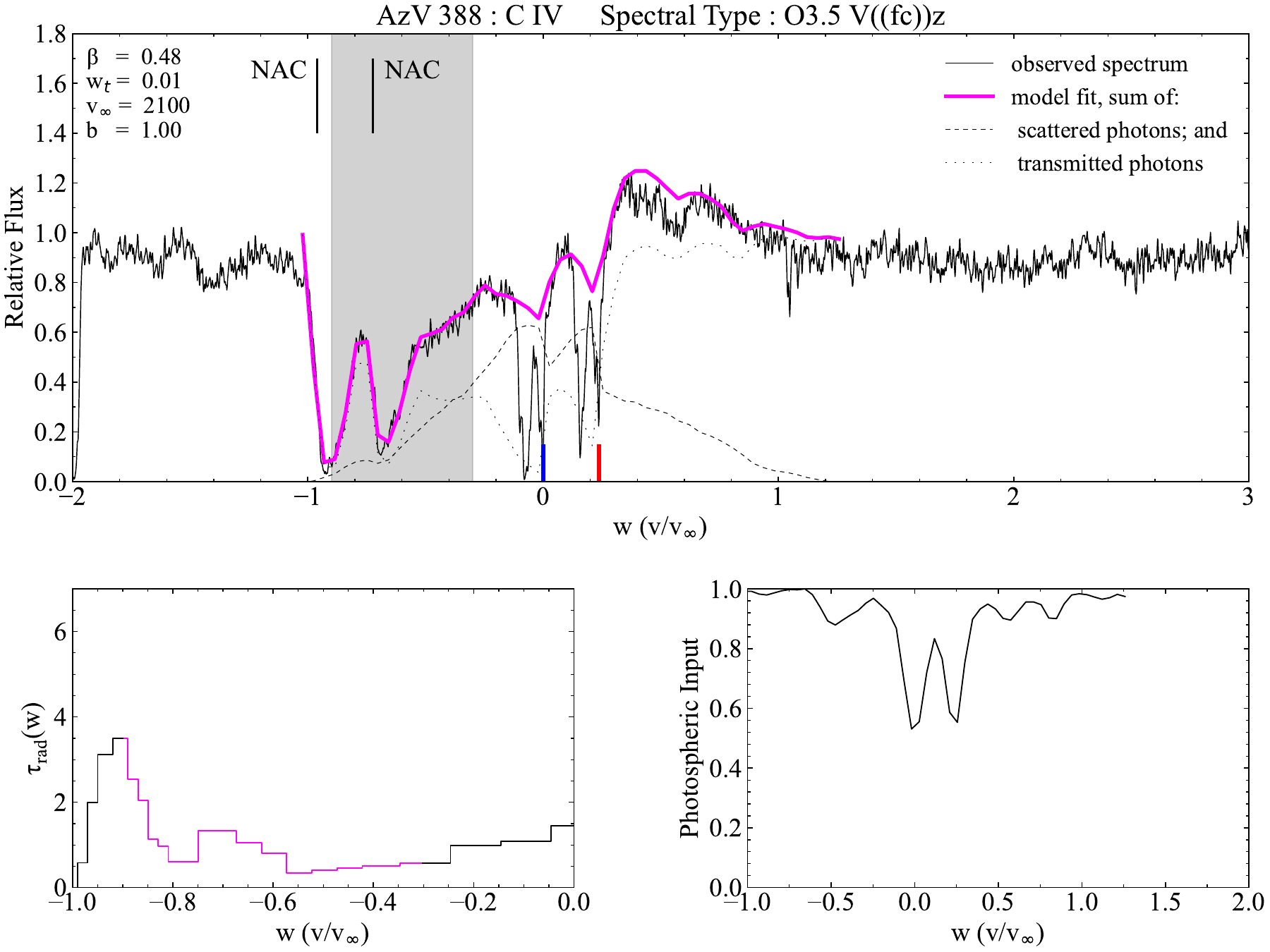} }
 \caption{SEI-derived model fits for the \ion{C}{IV} doublet feature for two observations, dated (a) 2009 December 21 and (b) 2015 October 26, of the SMC star AzV 388. The deep NAC features can be observed in each profile, as well as the fact that each broadens significantly between the first observation and the second.}
 \label{fig:388_c4_SEI}
\end{center}
    \end{figure*}

   \begin{figure*}
\begin{center}
 \subfloat[ ]{\includegraphics[width=3.34in]{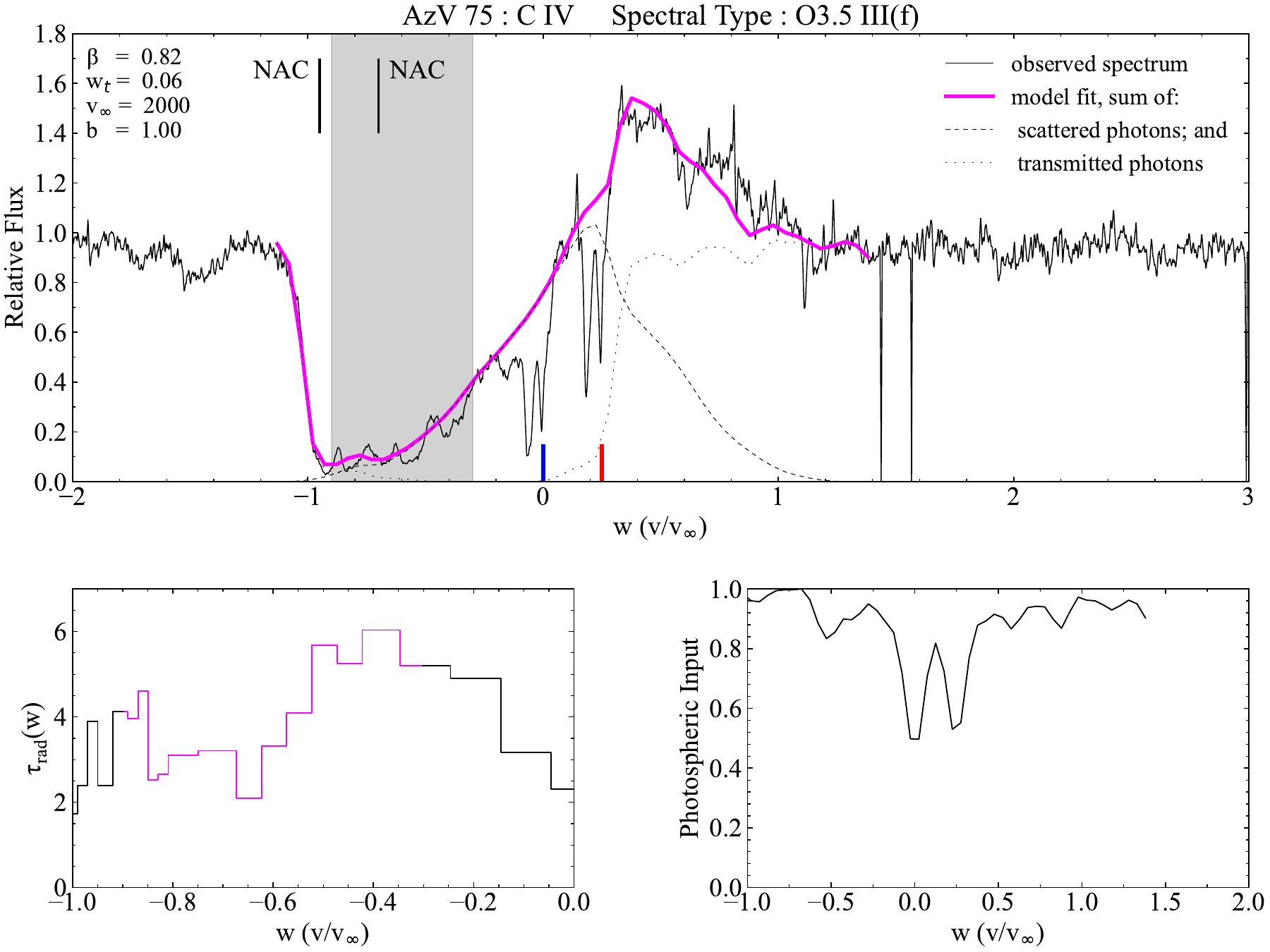} }
 \qquad
 \subfloat[ ]{\includegraphics[width=3.34in]{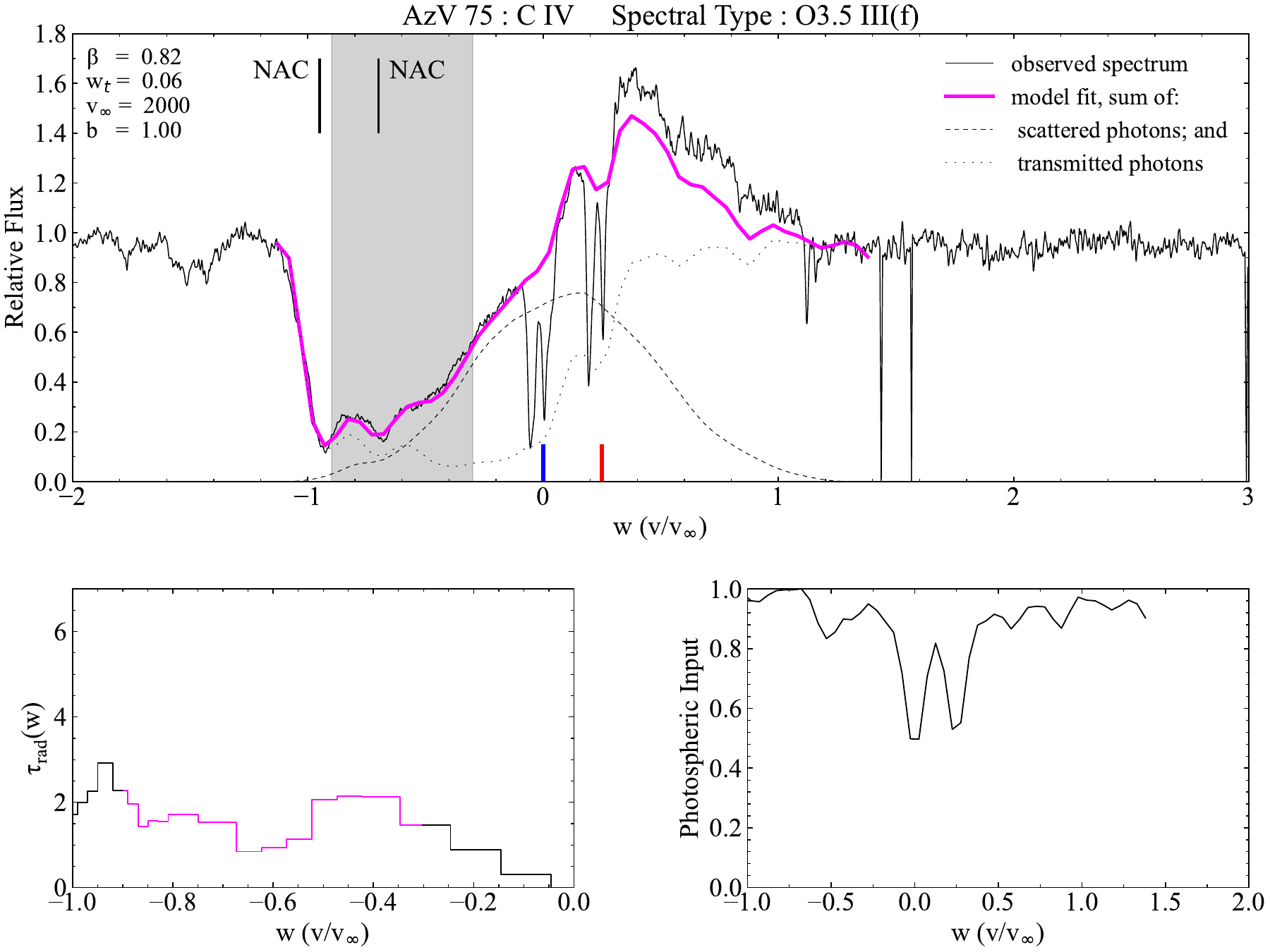} }
 \caption{SEI-derived model fits for the \ion{C}{IV} doublet feature for two representative observations, dated (a) 2012 September 11 and (b) 2013 March 15, of the SMC star AzV 75. These two observations show extremes of variation observed in average optical depth of the stellar wind.}
 \label{fig:av75_c4_SEI}
\end{center}
    \end{figure*}

   \begin{figure*}
\begin{center}
 \subfloat[ ]{\includegraphics[width=3.34in]{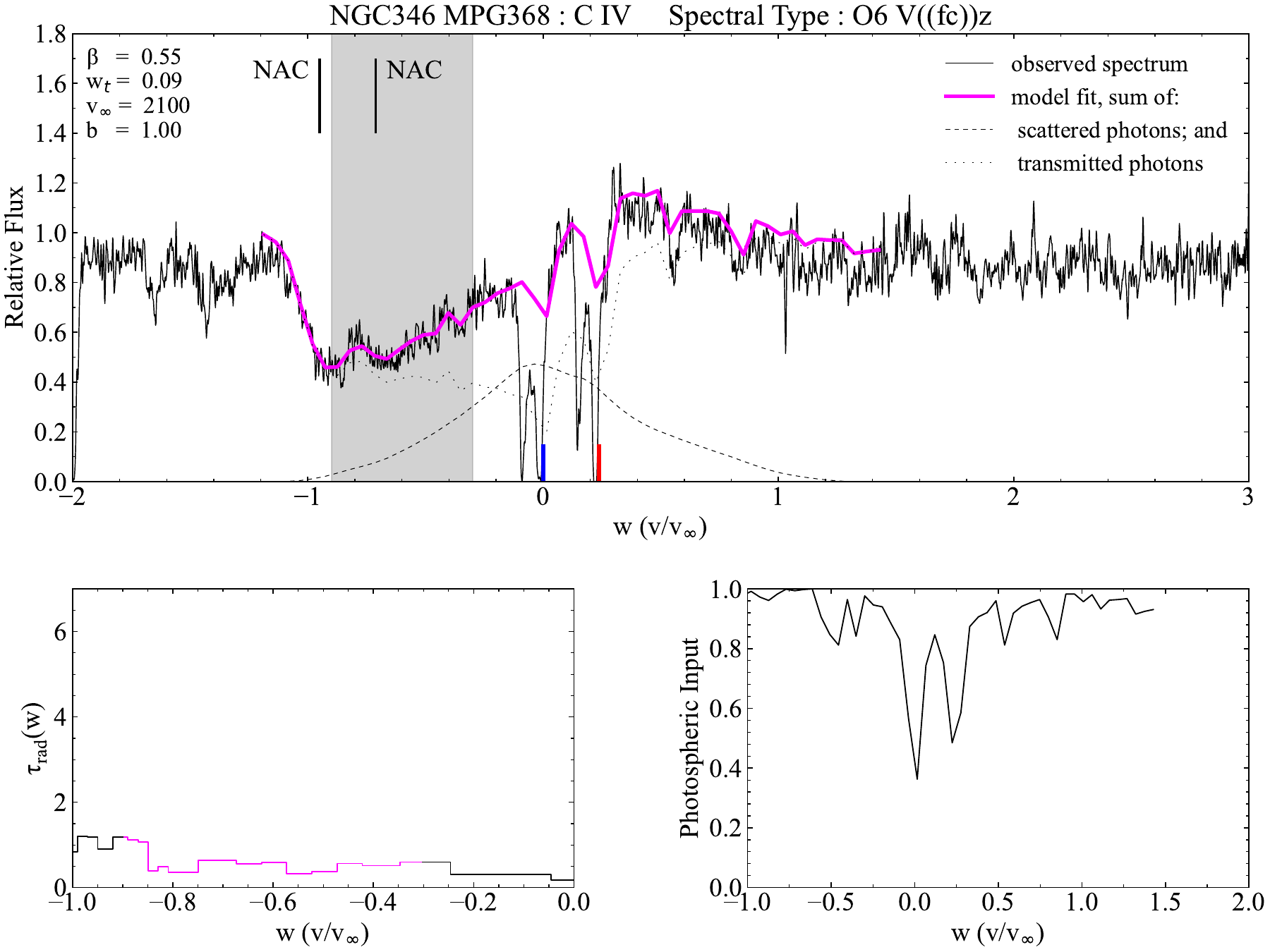} }
 \qquad
 \subfloat[ ]{\includegraphics[width=3.34in]{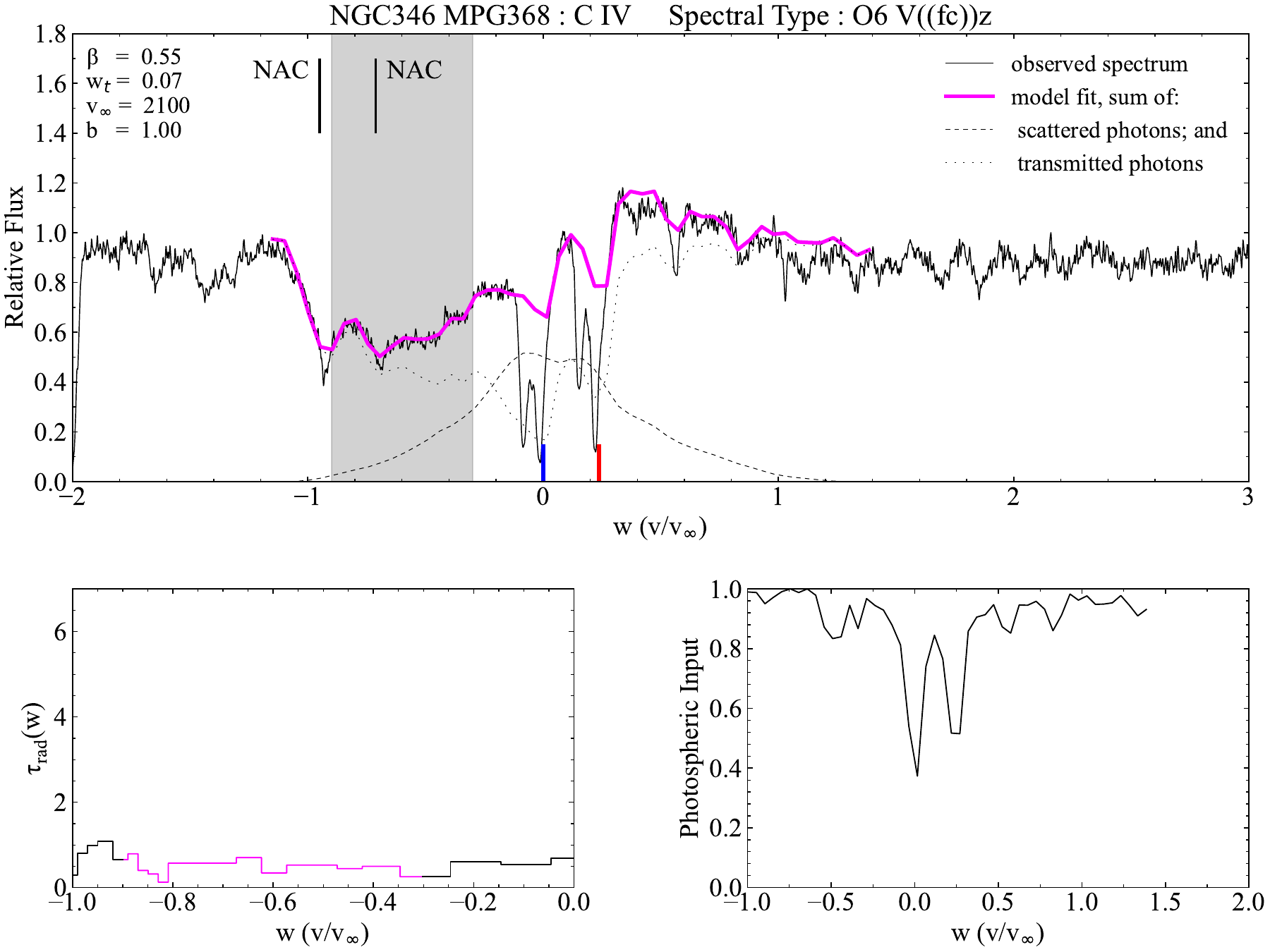} }
 \caption{SEI-derived model fits for the \ion{C}{IV} doublet feature for two observations, dated (a) 1998 August 08 and (b) 2020 July 03, of the SMC star NGC 346 MPG 368.}
 \label{fig:368_c4_SEI}
\end{center}
    \end{figure*}

   \begin{figure*}
\begin{center}
 \subfloat[ ]{\includegraphics[width=3.34in]{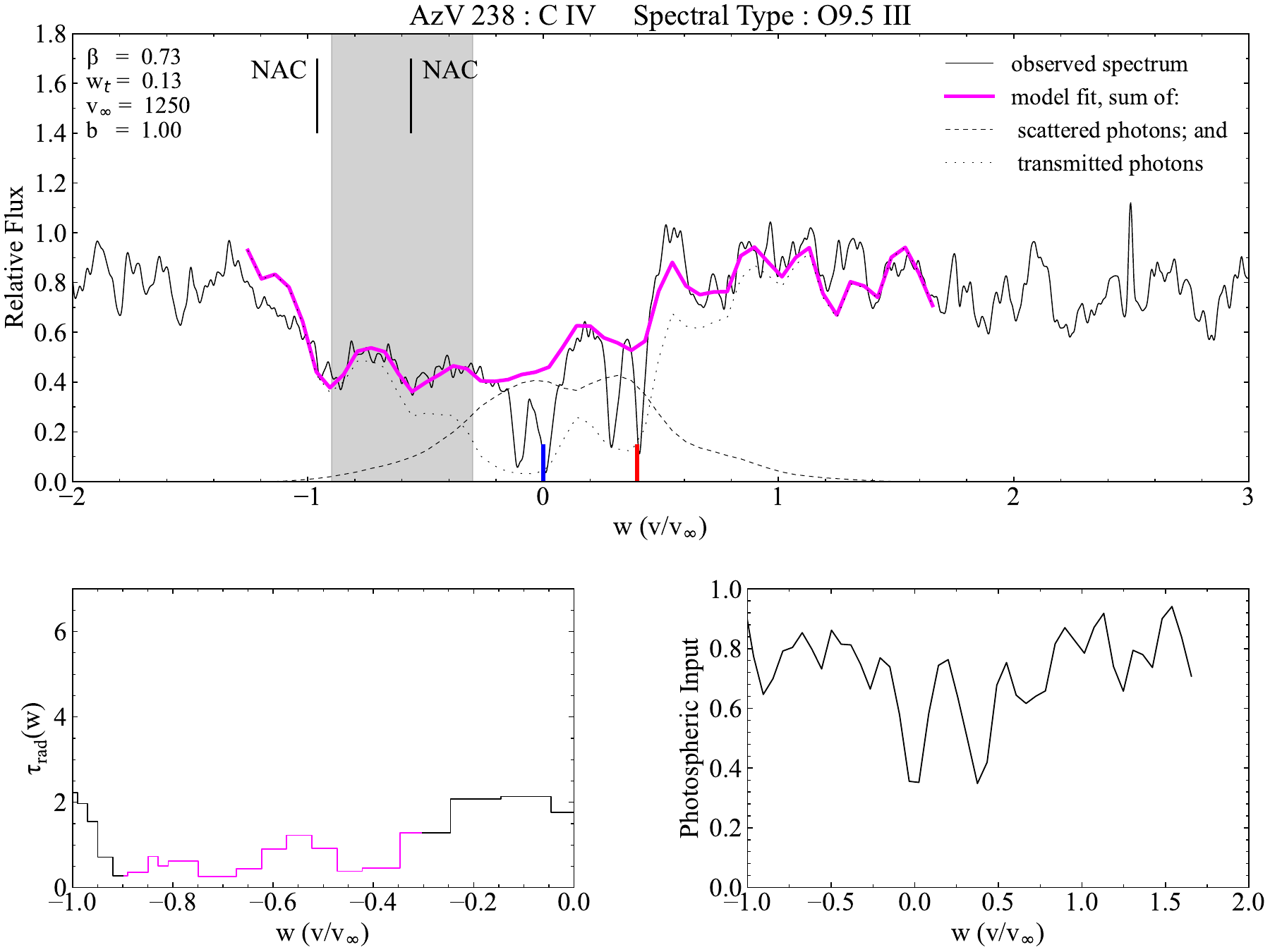} }
 \qquad
 \subfloat[ ]{\includegraphics[width=3.34in]{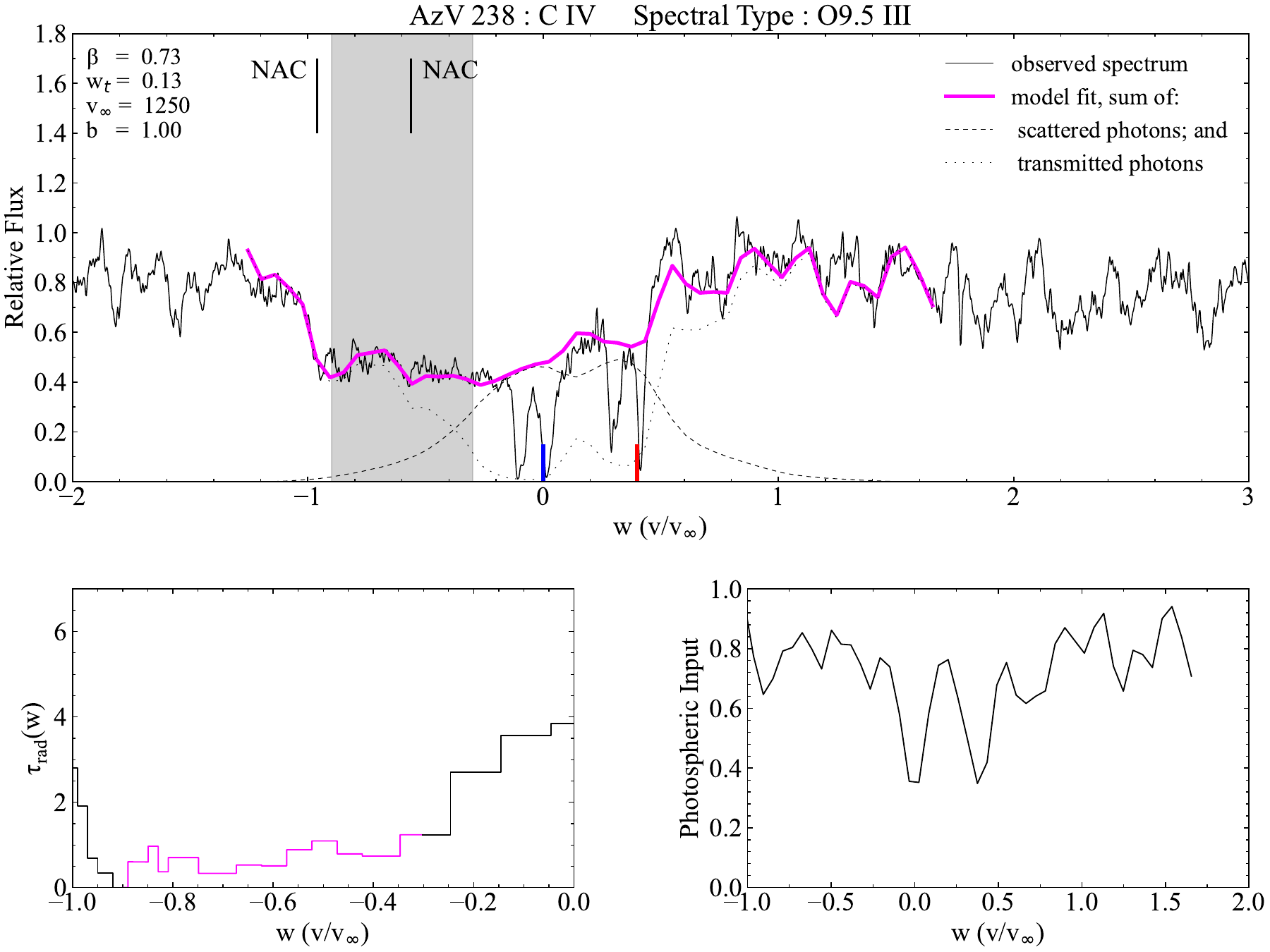} }
 \caption{SEI-derived model fits for the \ion{C}{IV} doublet feature for two observations, dated (a) 2023 March 11 and (b) 2023 April 03, of the SMC star AzV 238. The lack of significant emission features, despite the evident wind-formed absorption profile, can be noted.}
 \label{fig:238_c4_SEI}
\end{center}
    \end{figure*}

   \begin{figure*}
\begin{center}
 \subfloat[ ]{\includegraphics[width=3.34in]{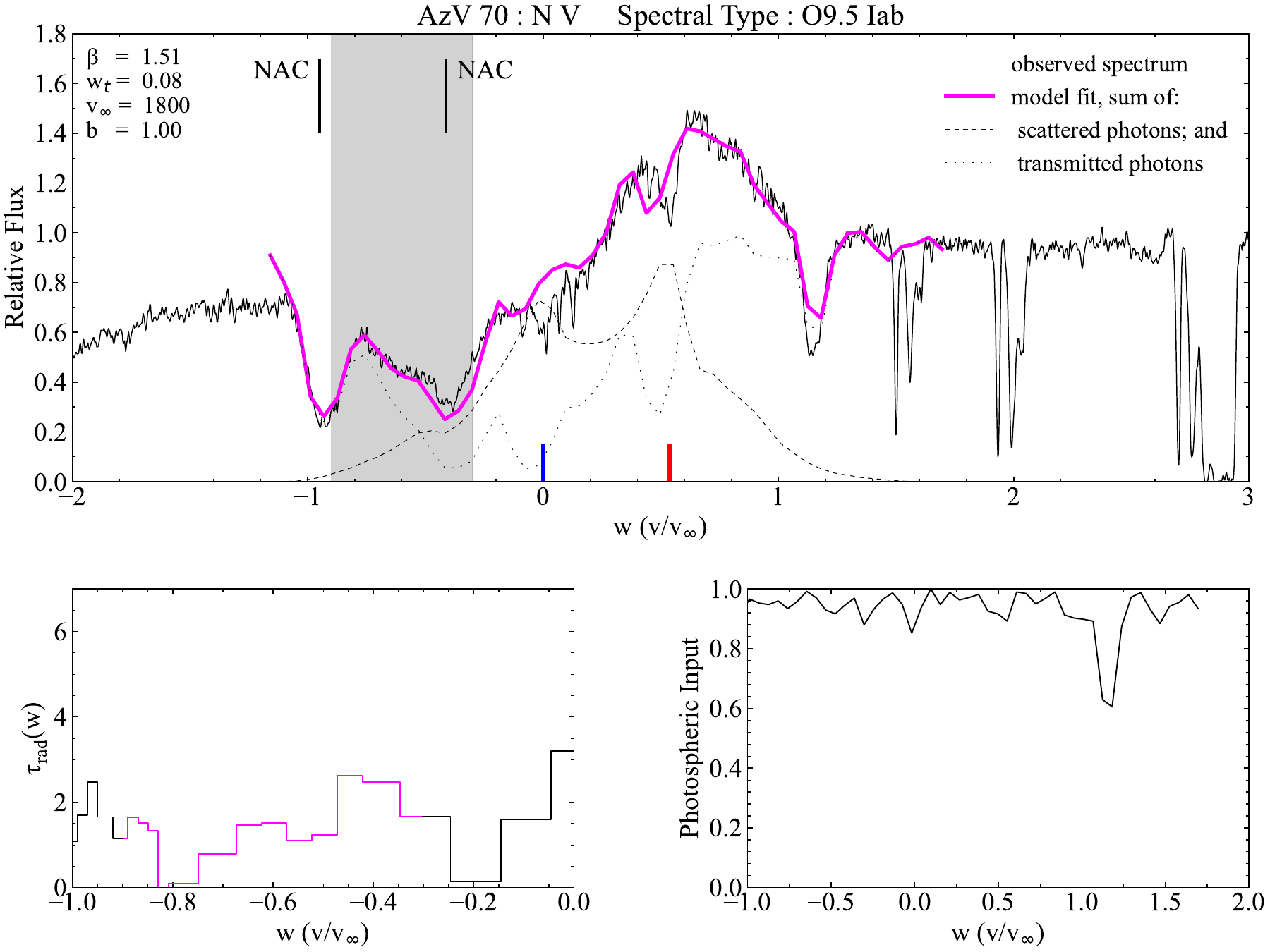} }
 \qquad
 \subfloat[ ]{\includegraphics[width=3.34in]{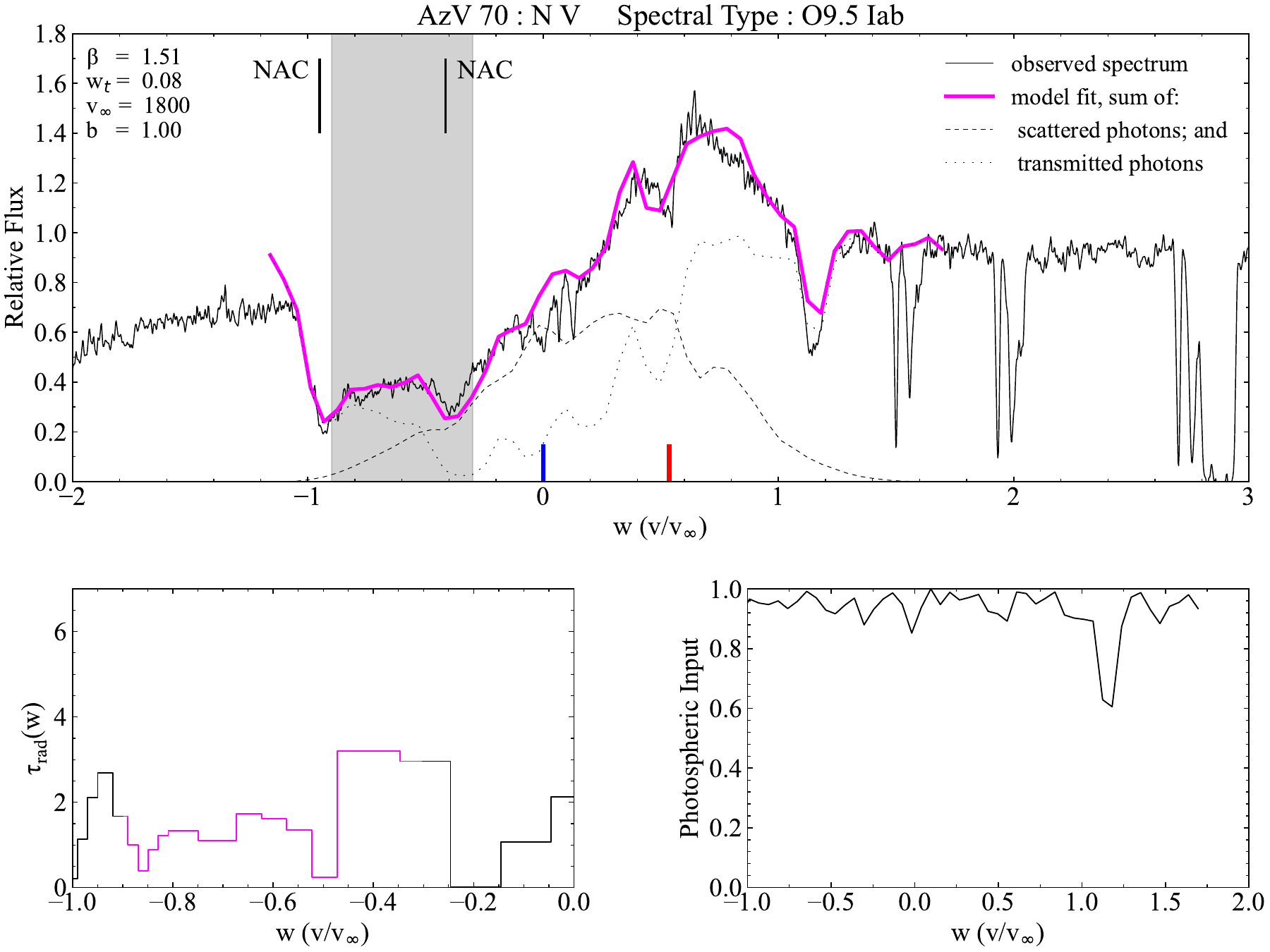} }
 \qquad
 \subfloat[ ]{\includegraphics[width=3.34in]{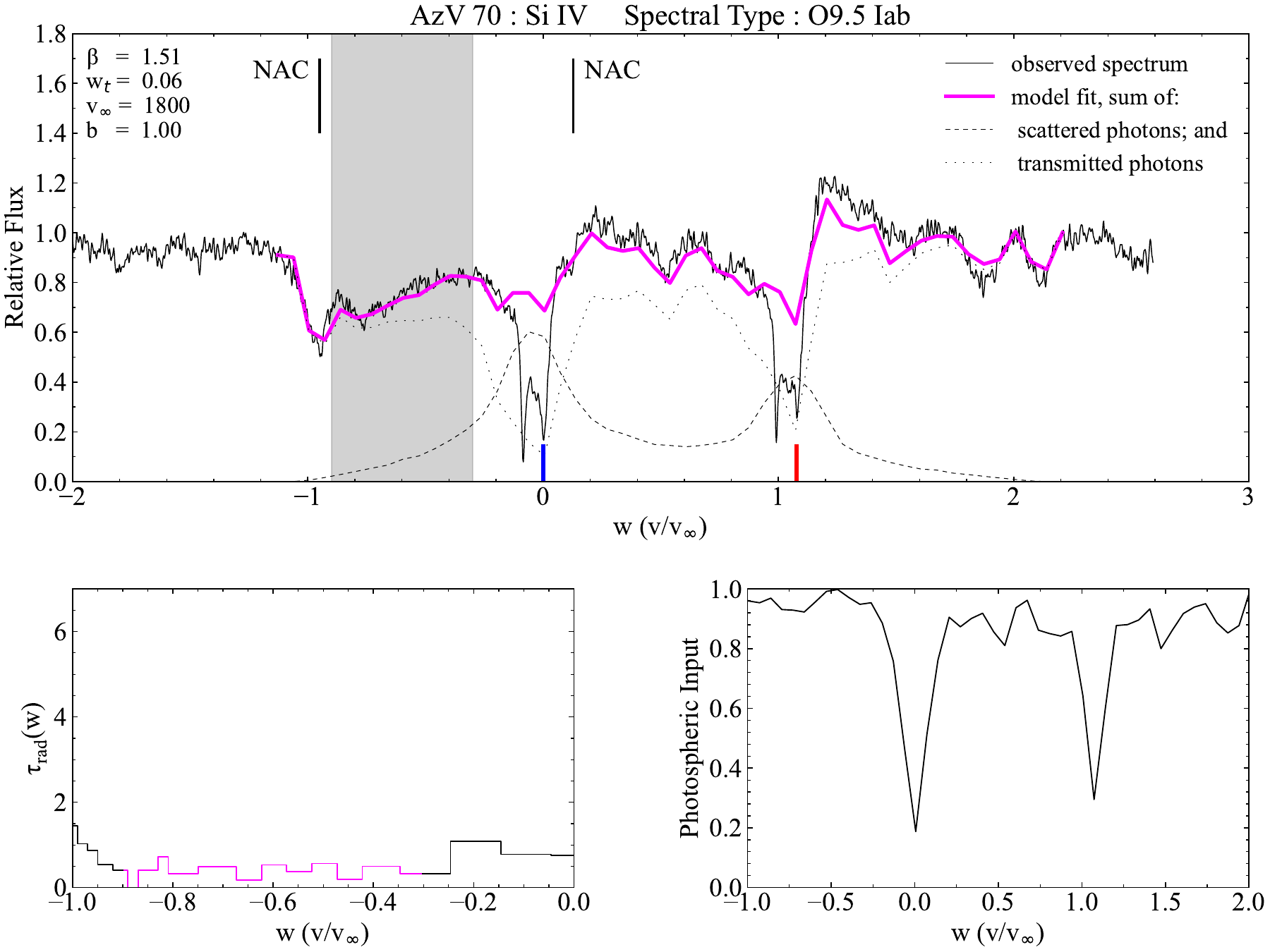} }
 \qquad
 \subfloat[ ]{\includegraphics[width=3.34in]{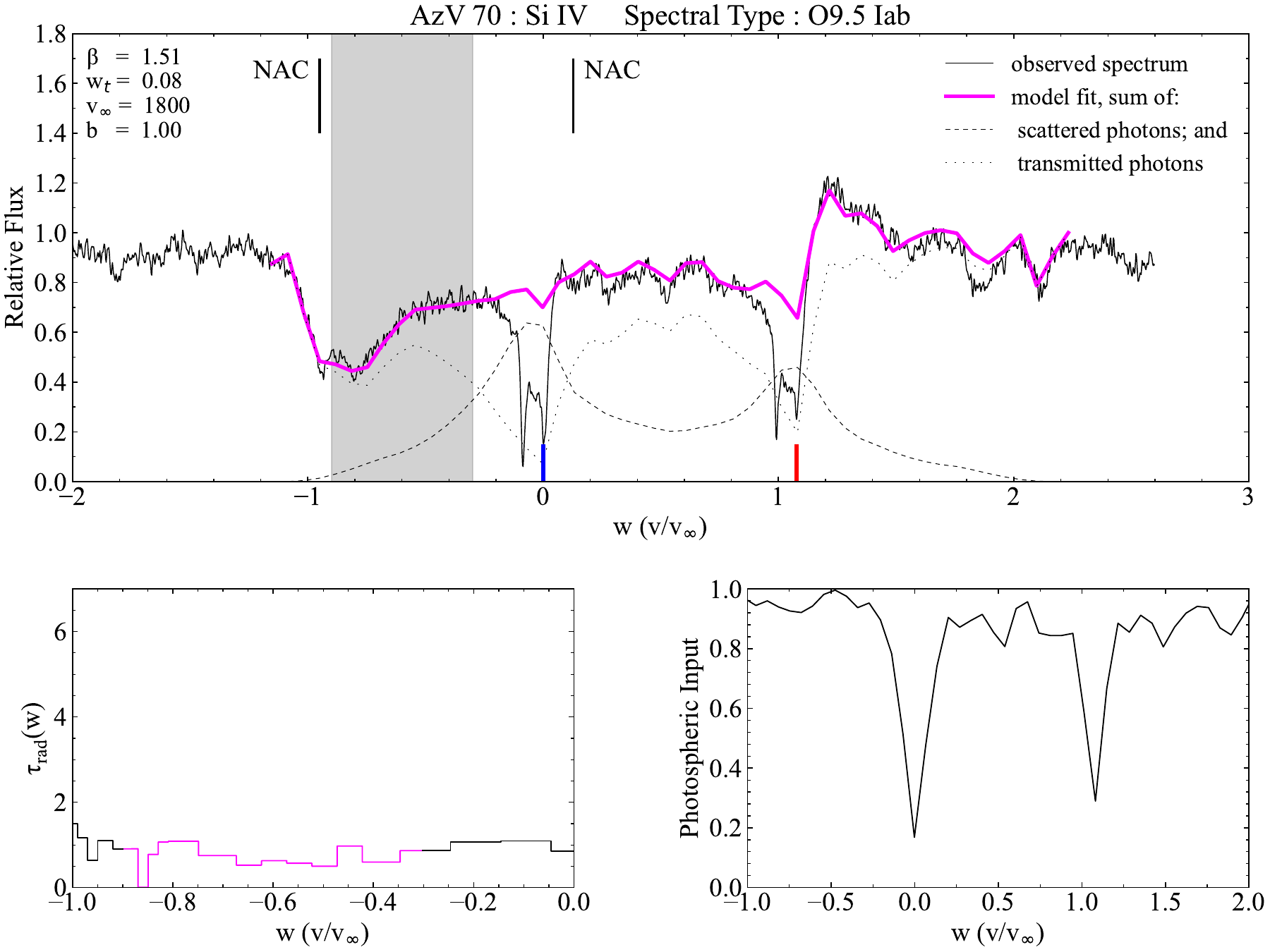} }
 \qquad
 \subfloat[ ]{\includegraphics[width=3.34in]{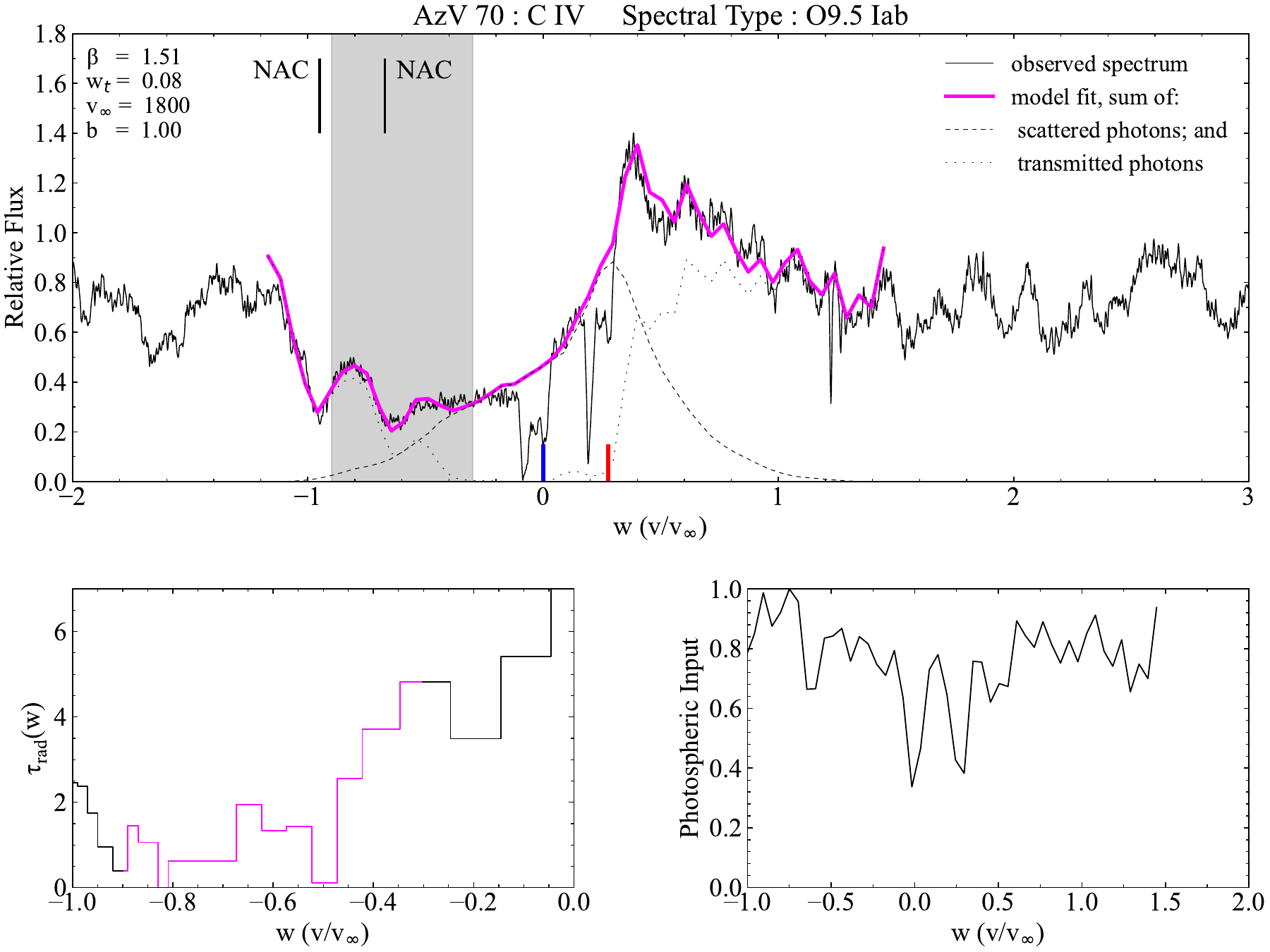} }
 \qquad
 \subfloat[ ]{\includegraphics[width=3.34in]{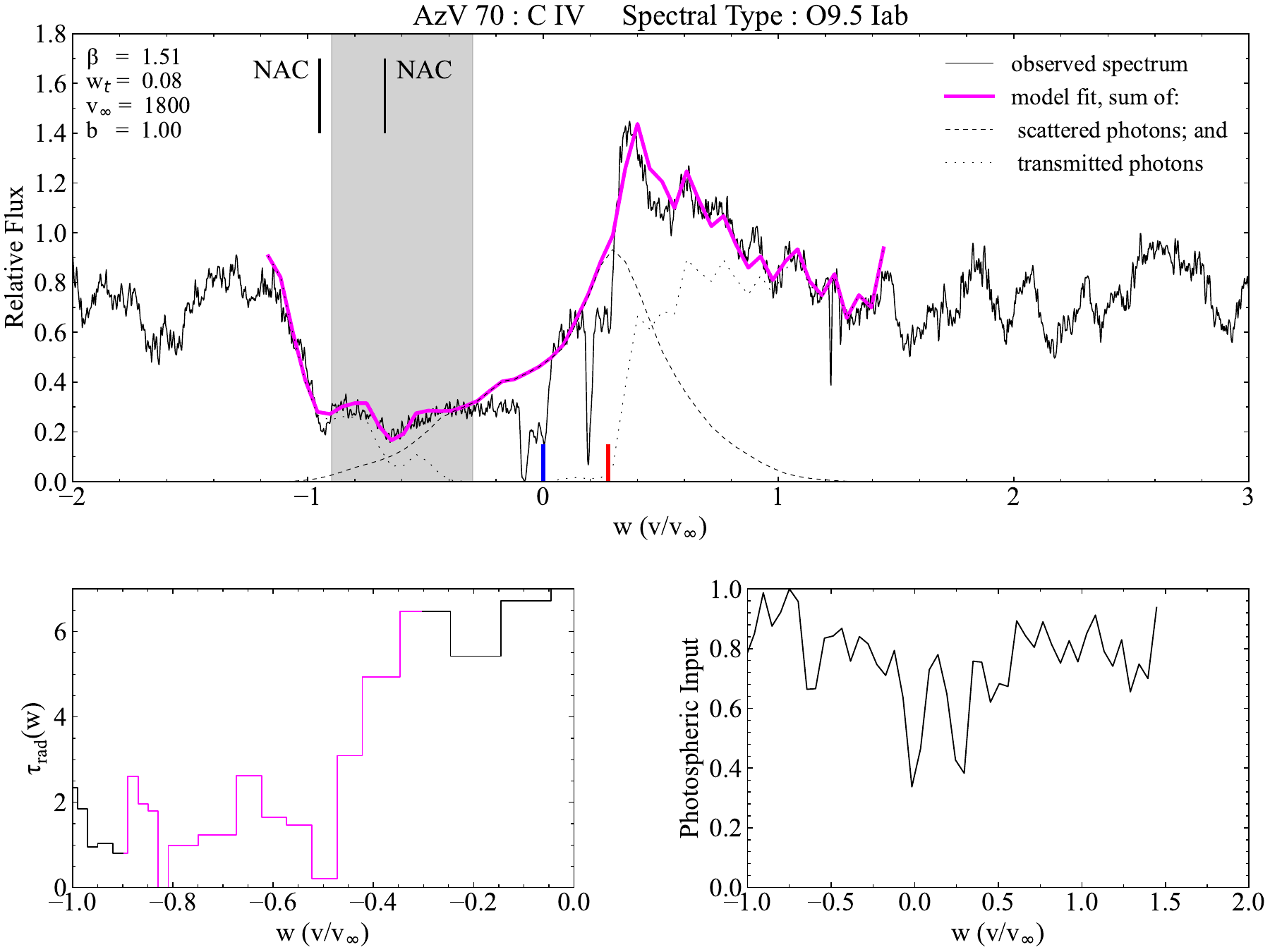} }
 \caption{SEI-derived model fits for: (a) and (b) the \ion{N}{V} doublet feature, (c) and (d) the \ion{Si}{IV} and (e) and (f) the \ion{C}{IV} doublet feature for two observations, dated 2020 June 28 and 2020 August 15, of the SMC star AzV 70.}
 \label{fig:70_si4_c4_SEI}
\end{center}
    \end{figure*}

   \begin{figure*}
\begin{center}
 \subfloat[ ]{\includegraphics[width=3.34in]{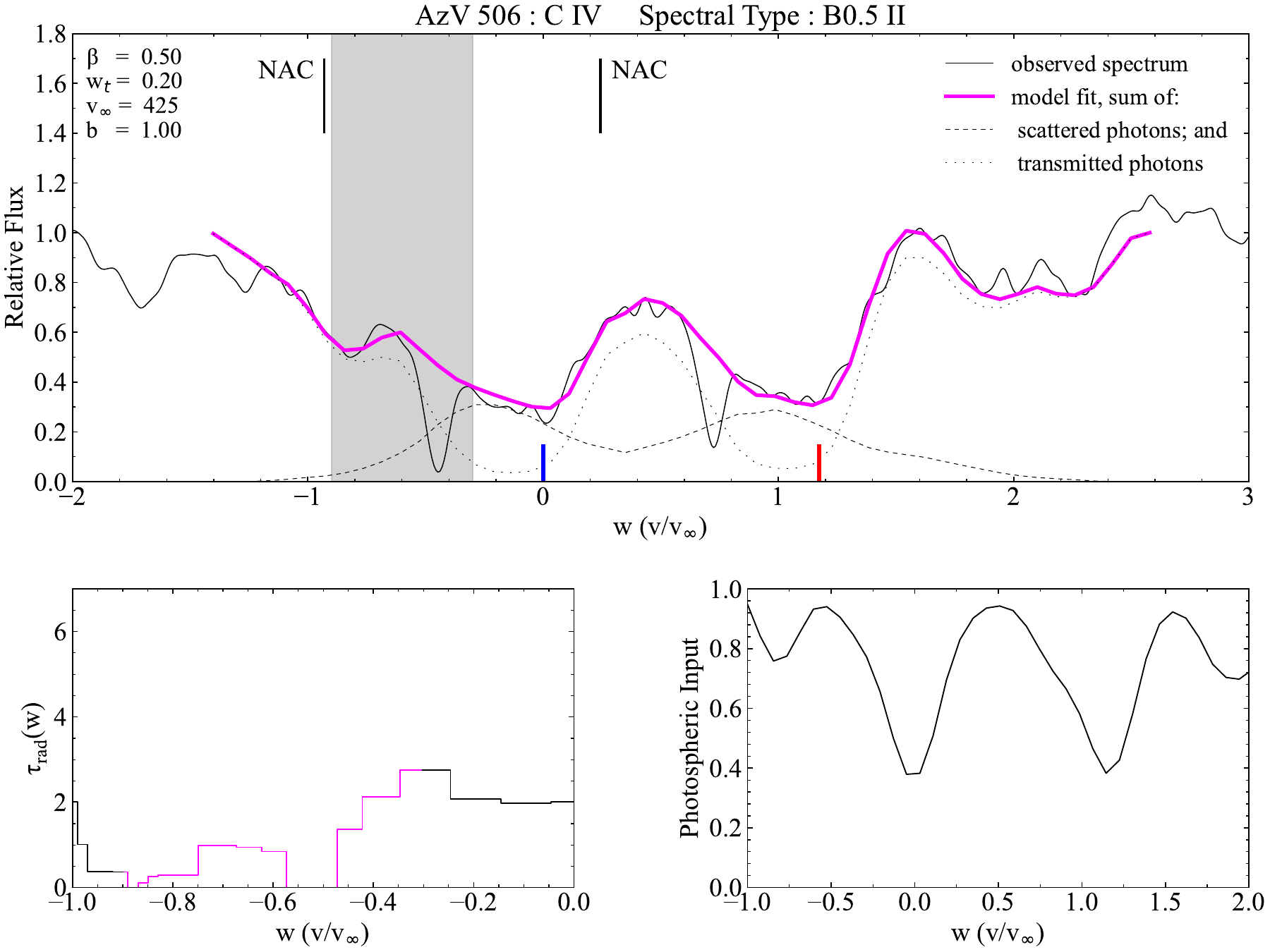} }
 \qquad
 \subfloat[ ]{\includegraphics[width=3.34in]{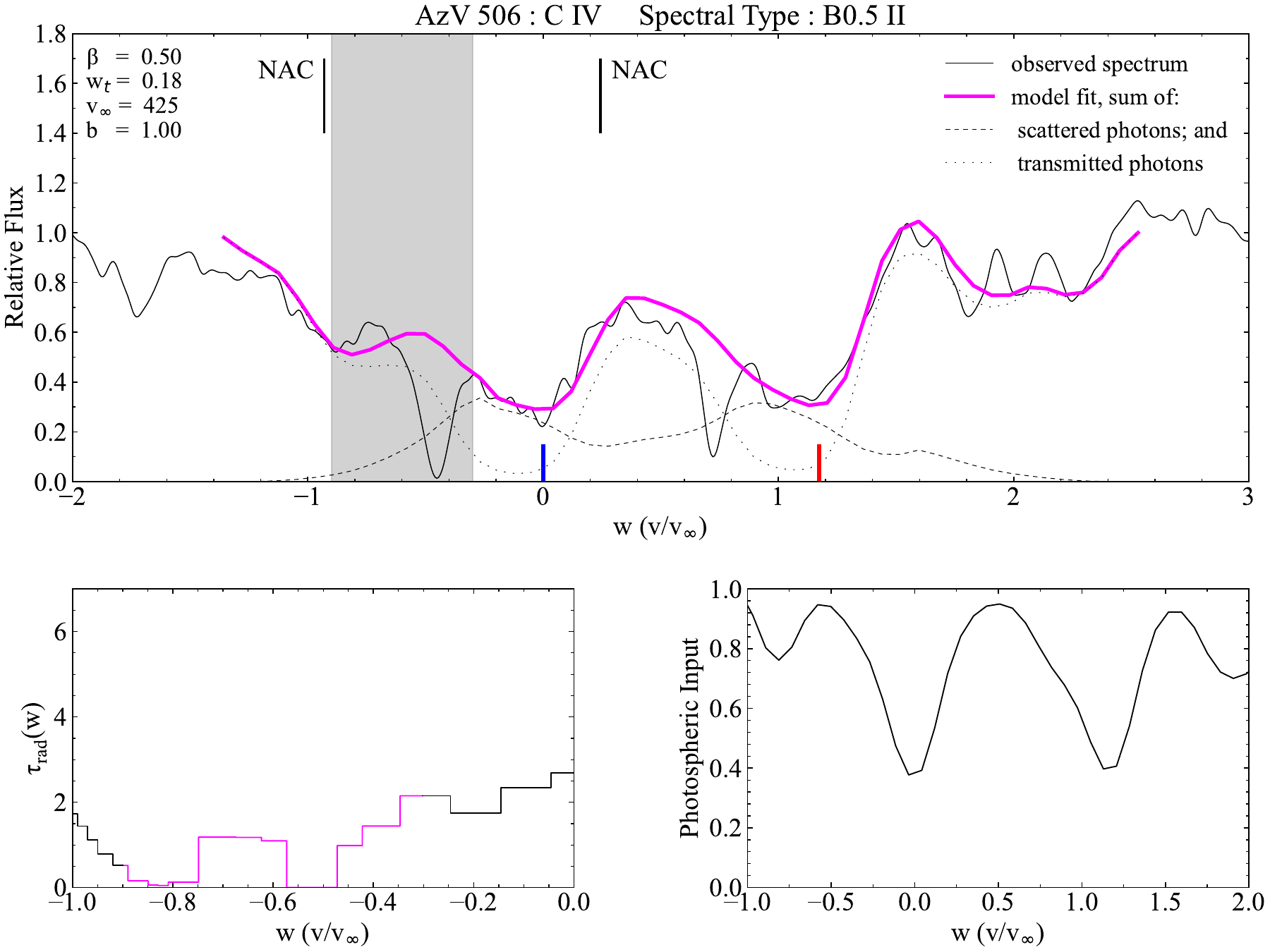} }
 \caption{SEI-derived model fits for the \ion{C}{IV} doublet feature for two observations, dated (a) 2023 March 11 and (b) 2023 April 03, of the SMC star AzV 506. There is a lack of significant emission features, despite some evidence of a wind-formed absorption profile.}
 \label{fig:506_c4_SEI}
\end{center}
    \end{figure*}

   \begin{figure*}
\begin{center}
 \subfloat[ ]{\includegraphics[width=3.34in]{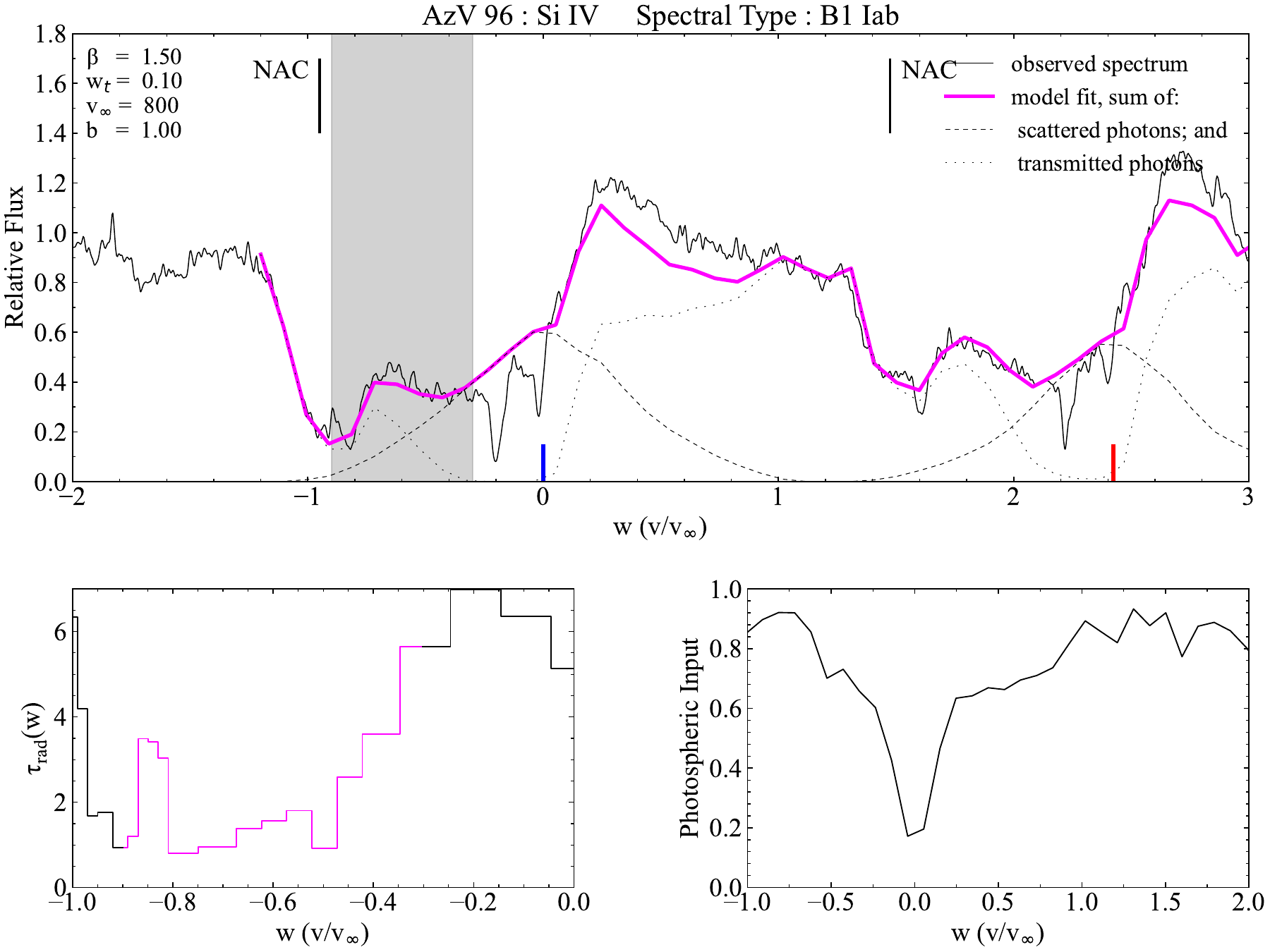} }
 \qquad
 \subfloat[ ]{\includegraphics[width=3.34in]{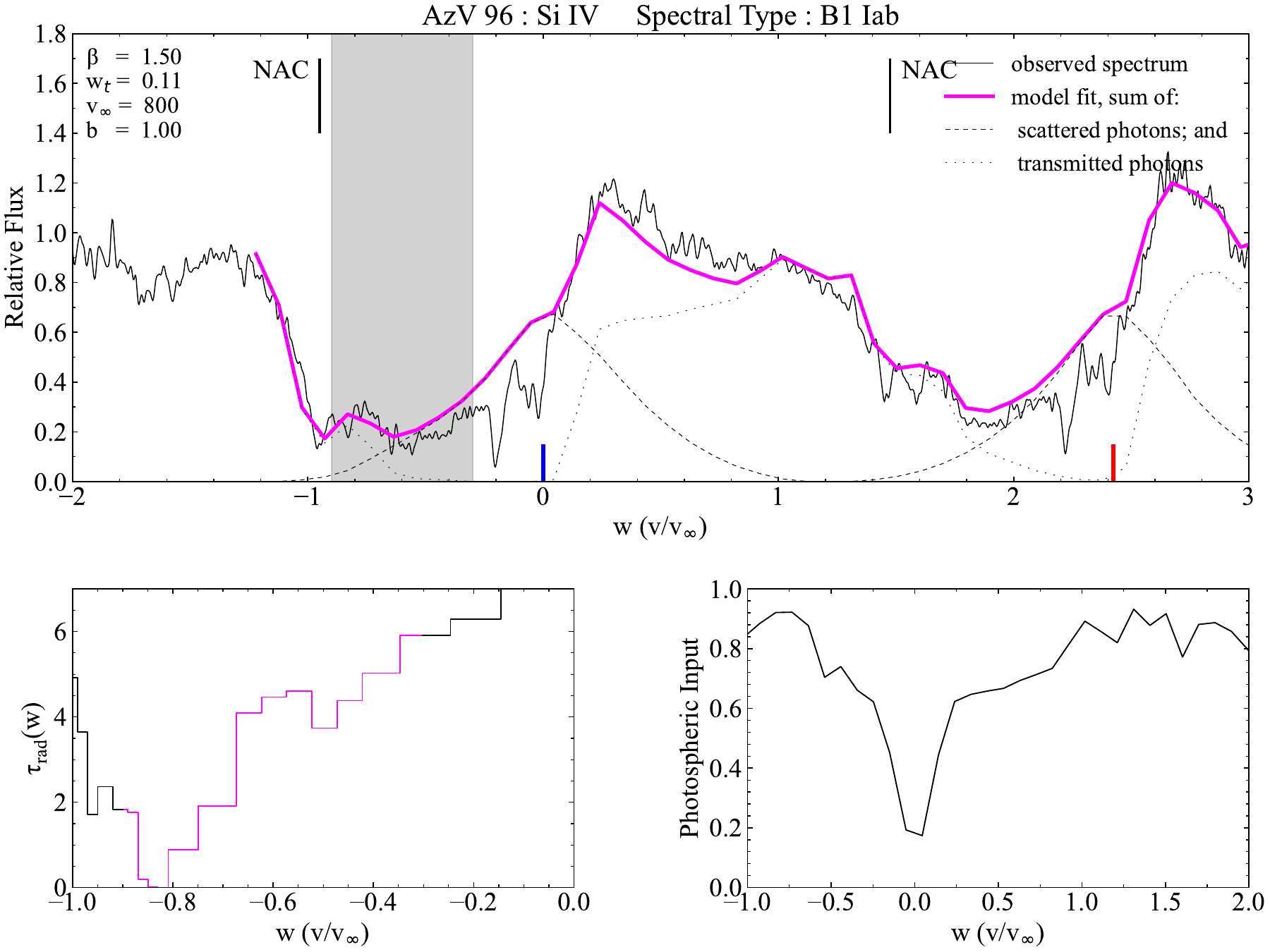} }
 \qquad
 \subfloat[ ]{\includegraphics[width=3.34in]{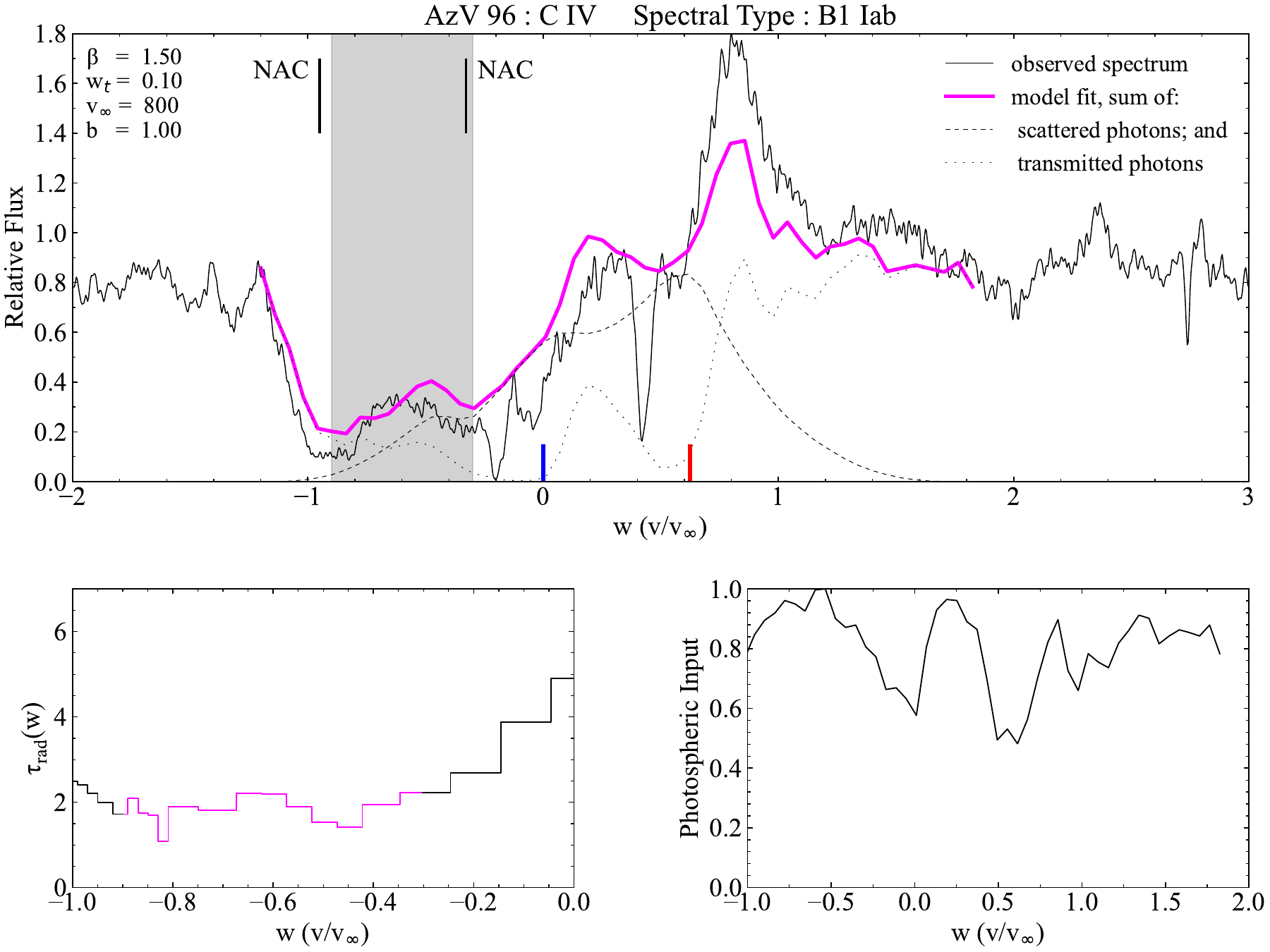} }
 \qquad
 \subfloat[ ]{\includegraphics[width=3.34in]{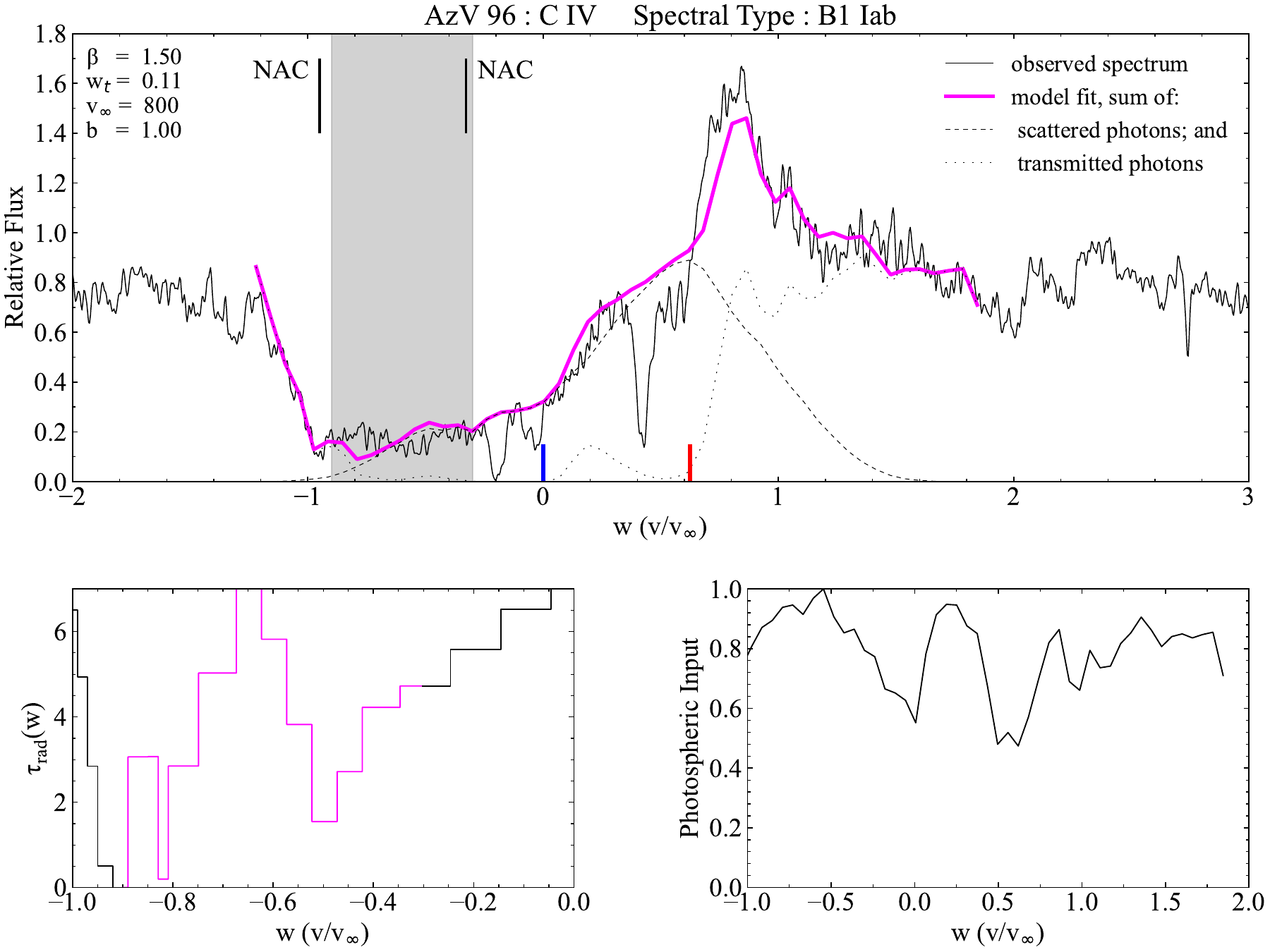} }
 \caption{SEI-derived model fits for (a) and (b) the \ion{Si}{IV} doublet feature and (c) and (d) the \ion{C}{IV} for two observations, dated 2022 February 26 and 2022 March 03, of the SMC star AzV 96.}
 \label{fig:96_si4_c4_SEI}
\end{center}
    \end{figure*}

   \begin{figure*}
\begin{center}
 \subfloat[ ]{\includegraphics[width=3.34in]{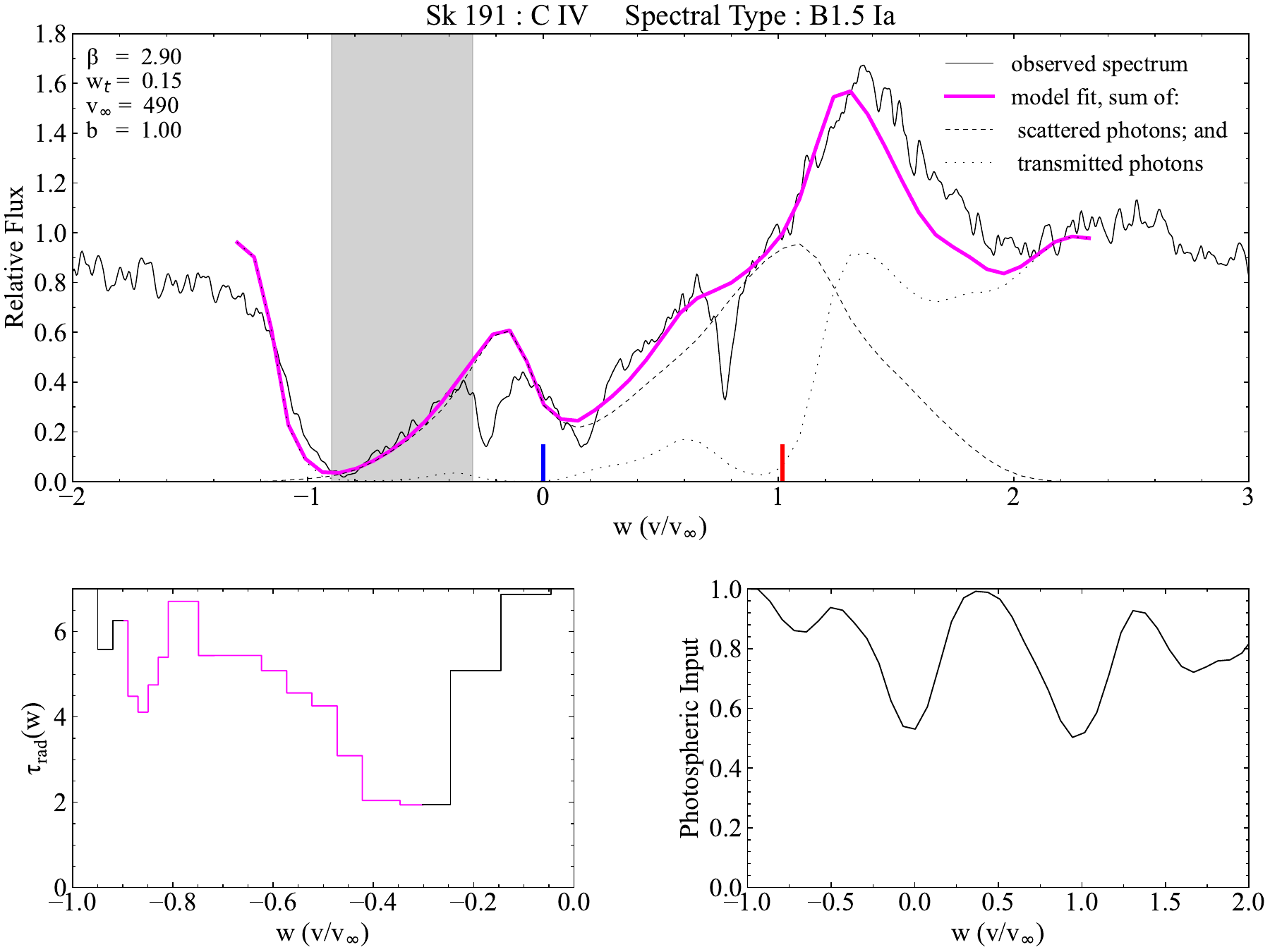} }
 \qquad
 \subfloat[ ]{\includegraphics[width=3.34in]{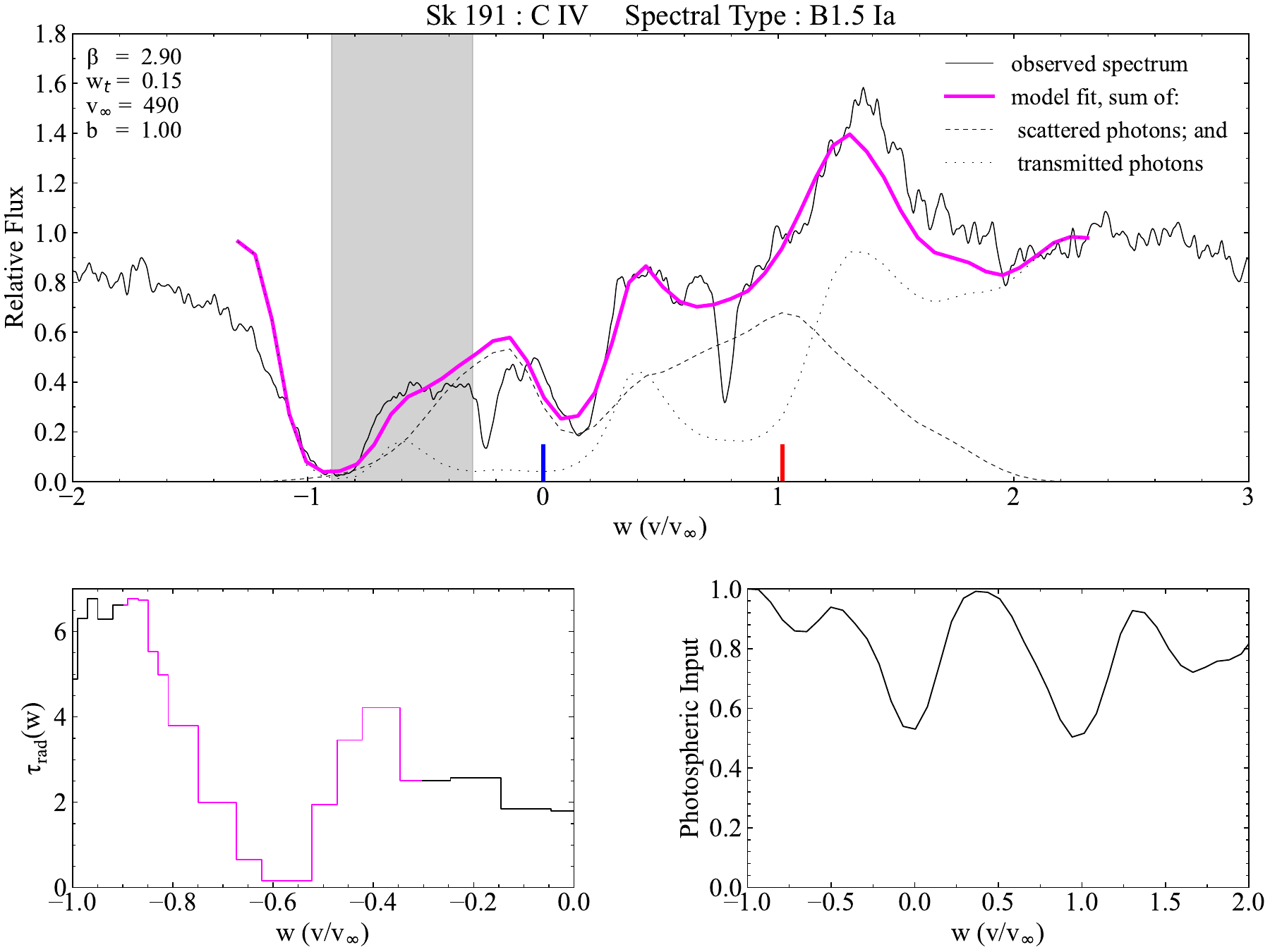} }
 \caption{SEI-derived model fits for the \ion{C}{IV} for two observations, dated (a) 2010 February 11 and (b) 2010 April 23, of the SMC star Sk 191.}
 \label{fig:191_c4_SEI}
\end{center}
    \end{figure*}

   \begin{figure*}
\begin{center}
 \subfloat[ ]{\includegraphics[width=3.34in]{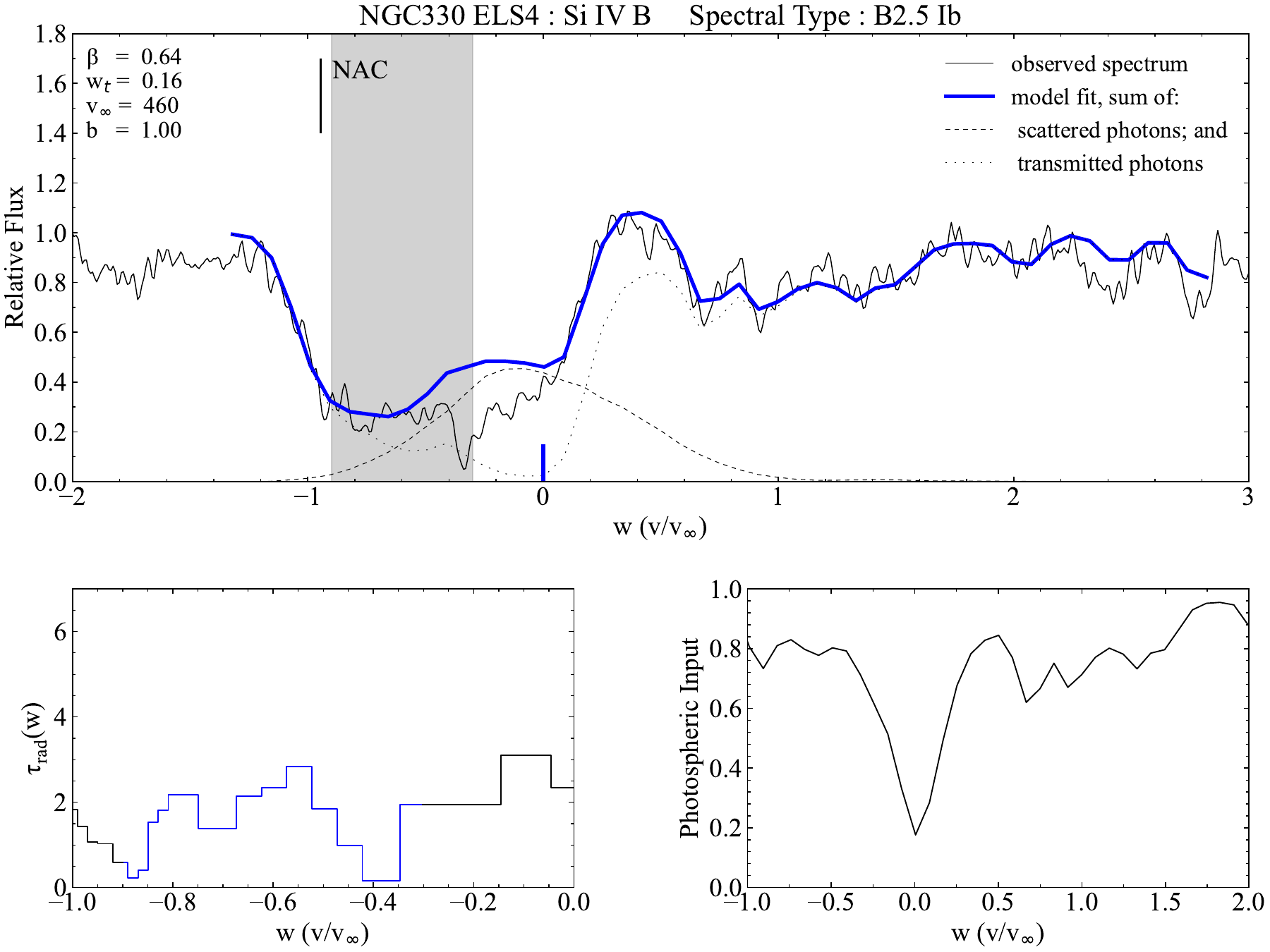} }
 \qquad
 \subfloat[ ]{\includegraphics[width=3.34in]{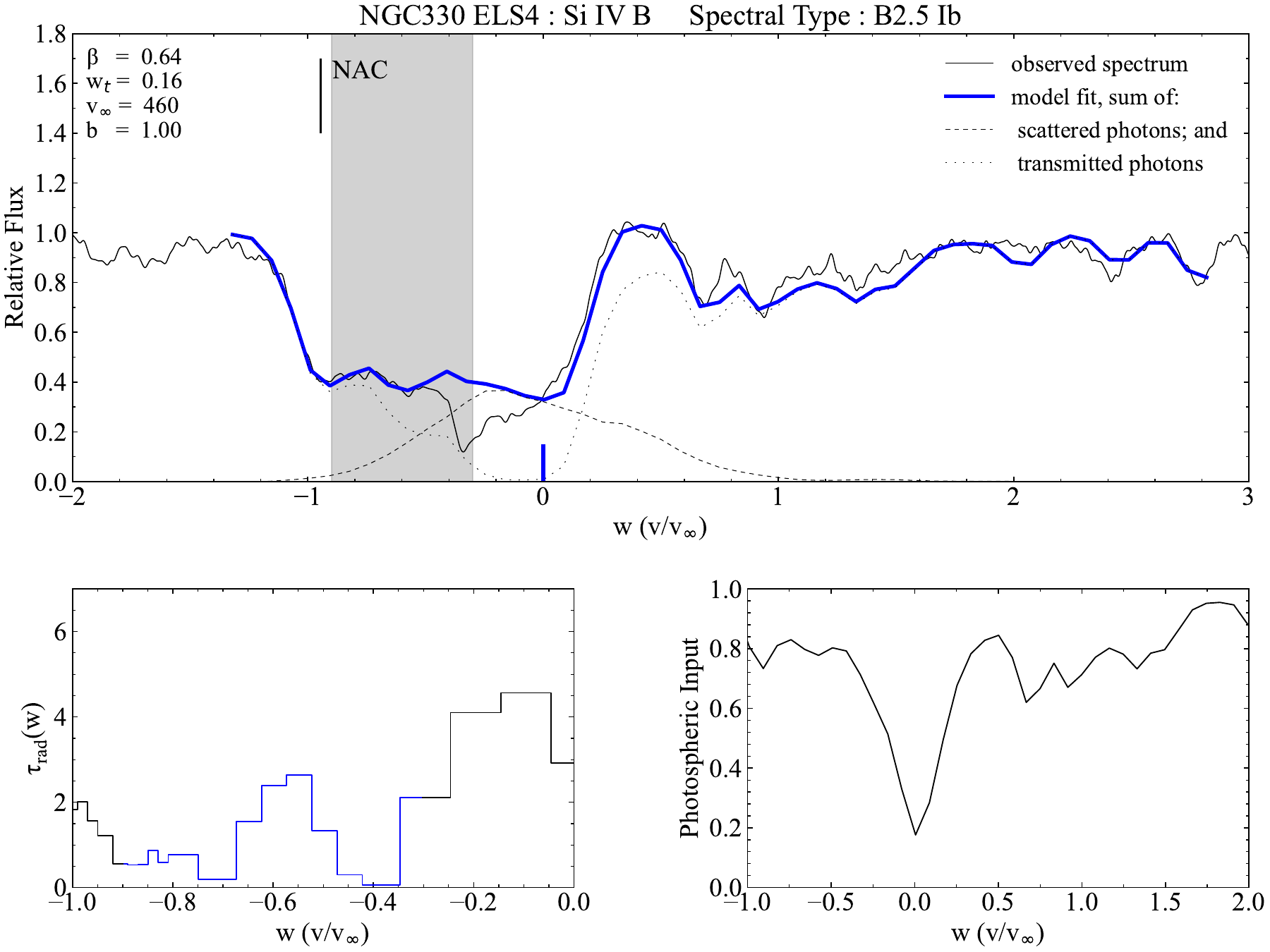} }
 \qquad
  \subfloat[ ]{\includegraphics[width=3.34in]{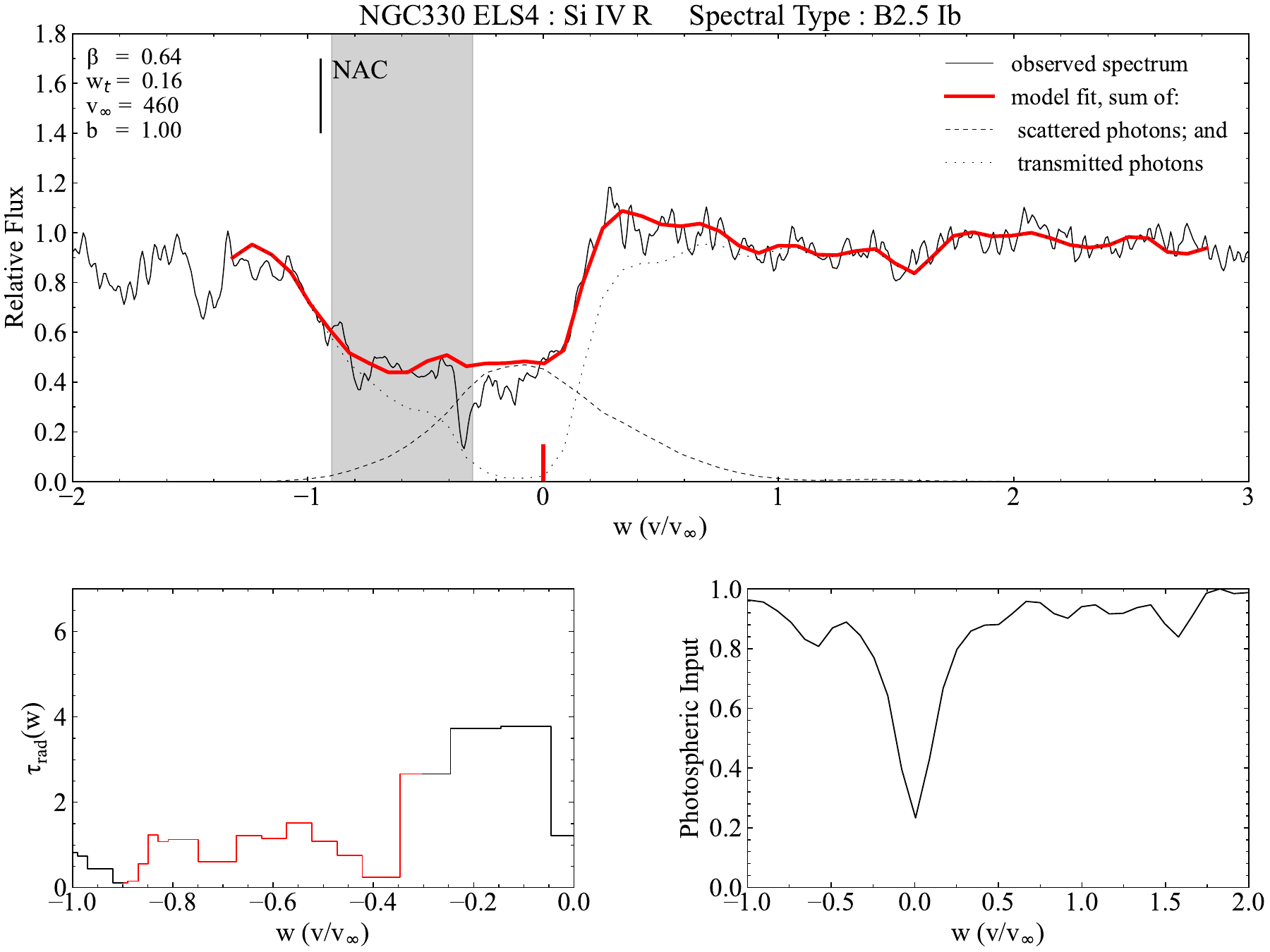} }
 \qquad
 \subfloat[ ]{\includegraphics[width=3.34in]{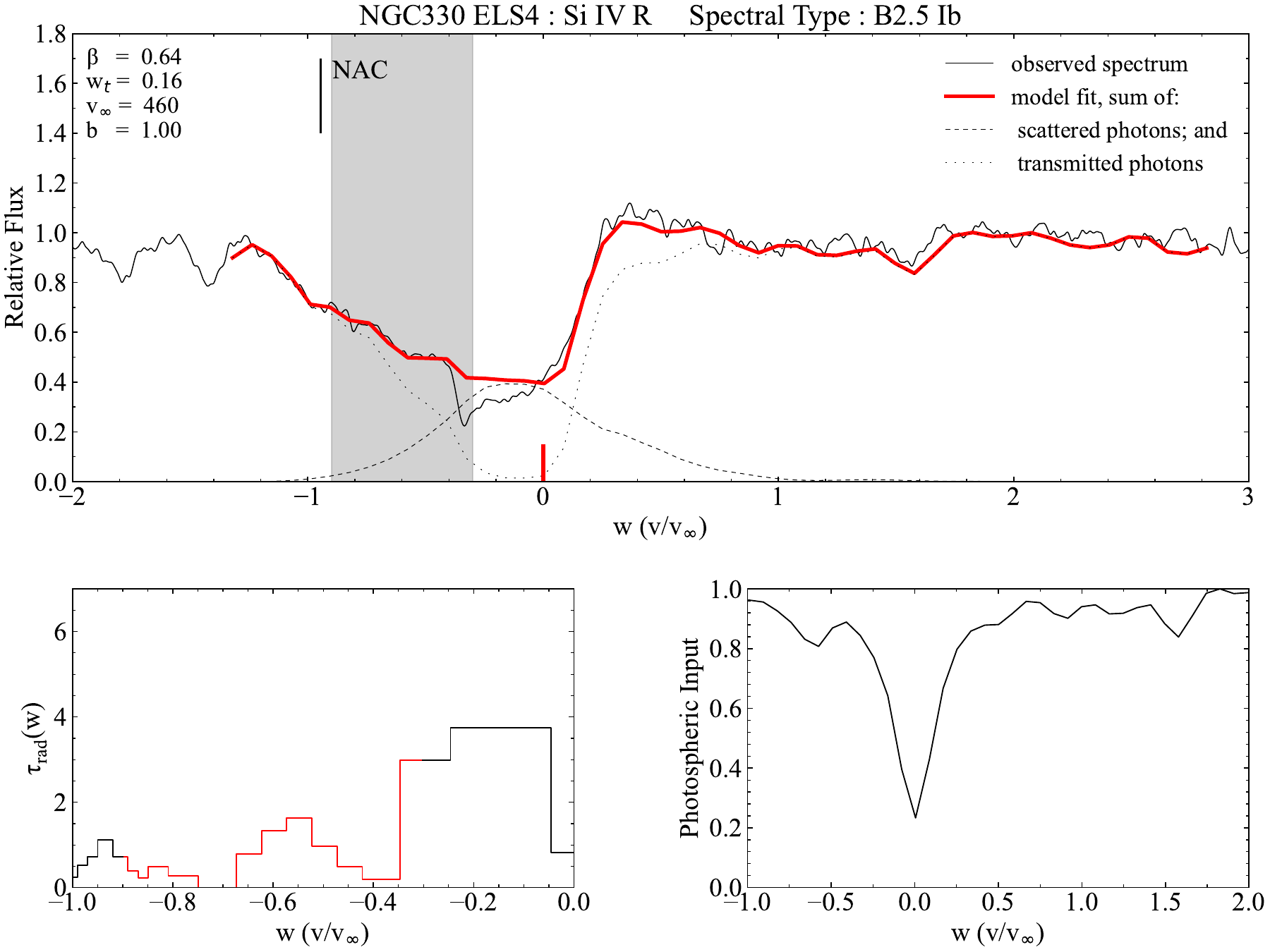} }
 \qquad
 \subfloat[ ]{\includegraphics[width=3.34in]{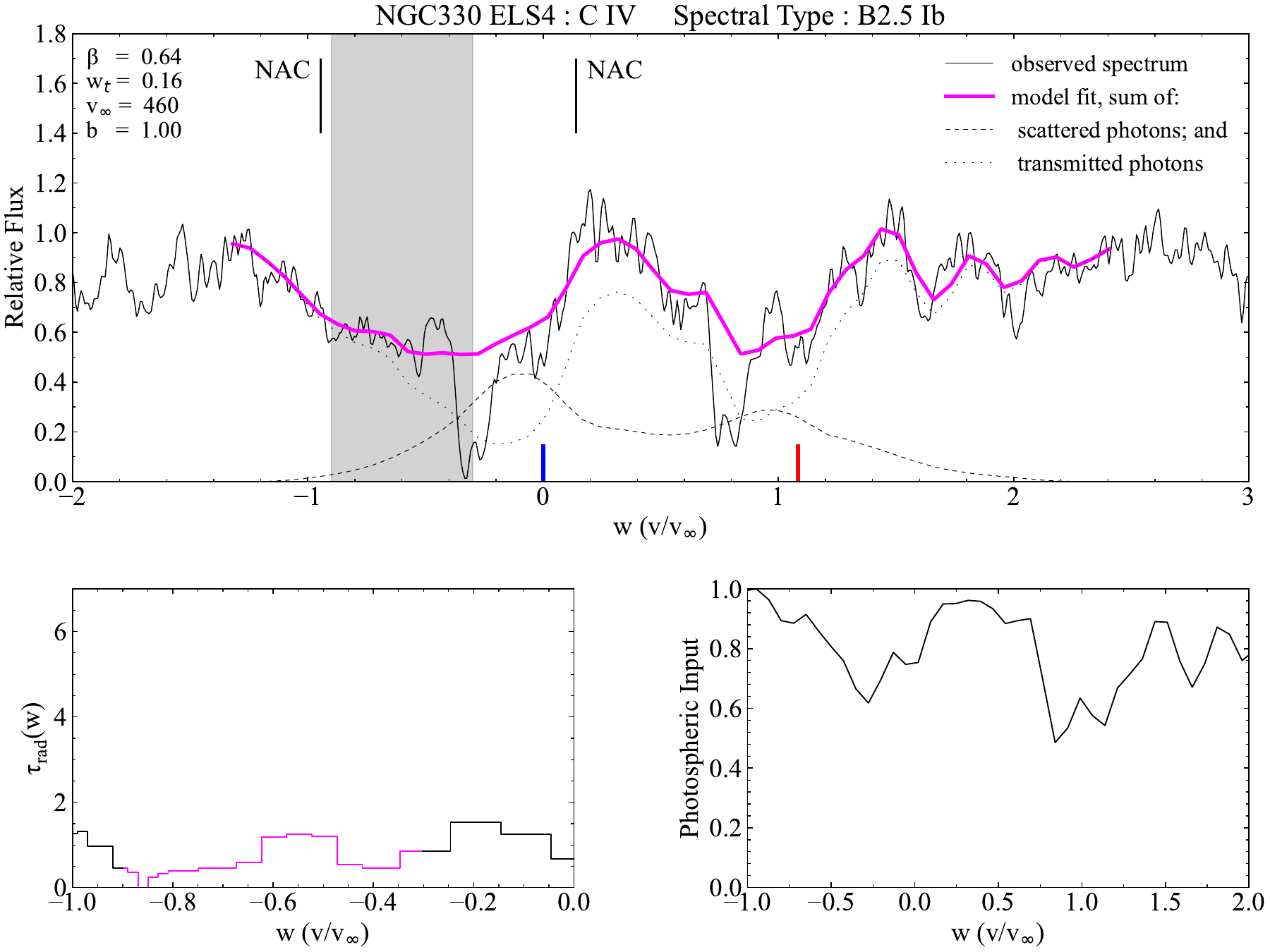} }
 \qquad
 \subfloat[ ]{\includegraphics[width=3.34in]{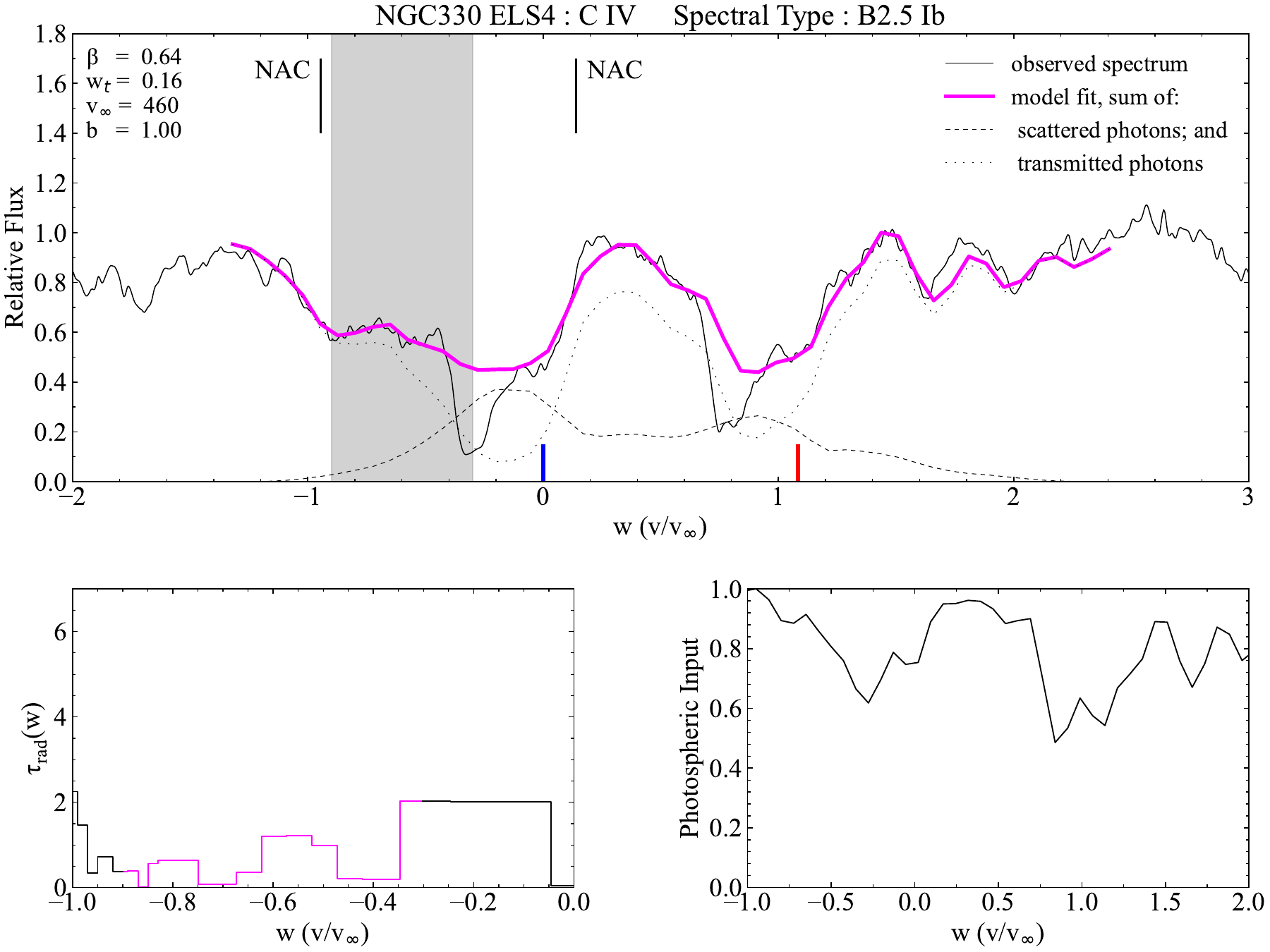} } \caption{SEI-derived model fits for (a) and (b) the blue component of the \ion{Si}{IV} resonance line doublet (effectively decoupled in this low-velocity wind example), (c) and (d) the red element of that doublet and (e) and (f) the \ion{C}{IV} doublet feature for two observations, in each case dated 2002 September 30 and 2009 January 08 respectively, of the SMC star NGC330 ELS 4.}
 \label{fig:3304_si4b_si4r_c4_SEI}
\end{center}
    \end{figure*}

   \begin{figure*}
\begin{center}
 \subfloat[ ]{\includegraphics[width=3.34in]{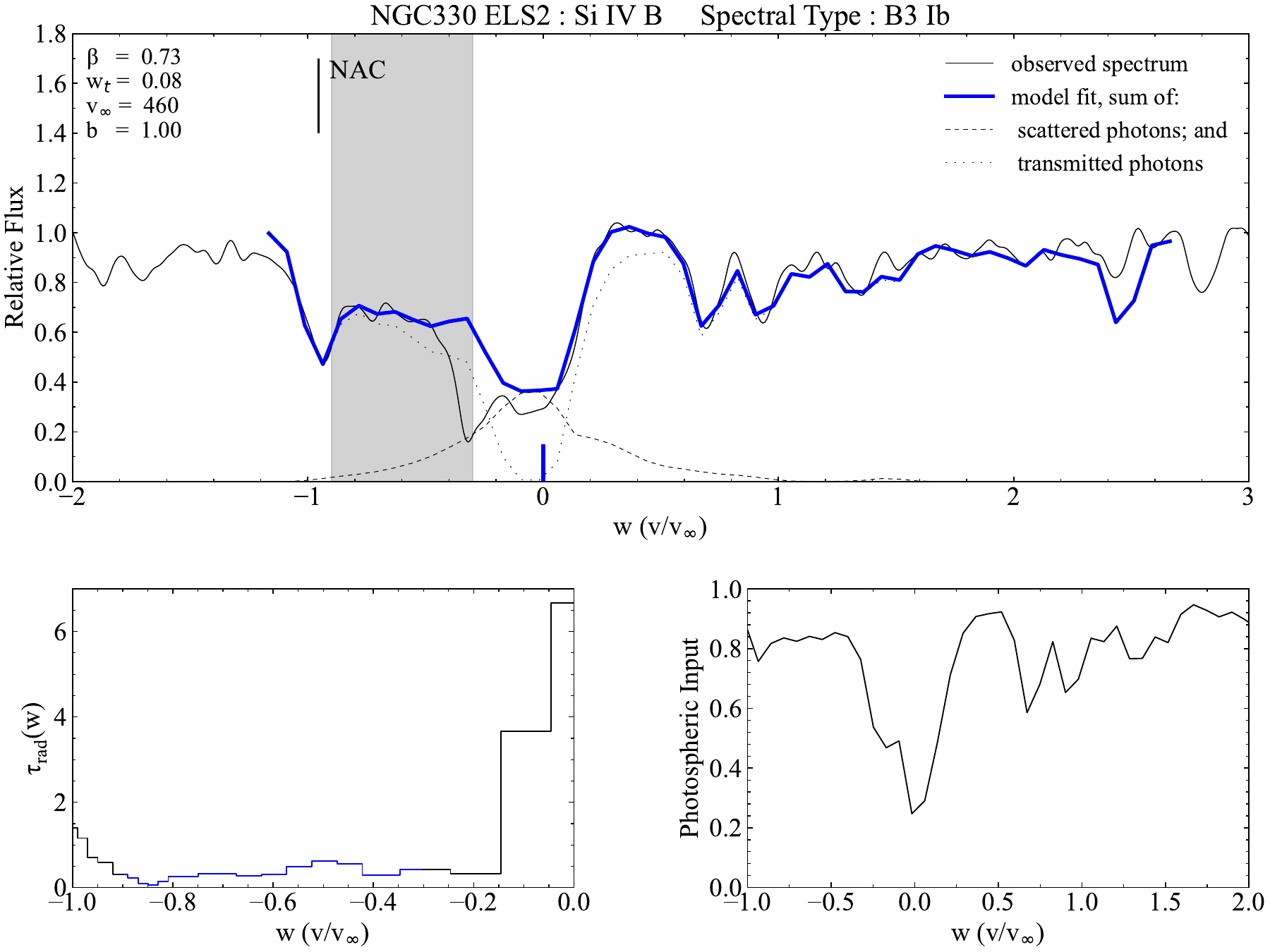} }
 \qquad
 \subfloat[ ]{\includegraphics[width=3.34in]{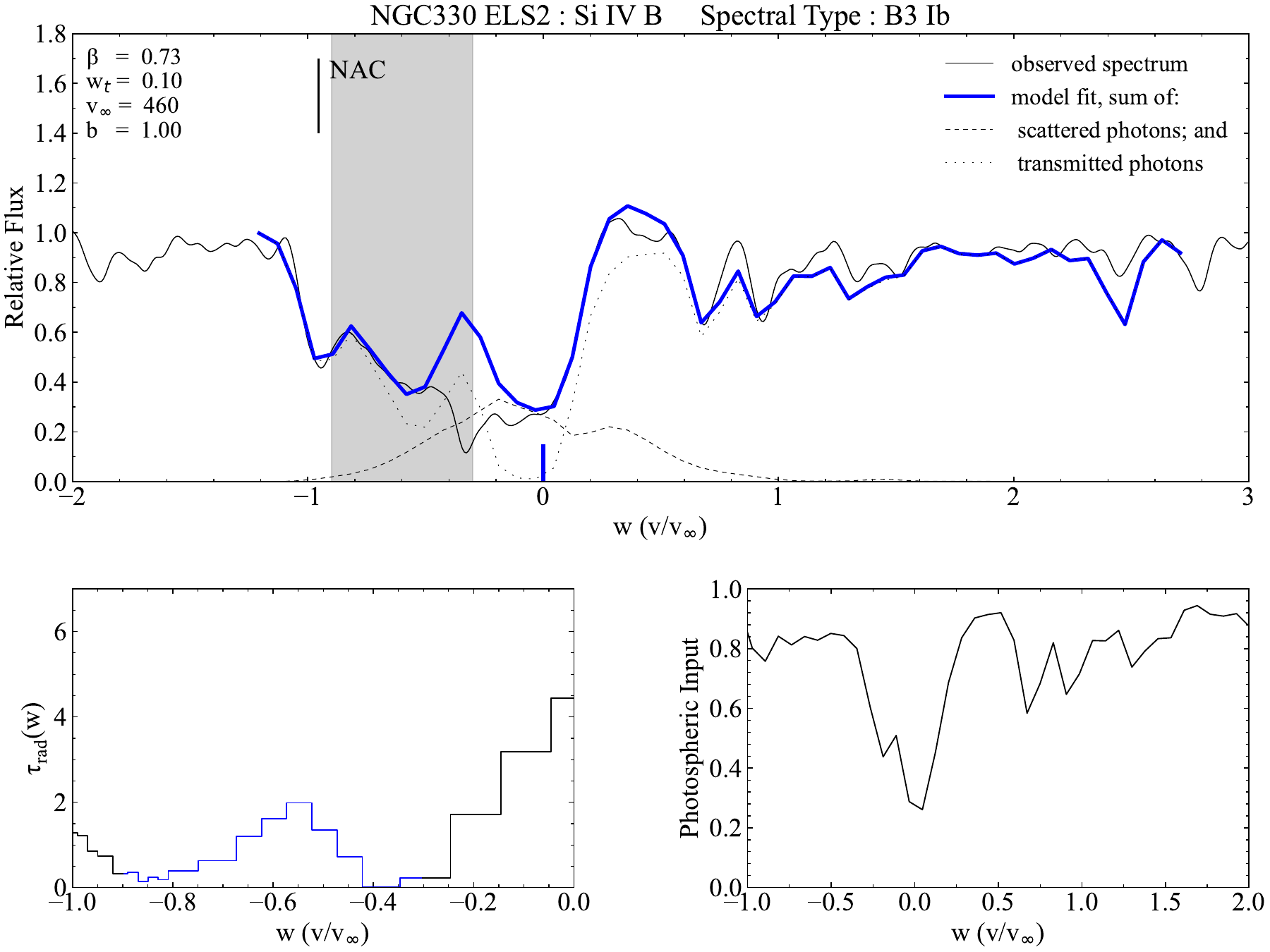} }
 \qquad
  \subfloat[ ]{\includegraphics[width=3.34in]{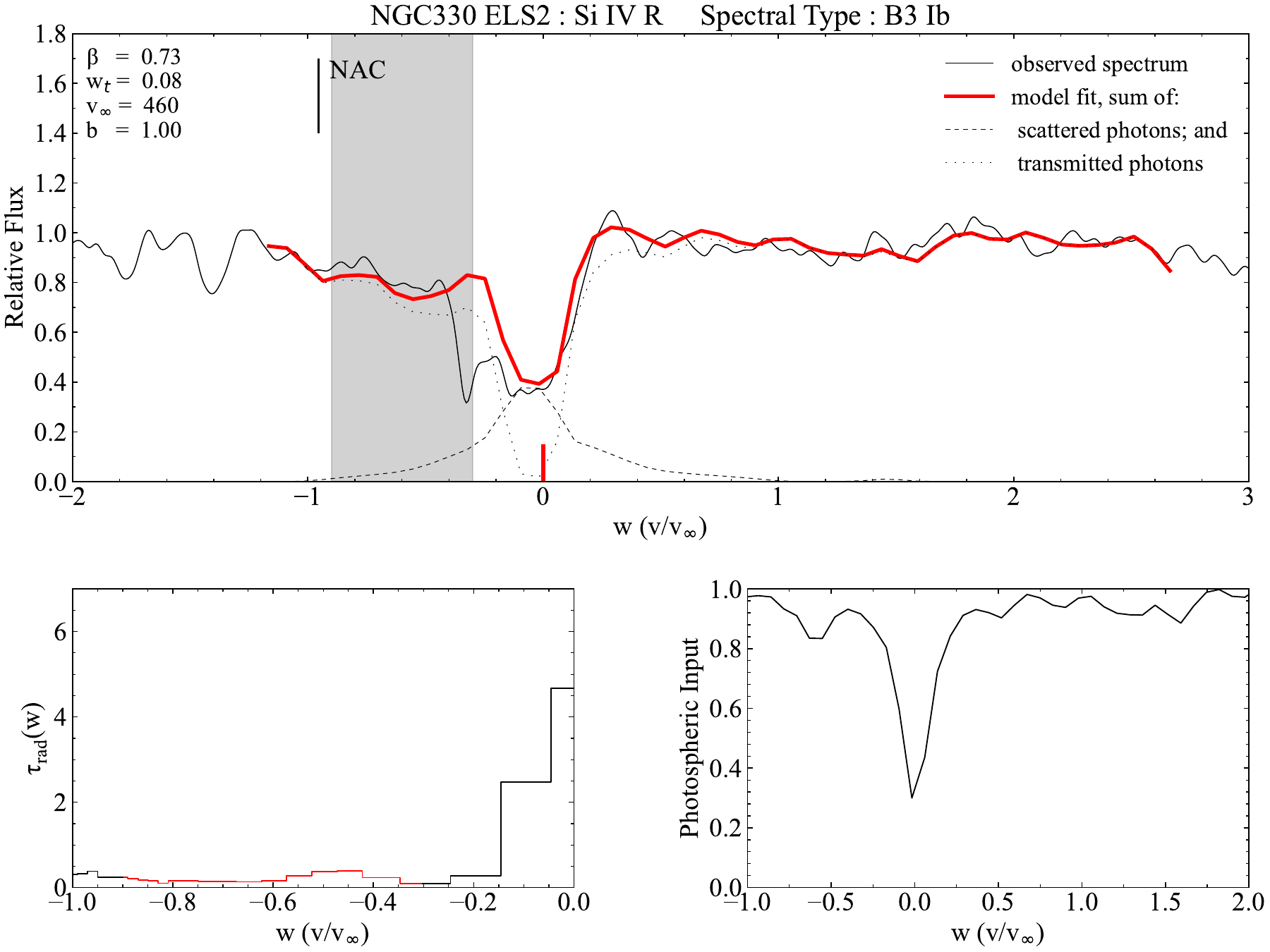} }
 \qquad
 \subfloat[ ]{\includegraphics[width=3.34in]{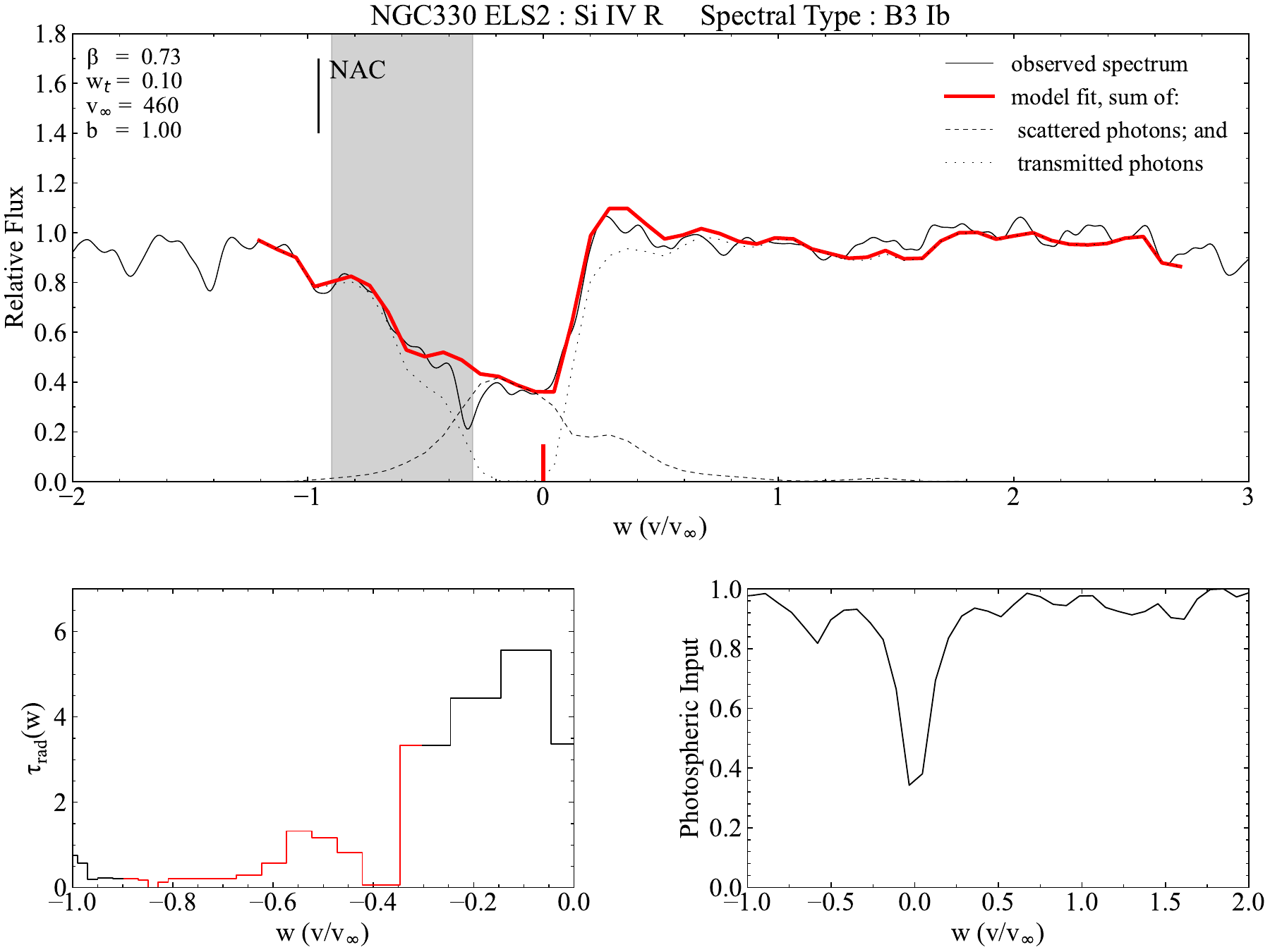} }
 \qquad
 \subfloat[ ]{\includegraphics[width=3.34in]{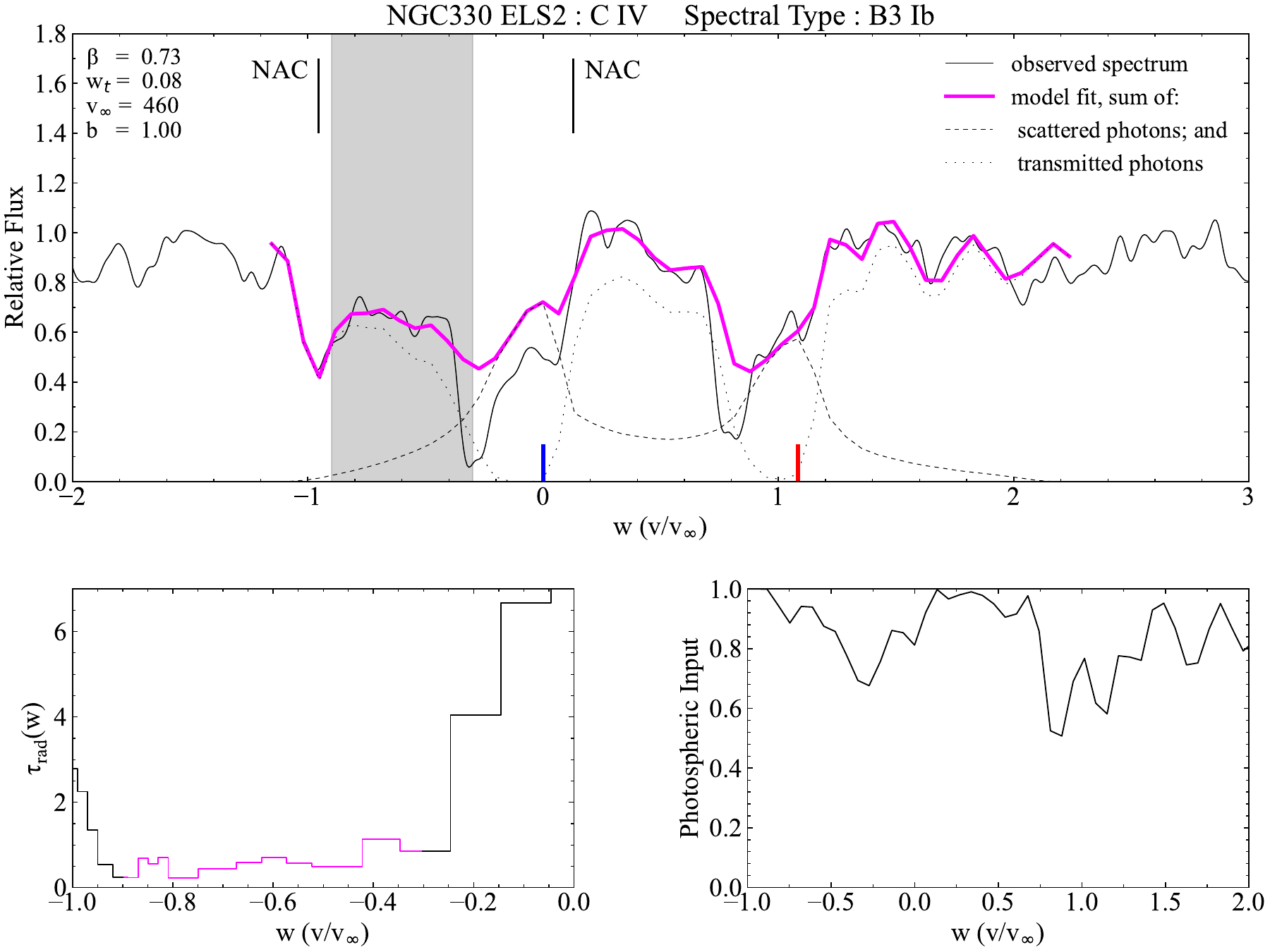} }
 \qquad
 \subfloat[ ]{\includegraphics[width=3.34in]{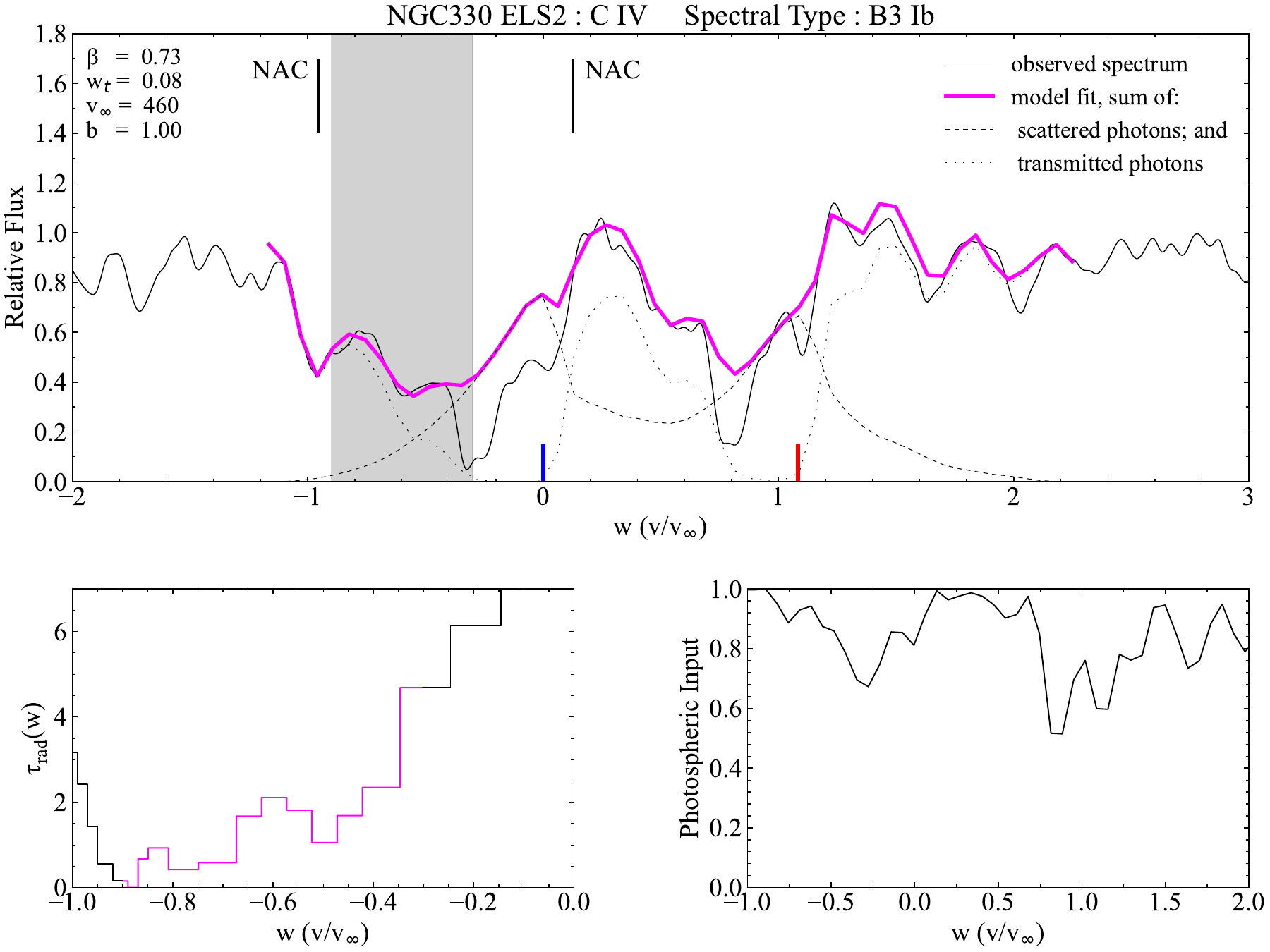} } \caption{SEI-derived model fits for (a) and (b) the blue component of the \ion{Si}{IV} resonance line doublet (effectively decoupled in this low-velocity wind example), (c) and (d) the red element of that doublet and (e) and (f) the \ion{C}{IV} doublet feature for two observations, in each case dated 2002 December 30 and 2003 January 10 respectively, of the SMC star NGC330 ELS 2. The considerable increase in the average optical depths of each feature in the 11 days from the first to the second observation can readily be seen. In this, and the preceding low-velocity wind example, the ISM lines severely affect a significant portion of the lower velocity part of the wind profile.}
 \label{fig:3302_si4b_si4r_c4_SEI}
\end{center}
    \end{figure*}

\clearpage

\section{Changes in Radial Optical Depths} \label{alltauradcomp}

The following plots compare the radial optical depths derived from the SEI model fits shown in Appendix \ref{allSEI} for each relevant species. Where multiple observations are available, for clarity, the plots are limited to showing 4 of those observations. Where 2 observations are plotted, the first observation's data are shown in dark violet and the second in light green, as in Appendix \ref{allplots}. Where 3 or 4 observations are plotted the same violet-blue-green-yellow sequence used elsewhere in this work is employed.

   \begin{figure*}
\begin{center}
 \qquad
 \subfloat[ ]{\includegraphics[width=3.34in]{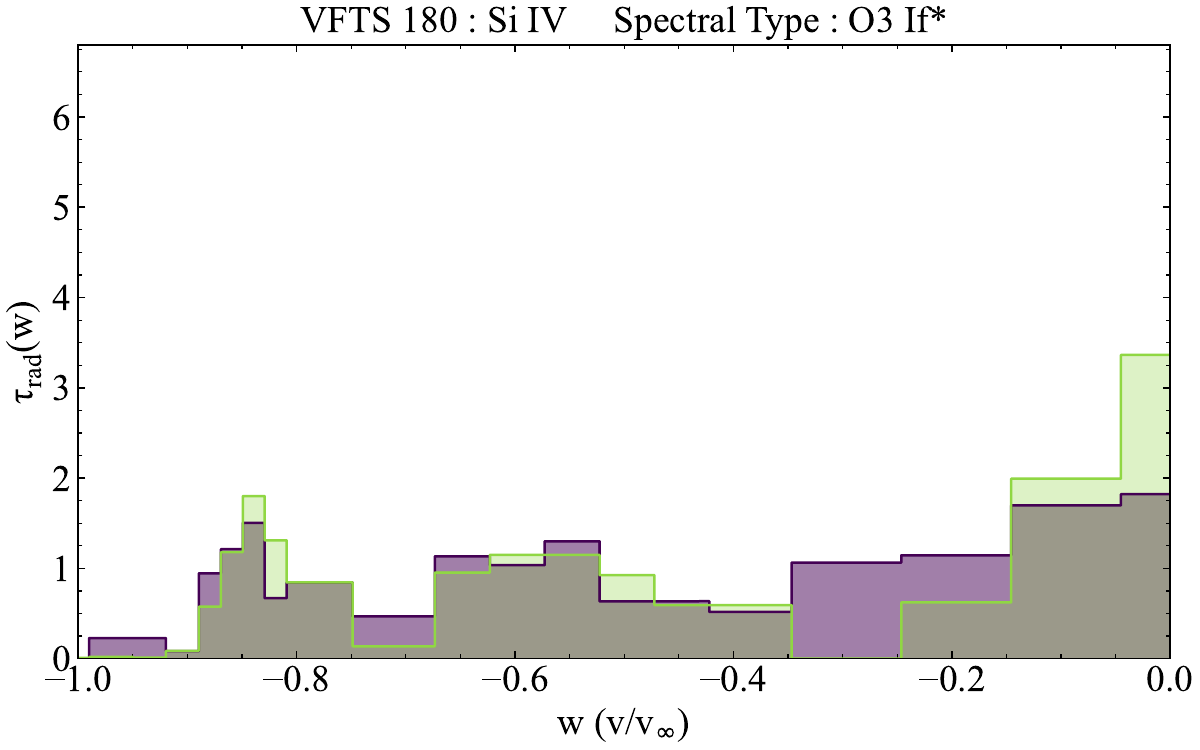} }
 \caption{SEI-derived radial optical depths of the \ion{Si}{IV} absorption profile, showing changes in those optical depths in the stellar wind of the LMC star VFTS 180 between two observations separated by $\Delta(t)= 164.637$ days.}
 \label{fig:180_tauradcomp}
\end{center}
    \end{figure*}

   \begin{figure*}
\begin{center}
 \qquad
 \subfloat[ ]{\includegraphics[width=3.34in]{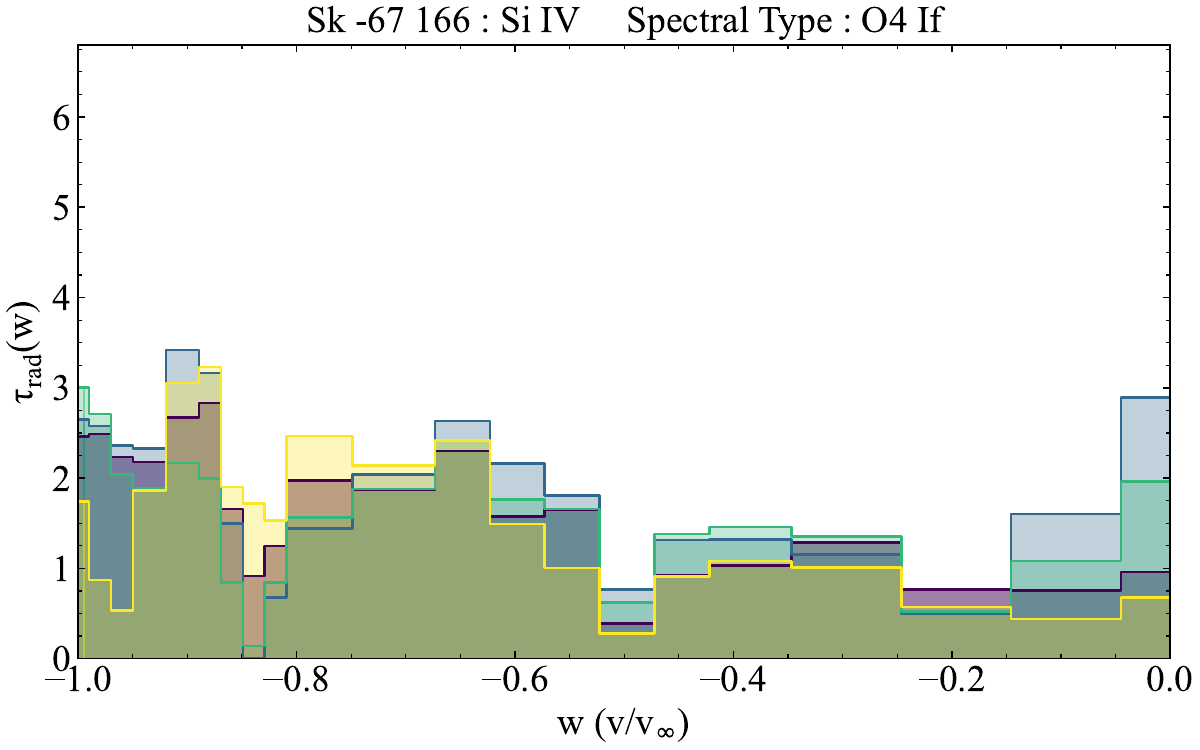} }
 \caption{SEI-derived radial optical depths of selected observations of the \ion{Si}{IV} absorption profile, showing changes in those optical depths in the stellar wind of the LMC star Sk -67 166. Elapsed time between each observation varies but is less than 2 days in each case.}
 \label{fig:67166_tauradcomp}
\end{center}
    \end{figure*}

   \begin{figure*}
\begin{center}
 \qquad
 \subfloat[ ]{\includegraphics[width=3.34in]{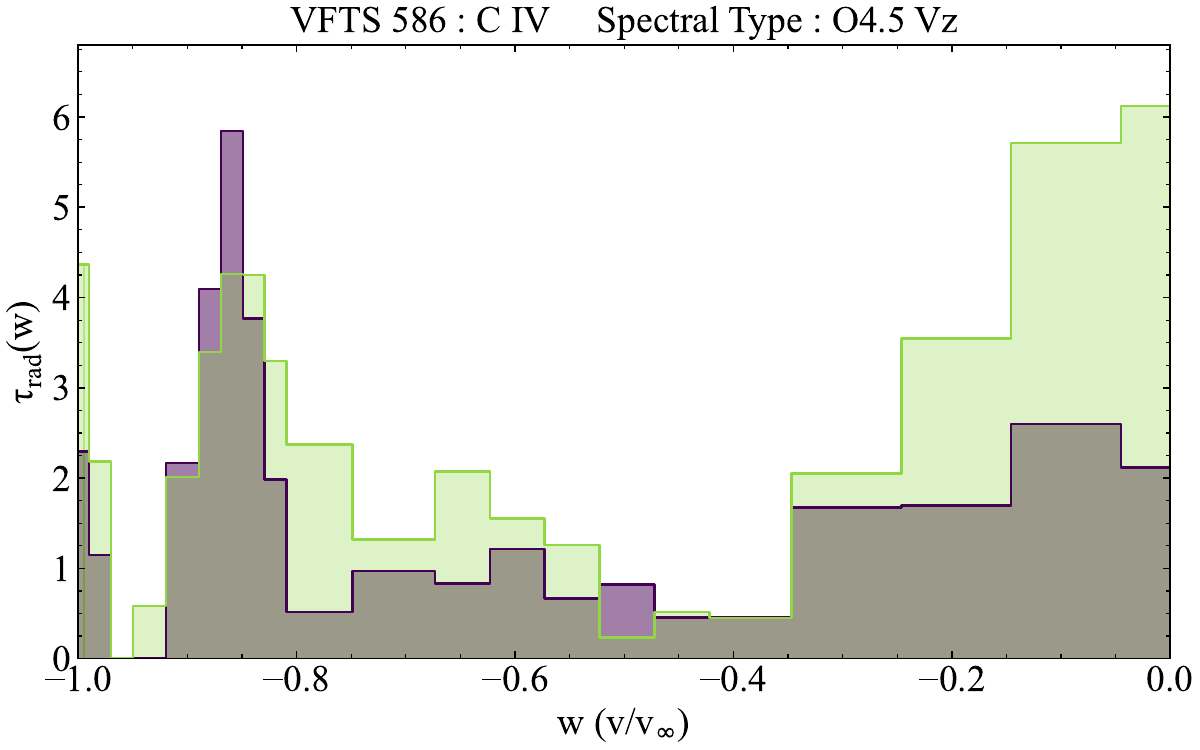} }
 \caption{SEI-derived radial optical depths of the \ion{C}{IV} absorption profile, showing changes in those optical depths in the stellar wind of the LMC star VFTS 586 between two observations separated by $\Delta(t)= 952.753$ days.}
 \label{fig:586_tauradcomp}
\end{center}
    \end{figure*}

   \begin{figure*}
\begin{center}
 \subfloat[ ]{\includegraphics[width=3.34in]{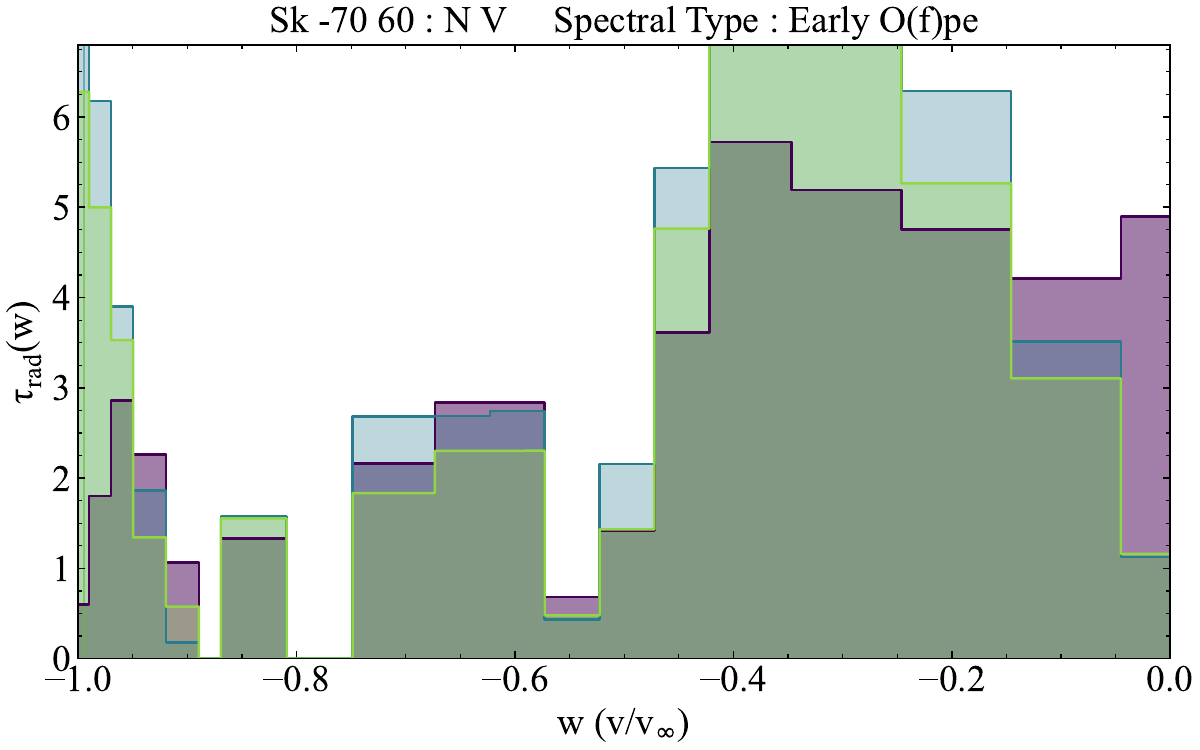} }
 \qquad
 \subfloat[ ]{\includegraphics[width=3.34in]{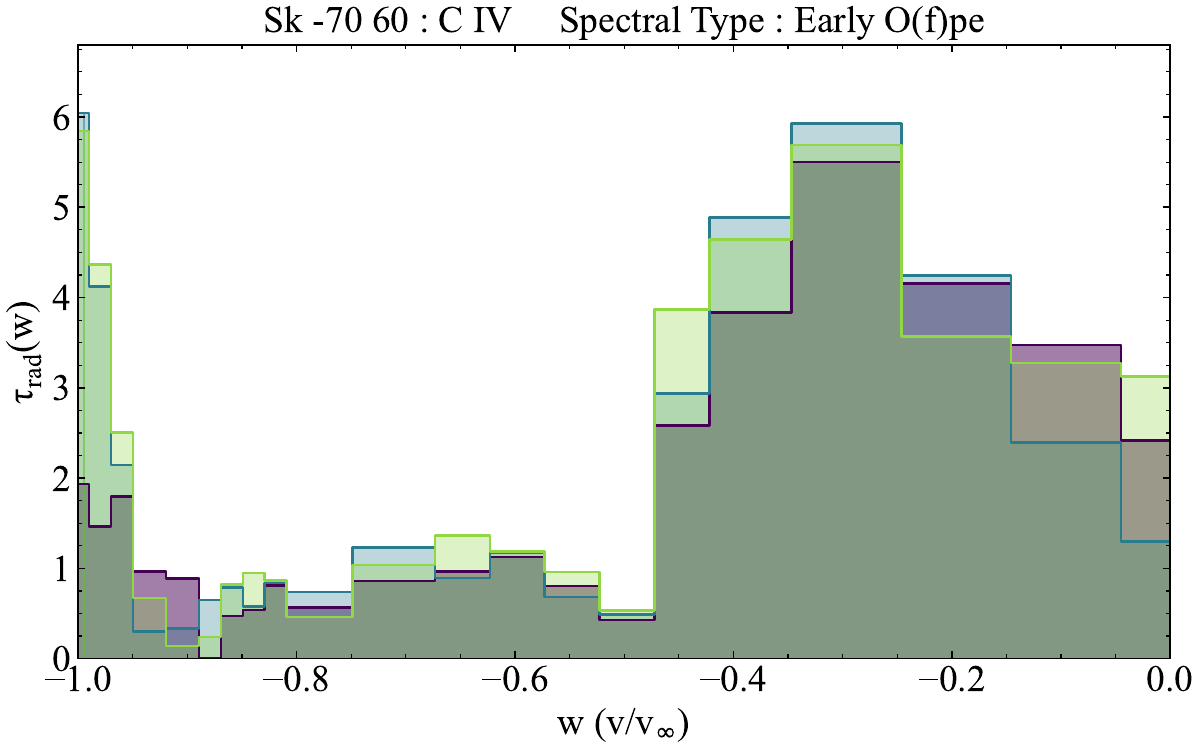} }
 \caption{SEI-derived radial optical depths of selected observations of the (a) \ion{N}{V} and (b) \ion{C}{IV} absorption profiles, showing changes in those optical depths in the stellar wind of the LMC star Sk -70 60 between three observations separated by a total of $\Delta(t)= 152.056$ days (the final two observations are separated by approximately 3 hours).}
 \label{fig:7060_tauradcomp}
\end{center}
    \end{figure*}

   \begin{figure*}
\begin{center}
 \qquad
 \subfloat[ ]{\includegraphics[width=3.34in]{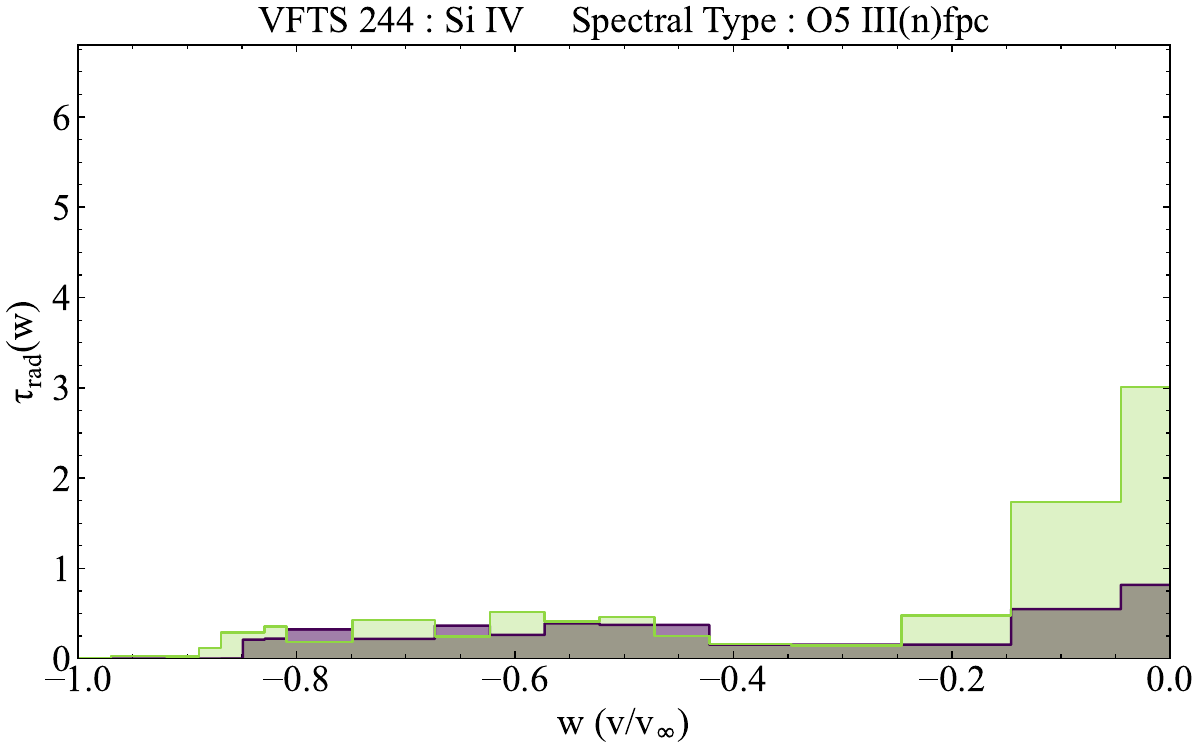} }
 \caption{SEI-derived radial optical depths of the \ion{Si}{IV} absorption profile, showing changes in those optical depths in the stellar wind of the LMC star VFTS 244 between two observations separated by $\Delta(t)= 1288.255$ days.}
 \label{fig:244_tauradcomp}
\end{center}
    \end{figure*}

   \begin{figure*}
\begin{center}
 \subfloat[ ]{\includegraphics[width=3.34in]{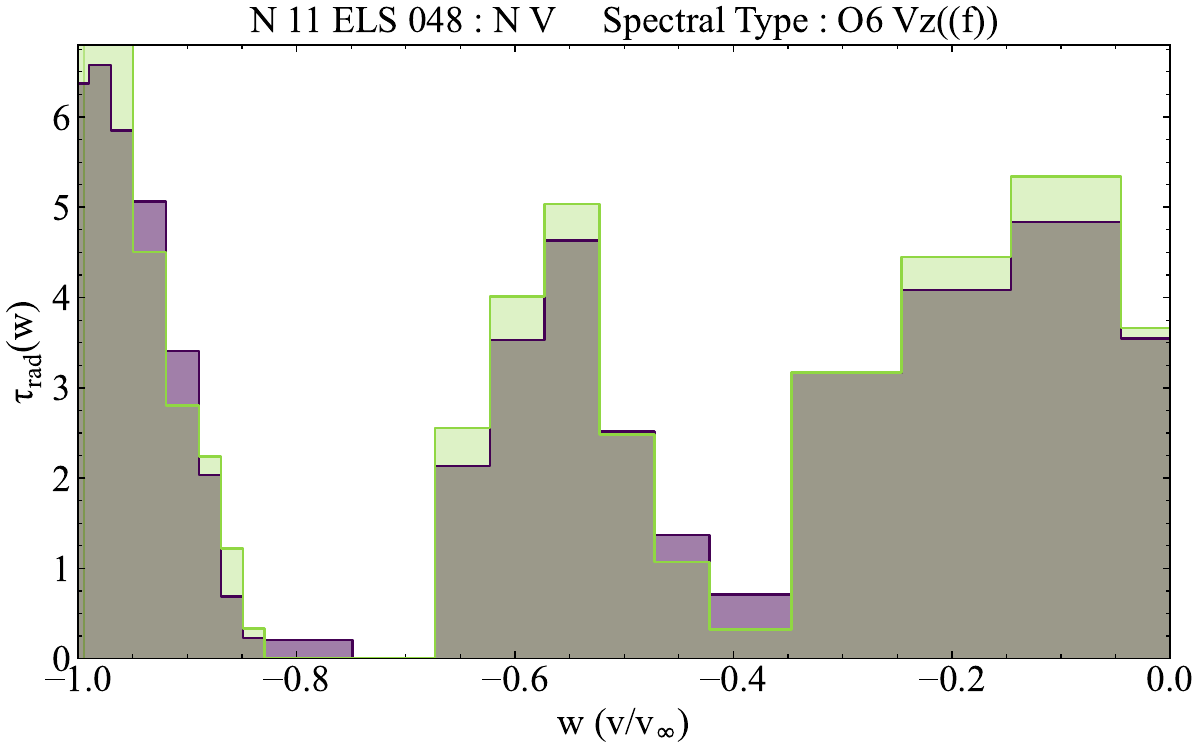} }
 \qquad
 \subfloat[ ]{\includegraphics[width=3.34in]{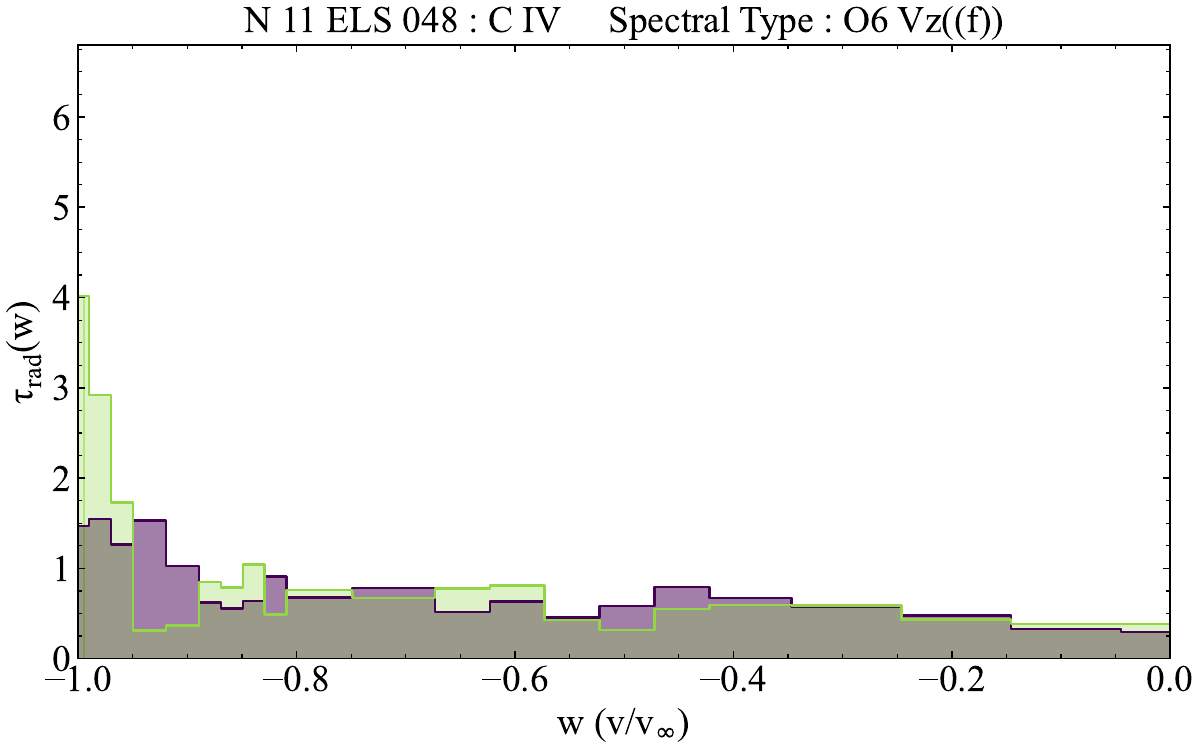} }
 \caption{SEI-derived radial optical depths of the (a) \ion{N}{V} and (b) \ion{C}{IV} absorption profile, showing changes in those optical depths in the stellar wind of the LMC star N11 ELS 048 between two observations separated by $\Delta(t)= 1.051$ days.}
 \label{fig:11048_tauradcomp}
\end{center}
    \end{figure*}

   \begin{figure*}
\begin{center}
 \subfloat[ ]{\includegraphics[width=3.34in]{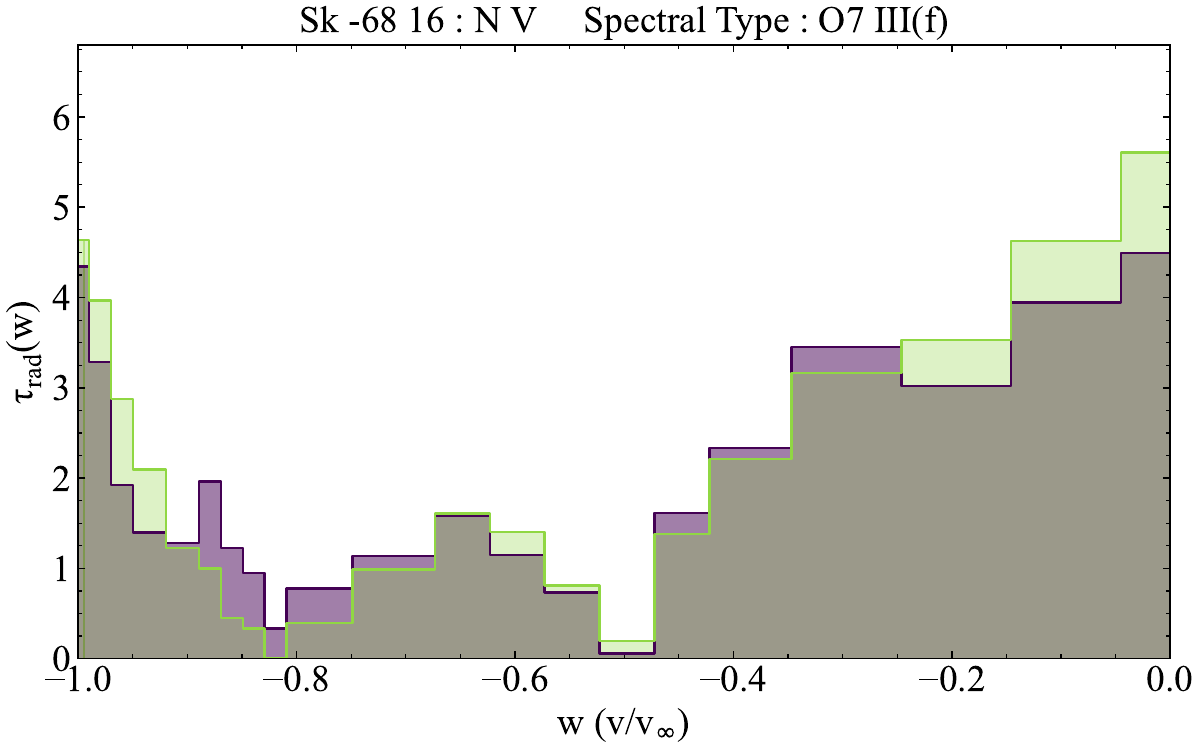} }
 \vspace*{-1.0 cm}
 \qquad
 \subfloat[ ]{\includegraphics[width=3.34in]{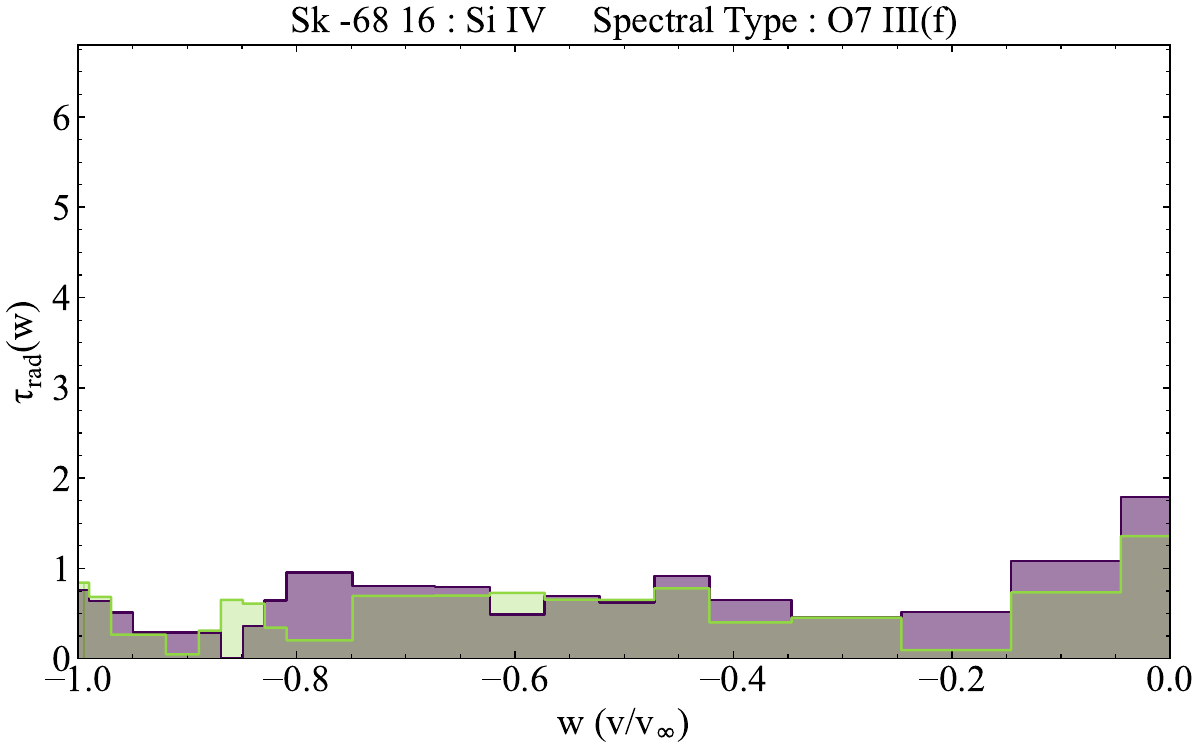} }
 \vspace*{1.0 cm} \qquad
 \subfloat[ ]{\includegraphics[width=3.34in]{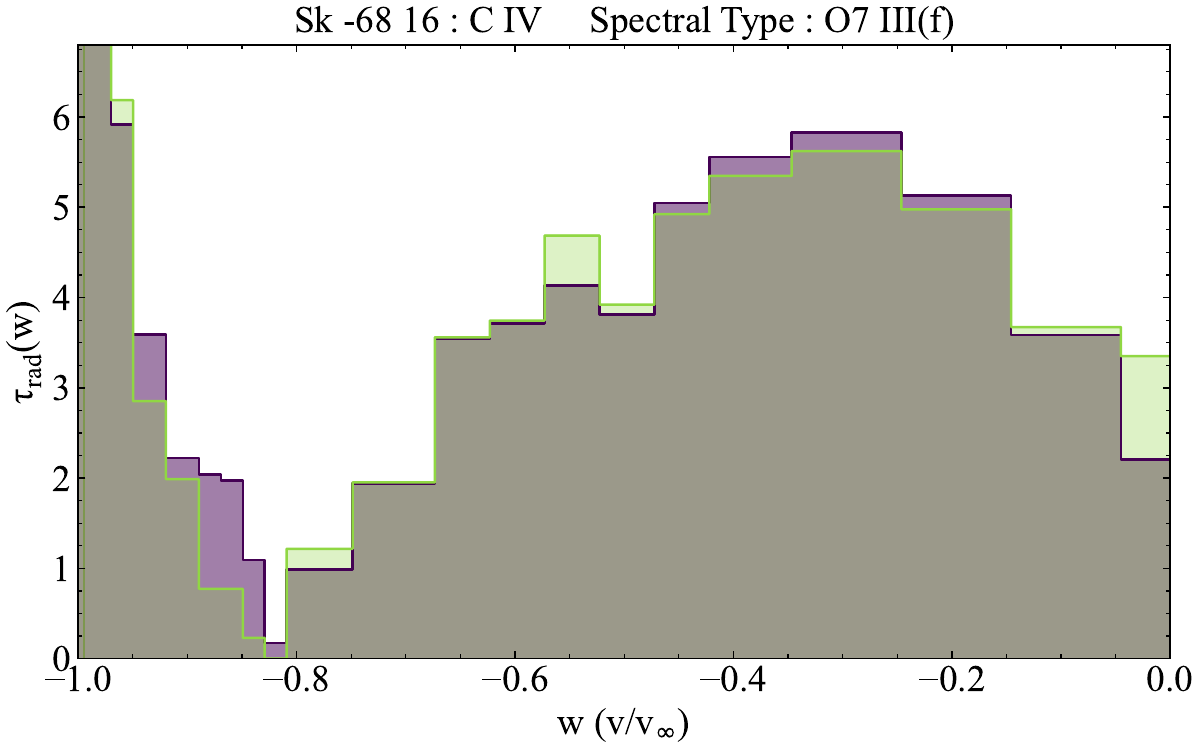} }
 \caption{SEI-derived radial optical depths of selected observations of the (a) \ion{N}{V}, (b) \ion{Si}{IV} and (c) \ion{C}{IV} absorption profiles, showing changes in those optical depths in the stellar wind of the LMC star Sk -68 16 between two observations separated by $\Delta(t)= 2.311$ days.}
 \label{fig:6816_tauradcomp}
\end{center}
    \end{figure*}

   \begin{figure*}
\begin{center}
 \qquad
 \subfloat[ ]{\includegraphics[width=3.34in]{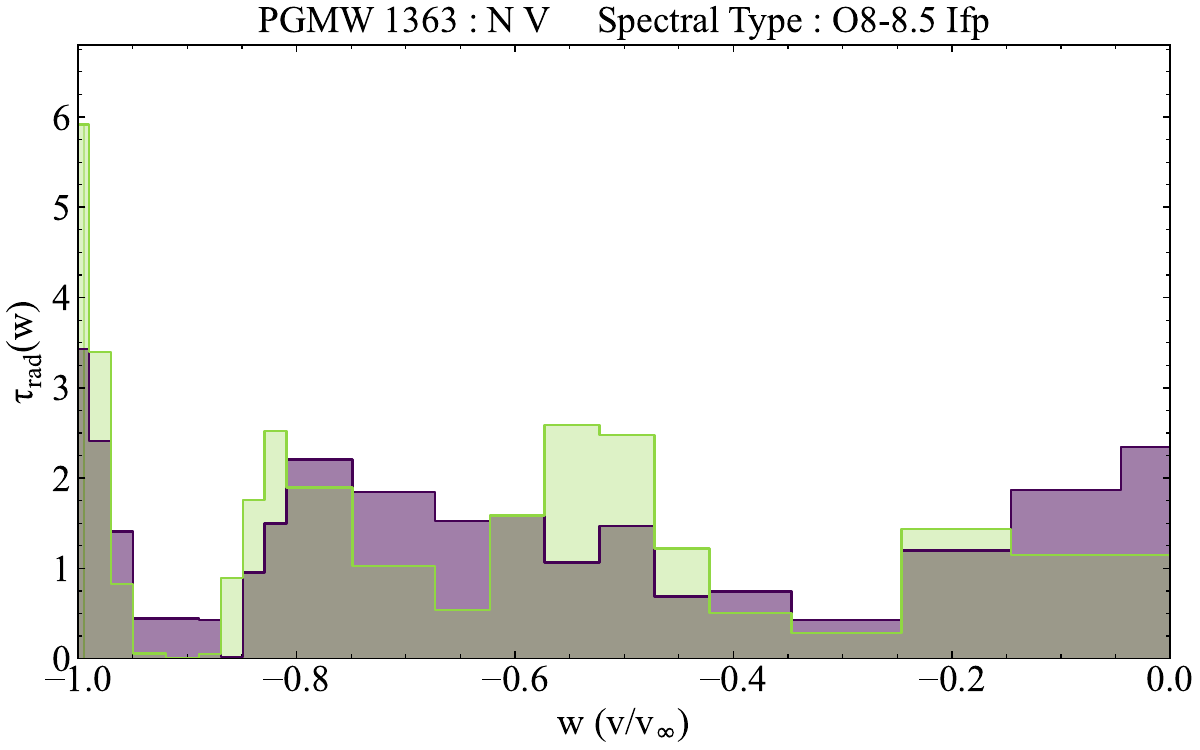} }
 \caption{SEI-derived radial optical depths of the \ion{N}{V} absorption profile, showing changes in those optical depths in the stellar wind of the LMC star PGMW 1363 between two observations separated by $\Delta(t)= 171.310$ days.}
 \label{fig:1363_tauradcomp}
\end{center}
    \end{figure*}

   \begin{figure*}
\begin{center}
 \qquad
 \subfloat[ ]{\includegraphics[width=3.34in]{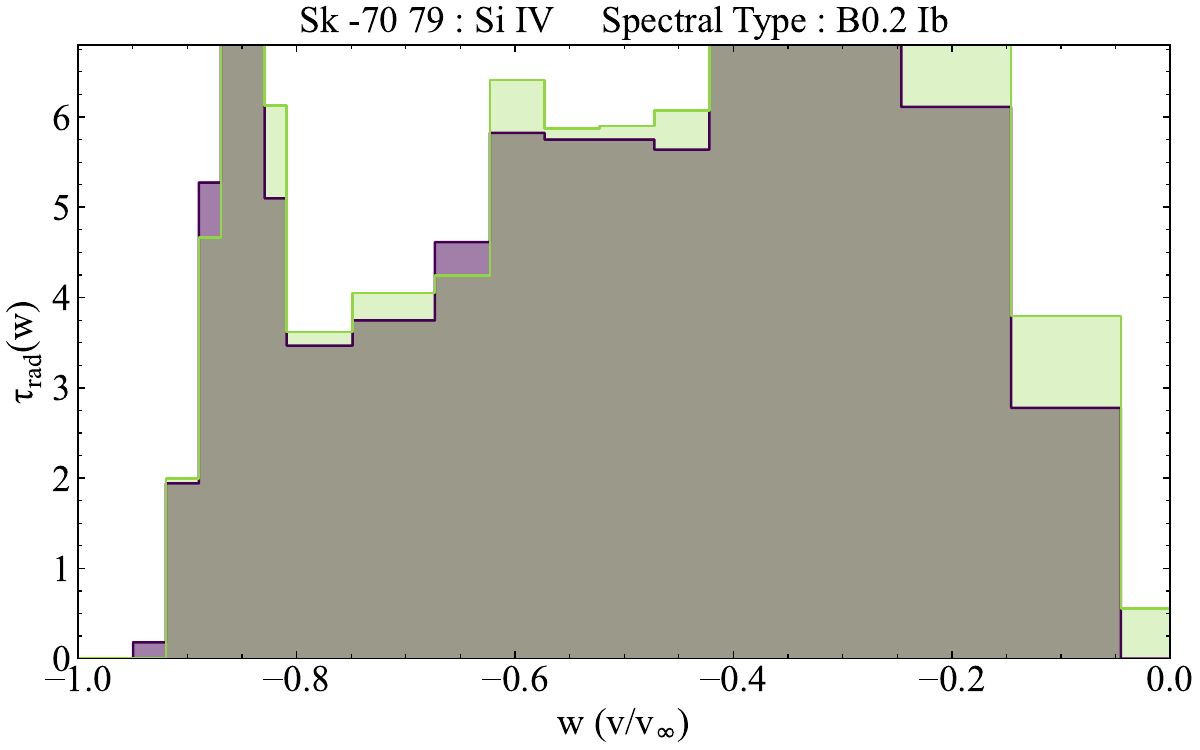} }
 \caption{SEI-derived radial optical depths of the \ion{Si}{IV} absorption profile, showing changes in those optical depths in the stellar wind of the LMC star Sk -70 79 between two observations separated by $\Delta(t)= 0.857$ days.}
 \label{fig:7079_tauradcomp}
\end{center}
    \end{figure*}

   \begin{figure*}
\begin{center}
 \qquad
 \subfloat[ ]{\includegraphics[width=3.34in]{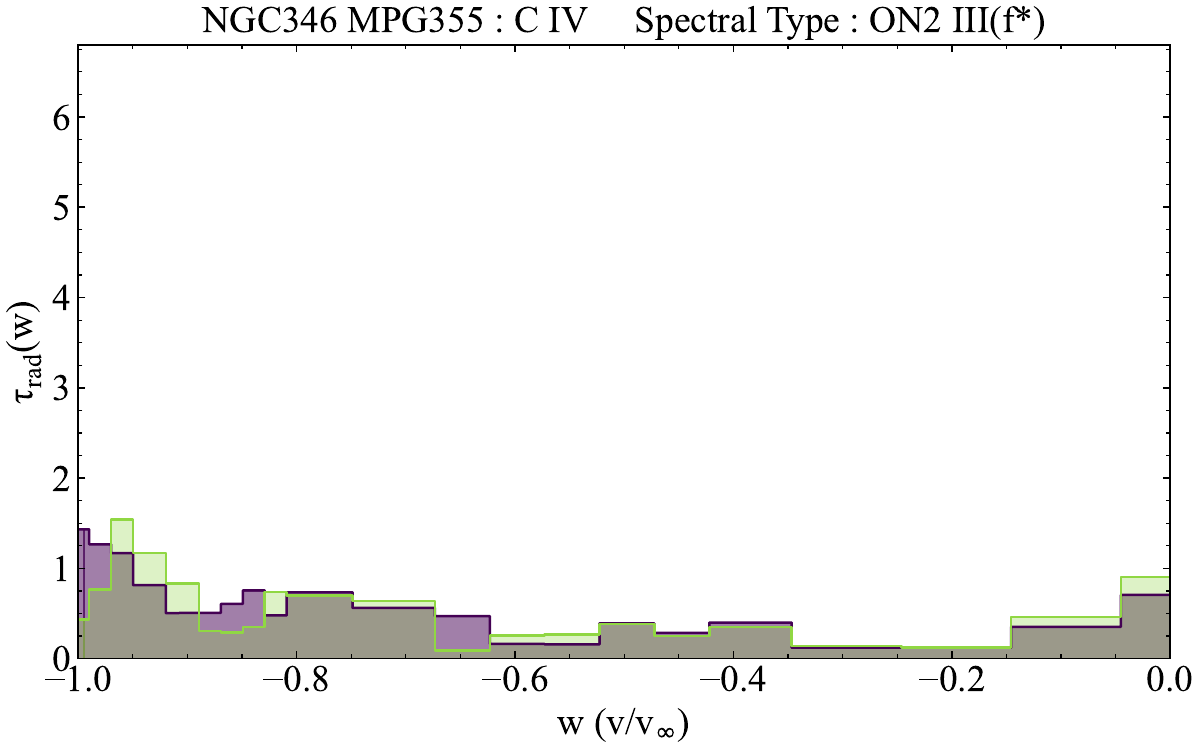} }
 \caption{SEI-derived radial optical depths of the \ion{C}{IV} absorption profile, showing changes in those optical depths in the stellar wind of the SMC star NGC 346 MPG 355 between two observations separated by $\Delta(t)= 8060.059$ days.}
 \label{fig:355_tauradcomp}
\end{center}
    \end{figure*}

   \begin{figure*}
\begin{center}
 \qquad
 \subfloat[ ]{\includegraphics[width=3.34in]{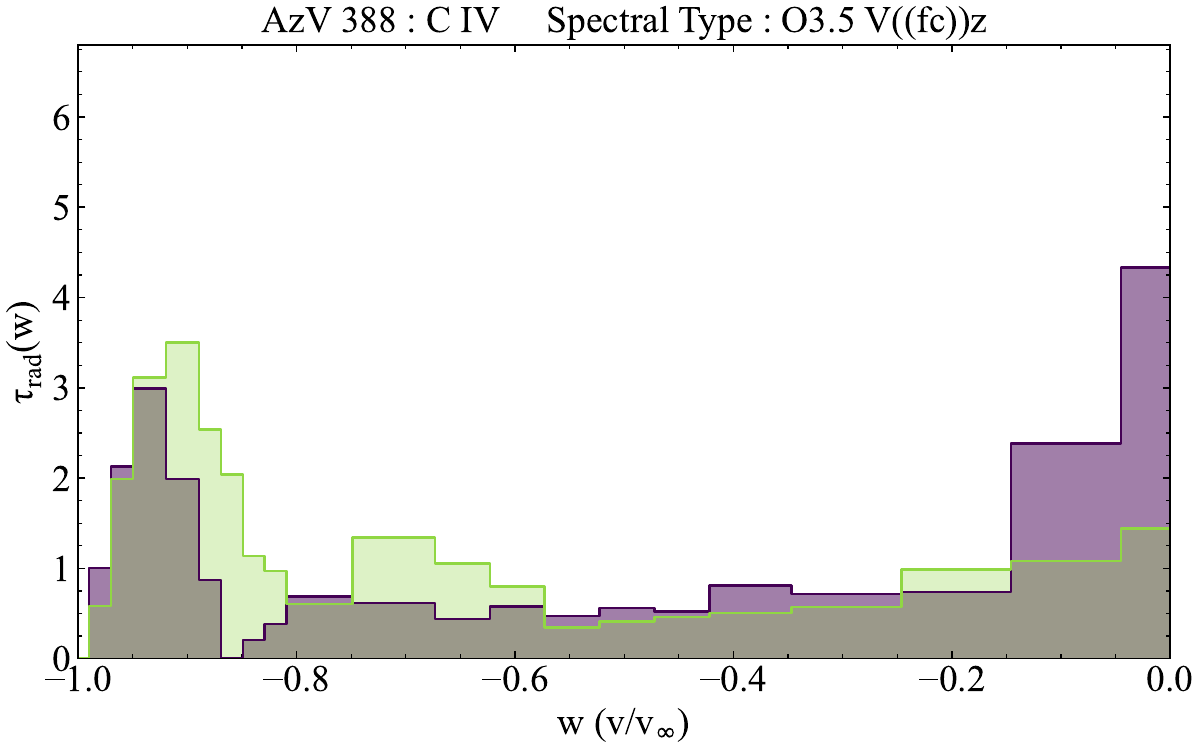} }
 \caption{SEI-derived radial optical depths of the \ion{C}{IV} absorption profile, showing changes in those optical depths in the stellar wind of the SMC star AzV 388 between two observations separated by $\Delta(t)= 2135.151$ days.}
 \label{fig:388_tauradcomp}
\end{center}
    \end{figure*}

   \begin{figure*}
\begin{center}
 \qquad
 \subfloat[ ]{\includegraphics[width=3.34in]{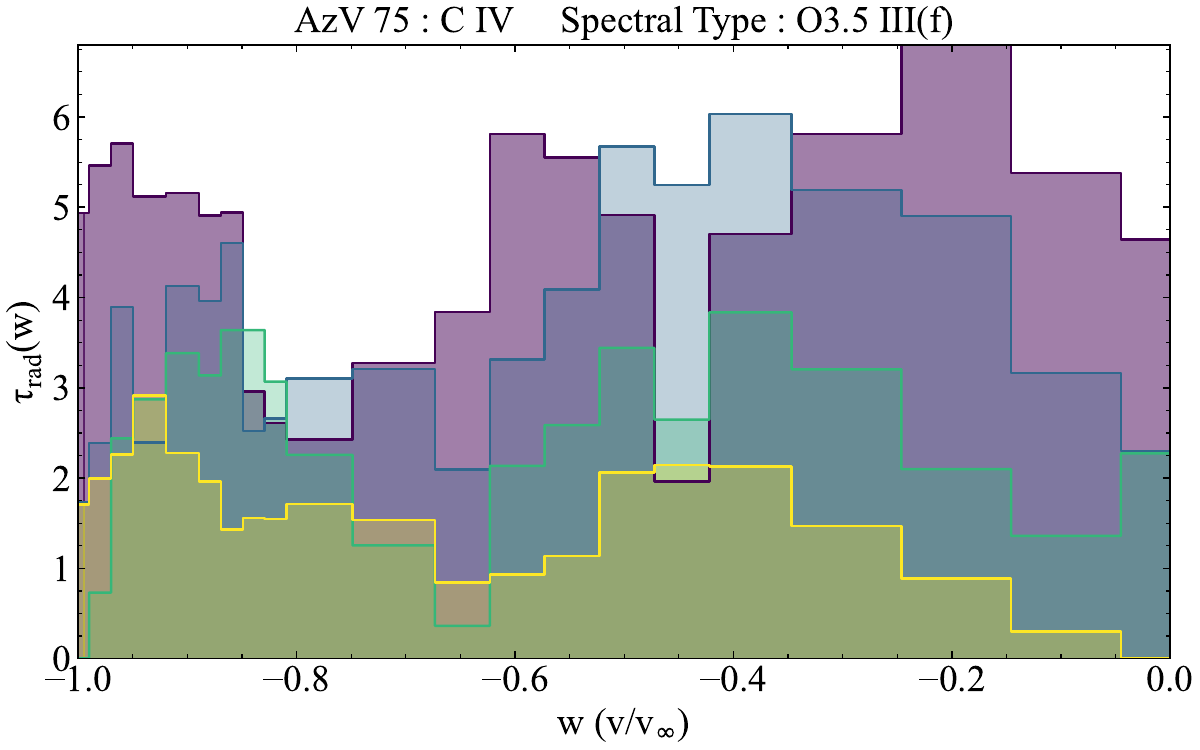} }
 \caption{SEI-derived radial optical depths of selected observations of the \ion{C}{IV} absorption profile, showing changes in those optical depths in the stellar wind of the SMC star AzV 75. Elapsed time between each selected observation varies but is of the order of several months up to approximately 2 years.}
 \label{fig:75_tauradcomp}
\end{center}
    \end{figure*}

   \begin{figure*}
\begin{center}
 \qquad
 \subfloat[ ]{\includegraphics[width=3.34in]{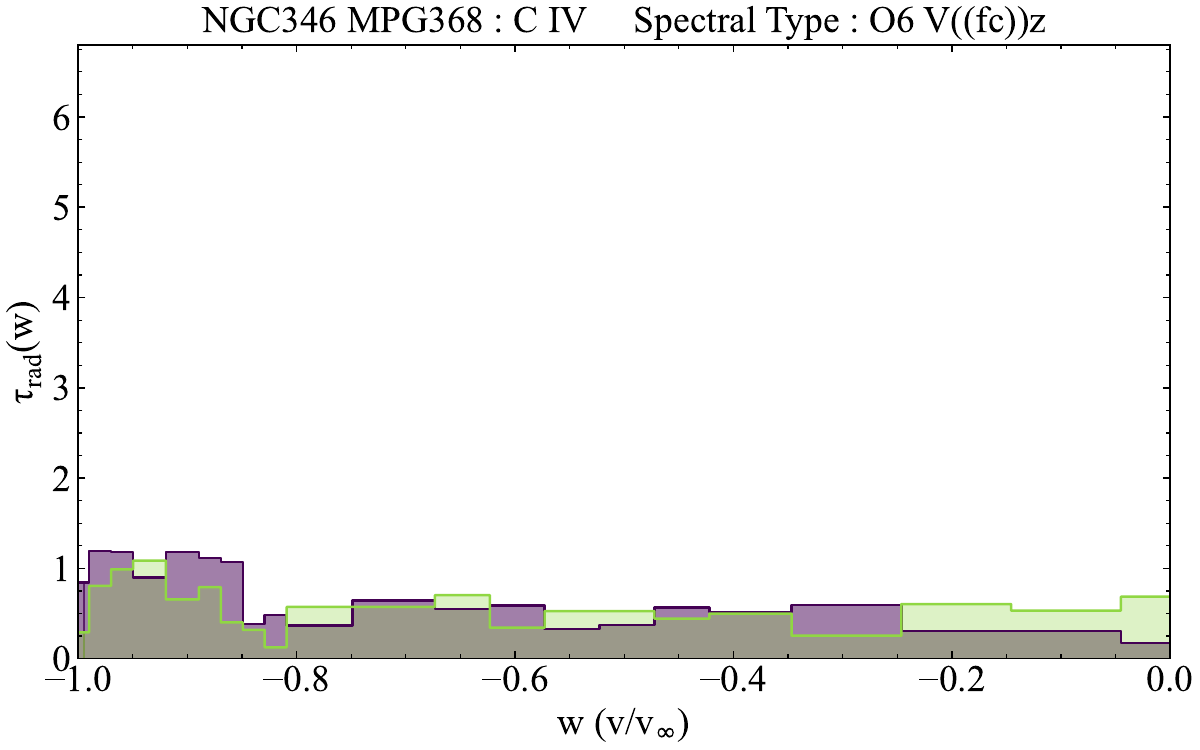} }
 \caption{SEI-derived radial optical depths of the \ion{C}{IV} absorption profile, showing changes in those optical depths in the stellar wind of the SMC star NGC 346 MPG 368 between two observations separated by $\Delta(t)= 8000.632$ days.}
 \label{fig:368_tauradcomp}
\end{center}
    \end{figure*}

   \begin{figure*}
\begin{center}
 \qquad
 \subfloat[ ]{\includegraphics[width=3.34in]{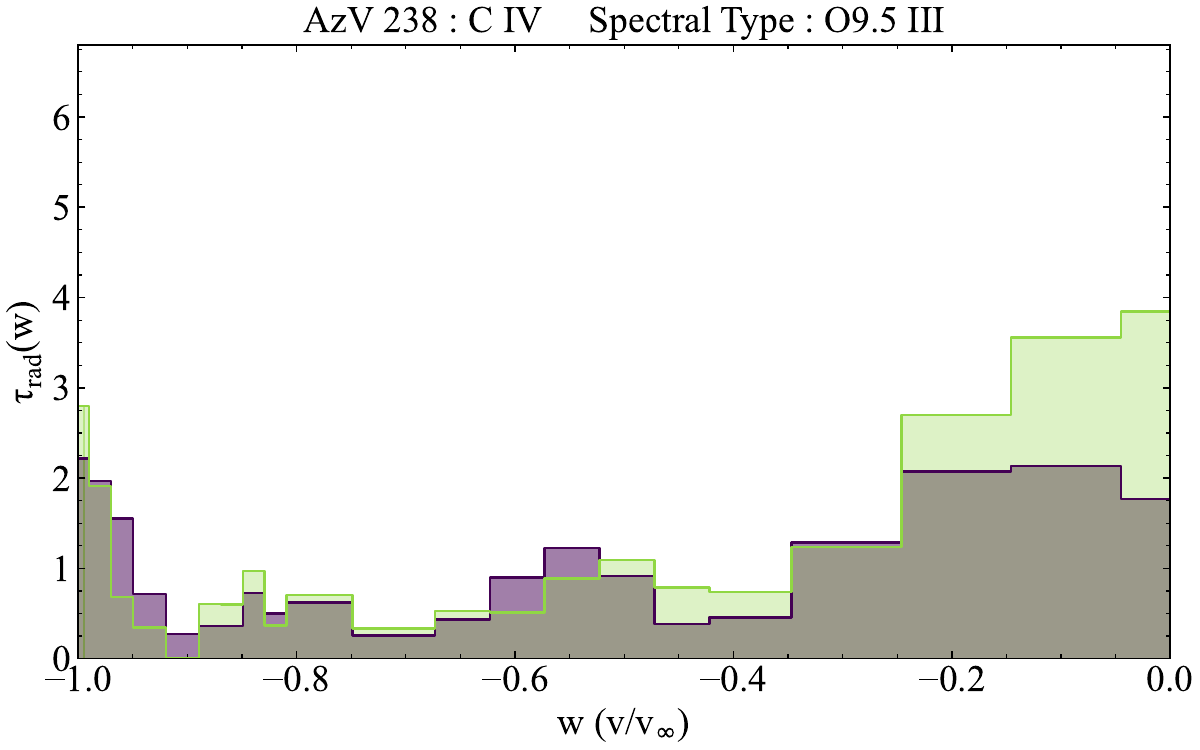} }
 \caption{SEI-derived radial optical depths of the \ion{C}{IV} absorption profile, showing changes in those optical depths in the stellar wind of the SMC star AzV 238 between two observations separated by $\Delta(t)= 23.434$ days.}
 \label{fig:238_tauradcomp}
\end{center}
    \end{figure*}

   \begin{figure*}
\begin{center}
 \subfloat[ ]{\includegraphics[width=3.34in]{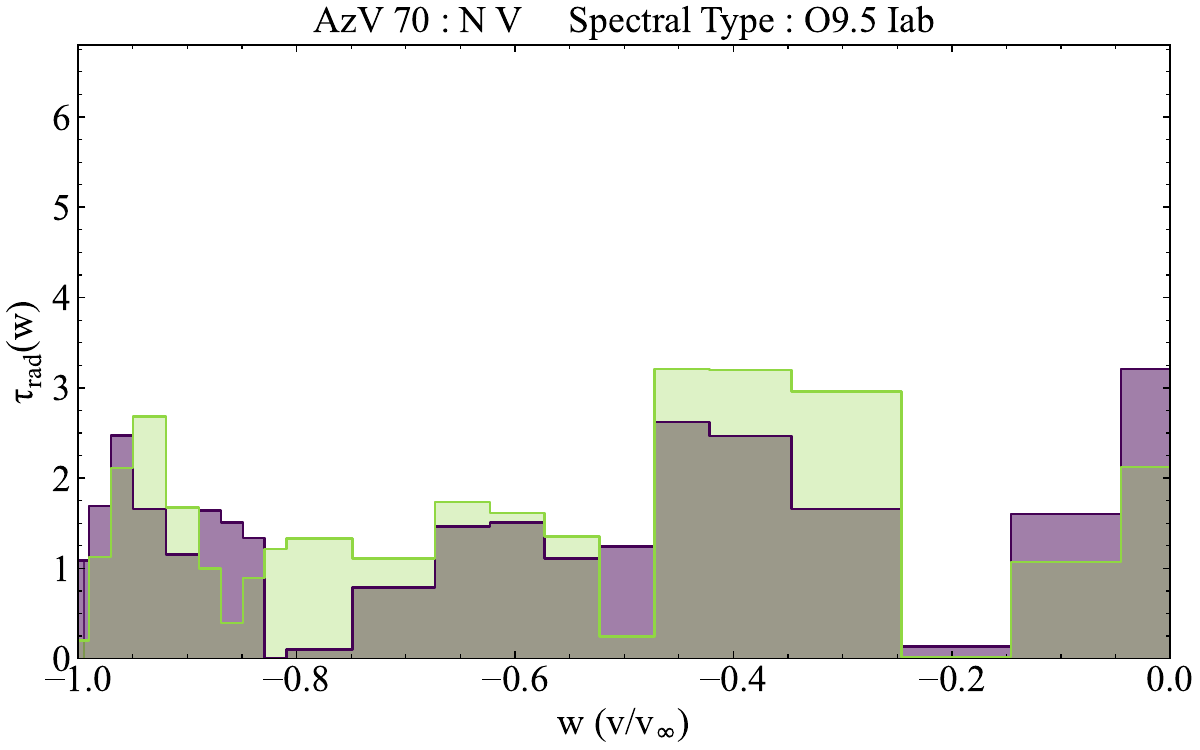} }
 \vspace*{-1.0 cm}
 \qquad
 \subfloat[ ]{\includegraphics[width=3.34in]{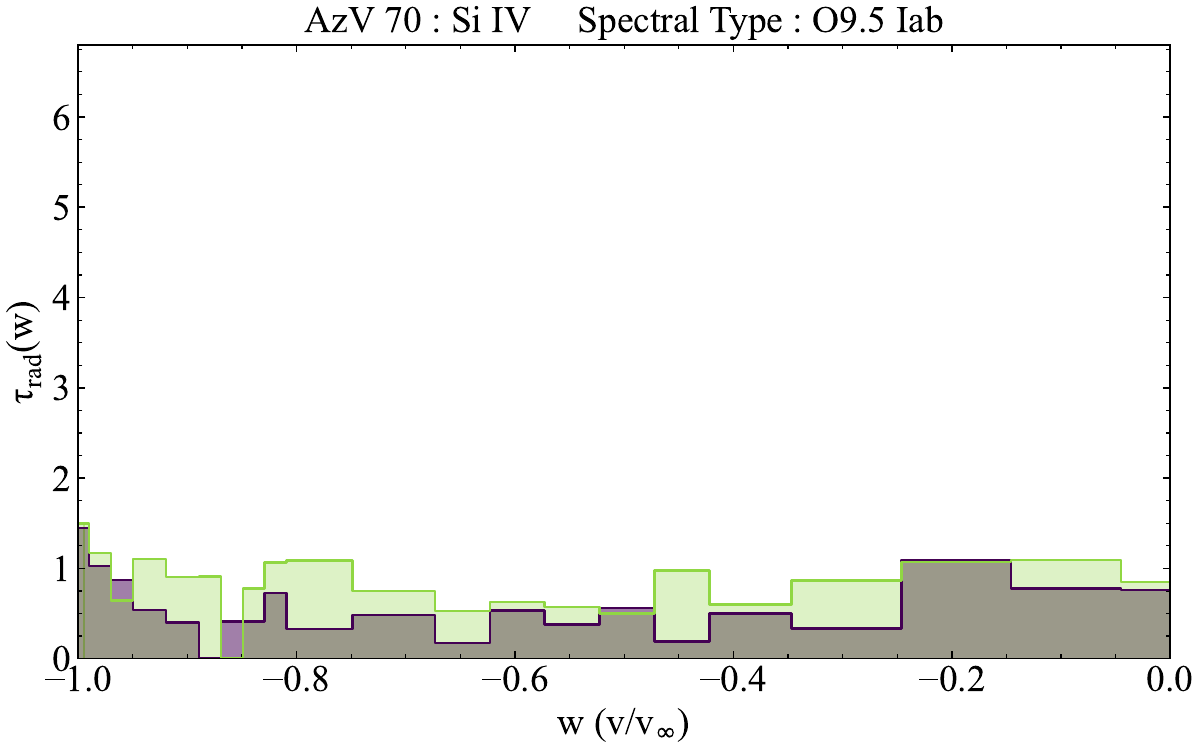} }
 \vspace*{1.0 cm} \qquad
 \subfloat[ ]{\includegraphics[width=3.34in]{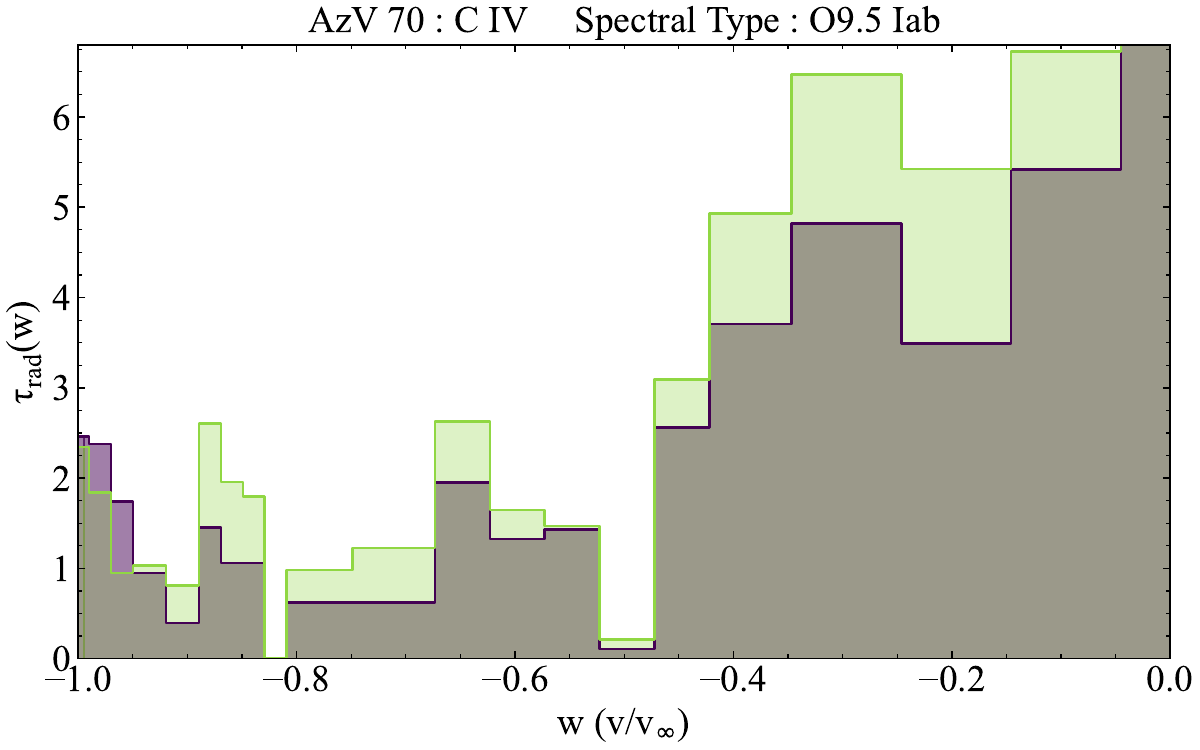} }
 \caption{SEI-derived radial optical depths of selected observations of the (a) \ion{N}{V}, (b) \ion{Si}{IV} and (c) \ion{C}{IV} absorption profiles, showing changes in those optical depths in the stellar wind of the SMC star AzV 70 between two observations separated by $\Delta(t)= 47.954$ days.}
 \label{fig:70_tauradcomp}
\end{center}
    \end{figure*}

   \begin{figure*}
\begin{center}
 \qquad
 \subfloat[ ]{\includegraphics[width=3.34in]{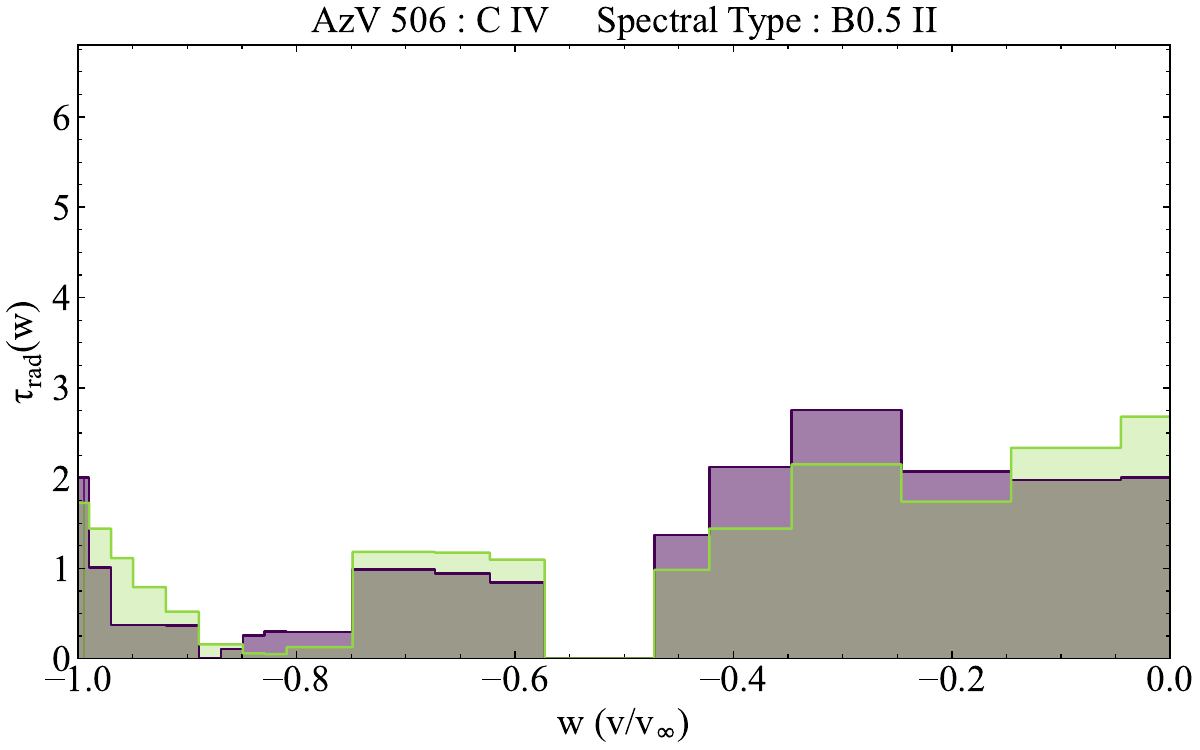} }
 \caption{SEI-derived radial optical depths of the \ion{C}{IV} absorption profile, showing changes in those optical depths in the stellar wind of the SMC star AzV 506 between two observations separated by $\Delta(t)= 6.135$ days.}
 \label{fig:506_tauradcomp}
\end{center}
    \end{figure*}

   \begin{figure*}
\begin{center}
 \subfloat[ ]{\includegraphics[width=3.34in]{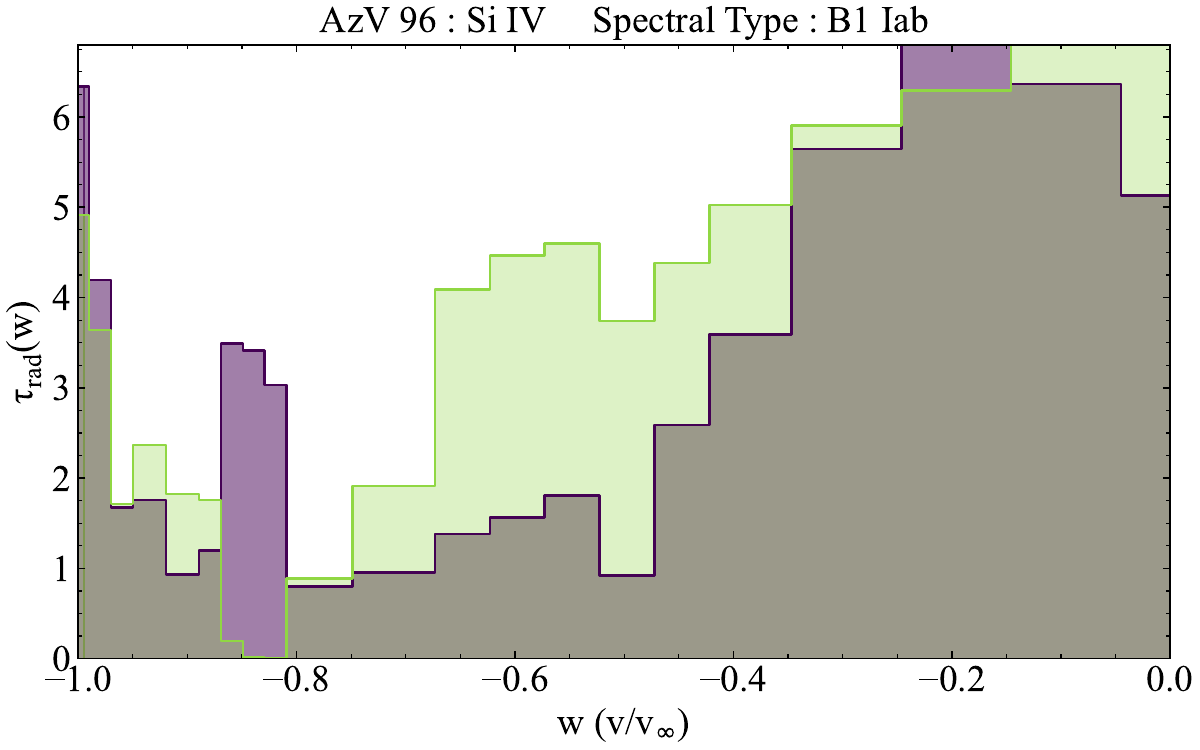} }
 \qquad
 \subfloat[ ]{\includegraphics[width=3.34in]{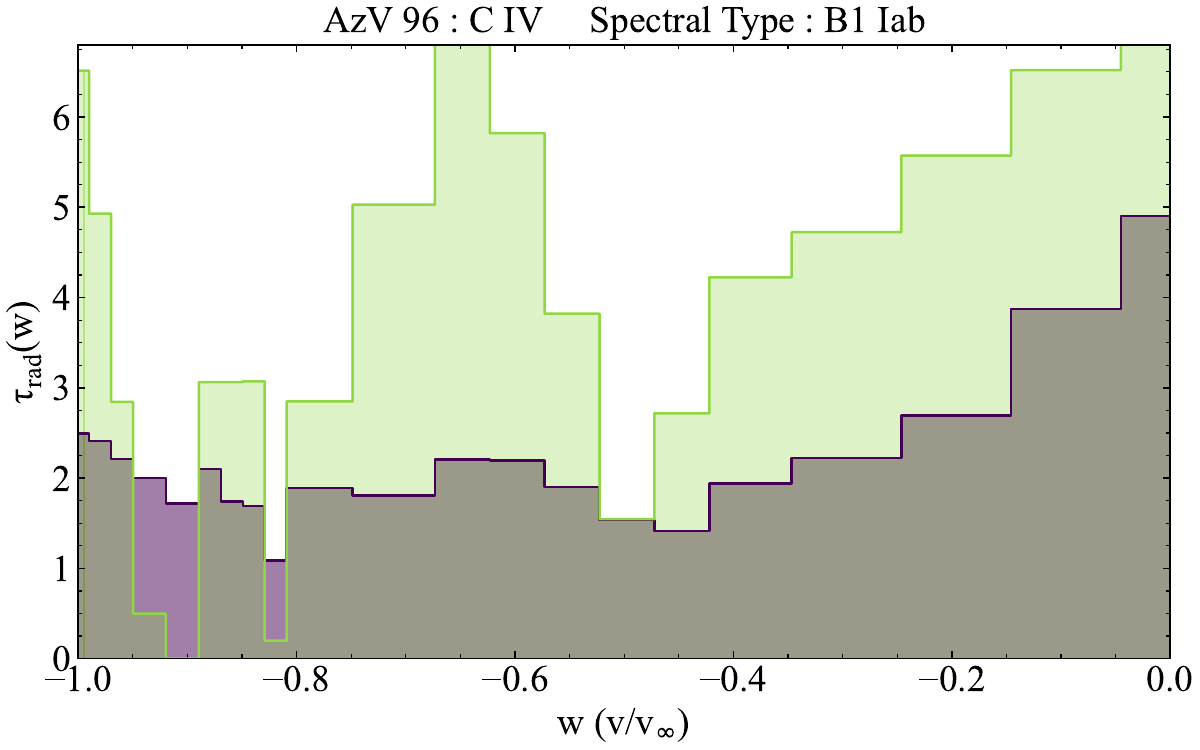} }
 \caption{SEI-derived radial optical depths of the (a) \ion{Si}{IV} and (b) \ion{C}{IV} absorption profiles, showing changes in those optical depths in the stellar wind of the SMC star AzV 96 between two observations separated by $\Delta(t)= 4.827$ days.}
 \label{fig:96_tauradcomp}
\end{center}
    \end{figure*}

   \begin{figure*}
\begin{center}
 \qquad
 \subfloat[ ]{\includegraphics[width=3.34in]{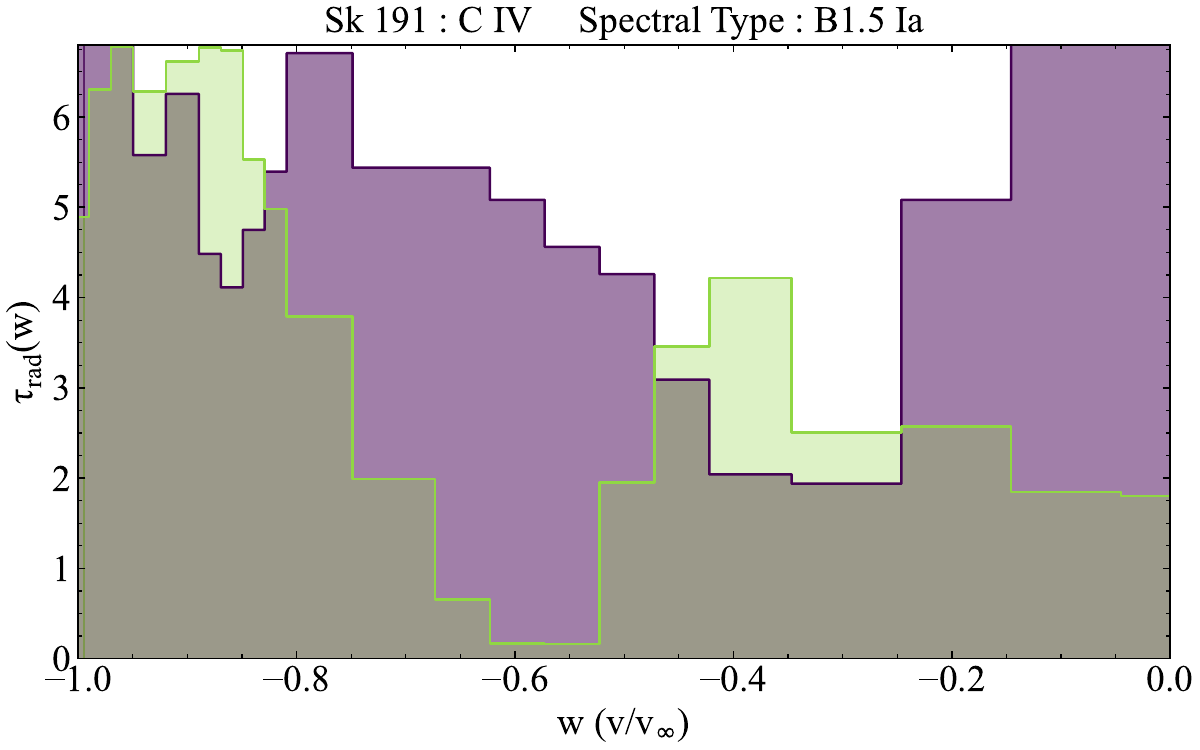} }
 \caption{SEI-derived radial optical depths of the \ion{C}{IV} absorption profile, showing changes in those optical depths in the stellar wind of the SMC star AzV 506 between two observations separated by $\Delta(t)= 71$ days.}
 \label{fig:191_tauradcomp}
\end{center}
    \end{figure*}

   \begin{figure*}
\begin{center}
 \subfloat[ ]{\includegraphics[width=3.34in]{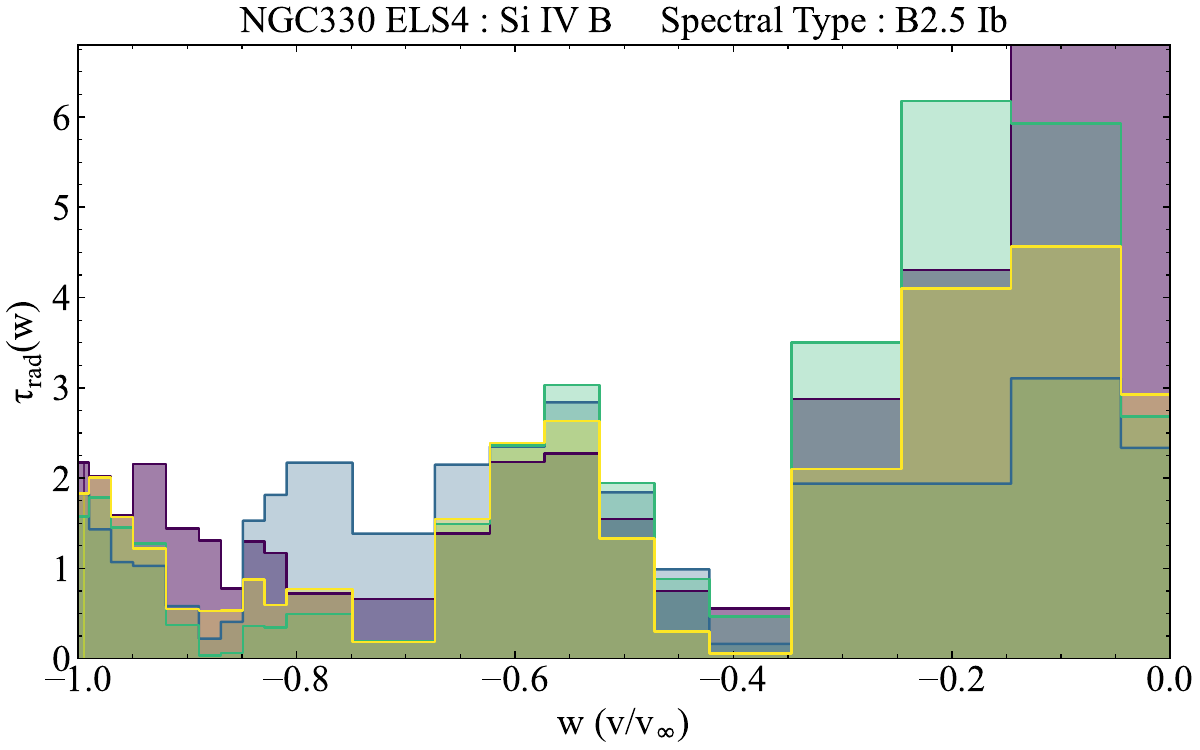} }
 \qquad
 \subfloat[ ]{\includegraphics[width=3.34in]{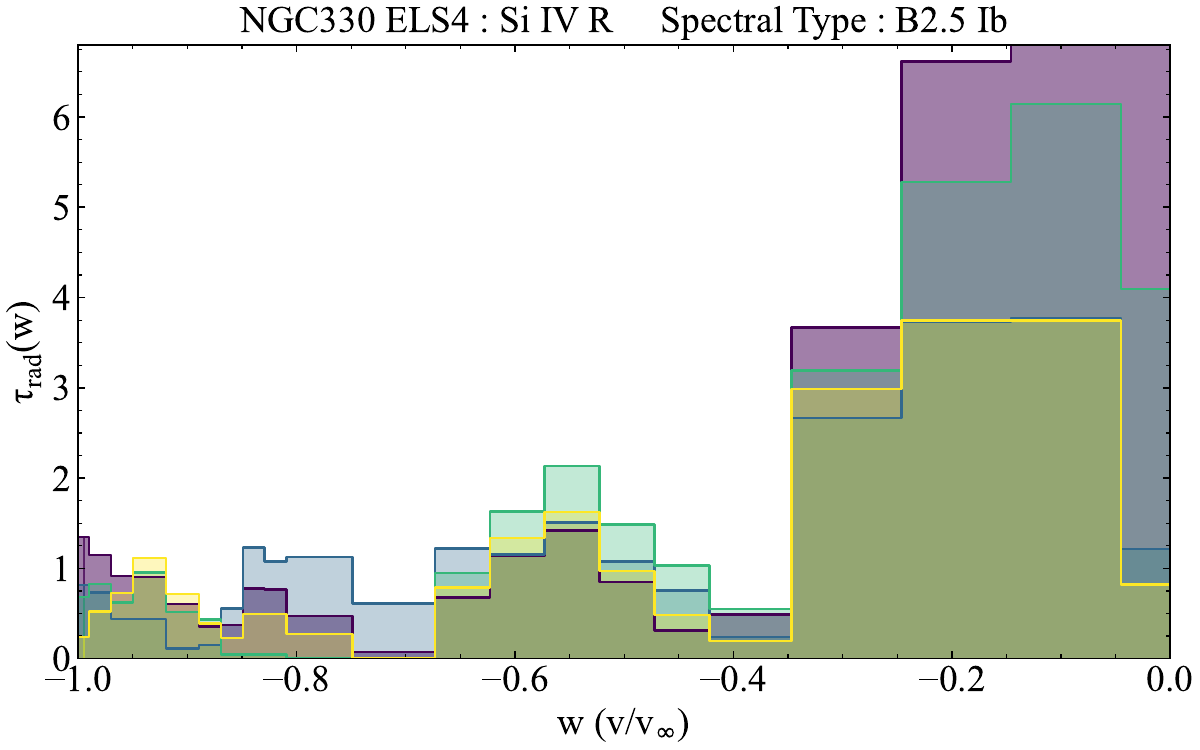} }
 \qquad
 \subfloat[ ]{\includegraphics[width=3.34in]{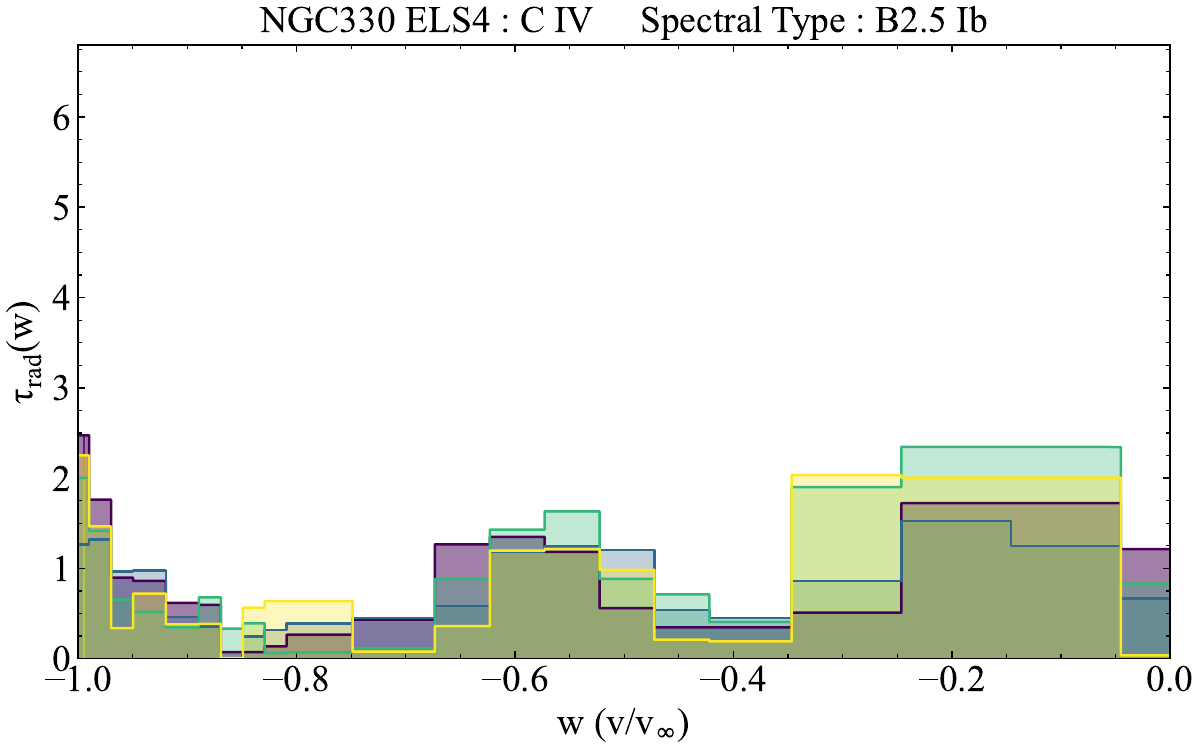} } \caption{SEI-derived radial optical depths of the (a) \ion{Si}{IVB}, (b) \ion{Si}{IVR} and (c) \ion{C}{IV} absorption profiles, showing changes in those optical depths in the stellar wind of the SMC star NGC330 ELS4. Overall elapsed time between first and last observation is $\Delta(t)= 2837.012$ days, however between Observations 1 and 2, $\Delta(t)= 340.840$ days and between Observations 3 and 4, $\Delta(t)= 4.864$ days.}
 \label{fig:3304_tauradcomp}
\end{center}
    \end{figure*}

   \begin{figure*}
\begin{center}
 \subfloat[ ]{\includegraphics[width=3.34in]{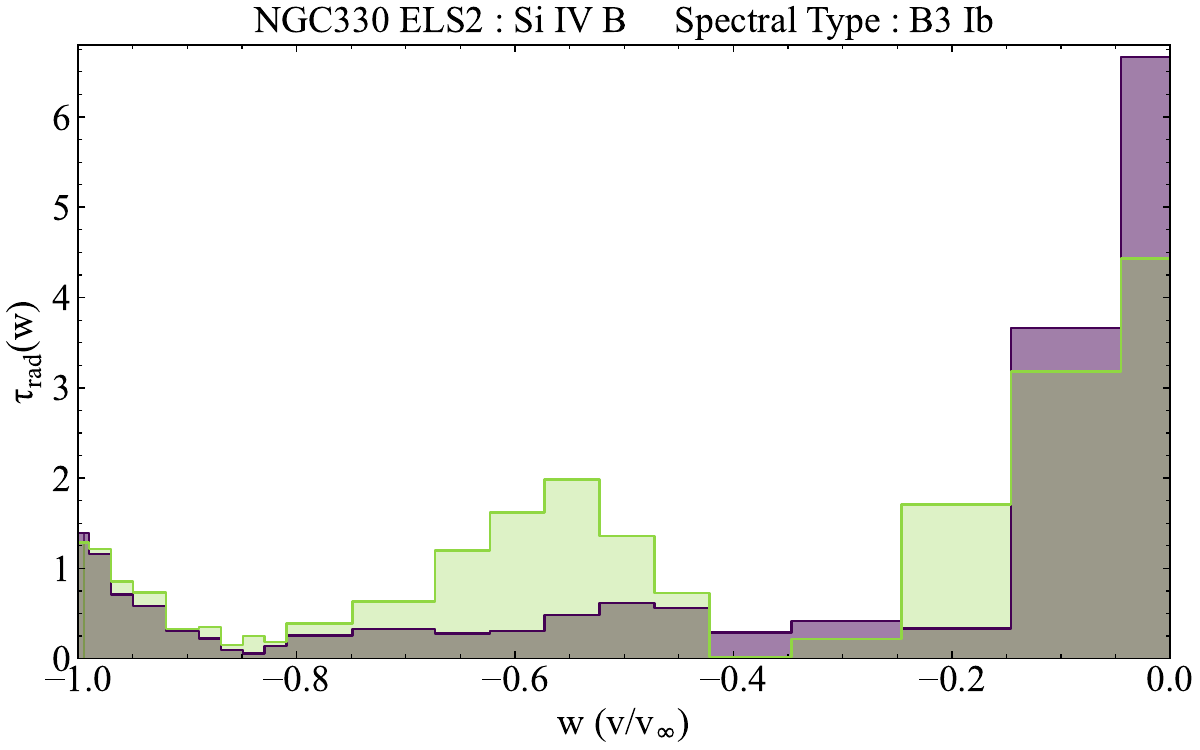} }
 \qquad
 \subfloat[ ]{\includegraphics[width=3.34in]{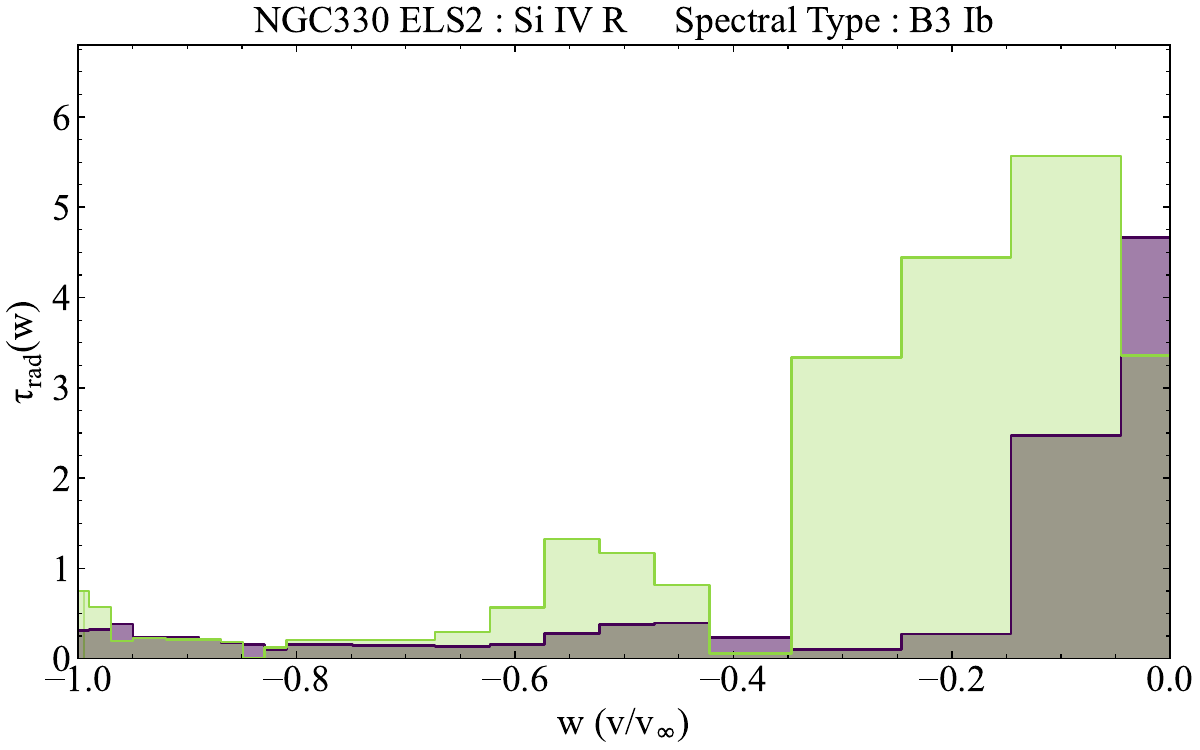} }
 \qquad
 \subfloat[ ]{\includegraphics[width=3.34in]{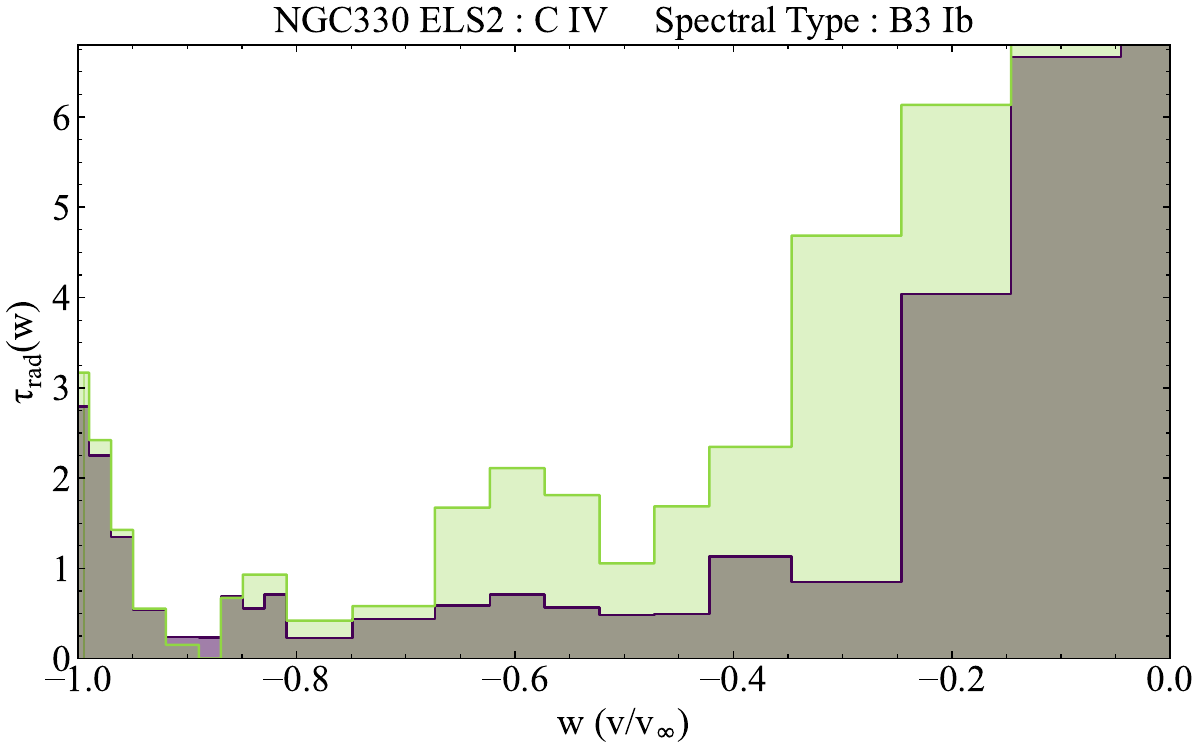} } \caption{SEI-derived radial optical depths of the (a) \ion{Si}{IVB}, (b) \ion{Si}{IVR} and (c) \ion{C}{IV} absorption profiles, showing changes in those optical depths in the stellar wind of the SMC star NGC330 ELS2 between two observations separated by $\Delta(t)= 10.967$ days.}
 \label{fig:3302_tauradcomp}
\end{center}
    \end{figure*}


\bsp	
\label{lastpage}
\end{document}